\documentclass[12pt]{osuthesis}
\usepackage{graphicx}
\usepackage{subfigure}
\usepackage{enumerate}
\usepackage{amsmath}
\usepackage{color}
\usepackage{colortbl}

\renewcommand{\thesubsubsection}{\Alph{subsubsection}}

\usepackage{epsfig,epsf,psfrag,amssymb,amsfonts,cite}
\usepackage{mathrsfs}
\usepackage{euscript}
\usepackage{multirow}
\usepackage{slashbox}
\suppressfloats %\nofiles

\title{{\bf NEW PHYSICS AT THE TEV SCALE}}
\formattedtitle{ NEW PHYSICS AT THE TEV SCALE}

\author{Shreyashi Chakdar}

\degreeone{\ssp Bachelor of Science in Physics (Honours)\\
        Lady Brabourne College, University of Calcutta\\
        Kolkata, India\\
        2006}
\degreetwo{\ssp Master of Science in Physics \\
        Indian Institute of Technology, Roorkee\\
        Roorkee, India\\
        2008}
\degreesought{DOCTOR OF PHILOSOPHY}
\degreedate{July, 2015}
\majorfield{Physics}
\begin{document}
\maketitle
%\makecopyright
\makeapproval{5} % Number denotes total number of required signatures, Committee+Dean
%\begin{approval}
%\input{approval}
%\end{approval}
\begin{acknowledge}
This Ph.D thesis is the result of six years of hard work and patience
under the inspiring supervision of my advisor Dr. Satya Nandi at the Physics Department
in Oklahoma State University. I am extremely grateful to him for his understanding,
endless support, patience and guidance during my Ph.D. His brilliant supervision encouraged me to always feel inspired. He was always more than a supervisor, there is a lot to say about him but in short he will always remain
as a very special person in rest of my career and life.
I would also like to thank Dr. Kaladi Babu, whom I was collaborating
with during my visit in KITP. I have learned a lot
from him about the cosmological aspects of our field. I am thankful to him for giving me time and guidance during my Ph.D. Much of this research would not have occurred without hard work and insight of my collaborators. For this I would like to thank Dr. K. Ghosh and Dr. S. K. Rai for helping me with computational tools. Through most of my graduate studies I have been financially supported by 
teaching assistantship in Physics Department, OSU. 
All the faculty and staff members at the Physics Department in OSU deserve many thanks for providing a friendly and effective research environment. 
Finally, I would like to thank my parents for always being there, showering me with unconditional love and always keeping faith in my abilities. This journey would not have been possible without the extensive support of my family and near friends. Their constant praise has encouraged
me throughout my studies to work hard and excel. More importantly, they contributed most to make me who I am today.
\footnote{Acknowledgements reflect the views of the author and are not endorsed by committee members or Oklahoma State University.}
\end{acknowledge}
\begin{abstract}{Shreyashi Chakdar}{Doctor of Philosophy}{Physics} % Creates abstract
    The Standard Model of particle physics is assumed to be a low-energy effective theory with new physics theoretically motivated to be around TeV scale. The thesis presents theories with new physics beyond the Standard Model in the TeV scale testable in the colliders. Work done in chapters 2, 3 and 5 in this thesis present some models incorporating different approaches of enlarging the Standard Model gauge group to a grand unified symmetry with each model presenting its unique signatures in the colliders. The study on leptoquarks gauge bosons in reference to TopSU(5) model in chapter 2 showed that their discovery mass range extends upto 1.5 TeV at 14 TeV LHC with luminosity of $100 fb^{-1}$. On the other hand, in chapter 3 we studied the collider phenomenology of TeV scale mirror fermions in Left-Right Mirror model finding that the reaches for the mirror quarks goes upto $750$ GeV at the $14$ TeV LHC with $~300 ~fb^{-1}$ luminosity.
In chapter 4 we have enlarged the bosonic symmetry to fermi-bose symmetry e.g. supersymmetry and have shown that SUSY with non-universalities in gaugino or scalar masses within high scale SUGRA set up can still be accessible at LHC with 14 TeV. In chapter 5, we performed a study in respect to the $e^+e^-$ collider and find that precise measurements of the higgs boson mass splittings upto $\sim 100$ MeV may be possible with high luminosity in the International Linear Collider (ILC).
In chapter 6 we have shown that the experimental data on neutrino masses and mixings are consistent with the proposed 4/5 parameter Dirac neutrino models yielding a solution for the neutrino masses with inverted mass hierarchy and large CP violating phase $\delta$ and thus can be tested experimentally. 
Chapter 7 of the thesis incorporates a warm dark matter candidate in context of two Higgs doublet model. The model has several testable consequences at colliders with the charged scalar and pseudoscalar being in few hundred GeV mass range.\\
This thesis presents an endeavor to study beyond standard model physics at the TeV scale with testable signals in the Colliders.

 \end{abstract}
\tableofcontents
\listoftables
\listoffigures
%\msp    %========  single space text.
        %========  For the final version, this command should not be used.

%\makeapproval{5}\addtocounter{placeholder}{-1}    %% Comment this in the final copy. This is for electronic copy
%\begin{nomenclature}
%	\input{notation}
%\end{nomenclature}

%====================== main  body of the dissertation ==========================
\pagenumbering{arabic}\setcounter{page}{1}
\chapter{Introduction} \label{chap:intro}

\section{The Standard Model}
\label{sect:standard_model}
The standard model \index{standard model|(} of particle physics, developed in the late '60's/ early 70's and based on the well known gauge symmetry $SU(3)_C \times SU(2)_L \times U(1)_Y$, contains our best formulation to date understanding the observed classification of elementary particles and their interactions. In the Standard model, all matter consist of a finite irreducible set of spin-1/2 particles denoted as fermions \index{fermion} which interact via the exchange of integral spin bosons\index{boson}. The bosons in the theory act as force carriers for the electro-weak and strong nuclear forces. The fermions are subdivided into classifications of leptons and quarks \index{lepton} \index{quark} based on their electric charge and ability to interact with strong nuclear force.

\begin{figure}[h]
\begin{center}
\includegraphics[width=8cm,height=8cm]{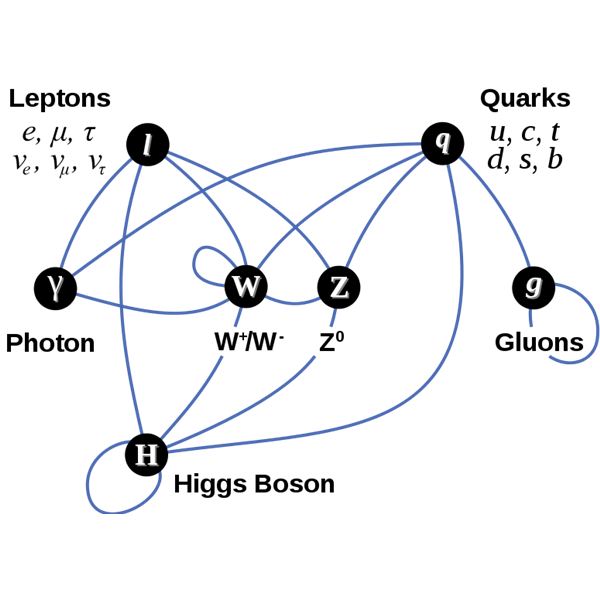}
\caption{Elementary particle structure in The Standard Model}
\end{center}
\end{figure}

\newpage

There are three flavors of leptons forming a progressive mass hierarchy in a doublet \index{lepton!flavor!doublets} arranged structure and consist of integral or zero electric charge (defined in units of the charge of the electron). Each charged lepton is also associated with a neutral particle denoted as a \index{neutrino} \index{lepton!neutrino|see{neutrino}} neutrino.
\begin{equation}
\label{eqn:lepton_flavor_doublets}
\begin{pmatrix} \nu_e       \\ e      \end{pmatrix} \qquad
\begin{pmatrix} \nu_{\mu}     \\ \mu  \end{pmatrix} \qquad
\begin{pmatrix} \nu_{\tau}    \\ \tau \end{pmatrix}
\end{equation}
%
%The three leptons, the \index{electron} \index{lepton!electron|see{electron}} electron, \index{muon} %\index{lepton!muon|see{muon}} muon, and \index{tau} \index{lepton!tau|see{tau}} tau each with negative %charge are taken as the base particles states while their charge conjugates the $e^{+}$, $\mu^{+}$, and %$\tau^{+}$ are denoted as their anti-particles states.
 The neutrinos are taken to be massless in the Standard Model, grouped into three generations corresponding to their associated leptons.  Within the standard model there exists no mechanism which in a direct fashion provides for horizontal mixing between the \index{lepton!family} \index{lepton!horizonal mixing} lepton families; as a result members of each family are assigned a \index{lepton!lepton number} quantum number $L_{\ell}$ corresponding to the lepton flavor of the particle.

%%%%%%%%%%%%%%%%%%%%%%%%
% Begin Table Block
%\begin{table}
%\begin{center}
%\begin{tabular}{|c|c|c| c c c|}
%\hline
%Lepton        & Mass           & Charge & $L_{e}$ & $L_{\mu}$ & $L_{\tau}$ \\
%\hline
%$e^{-}$      & $0.51 \mbox{ MeV}$     & $-1 e$ & 1       & 0         & 0 \\
%$\mu^{-}$    & $105.65 \mbox{ MeV}$   & $-1 e$ & 0       & 1         & 0 \\
%$\tau^{-}$   & $1777.03 \mbox{ MeV}$  & $-1 e$ & 0       & 0         & 1 \\
%\hline
%$\nu_{e}$    & $< 3 \mbox{ eV}$       & $0$    & 1       & 0         & 0 \\
%$\nu_{\mu}$  & $< 0.19 \mbox{ MeV}$   & $0$    & 0       & 1         & 0 \\
%$\nu_{\tau}$ & $< 18.2 \mbox{ MeV}$   & $0$    & 0       & 0         & 1 \\
%\hline
%\end{tabular}
%\caption{Lepton Properties}
%\label{table:lepton_properties}
%\end{center}
%\end{table}
% End Table Block
%%%%%%%%%%%%%%%%%%%%%%%%
The leptons do not experience a direct interaction with the strong color force.  All lepton interactions occur through \index{electroweak interactions} electro-weak interaction couplings and as such are a sensitive probe into the structure of the \index{weak current} weak currents.
%%%%%%%%%%%%%%%%%%%%%%%%%%%
% Begin Figure Block
%\begin{figure}
%\begin{center}
%\input{lepton_electro_weak_1_psfrags.tex}
%\subfigure[Electromagnetic Interaction]{%
%\includegraphics*[width=\textwidth/3]{lepton_electro_weak_1.eps}}
%\subfigure[Weak Neutral Current]{%
%\input{lepton_electro_weak_2_psfrags.tex}
%\includegraphics*[width=\textwidth/3]{lepton_electro_weak_2.eps}}
%\subfigure[Weak Charged Current]{%
%\input{lepton_electro_weak_3_psfrags.tex}
%\includegraphics*[width=\textwidth/3]{lepton_electro_weak_3.eps}}
%\caption{Primitive electro-weak lepton interactions}
%\label{fig:lepton_electro_weak}
%\end{center}
%\end{figure}
% End Figure Block
%%%%%%%%%%%%%%%%%%%%%%%%%%%
%\Omit{
All matter interactions can be broken down into a finite system of fundamental particles interacting with a set of four fundamental forces. 

In contrast to leptons, quarks are distinguished by their interactions via the strong color force and their fractional electric charge.  Strong force \index{strong force} \index{quark!confinement} binding and confinement lead quarks to form the fundamental substructure for all hadronic matter, either in the form of a color neutral three quark bound states that form the common baryons such as the proton and neutron, or in quark-antiquark bound state mesons such as the $\pi,K,\eta$, and $\rho$.  Free quarks are not accessible due to the requirements of color neutrality and strong force confinement at low energies.  Similar to the leptons there exists a generational hierarchy of distinct quark flavor doublets \index{quark!flavor!doublets} based on the masses of each quark and their associated quantum properties. Each generation consists of two quarks each with fractional electric charges equal to $-\frac{1}{3}$ and $\frac{2}{3}$ $\times$ the charge magnitude of the electron.  There are three quark generations we label as \textit{up}, \textit{down}, \textit{charm}, \textit{strange}, \textit{top}, \textit{bottom}.  They are arranged in flavor doublets as: \index{quark!up|see{up quark}} \index{quark!down|see{down quark}} \index{quark!strange|see{strange quark}} \index{quark!charm|see{charm quark}} \index{quark!top|see{top quark}} \index{quark!bottom|see(bottom quark)} \index{up quark} \index{down quark} \index{strange quark} \index{charm quark} \index{top quark} \index{bottom quark}
\begin{equation}
\label{eqn:quark_flavor_doublets}
\begin{pmatrix} u \\ d \end{pmatrix} \qquad
\begin{pmatrix} c \\ s \end{pmatrix} \qquad
\begin{pmatrix} t \\ b \end{pmatrix}
\end{equation}
% Begin Table Block
\begin{table}
\begin{center}
\begin{tabular}{|c|c|c| c c c|}
\hline
Lepton        & Mass           & Charge & $L_{e}$ & $L_{\mu}$ & $L_{\tau}$ \\
\hline
$e^{-}$      & $0.51 \mbox{ MeV}$     & $-1 e$ & 1       & 0         & 0 \\
$\mu^{-}$    & $105.65 \mbox{ MeV}$   & $-1 e$ & 0       & 1         & 0 \\
$\tau^{-}$   & $1777.03 \mbox{ MeV}$  & $-1 e$ & 0       & 0         & 1 \\
\hline
$\nu_{e}$    & $< 3 \mbox{ eV}$       & $0$    & 1       & 0         & 0 \\
$\nu_{\mu}$  & $< 0.19 \mbox{ MeV}$   & $0$    & 0       & 1         & 0 \\
$\nu_{\tau}$ & $< 18.2 \mbox{ MeV}$   & $0$    & 0       & 0         & 1 \\
\hline
\end{tabular}
\caption{Lepton Properties}
\label{table:lepton_properties}
\end{center}
\end{table}
% End Table Block
% Begin Table Block
\begin{table}
\begin{center}
\begin{tabular}{|c|c|c|c|c|}
\hline
Quark & Mass & Charge & B number & L number \\
\hline
u & $1.8-3 \mbox{ MeV/c}^2$         & $\phantom{-}\frac{2}{3} e$  & $\frac{1}{3}$  & 0 \\
d & $4.5-5.3 \mbox{ MeV/c}^2$         & $-\frac{1}{3} e$ & $\frac{1}{3}$ & 0 \\
c & $1.25-1.3 \mbox{ GeV/c}^2$   & $\phantom{-}\frac{2}{3} e$  & $\frac{1}{3}$ & 0 \\
s & $90-100 \mbox{ MeV/c}^2$      & $-\frac{1}{3} e$ & $\frac{1}{3}$ & 0 \\
t & $\approx 174 \mbox{ GeV/c}^2$ & $\phantom{-}\frac{2}{3} e$  & $\frac{1}{3}$ & 0 \\
b & $4.15-4.21 \mbox{ GeV/c}^2$     & $-\frac{1}{3} e$ & $\frac{1}{3}$ & 0 \\
\hline
%Lepton        & Mass           & Charge & $L_{e}$ & $L_{\mu}$ & $L_{\tau}$ \\
%\hline
%$e^{-}$      & $0.51 \mbox{ MeV}$     & $-1 e$ & 1       & 0         & 0 \\
%$\mu^{-}$    & $105.65 \mbox{ MeV}$   & $-1 e$ & 0       & 1         & 0 \\
%$\tau^{-}$   & $1777.03 \mbox{ MeV}$  & $-1 e$ & 0       & 0         & 1 \\
%\hline
%$\nu_{e}$    & $< 3 \mbox{ eV}$       & $0$    & 1       & 0         & 0 \\
%$\nu_{\mu}$  & $< 0.19 \mbox{ MeV}$   & $0$    & 0       & 1         & 0 \\
%$\nu_{\tau}$ & $< 18.2 \mbox{ MeV}$   & $0$    & 0       & 0         & 1 \\
%\hline
\end{tabular}
\caption{Quark Properties}
\label{table:quark_properties}
\end{center}
\end{table}
% End Table Block
%%%%%%%%%%%%%%%%%%%%%%%%
As with the leptons, each quark flavor has a corresponding anti-particle state leading to a total of 36 distinct particles.  These quarks have strong, weak, and electro-magnetic interactions. The various properties like mass, charge for the leptons and the quarks are listed in Tables  \ref{table:lepton_properties} and \ref{table:quark_properties} respectively.\\
%%%%%%%%%%%%%%%%%%%%%%%%%%%
% Begin Figure Block
\begin{figure}
\begin{center}
%\subfigure[Strong Color Exchange]{%
%\input{quark_strong_psfrags.tex}
%\includegraphics*[width=\textwidth/3]{quark_strong.eps}}
%\subfigure[Electro-magnetic]{%
%\input{quark_em_psfrags.tex}
%\includegraphics*[width=\textwidth/3]{quark_em.eps}}
%\subfigure[Weak Charged Current]{%
%\label{fig:quark_interactions_weak_charged}
%\input{quark_weak_1_psfrags.tex}
%\includegraphics*[width=\textwidth/3]{quark_weak_1.eps}}
%\subfigure[Weak Neutral Current]{%
%\label{fig:quark_interactions_weak_neutral}
%\input{quark_weak_2_psfrags.tex}
%\includegraphics*[width=\textwidth/3]{quark_weak_2.eps}}
%\caption{Primitive strong, E\&M, and weak quark interactions}
\label{fig:quark_interactions}
\end{center}
\end{figure}
% End Figure Block
%%%%%%%%%%%%%%%%%%%%%%%%%%%
%%%%%%%%%%%%%%%%%%%%%%%%

\index{standard model|)}
The non-abelian gauge symmetry for the Standard Model is known to be $SU(3)_C \times SU(2)_L \times U(1)_Y$. For Standard Model, SU(3) is unbroken and has eight massless gluons. The remaining group $SU(2)_L \times U(1)_Y$ spontaneously breaks down to $U(1)_{EM}$ producing three massive gauge bosons $W^\pm$ and $Z$. From the leftover $U(1)_{EM}$, we get a massless gauge boson known as photon.\\
Apart from these quarks and leptons and the gauge bosons, the theory also contains a fundamental particle called the Higgs Boson. Over the years, the Standard Model has successfully explained almost all experimental results and precisely predicted a wide variety of phenomena to be discovered later, one example of which will be the existence of top quark, which was discovered at the Fermilab Tevatron. In July 2012, a new boson with a mass of 125 GeV has been discovered at the Large Hadron Collider(LHC) in CERN, Geneva. The experimental study of its properties, so far, shows it is most likely to be the Standard Model Higgs Boson. This discovery seems to complete the validation of the Standard Model as all of the predicted particles and interactions in Standard Model have now been observed experimentally. \\
The Lagrangian based on the most general gauge theory $SU(3)_C \times SU(2)_L \times U(1)_Y$ can be written as
\begin{center}
$\mathcal{L} = L_k + L_f + L_s + L_y$
\end{center}
where $L_k$ contains the gauge boson kinetic terms, $L_f$ contains the fermionic kinetic terms, $L_s$ contains the scalar mass terms, kinetic energy terms as well as the self interactions and $L_y$ contains interactions between the fermions and the scalars. 
To summarize the standard (Weinberg-Salam) model, we gather together all the
ingredients of the Lagrangian. The complete Lagrangian is explicitly given by:

\begin{center}
\begin{tabular}{rll}
   $\mathcal{L} =$ 
               & $-\frac{1}{4} {\bf W}_{\mu\nu} \cdot {\bf W}^{\mu\nu}
                  -\frac{1}{4}       B_{\mu\nu}             B^{\mu\nu}$
               & $\left\{ \begin{array}{l}
                          \mbox{W}^{\pm}, \mbox{Z}, \gamma \mbox{ kinetic} \\
                          \mbox{energies and}                              \\
                          \mbox{self-interactions}
                          \end{array}
                  \right.$ 
               \\ & & \\
               & $\begin{array}{l}
                  +\bar{L}\gamma^{\mu}\left( i\partial_{\mu}
                                             -g\frac{1}{2}{\bf \tau} \cdot {\bf W}_{\mu}
                                             -g'\frac{Y}{2}B_{\mu}
                                      \right) L \\
                  +\bar{R}\gamma^{\mu}\left( i\partial_{\mu}
                                             -g'\frac{Y}{2}B_{\mu}
                                      \right) R \\
                  \end{array}$
               & $\left\{ \begin{array}{l}
                          \mbox{lepton and quark}          \\
                          \mbox{kinetic energies}          \\
                          \mbox{and their}                 \\
                          \mbox{interactions with}         \\
                          \mbox{W}^{\pm}, \mbox{Z}, \gamma \\
                          \end{array}
                  \right.$ 
               \\ & & \\
               & $+\left|\left( i\partial_{\mu}
                                -g\frac{1}{2}{\bf \tau} \cdot {\bf W}_{\mu}
                                -g'\frac{Y}{2}B_{\mu}
                         \right) \phi
                   \right|^2 - V(\phi)$
               & $\left\{ \begin{array}{l}
                          \mbox{W}^{\pm}, \mbox{Z}, \gamma, \mbox{ and Higgs} \\
                          \mbox{masses and}                                   \\
                          \mbox{couplings}                                    \\
                          \end{array}
                  \right.$
               \\ & & \\
               & $-(G_1\bar{L}\phi R 
                    +G_2\bar{L}\phi_c R
                    +\mbox{hermitian conjugate}).$ 
               & $\left\{ \begin{array}{l}
                          \mbox{lepton and quark}  \\
                          \mbox{masses and}        \\
                          \mbox{coupling to Higgs} \\
                          \end{array}
                  \right.$ \\
\end{tabular}
\end{center}

\noindent
where $\bf W_{\mu\nu}$, $B_{\mu\nu}$ are the gauge bosons field strength tensors and the constants g, g', $G_1$ and $G_2$ are respectively the weak and yukawa coupling constants. 
Here $L$ denotes a left-handed fermion (lepton or quark) doublet, and R denotes a
right-handed fermion singlet. The scalar doublet $\phi$ which breaks the $SU(2)_L \times U(1)_Y$ symmetry is written with gauge quantum numbers and Vacuum expectation value (VEV) as,
\begin{center}
\begin{eqnarray}
\Phi \sim (1,2,1) &\nonumber\\
\left < \Phi \right> = \frac{1}{\sqrt 2}{\begin{pmatrix} 0 \\ v \end{pmatrix}}
\end{eqnarray}
\end{center}

\section{Limitations of the Standard Model}

The Standard Model (SM) of particle physics has been remarkably successful. However there are several
reasons why it is widely believed that while working well in the low energy regimes which
have been investigated to date, the Standard Model does not present the full picture. There are also several evidences and hints that seem to suggest the presence of new physics ‘Beyond the Standard Model’ (BSM). 
It is the belief that there is physics ‘Beyond the Standard Model’ (BSM) which
motivates the continuation of large scale particle physics experiments, most notably the
Large Hadron Collider (LHC) at CERN which is colliding protons with a center-of-mass
energy of 8 and 13 TeV respectively.\\
The issues with the Standard Model can be roughly divided into two classes: experimental discrepancies and theoretical considerations that does not allow us to accept the Standard Model as the ultimate theory.
The work in this thesis will address some of these issues but by no means all that are presented below.\\
Standard Model can not explain the experimental discoveries such as non-zero neutrino masses, existence of dark matter and dark energy, strong CP problem, baryon asymmetry in the Universe to name a few in its original framework. On the other hand, there are theoretical drawbacks such as, the higgs mass being in the electroweak scale, parity violation, gauge coupling unification, quantization of the electric charges, mass hierarchy of the elementary particles, flavor problem, quantum gravity etc which cannot be accommodated in the Standard Model itself. Some of these issues are discussed in detail in the following sections:\\
\textbf{Neutrino masses}: In the Standard Model, neutrinos have exactly zero mass. This is a consequence of the Standard Model containing only left-handed neutrinos. With no suitable right-handed partner, it is impossible to add a renormalizable mass term to the Standard Model. Measurements of neutrino oscillations however indicate that neutrinos spontaneously change flavor, which implies that neutrinos have non-zero masses. These measurements only give the mass difference squares of the different flavors. The best constraint on the absolute mass of the neutrinos comes from precision measurements of tritium decay, providing an upper limit 2 eV electron anti-neutrinos, which makes them at least seven orders of magnitude lighter than the electron in the Standard Model. This necessitates an extension of the Standard Model, which not only needs to explain how neutrinos get their mass, but also why the mass is so small.\\
\textbf{Existence of dark matter and dark energy}: Cosmological observations tell us that the Standard Model explains only about $5\%$ of the energy present in the universe. About $26\%$ is the dark matter, which would behave just like other matter, but only interacts weakly (or does not) with the Standard Model fields. Yet, the Standard Model does not supply any fundamental particles that are good dark matter candidates. The rest ($69\%$) should be dark energy, a constant energy density for the vacuum and Standard Model does not include it also.\\
\textbf{Strong CP problem}: The Standard Model should contain a term that breaks CP symmetry relating matter to antimatter, in the strong interaction sector. Experimentally, however, no such violation has been found, implying that the coefficient of this term is very close to zero. This fine tuning is also considered unnatural.\\
\textbf{Matter-antimatter asymmetry}: The universe is made out of mostly matter. However, the Standard Model predicts that matter and antimatter should have been created in (almost) equal amounts if the initial conditions of the universe did not involve disproportionate matter relative to antimatter. Yet, no mechanism sufficient to explain this asymmetry exists in the Standard Model.\\
\textbf{Hierarchy problem}: The Standard Model introduces particle masses through spontaneous symmetry breaking caused by the Higgs field. Within the Standard Model, the mass of the Higgs gets some very large quantum corrections. These corrections are of the order of Planck's scale which is seventeen orders larger than the actual mass of the Higgs. This means that the bare mass parameter of the Higgs in the Standard Model must be fine tuned in such a way that almost completely cancels the quantum corrections. This level of fine-tuning is deemed unnatural by many theorists.There are also issues of Quantum triviality, which suggests that it may not be possible to create a consistent quantum field theory involving elementary scalar particles.\\
\textbf{Coupling unification}: If the strong and electroweak forces are to unify into a single gauge theory at a high energy scale then the gauge couplings must also unify. The particle content of a model
determines how the couplings ‘run’ with energy. The Standard model contains a semi-simple group in which the three gauge couplings almost unify at a scale of about $10^{16}$ GeV if suitable new set of particles are added, although this unification isn’t quite exact. This can be seen as a motivation for a theory with a larger group with additional particle content or new physics containing the Standard model as a subgroup which will modify the running of the couplings to make the unification more exact.\\
\textbf{Gravity}: A complete model for particle physics would be expected to describe all the fundamental
forces between particles, electroweak, strong and gravitational. However the Standard
Model does not include any gravitational force, and hence cannot be the full
fundamental model. At energy scales of order the Planck mass $M_P$, a theory of quantum
gravitation will be required to describe the interactions between particles. This shows
that the Standard Model will need to be replaced by an alternative theory at very
high energy scales, and it is reasonable to expect that a more complete model than the
Standard Model might be required even at energy scales only moderately higher than
those which have already been investigated in detail.\\\newpage

\section{Extensions of the Standard Model}

The work done in this thesis tries to address some of the limitations of the Standard Model. Theoretical progress towards these questions has been made basically along three directions:\\
(i) The first direction is enlarging the Standard Model gauge symmetry to a grand unified symmetry such as SU(5), SO(10) or some other extensions like the Left-Right symmetry. This explains the charge quantization and unification of the couplings, but there is no sign of proton decay so far. Work done in the chapters 2, 3, 5 in this thesis present some models in this direction.\\
(ii) In the second direction, the bosonic symmetry can be enlarged to fermi-bose symmetry, example is supersymmetry. This approach solves hierarchy problem, incorporates the unification of the three gauge couplings, and has good candidate for dark matter. But there is also no sign of superpartners at the Large Hadron Collider (LHC) yet. Work done in the chapter 4 in this thesis presents models in this direction.\\
(iii) The third direction is increasing the number of spatial dimensions, an example of this is the extra dimensions theories. Gravity can also be included into the mix as in the examples of supergravity and string theory. Experimentally, we have not seen any sign of Kaluza-Klein excitations so far in the Colliders.\\
Some other interesting avenues that can be persued includes explaining the non-zero neutrino masses and mixing. Work done in the chapter 6 presents some models in this neutrino sector.\\
Another important sector is cosmological model building which has been investigated in Chapter 7 in this thesis. Some aspects of these kind of models is the presence of dark matter candidates and description of their cosmological history. 
This thesis presents the completed works in this framework of ``Beyond the Standard Model Physics"  which reside in TeV scale and are thus testable at Colliders like the Large Hadron Collider (LHC) and the proposed International Linear Collider (ILC).

\chapter{Top $SU(5)$ Model: Baryon and Lepton Number Violating Resonances at the LHC}\label{chap:chap2}

\section{Introduction}

 The Standard Model (SM), based on the local
gauge symmetry $SU(3)_C\times SU(2)_L\times U(1)_Y$, is very
successful in describing all the experimental results below
the  TeV scale. It is an excellent effective field theory, but
it is widely believed not to be  the final theory.  
 Discovery of new particles is highly anticipated at 
the Large Hadron Collider (LHC). The most likely and reasonably
well motivated candidates are  supersymmetric particles, 
and extra $Z'$ boson. However, it is important to explore other alternatives or entirely new possibilities at the current and future LHC.

In the SM, we have fermions (spin 1/2) and scalars 
(Higgs fields)(spin 0) which do not belong to adjoint 
representations under the SM gauge symmetry. 
Can we also have TeV scale gauge bosons (spin 1) belonging
to the non-adjoint representations under the SM 
gauge symmetry? Can we achieve the (partial) grand unified theory at the TeV scale? Can we construct a renormalizable theory realizing such a possibility which can be  tested at the LHC? These are very interesting theoretical questions that we shall address in this work. Discovery of such gauge  bosons in the TeV scale at the LHC will open up a new window for our understanding of the fundamental theory describing the nature.

How can we construct a consistent  theory involving the massive vector bosons which do not belong to the adjoint representations under the SM gauge symmetry? If the massive vector bosons are not the gauge bosons of a symmetry group, there are some theoretical problems from the consistency of quantum field theory, for instance, the unitarity and renormalizability~\cite{Lee:1962vm}.
%On the other hand, the gauge bosons are massless if the gauge %symmetry
%is exact.
When the gauge symmetry is spontaneously
broken via the Higgs mechanism, the interactions of
the massive gauge bosons satisfy both the unitarity and the
renormalizability of the theory~\cite{Hooft,Lee}.
Thus, the massive vector bosons must be the gauge bosons arising from the spontaneous gauge symmetry breaking.
As we know, a lot of models with extra TeV scale gauge bosons have been proposed previously in the literature. However, those massive gauge bosons  either  belong to the adjoint representations or are singlets under the SM gauge symmetry~\cite{Hill:1991at,Hill:1993hs,Dicus:1994sw,
Muller:1996dj,Malkawi:1996fs,Erler:2002pr,Chiang:2007sf,PSLR}. For example, in the top color model~\cite{Hill:1991at,Hill:1993hs,Dicus:1994sw}, 
the colorons belong to the adjoint representation
of the $SU(3)_C$; in the top 
flavor model~\cite{Muller:1996dj,Malkawi:1996fs}, the extra $W'$ and $Z'$
bosons belong to the adjoint representation of the $SU(2)_L$, while
in the $U(1)'$ model~\cite{Erler:2002pr} or 
top hypercharge model~\cite{Chiang:2007sf}, 
the new $Z'$ boson is a singlet under the SM gauge symmetry.
In the Grand Unified Theories such as $SU(5)$ and
 $SO(10)$~\cite{Georgi:1974sy,Fritzsch:1974nn},
there are such kind of massive gauge bosons. However, their
masses have to be around the unification scale $\sim 10^{16}$ GeV 
to satisfy the proton decay constraints. 

Some years ago, TL and SN had  proposed
 a class of models where the gauge symmetry is  ${\cal G} \equiv 
\prod_i G_i \times SU(3)'_C \times SU(2)'_L \times U(1)'_Y$~\cite{Li:2004cj}.
The quantum numbers of the SM fermions and Higgs fields under
the $SU(3)'_C \times SU(2)'_L \times U(1)'_Y$ gauge symmetry are
the same as they have under the SM gauge 
symmetry $SU(3)_C \times SU(2)_L \times U(1)_Y$,
while they are all singlets under $\prod_i G_i$. Hence $\prod_i G_i$ is 
the hidden gauge symmetry. After the 
gauge symmetry ${\cal G}$ is spontaneously broken down to
the SM gauge symmetry at the TeV scale via Higgs mechanism,
some of the massive gauge bosons from the ${\cal G}$ breaking
do not belong to the adjoint representations under the SM
gauge symmetry. In particular, a
 concrete $SU(5) \times SU(3)'_C \times SU(2)'_L \times U(1)'_Y$
has been studied in detail.
However, the corresponding $(X_{\mu}, Y_{\mu})$ massive gauge 
bosons are meta-stable and behave like the stable heavy
quarks and anti-quarks at the LHC~\cite{Li:2004cj}.
Thus, an interesting question is whether we can construct the
$SU(5) \times SU(3)'_C \times SU(2)'_L \times U(1)'_Y$
models where the $(X_{\mu}, Y_{\mu})$ gauge bosons can decay and produce interesting signals at the LHC. By the way, the six-dimensional orbifold non-supersymmetric and supersymmetric $SU(5)$ and $SU(6)$ models with low energy gauge unification
have been constructed previously~\cite{Li:2001qs, Li:2002xw, Jiang:2002at}. However, there is no direct interactions between the  $(X_{\mu}, Y_{\mu})$ particles and the SM fermions.

As pointed out above, the top color model~\cite{Hill:1991at,Hill:1993hs,Dicus:1994sw}, 
top flavor model~\cite{Muller:1996dj,Malkawi:1996fs},
and top hypercharge model~\cite{Chiang:2007sf} have been constructed before. 
Because of the proton decay problem and quark CKM mixings,  etc,
the real challenging question is whether we can construct the top $SU(5)$ model as the unification of these models. Consequently,
we can explain the charge quantization for the third family, and 
 probe  the baryon and lepton number violating interactions involving the third family at the LHC. Such a model was proposed by us recently \cite{Chakdar:2012kd}, and its implications for LHC was briefly explored.  

In this work, we propose two such models: the minimal and the renormalizable top $SU(5)$ model where the 
$SU(5)\times SU(3)'_C \times SU(2)'_L \times U(1)'_Y$ gauge symmetry 
is broken down to the SM
gauge symmetry via the bifundamental Higgs fields at low energy. The first 
two families of the SM fermions are charged under $ SU(3)'_C \times SU(2)'_L \times U(1)'_Y$
while the third family is charged under $SU(5)$. 
In the minimal top $SU(5)$ model, we show that the quark CKM mixing matrix can be generated via dimension-five operators, and the proton decay problem can be solved by fine-tuning the coefficients of the higher dimensional operators at the order of $10^{-4}$. In the renormalizable top $SU(5)$ model, we can explain 
the quark CKM mixing matrix by introducing vector-like particles, and we do not have proton decay problem. In these models, the non-unification of the three SM couplings are remedied, because three SM couplings $g_3$, $g_2$, $g_1$ are now combinations of $(g_5, g_3')$, $(g_5, g_2')$,$(g_5, g_1')$, and need not be unified. Since the models have baryon and lepton number violating interactions, it might be useful in generating the baryon asymmetry of the Universe. In our models, since the third family quark lepton unification is at the TeV scale,  we can probe the new $(X_{\mu}, ~Y_{\mu})$ gauge bosons at the LHC through their decays  to the third family of the SM fermions. 

This chapter is organized as follows. In section 2.2, we discuss the two models and their formalism. In section 2.3, we discuss in detail the phenomenological implications of the models. These include the productions and decays of the X and Y gauge bosons at the LHC energies of 7, 8 and 14 TeV, their decay modes, and the signals for the final states. We also discuss the LHC reach for the masses of these particle for various LHC energies and luminosities. Section 2.4 contains the summary and conclusions.

\section{The Minimal and Renormalizable Top $SU(5)$ Models}
%%%%%%%%%%%%%%

We propose two non-supersymmetric top $SU(5)$ models 
where the gauge symmetry is 
$ SU(5)\times SU(3)'_C \times SU(2)'_L \times U(1)'_Y$.
The first 
two families of the SM fermions are charged under 
$ SU(3)'_C \times SU(2)'_L \times U(1)'_Y$
while the third family is charged under $SU(5)$. 
We denote the gauge fields
for $SU(5)$ and $SU(3)'_C \times SU(2)'_L \times U(1)'_Y $
as ${\widehat A}_{\mu}$ and ${\widetilde A}_{\mu}$, respectively,
and the gauge couplings for $SU(5)$, $SU(3)'_C$, $SU(2)'_L$
and $U(1)'_Y$ are $g_5$, $g'_3$, $g'_2$ and $g'_Y$,
respectively. The Lie algebra indices for the generators of
 $SU(3)$, $SU(2)$ and $U(1)$
are denoted by $a3$, $a2$ and $a1$, respectively, and
the  Lie algebra indices for the generators of
$SU(5)/(SU(3)\times SU(2)\times U(1))$ are denoted by ${\hat a}$.
After the $SU(5)\times SU(3)'_C \times SU(2)'_L \times U(1)'_Y $
gauge symmetry is broken down to the SM gauge symmetry
$SU(3)_C\times SU(2)_L \times U(1)_Y$,
 we denote the massless gauge fields for the
SM gauge symmetry as $A_{\mu}^{ai}$, and
the massive gauge fields as $B_{\mu}^{ai}$ and ${\widehat A}_{\mu}^{\hat a}$.
The gauge couplings for the SM gauge symmetry
$SU(3)_C$, $SU(2)_L$ and $U(1)_Y$ are $g_3$, $g_2$ and $g_Y$, respectively.

To break the $SU(5)\times SU(3)'_C \times SU(2)'_L \times U(1)'_Y $
gauge symmetry down to the SM gauge symmetry, we introduce
two bifundamental Higgs fields $U_T$ and $U_D$~\cite{Li:2004cj}. 
Let us explain our convention. We denote the first two family
quark doublets, right-handed up-type quarks, right-handed down-type 
quarks, lepton doublets, right-handed neutrinos, right-handed charged leptons,
and the corresponding Higgs field respectively
as $Q_i$, $U_i^c$, $D_i^c$, $L_i$, $N^c_i$, $E_i^c$, and $H$, as in the 
supersymmetric SM convention. We denote the third family 
SM fermions as $F_3$, $\overline{f}_3$, and $N_3^c$.
To give the masses to the third family of the SM fermions, we introduce
a $SU(5)$ anti-fundamental Higgs field $\Phi\equiv (H'_T, H')$. We also
 need to introduce a scalar field $XT$ if we require that the triplet
Higgs $H'_T$ have mass around the 
$SU(5)\times SU(3)'_C \times SU(2)'_L \times U(1)'_Y$ gauge symmetry breaking
scale. However, it is not necessary, and we will explain it in
the following. In addition, note that the neutrino PMNS mixings can be
generated via the right-handed neutrino Majorana mass mixings.
we propose two top $SU(5)$ models which can generate the mass for the possible
pseudo-Nambu-Goldston boson (PNGB) $\phi$ during the gauge symmetry
breaking and generate the quark CKM mixings.
In the minimal top $SU(5)$ model,  we consider the dimension-five
non-renormalizable operators and fine-tune some
coefficients of the higher dimensional operators
at the order $10^{-4}$ to suppress the proton decay.
In the renormalizable top $SU(5)$ model,  we introduce the additional vector-like 
particles. To give the PNGB mass, we introduce a scalar field $XU$ in the
$SU(5)$ anti-symmetric representation.
And to generate the quark CKM mixings while not to introduce
the proton decay problem, we introduce the vector-like fermionic particles 
$(Xf, ~Xf^c)$ and $(XD, ~XD^c)$. Note that the 
$SU(3)'_C \times SU(2)'_L \times U(1)'_Y $ gauge symmetry can be formally embedded
into a global $SU(5)'$ symmetry, and to do that, we introduce the vector-like particles
$(XL, ~XL^c)$ as well. The complete
particle content and the particle quantum numbers under 
$SU(5)\times SU(3)'_C \times SU(2)'_L \times U(1)'_Y $ gauge symmetry
are given in Table~\ref{tab:Content}.

\begin{table}[htb]
%\caption[]{The complete
%particle content and the particle quantum numbers under 
%$SU(5)\times SU(3)'_C \times SU(2)'_L \times U(1)'_Y $ gauge symmetry in the
%top $SU(5)$ model. Here, $i=1,~2$, and $k=1,~2,~3$.}
%\label{tab:Content}
\begin{center}
  \begin{tabular}{|c|c||c|c|}
   \hline
 Particles  & Quantum Numbers  & Particles & Quantum Numbers \\ \hline
 $Q_i$ & $({\bf 1}; {\bf {3}}, {\bf 2}, {\bf {1/6}})$  
& $L_i$ & $({\bf 1}; {\bf {1}}, {\bf 2}, {\bf {-1/2}})$  
\\\hline
 $U^c_i$ & $({\bf 1}; {\bf {\bar 3}}, {\bf 1}, {\bf {-2/3}})$  
& $N^c_k$ & $({\bf 1}; {\bf {1}}, {\bf 1}, {\bf {0}})$  
\\\hline
 $D^c_i$ & $({\bf 1}; {\bf {\bar 3}}, {\bf 1}, {\bf {1/3}})$  
& $E^c_i$ & $({\bf 1}; {\bf {1}}, {\bf 1}, {\bf {1}})$  
\\\hline
 $F_3$ & $({\bf 10}; {\bf {1}}, {\bf 1}, {\bf {0}})$  
& $\overline{f}_3$ & $({\bf {\bar 5}}; {\bf {1}}, {\bf 1}, {\bf {0}})$  
\\\hline
 $H$ & $({\bf 1}; {\bf {1}}, {\bf 2}, {\bf {-1/2}})$  
& $\Phi$ & $({\bf {\bar 5}}; {\bf {1}}, {\bf 1}, {\bf {0}})$  
\\\hline
 $U_T$ & $({\bf 5}; {\bf {\bar 3}}, {\bf 1}, {\bf {1/3}})$
& $U_D$ & $({\bf 5}; {\bf {1}}, {\bf { 2}}, {\bf -1/2})$
\\\hline \hline
$XT$ & $({\bf 1}; {\bf {\bar 3}}, {\bf 1}, {\bf {1/3}})$ &
$XU$ & $({\bf 10}; {\bf {1}}, {\bf 1}, {\bf {-1}})$ 
\\\hline 
 $Xf$ & $({\bf 5}; {\bf {1}}, {\bf 1}, {\bf {0}})$  
& $\overline{Xf}$ & $({\bf {\bar 5}}; {\bf {1}}, {\bf 1}, {\bf {0}})$  
\\\hline
$XD$ & $({\bf 1}; {\bf {3}}, {\bf 1}, {\bf {-1/3}})$  
& $\overline{XD}$ & $({\bf {1}}; {\bf {\bar 3}}, {\bf 1}, {\bf {1/3}})$  
\\\hline
$XL$ & $({\bf 1}; {\bf {1}}, {\bf 2}, {\bf {-1/2}})$  
& $\overline{XL}$ & $({\bf {1}}; {\bf {1}}, {\bf 2}, {\bf {1/2}})$  
\\\hline
\end{tabular}
%\end{center}
\caption{The complete
particle content and the particle quantum numbers under 
$SU(5)\times SU(3)'_C \times SU(2)'_L \times U(1)'_Y $ gauge symmetry in the
top $SU(5)$ model. Here, $i=1,~2$, and $k=1,~2,~3$.}
\label{tab:Content}
\end{center}
%\end{tabular}
\end{table}

To give the vacuum expectation values (VEVs)
 to the bifundamental Higgs fields $U_T$ and $U_D$, we consider the
following Higgs potential
\begin{eqnarray}
V &=& -m_{T}^2 |U_T^2|-m_{D}^2 |U_D^2| + \lambda_T |U_T^2|^2
+ \lambda_D |U_D^2|^2 + \lambda_{TD} |U_T^2| |U_D^2|
\nonumber\\ &&
+\left[ A_T \Phi U_T XT^{\dagger} + A_D \Phi U_D H^{\dagger} +
 {{y_{TD}}\over {M_{*}}} U^3_T U^2_D + {\rm H.C.}\right]~,~\,
\label{potential}
\end{eqnarray} 
where $M_*$ is a normalization mass scale.

A few remarks are in order. First, with $XT$ particle, 
the Higgs triplet $H'_T$ will have mass around the 
$SU(5) \times SU(3)'_C \times SU(2)'_L \times U(1)'_Y$
gauge symmetry breaking scale, as given by the above
$A_T$ term. However, it is still fine even if we do not
introduce the $XT$ field. Let us explain it in detail.
In our models, we have
two Higgs doublets $H$ and $H'$, which give
the masses to the first two families and
the third family of the SM fermions, respectively. Thus,
$H'_T$ will have mass around a few hundred GeV,
and it has interesting decay channels via Yukawa couplings, 
which will be discussed in the following.

Second, without the non-renormalizable $y_{TD}$ term,
we have global symmetry $U(5)\times SU(3)'_C \times SU(2)'_L \times U(1)'_Y $
in the above potential, and then we will have a PNGB $\phi$
during the $SU(5) \times SU(3)'_C \times SU(2)'_L \times U(1)'_Y$ gauge
symmetry breaking. To break the $U(5)$ global symmetry
down to $SU(5)$ and then give mass to $\phi$, we do need this non-renormalizable term.
Moreover, $M_*$ can be around the intermediate scale, for example,
1000~TeV. If we assume that all the high-dimensional operators are 
suppressed by the reduced Planck scale, {\it i.e.}, $ M_*=M_{\rm Pl}$, 
we can generate the $y_{TD}$ term by introducing the $XU$ field. 
The relevant Lagrangian is 
\begin{eqnarray}
-{\cal L} &=& \left(y_T U_T^3 XU + y_D \mu' U_D^2 XU^{\dagger} + {\rm H.C.} \right)
+ M^2_{XU} |XU|^2~,~\, 
\label{Eq-P-XU}
\end{eqnarray} 
where the mass scales
$\mu'$ and  $M_{XU} $ will be assumed to be around 1000~TeV. 
After we integrate
out $XU$, we get the needed high-dimensional operator
\begin{eqnarray}
V \supset- {{y_T y_D \mu'}\over {M_{XU}^2}}  U^3_T U^2_D~.~\,
\end{eqnarray} 
%Also, after $U_T$ and $U_D$
%obtain VEVs, we will have a tadpole term for $XU$. And then
%$XU$ will get a VEV as well. For the VEVs of $U_T$ and $U_D$
%around 1~TeV, the VEV of $XU$ is around 1 GeV. Thus, 
%our following discussions will not be affected due to the $XU$'s %VEV,
%and then we will neglect the $XU$'s VEV in the following.

We choose the following VEVs for the fields $U_T$ and $U_D$
\begin{eqnarray}
 <U_T> =  {v_T} \left(
  \begin{array}{c}
    I_{3\times3} \\
    0_{2\times3} \\
  \end{array}
  \right)~, \quad
 <U_D> =  {v_D} \left(
  \begin{array}{c}
    0_{3\times2} \\
    I_{2\times2} \\
  \end{array}
  \right)~,
\end{eqnarray}
where $I_{i\times i}$ is the $i\times i$ identity matrix, and
$0_{i\times j}$ is the $i\times j$ matrix where all the entries are
zero. We assume that $v_D$ and $v_T$ are in the 
TeV range so that the massive gauge bosons  have TeV scale masses.

From the kinetic terms for the fields
$U_T$ and $U_D$ , we obtain the mass terms for the gauge fields 
\begin{eqnarray}
\sum_{i=T, D} \langle (D_{\mu} U_i)^{\dagger} D^{\mu} U_i \rangle
&=& {1\over 2} v_T^2   \left( g_5 {\widehat A}_{\mu}^{a3} - 
g'_3 {\widetilde A}_{\mu}^{a3} \right)^2
+ {1\over 2} v_D^2  \left( g_5 {\widehat A}_{\mu}^{a2} - 
g'_2 {\widetilde A}_{\mu}^{a2} \right)^2
\nonumber\\ &&
+\left( {{v_T^2}\over 3} 
+ {{v_D^2}\over 2}  \right) 
\left( g_5^Y {\widehat A}_{\mu}^{a1} - 
g'_Y {\widetilde A}_{\mu}^{a1} \right)^2
\nonumber\\ &&
+ {1\over 2} g_5^2 \left(v_T^2  +v_D^2 \right) 
\left(X_{\mu} \overline{X}_{\mu}
+ Y_{\mu} \overline{Y}_{\mu}\right)
~,~\,
\label{massterm}
\end{eqnarray}
 where $g_5^Y \equiv {\sqrt 3} g_5/{\sqrt 5}$,
 and we define the complex fields
($X_{\mu}$, $Y_{\mu}$) 
with quantum numbers (${\bf 3}$, ${\bf 2}$, ${\bf {5/6}}$) 
from the gauge fields ${\widehat A}_{\mu}^{\hat a}$, similar to that
in the usual $SU(5)$ model~\cite{Georgi:1974sy}.

From the original gauge fields ${\widehat A}_{\mu}^{ai}$ and
${\widetilde A}_{\mu}^{ai}$ and from
Eq. (\ref{massterm}), we obtain the massless gauge bosons $A_{\mu}^{ai}$ and the
TeV scale massive gauge bosons $B_{\mu}^{ai}$ ($i=3, 2, 1$)
which are in the adjoint representations of the SM gauge
symmetry
\begin{eqnarray}
\left(
\begin{array}{c}
A_\mu^{ai} \\
B_\mu^{ai}
\end{array} \right)=
\left(
\begin{array}{cc}
\cos\theta_i & \sin\theta_i \\
-\sin\theta_i & \cos\theta_i
\end{array}
\right)
\left(
\begin{array}{c}
{\widehat A}_\mu^{ai} \\
{\widetilde A}_\mu^{ai}
\end{array} \right)
~,~\,
\end{eqnarray}
where $i=3, 2, 1$, and 
\begin{eqnarray}
\sin\theta_j \equiv {{g_5}\over\displaystyle {\sqrt {g_5^2 +(g'_j)^2}}}
~,~
\sin\theta_1 \equiv {{g^Y_5}\over\displaystyle 
{\sqrt {(g_5^{Y})^2 +(g_Y^{\prime})^2}}} ~,~\,
\end{eqnarray}
where $j=3, 2$.
We also have the massive gauge bosons
($X_{\mu}$, $Y_{\mu}$) and (${\overline{X}_{\mu}}$, ${\overline{Y}_{\mu}}$)
which are not in the adjoint representations of the SM gauge symmetry.
 So, the $SU(5)\times SU(3)'_C \times SU(2)'_L \times U(1)'_Y $
gauge symmetry is  broken down to the diagonal SM gauge symmetry
$SU(3)_C\times SU(2)_L \times U(1)_Y$, and the theory is
unitary and renormalizable. The SM gauge couplings
$g_j$ ($j=3, 2$) and $g_Y$ are given by
\begin{eqnarray}
{1\over {g_j^2}} ~=~ {1\over {g_5^2}} + {1\over {(g'_j)^2}}~,~
{1\over {g_Y^2}} ~=~ {1\over {(g^Y_5)^2}} + {1\over {(g'_Y)^2}}~.~\,
\end{eqnarray}

If the theory is perturbative,  the upper and lower
bounds on the gauge couplings $g_5$, $g'_3$, $g'_2$ and $g'_Y$ are 
\begin{eqnarray}
&& g_3 ~< ~g_5 ~< ~{\sqrt {4\pi}} ~,~ g_3 ~<~ g'_3 ~<~ {\sqrt {4\pi}} ~,~
\\ &&
g_2 ~<~ g'_2 ~<~
{{g_3 g_2}\over {\sqrt {g_3^2 -g_2^2}}}~,~ \\ &&
g_Y ~<~ g'_Y  ~<~
{{{\sqrt 3} g_3 g_Y}\over {\sqrt {3g_3^2 -5g_Y^2}}}~.~ \,
\end{eqnarray}
Note that the gauge coupling  $g_5$ for $SU(5)$  is naturally 
large at the TeV scale because
the beta function of $SU(5)$ is negative, {\it i.e.},
$SU(5)$ is asymptotically free.

\subsection{The Minimal Model}

We consider the minimal model first, where we do not 
introduce any extra (``$X$'') particles $XT$, $XU$, $Xf$,
$\overline{Xf}$, $XD$, $\overline{XD}$, $XL$, and
$\overline{XL}$. So the Higgs triplet $H'_T$ will be 
a few hundred GeV.
We introduce the non-renormalizable
operators to generate the quark CKM mixings. We also escape the 
proton decay problem by fine-tuning some coefficients
of the higher-dimensional operators.

The renormalizable SM fermion Yukawa couplings are
\begin{eqnarray}
-{\cal L} &=& y^u_{ij} U_i^c Q_j {\widetilde H} +
y^{\nu}_{kj} N_k^c L_j  {\widetilde H} + 
 y^d_{ij} D_i^c Q_j H + y^e_{ij} E_i^c L_j H 
\nonumber\\ &&
+y^u_{33} F_3 F_3 \Phi^{\dagger}
+ y^{de}_{33} F_3 {\overline f}_3 \Phi
+ y^{\nu}_{k3} N_{k}^c  {\overline f}_3 \Phi^{\dagger}
+ m^N_{kl}  N_{k}^c  N_{l}^c  + {\rm H.C.} ~,~\,
\end{eqnarray}
where $i/j=1,~2$, $k/l=1,~2,~3$, and $ {\widetilde H}=i\sigma_2 H^{\dagger}$
with $\sigma_2$ the second Pauli matrix.
Because the three right-handed neutrinos can mix among
themselves via the Majorana masses, we can generate
the observed neutrino masses and mixings.
In addition, we make a wrong prediction that the bottom Yukawa coupling is
equal to the tau Yukawa coupling at the low energy. 
We can easily avoid this problem by introducing
the high-dimensional Higgs field under $SU(5)$, which
is out of the scope of this paper.
In addition, the Yukawa terms between 
the triplet Higgs field $H'_T$ in $\Phi$ and
 the third family of the SM fermions are
 $ y^{de}_{33} t^c b^c H'_T$, $ y^{de}_{33} Q_3 L_3 H'_T$, 
and $y^{u}_{33} t^c \tau^c H^{\prime \dagger}_T$. So, we have
$(B+L)$ violating interactions as well.

To generate the quark CKM mixings, we consider the higher-dimensional
operators. The dimension-five operators are
\begin{eqnarray}
-{\cal L} &=& {1\over {M_*}}
\left(  y^d_{i3} D_i^c F_3 \Phi U_T^{\dagger}
+ y^{e}_{i3} E_i^c {\overline f}_3 H U_D 
+
y^d_{3i} {\overline f}_3  Q_i H U_T
+ y^{e}_{3i} F_3 L_i \Phi U_D^{\dagger} \right) + {\rm H.C.} 
 ~.~\,
\label{O-Dim-5}
\end{eqnarray}
And the dimension-six operators are
\begin{eqnarray}
-{\cal L} &=&  {1\over {M^2_*}}
\left(y^u_{i3} U_i^c F_3 {\widetilde H} U_T^{\dagger} U_D^{\dagger}
+ y_{i3}^{\prime d} D_i^c F_3 H U_T^{\dagger} U_D^{\dagger}
+ y^u_{3i} F_3 Q_i \Phi^{\dagger} U_T U_D
\right.\nonumber\\ &&\left.
+ y^{\prime d}_{3i} \overline{f}_3 Q_i \Phi U_T U_D \right) + {\rm H.C.} 
 ~.~\,
\label{O-Dim-6}
\end{eqnarray}
Interestingly, if we neglect the dimension-six operators in Eq.~(\ref{O-Dim-6}),
we will generate the down-type quark mixings and charged
lepton mixings via the dimension-five operators in Eq.~(\ref{O-Dim-5}). 
Thus, the quark CKM mixing matrix can be
realized via the down-type quark mixings. 
The  proton decay is not a problem since there
is no mixing between the top quark and up quark.
For example, if we assume that the Yukawa couplings
$y^d_{i3}$ and $y^d_{3i}$ are order one and the VEVs of
$U_T$ and $U_D$ are about 1 TeV, we get $M_* \sim 1000$~TeV to generate the correct CKM mixings.

However, if we introduce the above dimension-six operators
in Eq.~(\ref{O-Dim-6}),
proton decay can indeed arises due to the up-type quark mixings.
%Note that the neutrino masses and mixings can be realized
%via the right-handed neutrino mixings,
For simplicity,  we assume that $y^d_{i3}$ and $y^d_{3i}$ are 
order one, $M_* \sim 1000$~TeV, and
the other Yukawa couplings $y^{e}_{i3}$, $y^{e}_{3i}$,
$y^u_{i3}$, $y^u_{3i}$ are very small and of the the same order.
Noting that the dimension-six proton decay operators have two up
quarks, one down quark and one lepton, 
from the current proton decay constraints, we obtain that
the Yukawa couplings $y^{e}_{i3}$, $y^{e}_{3i}$,
$y^u_{i3}$, $y^u_{3i}$ are about $10^{-4}$.
Because $m_e/m_t \sim 10^{-5}$, our fine-tuning is 
 one order smaller and therefore is still acceptable. We would like to point
out that the tau lepton decays to electron and muon will
be highly suppressed due to the very small $y^{e}_{i3}$ and $y^{e}_{3i}$
in the minimal model.
 
\subsection{The Renormalizable Model}

In the renormalizable model, we assume that all the non-renormalizable
operators are suppressed by the reduced Planck scale.
Thus, we need to introduce all the particles in Table~\ref{tab:Content}.
However, there are two exceptions: (1) We do not have to introduce the $XT$ field
since the triplet Higgs field $H'_T$ can have mass around a few 
hundred GeV; (2) We do not have to introduce the vector-like particles
$(XL,~ \overline{XL})$ since the neutrino masses and mixings can arise
from the right-handed neutrino Majorana mass mixings. Then
both the tau lepton decays to electron/muon and the proton decays
to $\pi^0 e^+$ will be highly suppressed. 

The relevant renormalizable operators for the SM fermions are 
\begin{eqnarray}
-{\cal L} & = &
F_3 \overline{Xf} \Phi + N_k^c \overline{Xf} \Phi^{\dagger}+
\overline{XD} Q_i H + E_i^c XL H + \overline{Xf} XD U_T 
\nonumber\\ &&
+ \overline{f}_3 XD U_T
+ \overline{XD} Xf U_T^{\dagger} + D_i^c Xf U_T^{\dagger}
+ \overline{XL} \overline{Xf} U_D + \overline{XL}  \overline{f}_3 U_D
\nonumber\\ &&
+ Xf XL U_D^{\dagger} + Xf L_i U_D^{\dagger} 
+ \mu_{Xf3} \overline{f_3} Xf + \mu_{XDi} D_i^c XD + \mu_{XL_i} \overline{XL} L_i
\nonumber\\ &&
+ M_{Xf} \overline{Xf} Xf + M_{XD} \overline{XD} XD + M_{XL} \overline{XL} XL  + {\rm H.C.} 
~,~\,
\label{Complete-Operators}
\end{eqnarray}
where we neglect the Yukawa couplings for simplicity.
We assume that the mass terms  $M_{Xf}$, $M_{XD}$, and $M_{XL}$ are around 
1000~TeV, while the mass terms $\mu_{Xf3}$, $\mu_{XDi}$, and $\mu_{XL_i}$
are relatively small. This can be realized via rotations of the fields
since $Xf$, $XD$ and $\overline{XL}$  only couple to one linear
combinations of $\overline{Xf}/\overline{f}_3$, $\overline{XD}/D_i^c$,
$XL/L_i$, respectively. Because
the VEVs of $U_T$ and $U_D$ are around 1 TeV, the mixing terms from
$\overline{f}_3 XD U_T$, $D_i^c Xf U_T^{\dagger}$, 
$\overline{XL}  \overline{f}_3 U_D$, and $Xf L_i U_D^{\dagger}$ 
are small and negligible.

For the dimension-five operators in Eq.~(\ref{O-Dim-5}),
the $y^d_{i3}$ term can be generated from the above
renormalizable operators $D_i^c Xf U_T^{\dagger}$ and 
$F_3 \overline{Xf} \Phi$, the $y^e_{i3}$ term can be 
generated from the above renormalizable operators $E_i^c XL H$ and
$\overline{XL}  \overline{f}_3 U_D$, 
the $y^d_{3i}$ term can be generated from the above
renormalizable operators $\overline{f}_3 XD U_T$ and
 $\overline{XD} Q_i H$, and the $y^e_{3i}$ term can be 
generated from the above renormalizable operators $F_3 \overline{Xf} \Phi$ and
$Xf L_i U_D^{\dagger} $.

In addition, we can show that there are no up-type quark mixings
after we integrate out the vector-like particles. Let us explain the
point. The $SU(3)'_C\times SU(2)'_L \times U(1)'_Y$ gauge symmetry
can be formally embedded into a global $SU(5)'$ symmetry. Under
$SU(5)\times SU(5)'$, the bifundamental fields $U_T$ and $U_D$ form
$({\bf 5}, {\bf {\bar 5}})$ representation, the
vector-like particles $Xf$ and $\overline{Xf}$ respectively
form $({\bf 5}, {\bf { 1}})$ and $({\bf {\bar 5}}, {\bf 1})$ representations,
and the vector-like particles ($XD,~\overline{XL}$) and
($\overline{XD}, ~XL$)  respectively form $({\bf { 1}}, {\bf 5})$ and 
$({\bf { 1}}, {\bf {\bar 5}})$ representations. Because all these
fields are in the fundamental and/or anti-fundamental representations of $SU(5)$
and/or $SU(5)'$,  we cannot create the Yukawa interactions $10_f 10'_f 5_{H}$ or $10_f 10'_f 5_{H'}$ for the up-type quarks
after we integrate out the vector-like particles. 
Therefore, there is no proton decay problem.

%%%%%%%%%%%%%%%%%%%%%%%%%%%%%%%%%%%%%%%%%%%%%%%%%%%%%%%%%%
\section{Phenomenology and signals at the LHC}
In this section we discuss the production mechanism for the exotic gauge bosons in our model and focus on the $X_\mu$
and $Y_\mu$ vector bosons predicted in our model. These vector bosons carry both color and electroweak quantum numbers
and behave as leptoquarks as well as diquarks. As the gauge bosons have their origins in the gauge group $SU(5)$ which
unifies only the third generation, as far as its coupling to fermions is concerned, it couples only to the third generation quarks and 
leptons. However, it interacts with the gluon as well as to all the other electroweak gauge bosons of the SM which would help in 
producing these particles at collider experiments. As far as their production at hadron colliders is concerned the dominant contributions 
would come from the strongly interacting subprocesses and therefore one can neglect the sub-dominant contributions 
coming from electroweak gauge boson exchanges. Note that they will however be produced only through the exchange of electroweak 
gauge bosons at electron positron colliders such as the {\it International Linear Collider} (ILC) \cite{Djouadi:2007ik} or the CLIC 
\cite{Linssen:2012hp}, envisioned and proposed for the future. We restrict ourselves to the study of these gauge boson at the currently 
operational LHC at CERN and therefore only focus on the couplings of the $X_\mu$ and 
$Y_\mu$ vector bosons  with the gluons which would be relevant for its production at the LHC.  
The general form of the interaction 
can be derived from the Lagrangian given by \cite{blumlein:belyaev}
\begin{align}
{\mathcal L} = -\frac{1}{2}{\mathcal V}^{i\dagger}_{\mu\nu}{\mathcal V}_i^{\mu\nu} + M_V^2 V_\mu^{i\dagger} V^\mu_i
 -i g_s V_\mu^{i\dagger} {T^a_{ij}} V^j_\nu {\mathcal G}_a^{\mu\nu}
\label{eq:lagrangian}
\end{align} 
where $V\equiv X,Y$ and $T^a$ are the $SU(3)_c$ generators. The field strength tensors for the exotic vector fields $V_\mu$ and 
gluon $G_\mu^a$ are
\begin{align}
{\mathcal G}_a^{\mu\nu} &=\partial_\mu G_\nu^a - \partial_\nu G_\mu^a + g_s f^{abc} G_{\mu b} G_{\nu c}  \\ 
{\mathcal V}_i^{\mu\nu}  &= D_\mu^{ik} V_{\nu k} - D_\nu^{ik} V_{\mu k}
\end{align}
and the covariant derivative is defined as
\begin{align}
D_\mu^{ij} = \partial_\mu \delta^{ij} - i g_s T^{ij}_a G^a_\mu.
\end{align}
\begin{center}
\begin{figure}[!t]
\includegraphics[width=6.0in,height=1.5in]{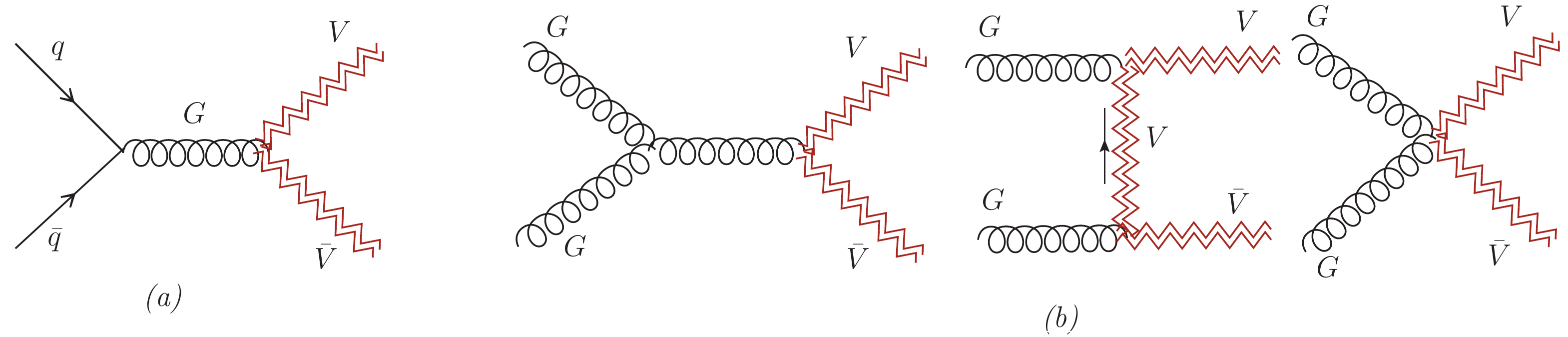}
\caption{The tree level Feynman diagrams which contribute to the pair production of the $X_\mu$ and $Y_\mu$ gauge bosons 
at the LHC, where both of them are denoted as $V_\mu$. The subprocesses that contribute are (a) $q\bar{q} \to V\bar{V}$ 
and (b) $G G \to  V\bar{V}$.  }
\label{fig:diagrams}
\end{figure}
\end{center}
Using the above Lagrangian we derive the Feynman rules for the interactions of the leptoquark gauge bosons $V\equiv X,Y$ with 
the gluon fields. These interactions then lead to the tree level Feynman diagrams as shown in 
Fig.(\ref{fig:diagrams}) which contribute to the pair production of these exotic particles at the LHC. 
\subsection{Calculation of cross sections}
\begin{figure}[t!]
\centering
\includegraphics[width=3.5in]{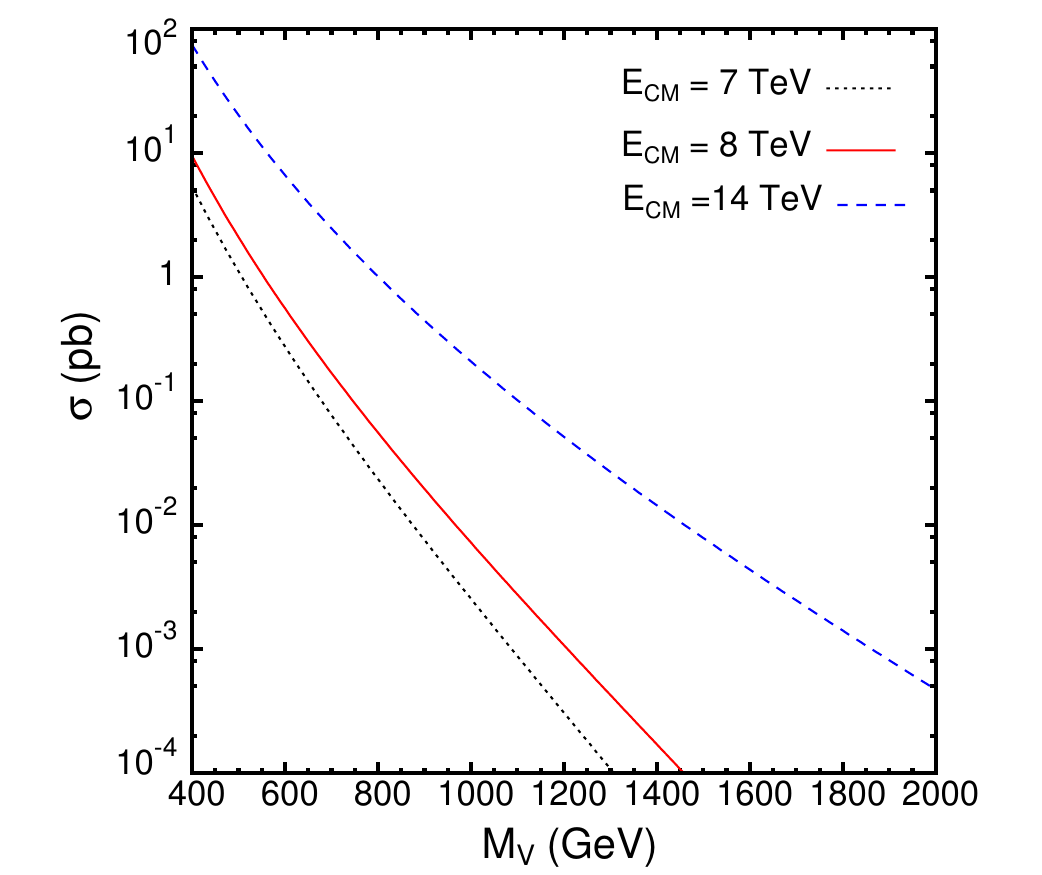}
\caption{\it The production cross sections for 
$p p \rightarrow V \bar{V}$ at the LHC as a function of 
leptoquark mass $M_V$ at center-of-mass energies, 
$E_{CM} = 7,8$ and $14$ TeV. We have chosen the scale as 
$Q=M_V$, the mass of the leptoquark.}
\label{prodfig}
\end{figure}

Using Feynman rules for the interaction vertices of the exotic gauge bosons with gluons derived 
from Eq.(\ref{eq:lagrangian}) we can write down the full spin and color averaged matrix amplitude square for the 
quark-antiquark annihilation subprocess  $q\bar{q} \to V \bar{V}$, (where $q\equiv u,d,c,s,b$ and $V\equiv X,Y$) as 
%\scriptsize
\begin{eqnarray*}
\overline{|{\mathcal M}|}^2_{q\bar{q}} = &&
\frac{g_s^4}{9 M_V^4 s^2} \left[-12 M_V^8-s^2 t (s+t)+4 M_V^6 (s+6 t)+2 M_V^2 s \left(2 s^2+3 s t+2 t^2\right) \right. \\ 
&& \left. -M_V^4 \left(17 s^2+20 s t+12 t^2\right)\right]
\end{eqnarray*}
while for the gluon induced  subprocess $GG \to V \bar{V}$, it is given by 
\begin{eqnarray*}
\overline{|{\mathcal M}|}^2_{GG} = && g_s^4
\left[\frac{9 M_V^4+4 s^2+9 s t+9 t^2-9 M_V^2 (s+2 t)}{24 s^2 \left(t-M_V^2\right)^2 \left(s+t-M_V^2\right)^2}\right]   \left[3 M_V^8+2 s^4-12 M_V^6 t+4 s^3 t  \right. \\
&& \left. +7 s^2 t^2+6 s t^3+3 t^4+M_V^4 \left(7 s^2+6 s t+18 t^2\right)-4 M_V^2 \left(s^3+2 s^2 t+3 s t^2+3 t^3\right)\right].
\end{eqnarray*}
%%%
Note that the Mandelstam variables $s$ and $t$ are defined in the parton frame of reference.  The
pair production cross section at the parton level is then easily obtained using the above 
expressions. To obtain the production cross section we convolute the parton level cross sections
$\hat{\sigma}(q_i\bar{q}_i\rightarrow V \bar{V})$ and $\hat{\sigma}(GG\rightarrow V \bar{V})$
 with the parton distribution functions (PDF).
%%%
\begin{equation}\label{prodcros}
\begin{split}
\sigma({pp \rightarrow V \bar{V}})=& 
            \left\{ \sum_{i=1}^{5}  \int dx_1 \int dx_2 
        ~\mathcal{F}_{q_i}(x_1,Q^2)\times \mathcal{F}_{\bar{q}_i}(x_2,Q^2) 
         \times \hat{\sigma}(q_i\bar{q}_i\rightarrow V \bar{V}) \right\} \\
  & + \int dx_1 \int dx_2 ~\mathcal{F}_g (x_1,Q^2) \times 
                           \mathcal{F}_g (x_2,Q^2) 
                  \times \hat{\sigma}(GG\rightarrow V \bar{V}), 
\end{split}
\end{equation}
%%%
where $\mathcal{F}_{q_i}$, $\mathcal{F}_{\bar{q}_i}$ and $\mathcal{F}_{g}$  
represent the respective PDF's for partons (quark, antiquark and gluons) in 
the colliding protons, while $Q$ is the factorization scale.  
In Fig.(\ref{prodfig}) we plot the leading-order production cross section 
for the process $p p \rightarrow V \bar{V}$  at center of mass energies 
of 7, 8 and 14 TeV  as a function of the leptoquark mass $M_V$.
%%%%
%%%%%
We set the factorization scale $Q$ equal to $M_V$,  
and have used the {\tt CTEQ6L1} PDF  \cite{Pumplin:2002vw}. As seen from the plot, we find that 
the pair production cross section for both the $X$ and $Y$ leptoquark 
gauge bosons are quite big for significantly large values of their mass 
even at the 7 and 8 TeV runs of LHC. Thus one expects severe bounds on such
particle masses from experimental data. In an earlier work \cite{Chakdar:2012kd}, we 
had studied specific signals from the pair production of $X_\mu$ at LHC and put 
expected limits on its mass. This work was also followed up by the CMS experimental group
which placed comparable limits on such leptoquark vector bosons \cite{Chatrchyan:2012sv}
using collision data from the 7 TeV run of the LHC. We note that as both the $X$ and $Y$
leptoquark gauge bosons have identical masses, any limits on one of them invariably leads
to a similar limit on the other. Thus it is important to explore all possible signals that 
come from the pair productions of these particles. In this work we extend our earlier 
study by looking at the different signals from the pair productions of such particles 
at LHC with center-of-mass energies of 8 TeV and 14 TeV. We note that at the 14 TeV run of LHC 
the production cross section for the leptoquark gauge bosons is significantly enhanced and 
would therefore improve the reach for such particle searches.
\subsection{Calculation of decays of the $X_\mu$ and $Y_\mu$ gauge bosons}
\begin{figure}[ht!]
\centering
\includegraphics[width=3.5in]{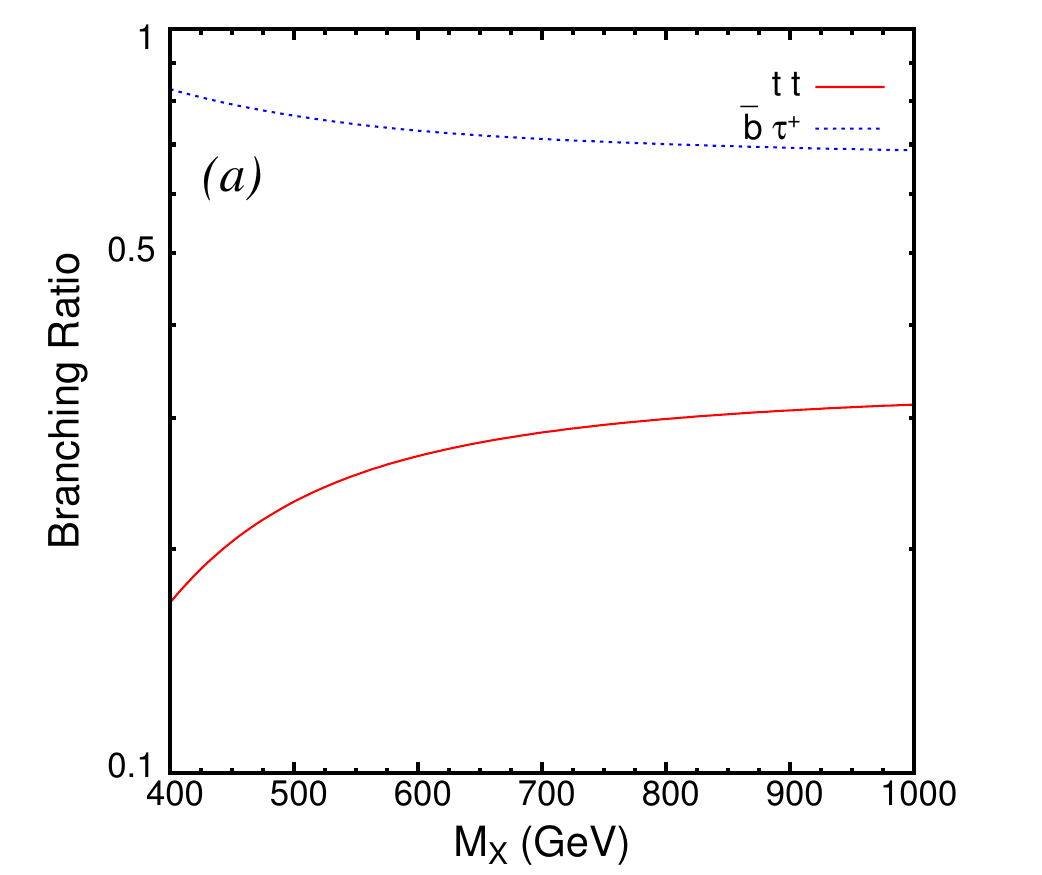}
\includegraphics[width=3.5in]{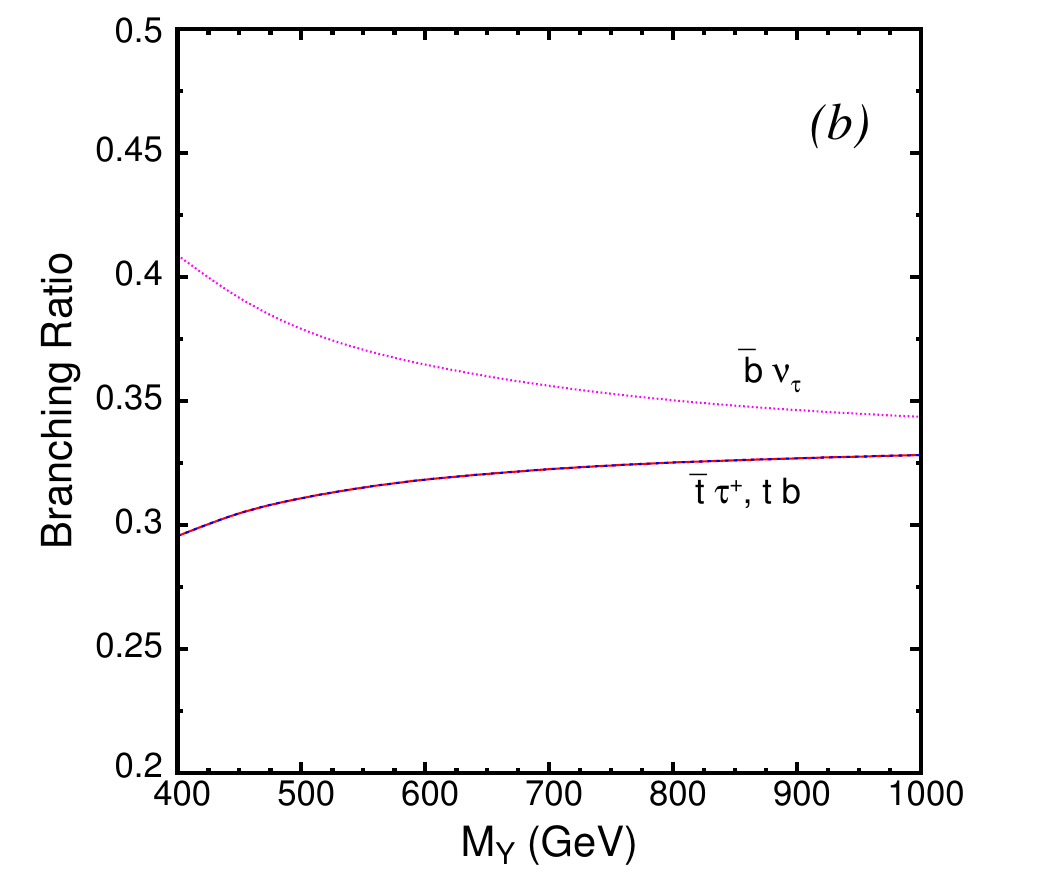}
\caption{\it Illustrating the decay branching fractions of the leptoquark gauge bosons (a) $X_\mu$ 
and (b) $Y_\mu$ as a function of their mass.}
\label{decayfig}
\end{figure}

To study the possible signals for the leptoquark gauge bosons, we need to know their decay 
properties. Since the third family of fermions is only charged under the gauge group $SU(5)$,
these leptoquark gauge bosons which come from the $SU(5)$ gauge fields are only coupled
to the third generation fermion fields. The interaction Lagrangian of the leptoquark gauge bosons
$X_\mu$ and $Y_\mu$ with the third generation fermions is given by \cite{Langacker:1980js},
\begin{align}
\mathcal{L}_G = & \frac{g_5}{\sqrt{2}} \bar{X}_\mu^\alpha \big[ \bar{b}_{R\alpha}\gamma^\mu \tau_R^+ +
\bar{b}_{L\alpha}\gamma^\mu \tau_L^+ + \epsilon^{\beta\gamma}_{\alpha} \bar{t}^c_{L\gamma} 
\gamma^\mu t_{L\beta} \big]  \nonumber \\ 
+ & \frac{g_5}{\sqrt{2}} \bar{Y}_\mu^\alpha \big[- \bar{b}_{R\alpha}\gamma^\mu \nu_R^c -
\bar{t}_{L\alpha}\gamma^\mu \tau_L^+ + \epsilon^{\beta\gamma}_{\alpha} \bar{t}^c_{L\gamma} 
\gamma^\mu b_{L\beta} \big] + H.C.
\label{eq:decaylag}
\end{align}
Using the above interaction Lagrangian, we can calculate the explicit decay modes of the leptoquark gauge bosons,
where $X_\mu$ decays to a top quark pair ($tt$) or anti-bottom quark + positively charged tau lepton ($\bar{b}\tau^+$) while $Y_\mu$ has three decay modes to anti-bottom quark + a tau-neutrino ($\bar{b}\nu_\tau$), anti-top quark + positively charged tau ($\bar{t}\tau^+$) or top quark + bottom quark ($tb$).
The partial decay width for each mode calculated using Eq.(\ref{eq:decaylag}) is then given by
\begin{align}
 \Gamma (X \to t t~~~) &= \frac{g_5^2 M_X}{24\pi} \left(1-\frac{m_t^2}{M_X^2} \right) 
                                          \left(1 -\frac{4m_t^2}{M_X^2} \right)^{1/2} \nonumber \\
 \Gamma (X \rightarrow \bar{b} \tau^+) &= \frac{g_5^2 M_X}{12\pi}   \\ \nonumber \\
%%%
 \Gamma (Y \to \bar{t} \tau^+) &= \frac{g_5^2 M_Y}{24\pi} \left(1-\frac{m_t^2}{M_Y^2}\right)^2 
                                                       \left(2 +\frac{m_t^2}{M_Y^2} \right) \nonumber \\
 \Gamma (Y \to \bar{b} \nu_\tau) &=\frac{g_5^2 M_Y}{12\pi}  \nonumber \\
 \Gamma (Y \to t b~) &=  \frac{g_5^2 M_Y}{24\pi} \left(1-\frac{m_t^2}{M_Y^2} \right)^2 
                                                      \left (2 +\frac{m_t^2}{M_Y^2} \right)
\end{align}
where $g_5$ is the $SU(5)$ gauge coupling and we have only kept the top quark mass ($m_t$) and neglected the other fermion masses.
We plot the branching fractions of the leptoquark gauge bosons as shown in Fig.(\ref{decayfig}). It is interesting to note that while the 
$X_\mu$ decays dominantly to $\bar{b}\tau^+$, it also has a substantial branching fraction to a pair of 
same sign top quarks. For very large values of the mass $M_X$ of the $X_\mu$, when the mass of the top
quark can be neglected, we find that $\frac{\Gamma(X\to \bar{b}\tau^+)}{\Gamma(X\to tt)} \simeq 2$. 
For the $Y_\mu$ leptoquark gauge boson we find that for smaller values of its mass it has the dominant 
decay fraction to $\bar{b}\nu_\tau$ while its decay to $\bar{t}\tau^+$ and $tb$ are equal. But for $M_Y$
quite large such that the top quark mass may be neglected, all $Y_\mu$ decay modes have the same branching
probability of 1/3.
%%%%
%%%%%
With the knowledge of the decay modes of the leptoquark gauge bosons and the branching fractions for the
decays we can now analyze all the different final states that we expect from the pair production of these 
leptoquarks at the LHC.

\subsection{Signals at the LHC}
In Ref.\cite{Chakdar:2012kd} we studied the signals for the pair production of the $X_\mu$ leptoquark gauge bosons
and their subsequent decays into the dominant mode $\bar{b}\tau^+$ at LHC center of mass energies of 7 TeV and 8 TeV. 
The final state signal was $b\bar{b}\tau^+\tau^-$ with all the four particles being detected in the respective flavor tagged
mode. It was observed that the signal stands out as resonances in the invariant mass distribution of the $\tau$ lepton paired 
with the $b$ jets against the continuum SM background, provided all the four final state particles carried significant transverse 
momenta. Using this signal a phenomenological prediction on the LHC reach was made on the mass of the $X_\mu$ which 
has subsequently been estimated as 760 GeV at 95 \% C.L. by the CMS Collaboration \cite{Chatrchyan:2012sv} at LHC with 
7 TeV center of mass energy. As our model predicts another decay mode (to top quark pairs) for the $X_\mu$ gauge boson, 
where the $X_\mu$ behaves as a {\it diquark}, carrying quantum numbers of two quarks, it is of extreme importance to 
be able to highlight this characteristic which distinguishes this particle from the usual leptoquark particles. Establishing the existence of both decay modes  is needed to show that these interactions are both baryon and lepton number violating. It is also worth 
pointing out that a similarly massive $Y_\mu$ in the spectrum which couples as strongly to the gluons as the $X_\mu$ will
also be produced with similar rates and needs to be studied in tandem with the production of the $X_\mu$ particles at the 
LHC.

We now consider all decay modes of both the $X_\mu$ and $Y_\mu$ and discuss final states which is then studied against the 
SM backgrounds. For the pair production process of $X_\mu$ gauge bosons,
%$$ p p \to X \bar{X}$$
where $X \to \bar{b}\tau^+,tt$ we have the following different final states given as
\begin{align}
p p \longrightarrow X \bar{X} \longrightarrow  \bar{b}\tau^+ b \tau^- ,~ t t b \tau^- ,~ \bar{b}\tau^+ \bar{t} \bar{t},
                                               ~  t t \bar{t} \bar{t}.  
\label{eq:xdecay}
\end{align}
The top quark would further decay, either semileptonically or hadronically to give multi-lepton and high jet multiplicity final states.
For our purposes, if we assume that the top quarks could be reconstructed with some reasonable efficiency in either modes, we can
just focus on the above mentioned final state signal.  Similarly for the pair production of the  $Y_\mu$ gauge bosons,
$$ p p \to Y \bar{Y} $$ 
where $Y \to \bar{b} \nu_\tau, \bar{t}\tau^+, t b$ we get the following set of final states given as
\begin{align}
p p \longrightarrow Y \bar{Y} \longrightarrow & ~ b \bar{b}{E}_T, ~\bar{b} t \tau^- {E}_T, ~\bar{t} b \tau^+{E}_T, ~\bar{b}\bar{b}\bar{t}{E}_T,~ b b t {E}_T,   
                                             \longrightarrow  & ~ t \bar{t}\tau^+\tau^-,~\bar{t}\bar{t}\bar{b}\tau^+,~ t t b \tau^-, ~t \bar{t} b \bar{b}.
\label{eq:ydecay}
\end{align}

Note that both the $X_\mu$ and $Y_\mu$ gauge boson productions at the LHC leads to a rich range of diverse final states which lead to 
many multi-particle signals and would lead to distinct resonances in the invariant mass distributions in some pairs corresponding to 
the mass of the $X_\mu$ and $Y_\mu$ states. Notably we find that each particular event rate is fixed once the model parameters have been
fixed, which in our case is the mass of the leptoquark gauge bosons while its coupling strength to the gluons 
has been fixed to be the 
strong coupling constant. Thus the success of the model is not dependent on an observation in only one 
particular final state but that observation needs to be complemented simultaneously in various other channels as listed above in Eqs.(\ref{eq:xdecay}) and (\ref{eq:ydecay}). Thus the study on all simultaneous channels 
deserves merit as it will be able to confirm or falsify the model in question.   

We now consider different final states and analyze the signals against the SM background. As we expect that the new gauge 
bosons when produced on-shell will decay to specific final state products, this would lead to a bump in the invariant mass distribution of 
the decay products. Keeping this in mind, it is instructive to first consider the most likely signals where the resonances would be observable.
Based on the decay channels and final states listed in Eqs.(\ref{eq:xdecay})--(\ref{eq:ydecay}) one 
should consider the ($b\tau$) mode for the
$X_\mu$ gauge bosons while the ($tb$ and $t\tau$) mode looks the more promising for the $Y_\mu$ resonance searches. The other modes either involve neutrinos or more than a single top
quark in the final state, which further decays either semileptonically or hadronically and therefore makes 
it more tasking to reconstruct the leptoquark gauge boson mass. 
However, we must emphasize that for measuring the electric charge of these gauge bosons one definitely requires that the $X_\mu$ resonance is observed in the invariant mass distribution of 
same-sign top quark pair ($tt$) while the $Y_\mu$ resonance is observed in the ($t\tau^-$) 
final state or its charge conjugate mode. Notwithstanding the fact that reconstructing the $tt$
state would be challenging, it would definitely lead to a very interesting observation. Final states
involving $b$ jets require measuring the $b$ jet charge which looks to be more difficult and 
hence not a desired mode to get information on the charge of the exotic gauge bosons. 

%%%
\begin{table}[!h]
\begin{center}
\begin{tabular}{|c|c|c|c|}
\hline ${\mathcal Signal}$& ${\mathcal SM}$ & ${\mathcal Signal}$& ${\mathcal SM}$  \\ 
\hline  $2b\tau^+\tau^-$ & $2b\tau^+\tau^-; ~ 2j\tau^+\tau^-;~ b j\tau^+\tau^-$  
       &  $ttb\tau^-,~\bar{t}\bar{t}b\tau^+$ & -- \\ 
\hline  $tt\bar{t}\bar{t}$ & $tt\bar{t}\bar{t}$ 
       &  $t\bar{t}\tau^+\tau^-$ & $t\bar{t}\tau^+\tau^-$ \\
\hline  $2b t\bar{t}$ & $2b t\bar{t}; ~ 2j t\bar{t}; ~ b j t\bar{t}$ 
       &  $2b {E}_T$ & $2b {E}_T; ~ 2j {E}_T; ~ j b {E}_T$ \\
\hline $b t \tau^- {E}_T$ & $b t \tau^- {E}_T; ~ j t \tau^- {E}_T$ 
       & $b \bar{t} \tau^+{E}_T$ & $b \bar{t} \tau^+ {E}_T;~ j\bar{t} \tau^+{E}_T$ \\
\hline  $2b t {E}_T$ & $ 2j t{E}_T; ~ b j t {E}_T$   
       &  $2b\bar{t}{E}_T$ & $2j\bar{t}{E}_T; ~bj\bar{t}{E}_T$ \\
\hline
\end{tabular}
\caption{Illustrating the final state signals and the corresponding SM background 
subprocesses. Note that ${E}_T$ for the SM subprocesses represents one or more neutrinos 
in the final state.}
\label{tab:sigbkg}
\end{center}
\end{table}
%%%%%%
In Table \ref{tab:sigbkg} we list the relevant SM background subprocesses  that 
we have considered for each set of final states for the signal. Note that we do not make a distinction
between the $b$ and $\bar{b}$ but we distinguish between a $\tau^+$ and $\tau^-$ by assuming exact 
charge measurement will be possible. We also distinguish between a top quark and anti-top quark assuming
that they will be reconstructed with their respective charge identifications from its semileptonic decay modes.
We associate an efficiency factor of $\varepsilon_t$ with this reconstruction. For final state signals not involving neutrinos we have not considered SM subprocesses with ${E}_T$ as they will involve  
extra electroweak vertices which suppress the contributions and further requirements on missing 
transverse momenta would make these contributions too small to take into further consideration.  
We highlight the above mentioned invariant mass distributions in our model for a few choices 
of the $X_\mu$ and $Y_\mu$ gauge boson masses considered at two different center of mass energies for the LHC. We focus our attention to the recently concluded 8 TeV run and the
proposed upgrade in energy of 14 TeV for the LHC. As a current limit of 760 GeV exists on the
leptoquark gauge boson mass from the CMS analysis \cite{Chatrchyan:2012sv} we choose 
a mass of 800 GeV to show the distributions at the 8 TeV run of LHC while a larger mass of 
1 TeV is chosen to highlight the signal distributions at the 14 TeV run. We note that there are 
more than one set of final states where a particular resonance could be observed in the invariant
mass distributions and so we consider the scenario where we look at a few definite invariant mass
distributions in individual final state modes listed in Eqs.(\ref{eq:xdecay}) and (\ref{eq:ydecay}). 
We list below the pair of final
state particles for which the invariant mass distribution is considered, motivated by favored modes for 
reconstructing the mass and the charge of the $X_\mu$ and $Y_\mu$ gauge bosons.

\begin{itemize}
\item [{\bf (C1)}] Invariant mass distribution of $b\tau^-$ coming from the final states $\bar{b}\tau^+ b \tau^- ,~ t t b \tau^-$. This is the most favorable mode for reconstructing the $X_\mu$ resonance.
\item [{\bf (C2)}] Invariant mass distribution of same sign top quark pair $tt$ coming from the final states
$ttb\tau^-,~tt\bar{t}\bar{t}$. The reconstruction of the leptoquark mass in this mode, although difficult, is
essential in measuring the charge of the $X_\mu$. 
\item [{\bf (C3)}] Invariant mass distribution of $tb$ coming from the final states $t\bar{t}b\bar{b},~ttb\tau^-,~t\bar{b}\tau^-{E}_T,~bbt{E}_T$. This is one of the 
favorable modes for reconstructing the $Y_\mu$ resonance.
\item [{\bf (C4)}] Invariant mass distribution of $t\tau^-$ coming from the final states $t\bar{t}\tau^-\tau^+,~tt b\tau^-,~\bar{b}t\tau^-{E}_T$. This mode is essential to measure the charge of the $Y_\mu$. 
Note that the $t\tau^-$ resonance corresponds to the charge conjugate mode of $Y_\mu$.   
\end{itemize}

We shall now discuss the signal and the associated SM backgrounds for the 
list of resonances given by {\bf C1--C4}. Note that the signal subprocesses which contribute 
to give a $b\tau^-$ final state as listed in {\bf (C1)} come from both $X_\mu$ and $Y_\mu$
pair productions. However the resonant distribution only happens for the $X_\mu$ production
while the $Y_\mu$ contribution acts to smear out the resonance although it does contribute 
in enhancing the signal over the SM background. A further smearing effect would come if the 
$t\bar{b}\tau^-{E}_T$ signal is included. But we can reject that contribution by demanding 
that we don't include events with large missing transverse momenta in the final state when 
reconstructing the $b\tau^-$ invariant mass. As discussed in Ref.\cite{Chakdar:2012kd} 
the dominant background for the resonant signal in the $b\tau$ channel comes from 
$pp\to 2b2\tau,4b,2j2b,2j2\tau,4j,t\bar{t}$ where $j=u,d,s,c$ when we consider the 
signal coming from the pair production of $X_\mu$ which then decay in the $b\tau$ mode 
to give a $2b\tau^+\tau^-$ final state.  The light jet final 
states in the SM can be mistaged as $\tau$ or $b$ jets and thus form a significant source for 
the background due to the large cross sections at LHC, as they are dominantly produced through 
strong interactions. Guided by previous analysis \cite{Chakdar:2012kd}, we 
note that a very strong requirement on the transverse momenta for the $b$ jet and the $\tau$
lepton is very helpful in suppressing the SM background. The SM background has been 
estimated using {\tt Madgraph 5} \cite{mad5}. In this analysis we further restrict the number of 
SM background sub-processes that contribute to the final state with $b\tau^-$ by 
demanding that the tau charge is measured. Therefore we neglect the 
contributions coming from jets that fake a tau. For example, when we consider 
the final state as $2b\tau^+\tau^-$ and demand that the tau lepton is tagged as 
well as its charge measured, we include $pp\to 2b2\tau,2j2\tau,t\bar{t}$ as the 
dominant SM processes for the background.

We have used two values for the leptoquark gauge boson ($V\equiv X,Y$) masses, $M_V=800$ 
GeV at LHC with center of mass energy 8 TeV and $M_V=1$ TeV at LHC with center of mass 
energy 14 TeV to highlight the signal cross sections and differential distributions for 
invariant mass. We set the factorization and renormalization scale ($Q=M_Z$) to the mass of the $Z$ boson and
also use the strong coupling constant value of $\alpha_s$ evaluated at the $Z$ boson mass.  
Note that we have evaluated the individual signals as listed in Eqs.(\ref{eq:xdecay}) and (\ref{eq:ydecay}) 
against their specific backgrounds independently. 
We have assumed in our analysis that the top quark and the anti-top quark are reconstructed
with good efficiencies which we can parameterize as $\varepsilon_t$. Note that we have used the 
following efficiencies for $b$ and $\tau$ tagging, $\epsilon_b=\epsilon_\tau=0.5$ while we assume a mistag rate 
for light jets to be tagged as $b$ jets as 1\% and $c$ jets tagged as $b$ jets to be 10\%. All our results here
are done at the parton level and therefore to account for the detector resolutions for energy measurement 
of particles, we have used a Gaussian smearing of the jet and $\tau$ energies with an energy resolution given 
by $\Delta E/E = 0.8/\sqrt{E~(GeV)}$ and $\Delta E/E = 0.15/\sqrt{E~(GeV)}$ respectively when analyzing 
the signal events. 

In Table \ref{tab:cuts} we list the kinematic selection cuts on the events. As the primary decay modes 
of the heavy leptoquark gauge bosons will have very large transverse momenta we put strong cuts on them. 
This helps in suppressing the SM background while it does not have any significant effect on the signal 
events. The cuts on ${E}_T$ applies only to final
states with neutrinos in the decay chain while the $\Delta R_{ij}$ cut is on any pair of visible particles. 
The invariant mass cut $M_{jj}$ is on any pair of jets in the final state.
 %%%
\begin{table}[!t]
\begin{center}
\begin{tabular}{|c|c|c|}
\hline {\bf Variable} &Cut ${\mathcal C}_1$ at 8 TeV & Cut ${\mathcal C}_2$ at 14 TeV \\ 
\hline $p_T^{\tau,b,j}$ & $>80$ GeV & $>200$ GeV \\ 
\hline ${E}_T$ & $>100$ GeV & $>200$ GeV \\ 
\hline $|\eta|$ & $<2.5$  & $<2.5$ \\ 
\hline $\Delta R_{ij}$ & $>0.4$ & $>0.4$ \\
\hline $M_{jj,\tau^+\tau^-}$ & $>5$ GeV & $>5$ GeV \\
\hline
\end{tabular}
\caption{\it Two different set of cuts, ${\mathcal C}_1$ at LHC with $\sqrt{s}=8$ TeV and 
${\mathcal C}_2$ at LHC with $\sqrt{s}=14$ TeV, imposed on the final states listed in 
Eqs.(\ref{eq:xdecay})--(\ref{eq:ydecay}) where the cuts on ${E}_T$ applies only to final
states with neutrinos in the decay chain.} \label{tab:cuts}
\end{center}
\end{table}
%%%%%%

\begin{figure}[ht!]
\centering
\includegraphics[height=2.2in]{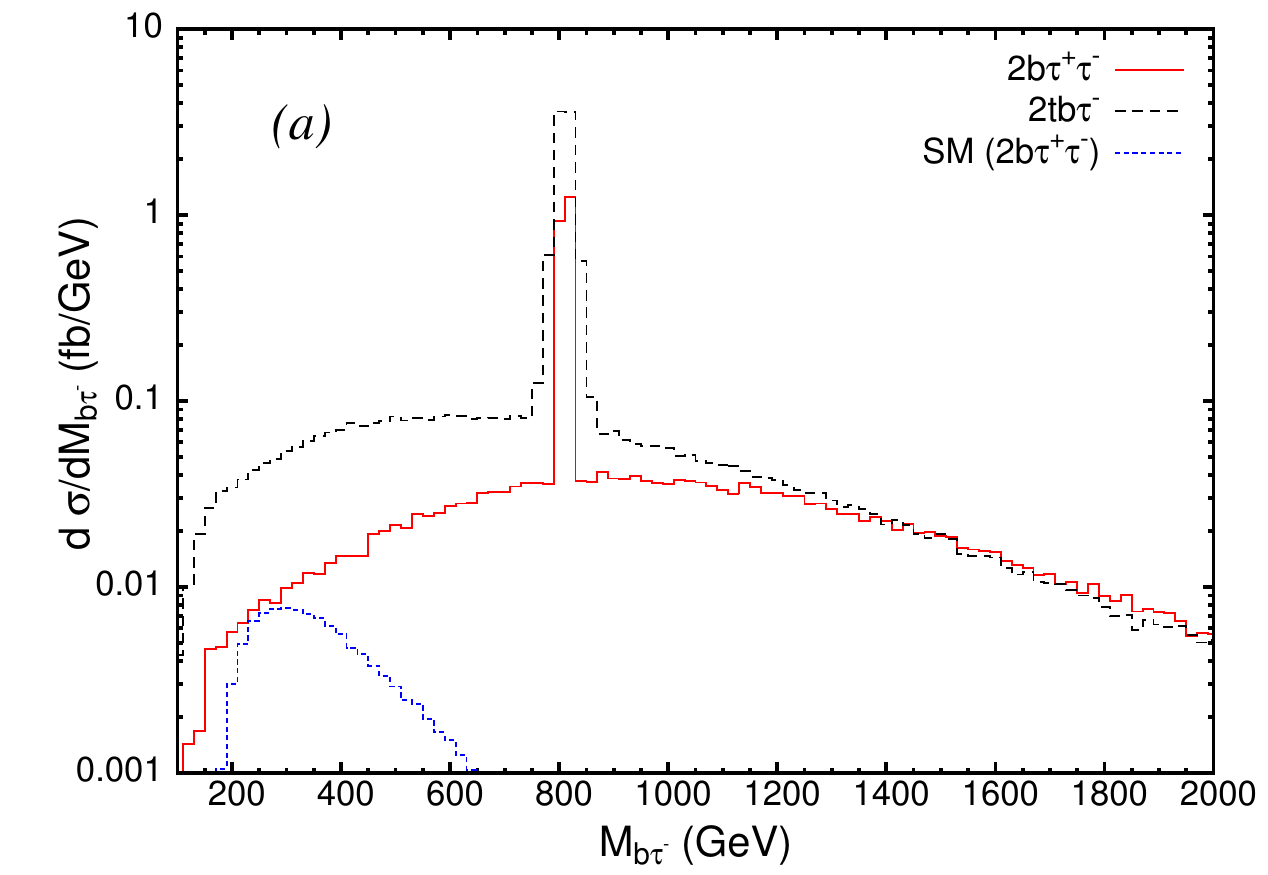}
\includegraphics[height=2.2in]{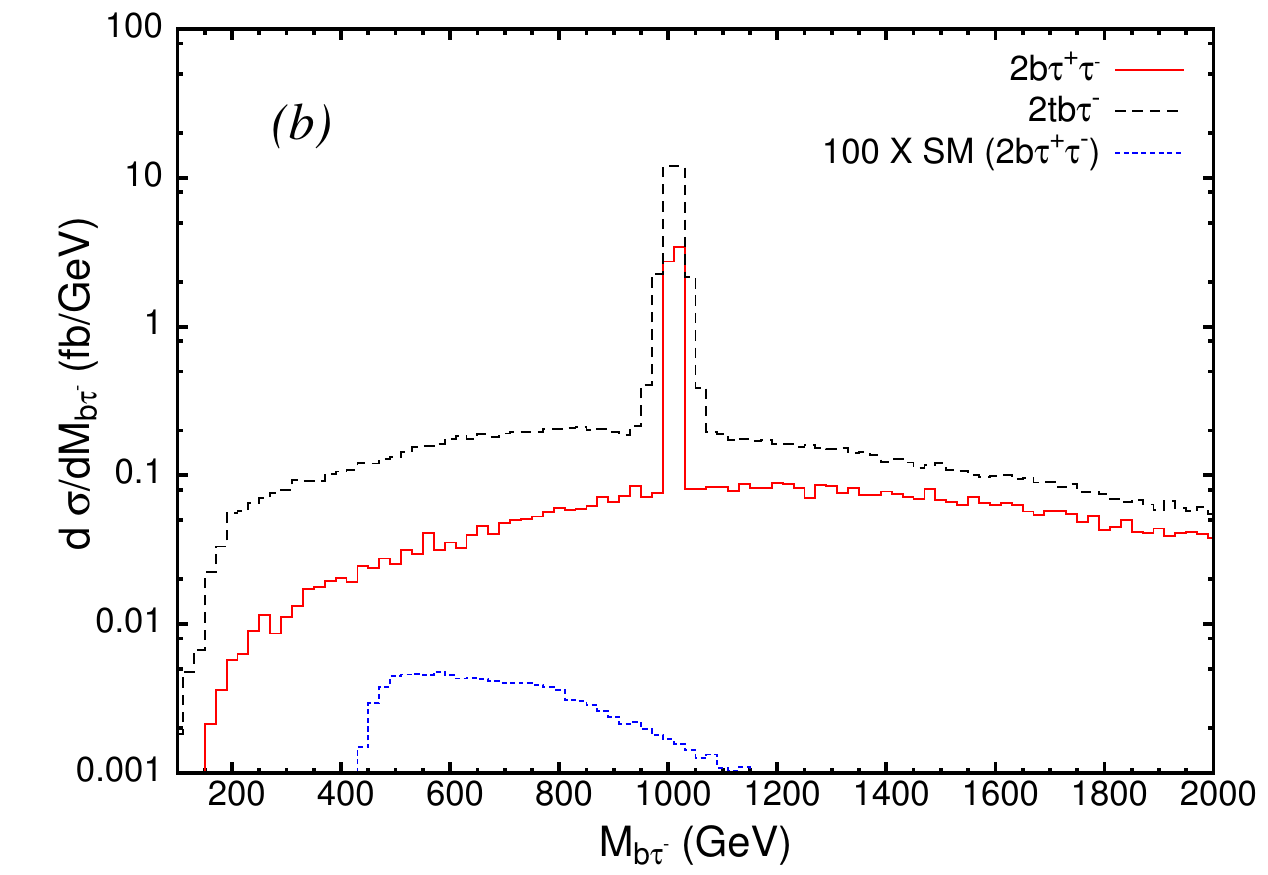}
\caption{\it Invariant mass distribution of $b\tau^-$ for the signal and SM background for two 
different choices of leptoquark gauge boson mass, (a) $M_V = 800$ GeV considered at LHC with 
$\sqrt{s}=8$ TeV and (b) $M_V = 1$ TeV considered at LHC with $\sqrt{s}=14$ TeV.}
\label{fig:btau}
\end{figure}

With the above set of kinematic selection on the final state events we evaluate the signal
cross sections and the corresponding SM background given in Table \ref{tab:sigbkg}. We 
first consider the resonance given by {\bf (C1)} and show the invariant mass distribution of $b\tau^-$ 
in Fig. \ref{fig:btau}. We must point out here that the $\tau^-$ is paired with the $b$ jet which has the 
leading transverse momenta in case there exist more than one tagged $b$ jets. 
After including the efficiency factors $\epsilon_b$ and $\epsilon_\tau$ associated with tagging the 
$b$ and $\tau$ jets and mistag rates, we estimate the signal cross section in the 
$2b\tau^+\tau^-$ mode as $4.23 ~fb$ for $M_{X,Y}=800$ GeV at LHC with $\sqrt{s}=8$ TeV and 
$12.05~fb$ 
%%%
%%%
for $M_{X,Y}=1$ TeV at LHC with $\sqrt{s}=14$ TeV. In Fig.\ref{fig:btau} we plot the invariant mass 
distribution for the signal. The dominant SM background are given by the following subprocesses, 
$\sigma (2b\tau^+\tau^-) \simeq 1.8 ~fb, ~\sigma (2c\tau^+\tau^-) \simeq 1.6 ~fb$ and 
$\sigma (2j\tau^+\tau^-) \simeq 167.6 ~fb$ which after including the efficiency factors, mistag rates 
is added to give $0.119~fb$. This is plotted in Fig.(\ref{fig:btau}) as ``SM $(2b\tau^+\tau^-)$". The 
corresponding SM background at 14 TeV center of mass energy is much more suppressed 
($\sim 0.002 ~fb$) because of the strong requirement on the transverse momenta of the jets and the 
charged tau leptons. The signal is clearly seen to stand out as resonance and one therefore expects 
this particular mode to be very  favorable in searching for the $X_\mu$ resonance by suppressing 
the SM background by demanding $\tau$ lepton charge identification which gets rid of the large all 
jet background.  Another mode for the $b\tau^-$ resonance which has completely negligible SM 
background, is for the final state $ttb\tau^-$. There are two different sources for the signal in this 
case, one which corresponds to the final states coming from the $X\bar{X}$ pair production while 
the other from the $Y\bar{Y}$ pair production. As the $Y\bar{Y}$ contribution does not lead to a 
resonance in the $b\tau^-$ mode, it will act to smear out the resonance as compared to that seen 
for the $2b\tau^+\tau^-$ final state. This is evident in Fig.(\ref{fig:btau}) where the width of the 
resonance is seen to spread out in more invariant mass bins for the $ttb\tau^-$  final state.  
Assuming a top reconstruction with an efficiency of $\varepsilon_t$ we find that the signal cross 
section from $X\bar{X}$ for $M_X=800~(1000)$ GeV at LHC with $\sqrt{s}=8~(14)$ TeV is 
$8.19 ~(28.23) \times \varepsilon_t^2~fb$ while the signal cross section from $Y\bar{Y}$ for 
$M_Y=800~(1000)$ GeV at LHC with $\sqrt{s}=8~(14)$ TeV is 
$4.04 ~(13.8) \times\varepsilon_t^2~fb$. Note that the $\tau$ and $b$ tagging 
efficiencies have been already included.
In Fig.(\ref{fig:btau}) we have assumed $\varepsilon_t=1$ for illustration purposes. Therefore the 
efficacy of the signal with the same sign top pairs in the final state is dependent on the inherent purity 
of the top reconstruction at experiments.

\begin{figure}[ht!]
\centering
\includegraphics[height=2.2in]{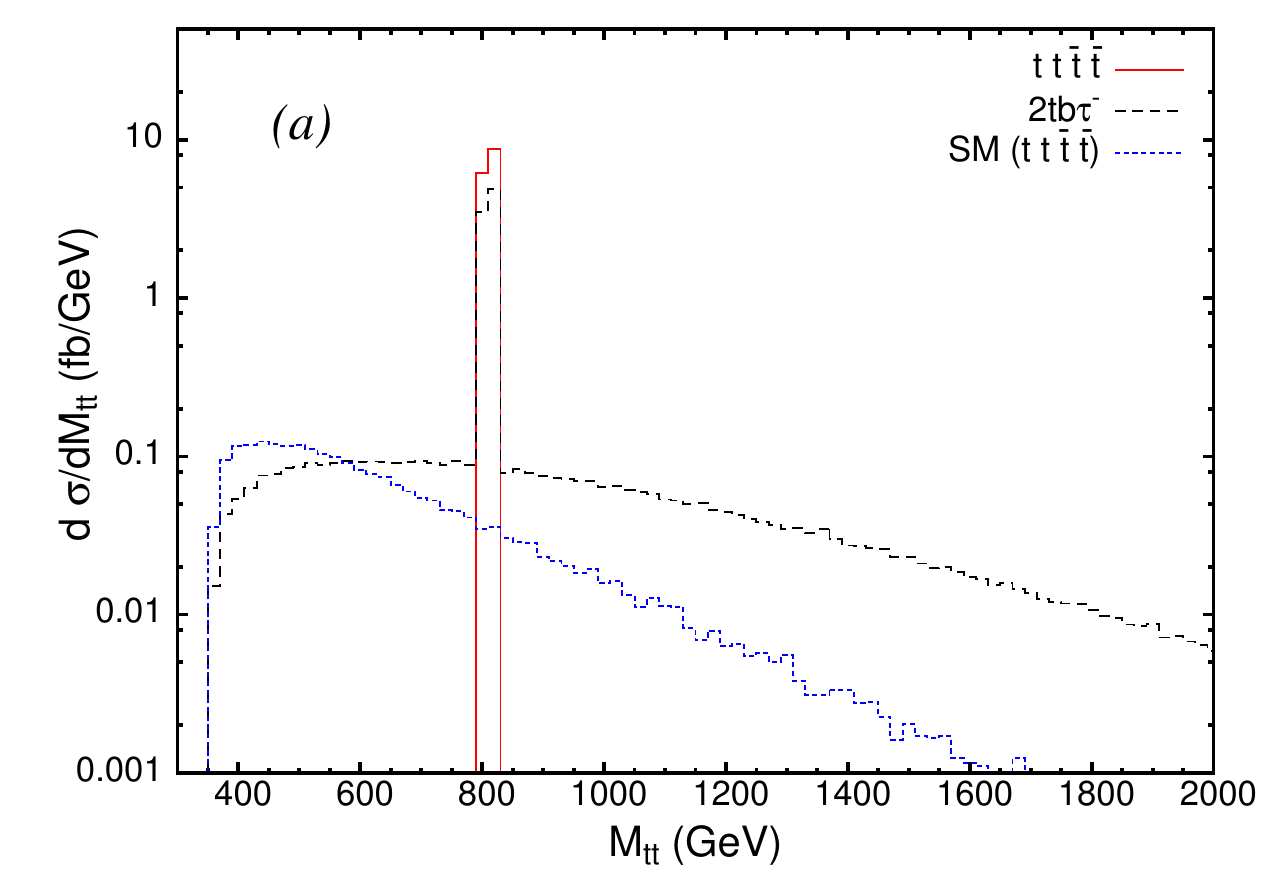}
\includegraphics[height=2.2in]{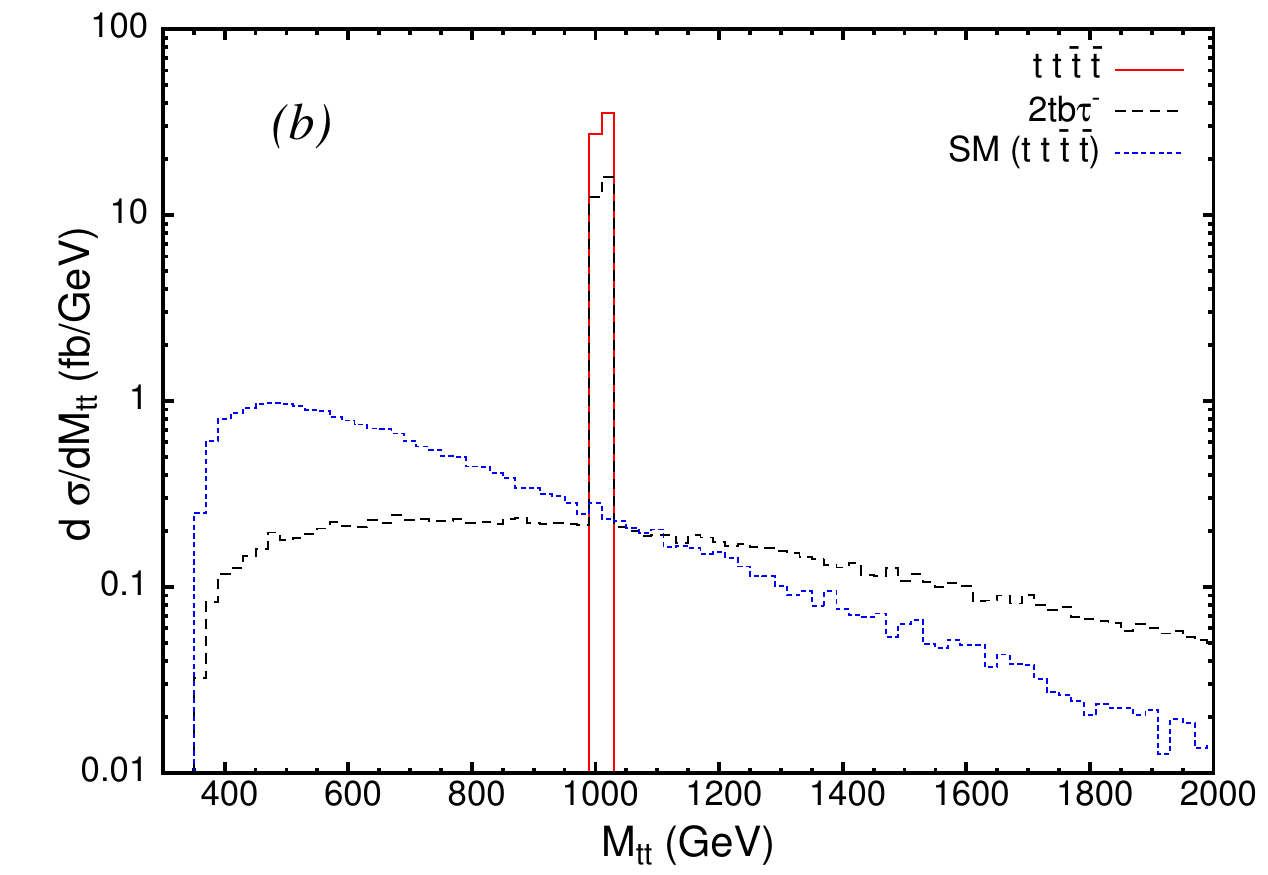}
\caption{\it Invariant mass distribution of same sign top pair $tt$ for the signal and SM background 
for two different choices of leptoquark gauge boson mass, (a) $M_V = 800$ GeV considered at LHC 
with $\sqrt{s}=8$ TeV and (b) $M_V = 1$ TeV considered at LHC with $\sqrt{s}=14$ TeV.}
\label{fig:tt}
\end{figure}

We now consider the resonance given by {\bf (C2)} and show the invariant mass distribution of 
the same sign top pair $tt$ in Fig. (\ref{fig:tt}). As pointed out earlier, this 
mode is necessary to measure the charge of the $X_\mu$ leptoquark gauge boson 
mass. A resonant bump in the same sign top pair invariant mass distribution would be a clear 
indication of a particle decaying into two same sign top quarks and therefore 
give a strong indication that the particle carries 4/3 electric charge and has quantum numbers of a
diquark.  
%%%
%%% 
The signal is again considered for two different set of final states, both of which show an invariant 
mass peak in the same sign top quark pair. In the $t\bar{t}t\bar{t}$ final state the signal cross section
comes solely from the pair production of the $X\bar{X}$ gauge bosons. As we have assumed a 
reconstruction efficiency for the top quarks as $\varepsilon_t$, the cross section for 
$M_X=800~(1000)$ GeV at LHC with $\sqrt{s}=8~(14)$ TeV is $14.93~(62.81) \times 
\varepsilon_t^4 ~fb$. The SM background for the same subprocess is 
$2.31~(24.34) \times \varepsilon_t^4 ~fb$ at LHC with $\sqrt{s}=8~(14)$ TeV. 
Although the strength of the signal crucially depends on the reconstruction 
efficiency, even a low efficiency in the long run will lead to a very important 
observation provided similar resonances are observed in the $b\tau^-$  or $b\tau^+$ final 
states. The other final state which shows a bump in $tt$ invariant mass is $ttb\tau^-$  
and its strength was already discussed for Fig.(\ref{fig:btau}). Note that again 
the $Y\bar{Y}$ contribution does not help the resonance, but is effective in 
enhancing the signal in this mode.

%%%
\begin{figure}[ht!]
\centering
\includegraphics[height=2.2in]{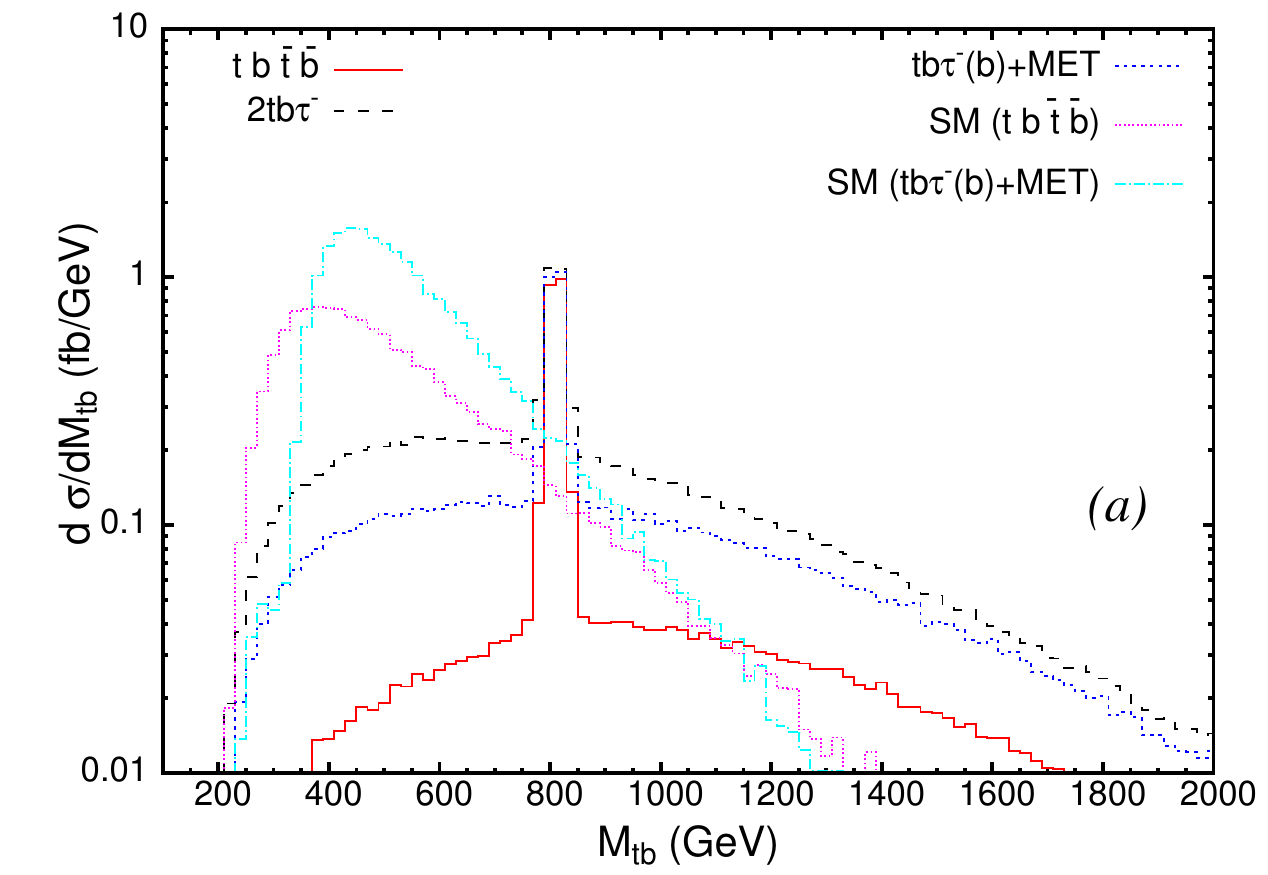}
\includegraphics[height=2.2in]{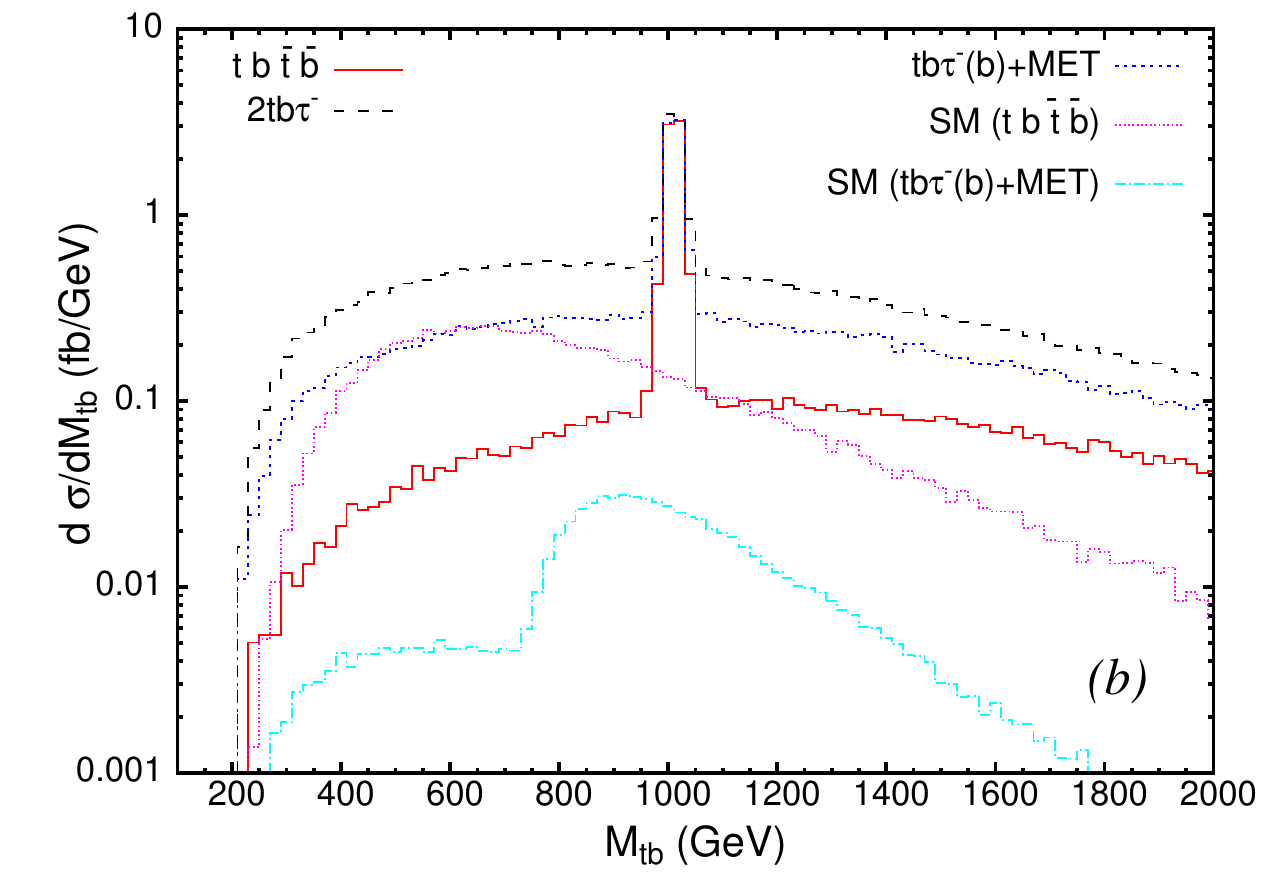}
\caption{\it Invariant mass distribution of $tb$ for the signal and SM background for two 
different choices of leptoquark gauge boson mass, (a) $M_V = 800$ GeV considered at LHC with 
$\sqrt{s}=8$ TeV and (b) $M_V = 1$ TeV considered at LHC with $\sqrt{s}=14$ TeV.}
\label{fig:tb}
\end{figure}
We now look at the final states which correspond to resonant signals for the $Y$ gauge boson. 
We therefore consider the resonance given by {\bf (C3)} and show the invariant mass distribution of 
the top-bottom pair $tb$ in Fig.(\ref{fig:tb}). Note that one of the dominant decay mode for the 
$Y_\mu$ gauge boson gives neutrinos in the final states that leads to large 
missing transverse energy (MET) and is not suitable to reconstruct the $Y_\mu$ mass. However, 
allowing one $Y$ to decay in the neutrino mode still allows reconstruction of 
the other in the visible decay modes of $tb$ and $t\tau$. A large MET in the 
final state also helps in suppressing large contributions to the SM background 
through all hadronic final states which proceed through strong interactions. In Fig.(\ref{fig:tb}) 
we consider four different final state signals which lead to a resonance in the $tb$ invariant 
mass, namely $t\bar{t}bb,~ttb\tau^-,~tb\tau^-{E}_T$ and $tbb{E}_T$. The 
signal $ttb\tau^-$ remains the same as discussed for Fig.(\ref{fig:btau}) with the only difference 
being that the contribution coming from the $X\bar{X}$ pair production now acts 
to smear out the resonance in $tb$ invariant mass distribution coming from the $Y_\mu$. This is
the cleanest mode with practically no SM background, although depending on the reconstruction of 
the top quarks. The signal cross section for the $t\bar{t}bb$ final state comes 
from the $Y_\mu$ pair production and for $M_Y=800~(1000)$ GeV at LHC with $\sqrt{s}=8~(14)$ 
TeV is $4.04~(13.75)\times \varepsilon_t^2 ~fb$. The SM background at LHC with $\sqrt{s}=8~(14)$ 
TeV for the signal comes dominantly from three subprocesses with $\sigma(t\bar{t}bb) 
\sim 54.4~(34.1)~fb, ~\sigma(t\bar{t}cc) \sim 55.1~(34.4)~fb$ and $\sigma(t\bar{t}jj) \sim 
10.14~(7.45)~pb$. The stronger cuts at the 14 TeV run is responsible for the 
relatively smaller numbers for the SM background for the higher energy run. Note that after 
including the tagging efficiencies and misstag rates, the corresponding SM background for the 
$t\bar{t}bb$ final state comes out to be $15.18~(9.65)\times \varepsilon_t^2 ~fb$. Although the
SM backgrounds are large in this case, the differential cross section is seen to fall rapidly for 
larger values of the invariant mass. Therefore, a strong cut on the $tb$ invariant mass will be useful
to suppress the background further. For the two final states involving missing transverse energy, we 
have combined their contribution in Fig.(\ref{fig:tb}) under the signal ``$tb\tau^-(b)+MET$". We find that
the SM background for $tbb{E}_T$ is completely negligible. The large contribution to the 
background comes from the $\sigma(tb\tau^-{E}_T) \sim 88.7~(3.05)~fb$ 
at LHC with $\sqrt{s}=8~(14)$ TeV, while the 
$tc\tau^-{E}_T$ and $tj\tau^-{E}_T$ are much suppressed due to the small CKM 
mixings between the first two generation quarks and the top quark. 
The SM background after including the efficiency factors is then given as 
$22.16~(0.75) \times \varepsilon_t~fb$, while the signal for $M_Y=800~(1000)$ GeV at the two center
of mass energies is $\sigma(tbb{E}_T)=4.21~(13.21) \times \varepsilon_t~fb$ and 
$\sigma(tb\tau^-{E}_T)=4.21~(13.20) \times \varepsilon_t~fb$. Note that $tbb{E}_T$ 
is the one which gives a resonant signal while $tb\tau^-{E}_T$ gives a continuum in the $tb$ 
invariant mass distribution because the $t$ and $b$ come from different 
$Y_\mu ~(Y\to \bar{b}\nu_\tau, ~\bar{Y}\to t\tau^-)$. This can be seen in Fig.(\ref{fig:tb}) where the 
large signal contribution in the $tb\tau^-{E}_T$ channel is spread out in the invariant mass distribution.
Therefore it is instructive to put a $\tau$ veto on the signal with missing transverse momenta when looking
at the invariant mass distribution in $tb$. Again for illustrative purposes we have chosen $\varepsilon_t=1$.    
 
\begin{figure}[t!]
\centering
\includegraphics[height=2.2in]{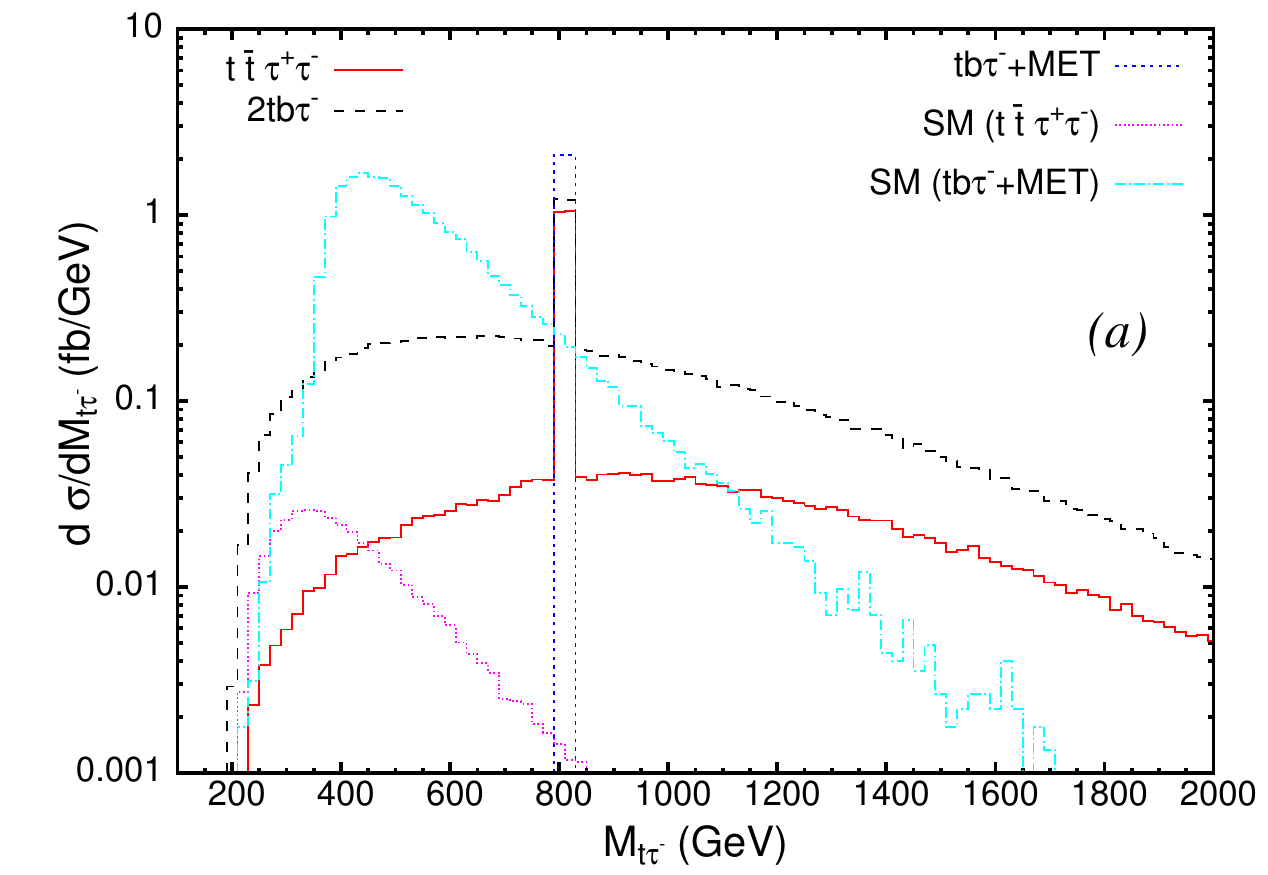}
\includegraphics[height=2.2in]{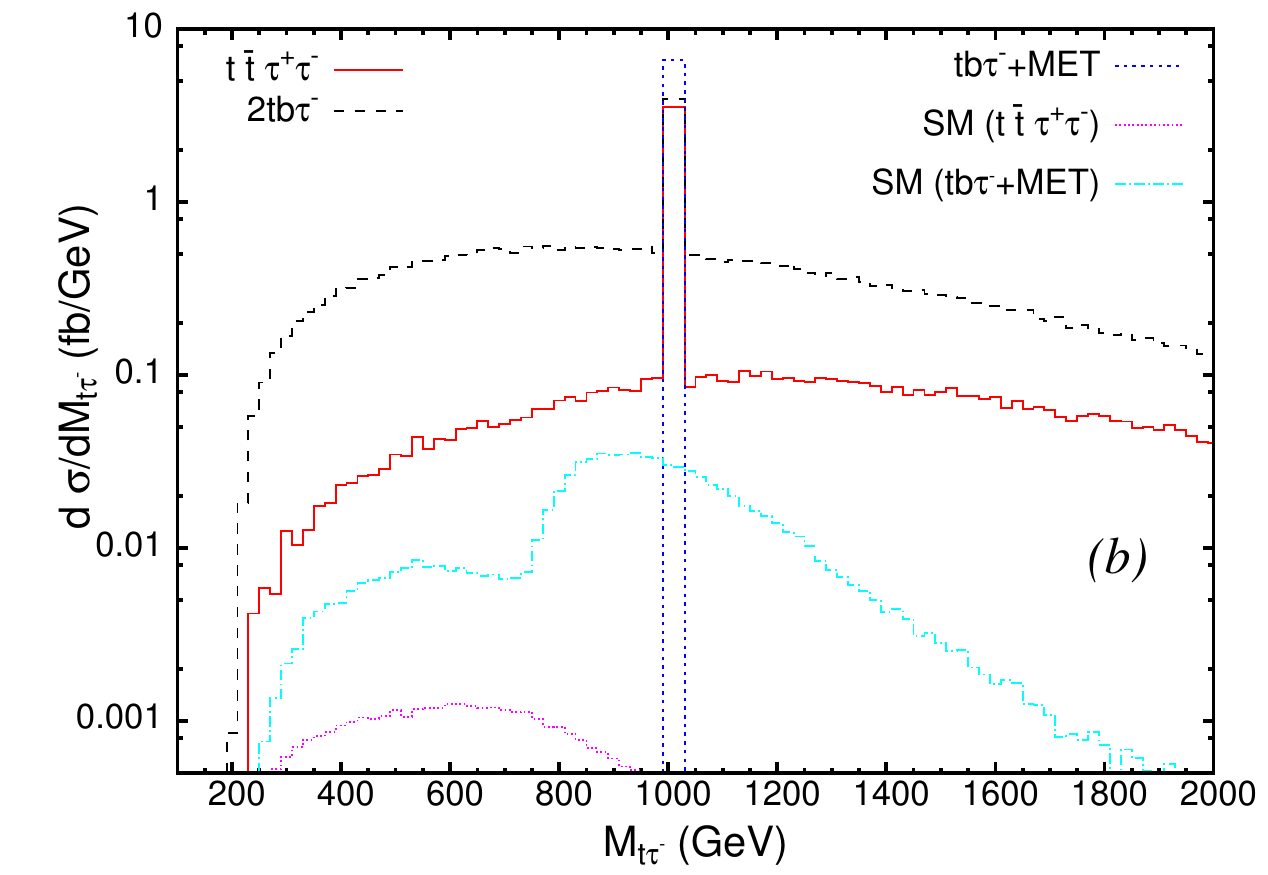}
\caption{\it Invariant mass distribution of $t\tau^-$ for the signal and SM background for two 
different choices of leptoquark gauge boson mass, (a) $M_V = 800$ GeV considered at LHC with $\sqrt{s}=8$ 
TeV and (b) $M_V = 1$ TeV considered at LHC with $\sqrt{s}=14$ TeV.}
\label{fig:ttau}
\end{figure}

We finally consider the resonance given by {\bf (C4)} which again is essential in measuring the charge of the 
$Y_\mu$ gauge boson. To measure the charge one requires the charge measurement of the $\tau$ lepton as
well as the reconstruction of the top quark in its semileptonic channel. We therefore show the invariant mass
distribution in the reconstructed top quark and charged tau lepton pair $(t\tau^-)$ in Fig.(\ref{fig:ttau}) which
corresponds to a resonance for the charge conjugated field of $Y_\mu$.  
%%%
%%%
The signal is obtained from three different set of final states given by 
$t\bar{t}\tau^+\tau^-, ~ tb\tau^-{E}_T$ and $2tb\tau^-$. As discussed before the  $2tb\tau^-$ 
contribution is found to have negligible SM background but the contribution from the $X\bar{X}$ 
production to the $t\tau^-$ invariant mass distribution itself acts as a background for the resonant signal from
the $Y\bar{Y}$ production. The $t\bar{t}\tau^+\tau^-$ signal comes solely from the $Y_\mu$ pair production 
and we find that with the proper charge identification of the $\tau$ leptons, we can ignore contributions from
SM background processes such as $t\bar{t}jj$. The signal cross section in this mode is found to be 
$4.04~(13.81) \times \varepsilon_t^2~fb$ for $M_Y=800~(1000)$ GeV at LHC with $\sqrt{s}=8~(14)$ TeV.
The SM background is quite suppressed at both center of mass energy values, given by 
$0.35~(0.04) \times \varepsilon_t^2~fb$. The $tb\tau^-{E}_T$ signal discussed for the $tb$
resonance in Fig.(\ref{fig:tb}) was found to give a continuum distribution in the $tb$ invariant mass.  However 
it leads to a resonance in the $t\tau^-$ invariant mass distribution as $\bar{Y}\to t\tau^-$. The event rates
are the same as before but one can clearly see a distinct resonance confined to a few bins in the 
invariant mass distribution of $t\tau^-$ in Fig.(\ref{fig:ttau}) for the $tb\tau^-{E}_T$ signal. 
The large SM background for this mode can again be suppressed with a significantly strong cut on the 
$t\tau^-$ invariant mass.

\subsection{LHC sensitivity to the $X_\mu$ and $Y_\mu$ gauge bosons}
As evident from our analyses of the resonant signals for the $X_\mu$ and $Y_\mu$ gauge 
bosons in our models, the LHC would be able to see the signals in various different channels for 
significantly large values of their mass.
A single channel analysis in the $b\tau$ mode relevant for $X_\mu$ search was considered for 
its search at the 7 TeV run of LHC \cite{Chakdar:2012kd,Chatrchyan:2012sv} while another 
experimental study relevant for the $Y_\nu$ search in the $bb{E}_T$ channel has been 
done by the CMS Collaboration \cite{Chatrchyan:2012st}. Here we do a more expansive sensitivity 
reach at the LHC for these gauge bosons that can be obtained at different integrated luminosities. 
For the top decaying semileptonically to $b\ell^+\nu_\ell$ where $\ell=e,\mu$ the events will be 
at most, or less than $\sim 22\%$ of the reconstructed top events. While it would be $\sim 66\%$ in 
the hadronic decay mode. Thus it gives a clear demarcation on the event rate we specify for the final 
states involving the top and anti-top quarks that would lead to any signal events to reconstruct 
the tops. 

For the sensitivity analysis we define the signal to be observable if the lower limit on the
signal plus background is larger than the corresponding upper limit on the 
background \cite{Sayre:2011ed} with statistical fluctuations
\begin{align*}
L (\sigma_s + \sigma_b) - N \sqrt{L (\sigma_s + \sigma_b)} \geq L\sigma_b + N\sqrt{L \sigma_b}
\end{align*}
or equivalently,
\begin{align}
\sigma_s \geq \frac{N}{L} \left[N + 2\sqrt{L\sigma_b}\right],
\label{eq:conflev}
\end{align}
where $L$ is the integrated luminosity, $\sigma_s$ is the signal cross section, and $\sigma_b$ is the
background cross section. The parameter $N$ specifies the level or probability of discovery.
We take $N = 2.5$, which corresponds to a $5\sigma$ signal. For $\sigma_b  \gg \sigma_s$, this 
requirement becomes similar to
\begin{align}
\mathcal{S} = \frac{N_s}{\sqrt{N_b}} = \frac{L\sigma_s}{\sqrt{ L\sigma_b}} \geq 5 ,
\end{align}
where $N_s$ is the number of events for the signal, $N_b$ is the number of events for the background, 
and  $\mathcal{S}$ equals the statistical significance. 

%%%
\begin{table}[!t]
\begin{center}
\begin{tabular}{|c|c|c|c|}
\hline 
${\mathcal Final ~State}$ &  $\sigma_{{\mathcal SM}}~(fb)$ & ${\mathcal Final ~State}$
& $\sigma_{{\mathcal SM}}~(fb)$  \\ 
\hline  $2b\tau^+\tau^-$ &   0.12 ~(0.002)
       &  $ttb\tau^-,~\bar{t}\bar{t}b\tau^+$ & -- \\ 
\hline  $tt\bar{t}\bar{t}$ & 2.31~(24.34) 
       &  $t\bar{t}\tau^+\tau^-$ & 0.35~(0.04) \\
\hline  $2b t\bar{t}$ & 15.18~(9.65)
       &  $2b{E}_T$ & 25.06~(3.83) \\
\hline $b t \tau^- {E}_T$ &  22.16~(0.75) 
       & $b \bar{t} \tau^+{E}_T$ &  22.16~(0.75) \\
\hline  $2b t {E}_T$ & 0.003~(0.001)  
       &  $2b\bar{t}{E}_T$ &  0.001~0.0006)\\
\hline
\end{tabular}
\caption{\textit{The combined SM cross sections estimated at parton level using MadGraph 5 for the 
different final state signals at LHC with $\sqrt{s}=8$ TeV and $\sqrt{s}=14$ TeV. The 14 TeV values
are given in parenthesis. Note that the cross sections given satisfy the kinematic cuts listed
in Table \ref{tab:cuts} and all tagging efficiencies and misstag rates are included.} } 
\label{tab:cross sections}
\end{center}
\end{table}
%%%%%%
 In Table \ref{tab:cross sections}, we have calculated the SM background for the different final states that 
we have considered for the signal coming form the pair productions of the $X_\mu$ and 
$Y_\mu$ gauge bosons. The cross sections shown in Table \ref{tab:cross sections} are obtained after passing the events through the 
kinematic selection conditions given in Table \ref{tab:cuts}. In most cases the SM 
backgrounds are quite small and would remain negligible even with an integrated luminosity 
of 100 fb$^{-1}$. Note that as the top reconstruction would require sufficient events after it has 
decayed, we need much larger cross sections for the final states involving top 
quarks. To use Eq.(\ref{eq:conflev}), we require the background events to be 
sufficiently large such that the fluctuations to a Gaussian distribution could be applied. We find that the best reaches are obtained 
 for  the  $bb{E}_T, ~b\bar{t}\tau^+{E}_T$ and $bt\tau^-{E}_T$
final states. For the $bb{E}_T$ final state at LHC with $\sqrt{s}=8$ TeV, the 
signal cross section for a  $5\sigma$ sensitivity must be greater than $8.54,~5.91,~4.78~fb$
for an integrated luminosity of $L=10,~20,~30~fb^{-1}$ respectively. This 
corresponds to the mass reach of $M_Y=737,~772,~793$ GeV respectively. With the 
higher center of mass energy option for LHC with $\sqrt{s}=14$ TeV, the 
signal cross section for a  $5\sigma$ sensitivity must be greater than $1.995,~1.041,~0.586~fb$
for an integrated luminosity of $L=30,~100,~300~fb^{-1}$ respectively. These lead 
to a mass reach of $1325,~1440,~1545$ GeV respectively. For the  other channels involving the
top quark in the final state, we assume the reconstruction efficiency for the top quark 
$\varepsilon_t \simeq 0.5$ which includes the event loss from kinematic cuts after the top decays.
Adding the contributions for $b\bar{t}\tau^+{E}_T$ and $bt\tau^-{E}_T$ 
we find that at the 8 TeV run of LHC, the mass reach is $770,~795$ GeV for an 
integrated luminosity of $L=20,~30~fb^{-1}$ respectively while at the 14 TeV run 
of LHC, where we use the high luminosity options of $200~fb^{-1}$ and 
$300~fb^{-1}$, the $5\sigma$ sensitivity comes out to be about $1650$ GeV and $1690$ GeV respectively. 
Inspired by this work, CMS collaboration has searched for the third generation leptoquarks in the "$b\tau$" mode and set a limit with vector leptoquarks with masses 760 GeV at $\sqrt s $ = 7 TeV with 95$\%$ CL. 

%For the other final states where the SM background is completely %negligible, even with high 
%integrated luminosities available from collision data, we can set %a criteria of observing a minimum 
%number of events in that mode for a discovery. Setting that %number as 10 events leads to 
%the following sensitivity for the different final states.

%{\bf Shreyashi: Please run the programs for the final states %listed in Table \ref{tab:crosssections} 
%using the usual tagging efficiencies and an additional efficiency %factor of 0.5 for a top quark 
%in the final state and find out the mass values which give 10 %events for different choices 
%of luminosity. Set the Q in the $\alpha_s$ running and also in %the PDF calling to the mass of the
%Z boson. Make a similar table as above and list the reach values %for the different final states.}

%%%%%%%%%%%%%%%%%%%%%%%%%%%%%%%%%%%%%%%%%%%%%%%%%%%%%%%%%%

\section{Summary and Conclusions}

Although the Standard Model, based on local gauge symmetries, accidentally conserve baryon and lepton numbers, there is no fundamental reason for the baryon and lepton numbers to be exact symmetries of Nature. In fact, Grand Unification, unifying quarks and leptons, naturally violate baryon and lepton number. The remarkable stability of the proton dictate that the masses of these leptoquark and diquark gauge bosons to be at the $10^{16}$ GeV scale. However, baryon and lepton number violating interaction involving only the 3rd family of fermions is not much constrained experimentally. Inspired by the topcolor, topflavor and top hypercharge models, we have a top-GUT model where only the third family of fermions are unified in an $SU(5)$ with the symmetry breaking scale at the TeV. These models give baryon and lepton number violating gauge interactions which involve only the third family, and with interesting resonant signals at the LHC. 

We have  proposed two models, the minimal and renormalizable top $SU(5)$ where the 
$SU(5)\times SU(3)'_C \times SU(2)'_L \times U(1)'_Y$ gauge symmetry 
is broken down to the Standard Model (SM)
gauge symmetry via the bifundamental Higgs fields at low energy. The first 
two families of the SM fermions are charged under $ SU(3)'_C \times SU(2)'_L \times U(1)'_Y$
while the third family is charged under $SU(5)$. 
In the minimal top $SU(5)$ model, we showed that the quark CKM mixing
matrix can be generated via dimension-five operators, and the
proton decay problem can be solved by fine-tuning the coefficients of the
high-dimensional operators at the order of $10^{-4}$.
In the renormalizable top $SU(5)$ model, we introduced additional vector-like fermions whose renormalizable interactions with the SM particles generate these dimension 5 interactions and  we can explain the quark CKM mixing matrix by introducing the vector-like particles, and also there is no proton decay problem.
We have discussed the phenomenology of the models in details 
looking for the resonant signals for the baryon and lepton number violating leptoquark as well as diquark gauge bosons at the LHC,
as well as the various final state arising from the productions and decays of these heavy gauge bosons. We have also calculated the corresponding SM backgrounds. We find that a $5\sigma$ signal can be observed for a mass leptoquark / diquark of about 770/800 GeV at the 8 TeV LHC with luminosity of $20 fb^{-1}$/$30 fb^{-1}$. The mass reach extends to about 1450 TeV for 14 TeV LHC with a luminosity of $100 fb^{-1}$.

%\begin{acknowledgments}

%This research was supported in part by the Natural Science Foundation of China 
%under grant numbers 10821504, 11075194, and 11135003, 
%and by the United States Department of Energy Grant Numbers DE-FG03-95-Er-40917, %DE-%FG02-04ER41306.The work of S.K.R was partially supported by funding available from the 
%Department of Atomic Energy, Government of India, for the Regional Center for Accelerator 
%based Particle Physics, Harish-Chandra Research Institute.

%\end{acknowledgments}

%%%%%%%%%%%%%%%%%%%%%%%%%%%%%%%%%%%%%%%%%%%%%%%%%%%%%%%%%%%%%%%%%%%%%%

\chapter{New fermions in the framework of Left-Right Mirror Model}\label{chap:chap3}
\section{Introduction}
The non-conservation of parity $P$ (the left-right asymmetry of elementary particles) is well incorporated in the Standard Model (SM) of particle physics. However, it has been considered as an unpleasant feature of the model. One possible way to understand the left-right asymmetry of elementary particles is to enlarge the SM into a left-right (LR) symmetric structure and then, by some spontaneously breaking mechanism, to recover the SM symmetry structure. For instance, in  {left-right} symmetric models \cite{LRmodels}, $SU(2)_R$ interactions are introduced to maintain parity invariance at high energy scales. The symmetry group $SU(2)_L\otimes SU(2)_R \otimes U(1)_{B-L}$ of LR symmetric models can be a part of a grand unified symmetry group such as 
$SO(10)$ \cite{Fritzsch:1974nn} or $E_6$ \cite{GUT} or  superstring inspired models \cite{string}. In the framework of LR symmetric SM, the SM left-handed fermions are placed in the $SU(2)_L$ doublets as they are in the SM while the SM right-handed fermions (with the addition of right-handed neutrinos for the case of leptons) are placed in the $SU(2)_R$ doublets. Subsequently, the LR symmetry is spontaneously broken down to the SM electroweak symmetry using suitable Higgs representations. There are different variants of LR symmetric models have been proposed in the literature \cite{LR1,LR2,LR3,LR4}.

Another interesting solution to the non-conservation of parity in the SM was proposed in a classic paper
 \cite{LY} by Lee and Yang. They postulated the existence of additional (mirror) fermions of opposite chirality to the SM ones to make the world left-right symmetric at high energies. The advantages of models with mirror fermions to solve some problems in particle physics have already been discussed in the literature. For instance, the existence of mirror neutrinos can naturally explain the smallness of neutrino mass via a see-saw like mechanism \cite{neutrino,Foot:1995pa,Berezhiani:1995yi}. Moreover, it can also be useful for the Dark Matter problem \cite{Berezhiani:1995yi}, neutrino oscillations as well as different neutrino physics anomalies such as solar neutrino deficit and atmospheric neutrino anomaly \cite{Foot:1995pa}. On the other hand, mirror fermions can provide a solution to the strong CP problem if the parity symmetry is imposed \cite{strongCP1,strongCP2}. Finally, the existence of mirror particles appear naturally in many extensions of the SM, like GUT and string theories \cite{string_GUT}. The masses of these mirror particles, though unknown, are not experimentally excluded to be at or below the TeV scale. The agreement of the models with mirror fermions with electroweak precision data, Higgs rate etc have been studied in Ref.~\cite{EWP}. Therefore, it is important to study the phenomenological consequences of the mirror particles in the context of collider experiments, in particular at the  {Large Hadron Collider} (LHC). In this work, we have investigated the phenomenology of mirror particles in the context of a particular variant of LR symmetric mirror model (LRMM), their associated final state signals, and the discovery potential at the LHC.

In the LRMM we propose in this work, the SM gauge group ($G_{SM}=SU(3)_C\otimes SU(2)_L \otimes U(1)_Y$) is extended to $G_{LR}=SU(3)_C\otimes SU(2)_L\otimes SU(2)_R \otimes U(1)_{Y^{\prime}}$ together with a discrete $Z_2$ symmetry. The SM particle spectrum is also extended to include mirror particles and a real scalar Higgs singlet under both $SU(2)_L$ and $SU(2)_R$. For the fermion sector, the right-handed (left-handed) components of mirror fermions transform as doublets (singlets) under $SU(2)_R$. The SM fermions are singlets under $SU(2)_R$, whereas  doublets  under $SU(2)_L$. Similarly there are mirror singlet fermions corresponding to the SM singlet fermions. Since the fermion representations are exactly mirror symmetric, all triangle anomalies are exactly canceled with respect to the entire gauge symmetry, the model is anomaly free. Because of even number of doublets, there is also no gravitational anomaly.
The SM fermions are even under the $Z_2$ symmetry, whereas,  {the} corresponding mirror fermions are odd. Therefore, any mass mixing  between SM charged fermions and with mirror partners are forbidden by the $Z_2$ symmetry. The spontaneous symmetry breaking (SSB), $G_{LR} \to G_{SM}$ is realized by introducing a mirror Higgs doublet which is singlet under $SU(2)_L$ and doublet under $SU(2)_R$. Subsequently the SSB, $G_{SM} \to SU(3)_C \otimes U(1)_{EM}$ is achieved via the SM Higgs doublet which is doublet under $SU(2)_L$ and singlet under $SU(2)_R$.
After the SSB, the gauge boson sector of LRMM contains the usual SM gauge bosons (gluon, $W^\pm$ bosons, $Z$ boson and photon) along with the mirror partners of $W^\pm$ and $Z$-boson. The non zero vacuum expectation value (VEV) for a singlet scalar breaks the $Z_2$ symmetry and gives rise to mixing between the SM and mirror fermions.

The parity symmetry in LRMM determines the ratio among the charged mirror fermion masses from the SM charged fermion mass spectrum. In particular, the ratio of the SM fermion mass and the corresponding mirror fermion mass is given by $ {\cal O}(1)\frac{v}{\hat v}$, where $ {\cal O}(1)$ is an order one number, $v\sim 250$ GeV and $\hat v$ are the VEV's for the SM Higgs and mirror Higgs respectively. Connecting the model for generating tiny neutrino masses $\simeq 10^{-11}$ GeV gives $\hat v \sim 10^{7}$ GeV. This gives  TeV scale masses, or few hundred GeV masses for the mirror partners of electron, up and down quarks, namely ${\hat e},~{\hat u}~{\rm and}~{\hat d}$. This makes the model testable at the ongoing LHC and proposed linear electron-positron collider experiments. 
%Apart from the collider phenomenological motivation, $\hat v \sim 10^{7}$ GeV is also motivated from the tiny neutrino mass generation which we have %discussed in the text.

Different variants of LR symmetric mirror models have been proposed and studied in the literature in different contexts. For example, in Ref.~\cite{Hung:2006ap}, the SM particle content have been extended to include mirror fermions and tiny neutrino mass have been explained via see-saw mechanism. In this model, the gauge group is the SM gauge group and for each SM left (right) handed $SU(2)_L$ doublet (singlet) there is right (left) handed mirror doublet (singlet). Therefore, both the SM (left handed) and mirror (right handed) neutrinos in this model transform as a doublet under $SU(2)_L$. As a result, triplet Higgs fields are required in this model for the Majorana mass terms. However, in our model the gauge structure is different and in addition to doublet neutrinos, we have singlet left handed and right handed neutrinos. Therefore, triplet Higgs is not required in our model. Another class of mirror models have been proposed in Ref.~\cite{strongCP1} as a solution to the strong CP problem. The gauge group and particle content for our model is somewhat similar to the model in Ref.~\cite{strongCP1}. However, our model includes a singlet scalar and the gauge symmetry is supplemented by an additional discrete $Z_2$ symmetry. This modifications give rise to TeV scale mirror fermions and make our model testable at the collider experiments.

One of the major goals of the LHC experiment is to find new physics beyond the SM. The LHC is a proton-proton collider and thus, the collision processes are overwhelmed by the QCD interactions. Therefore, in the framework of LRMM, the new TeV scale colored particles, namely ${\hat u}~{\rm and}~{\hat d}$ quarks will be copiously pair produced at the LHC. After being produced, ${\hat u}~{\rm and}~{\hat d}$ quarks will decay to the SM particles giving rise to interesting signatures at the LHC. The TeV scale mirror quarks   {are found to} decay into a $Z/W$ boson or a Higgs boson in association with a SM quark.  {This leads to new fermionic resonances as well as  new physics signals in  
two SM gauge bosons + two jet final states. Note that the gauge bosons could be either $Z$ or $W$ and 
the highlight of the signal would be the presence of a clear resonance in the jet+$Z$ and jet+$W$ 
invariant mass distributions. Such a resonance will stand out against any SM background in these final 
states. In this paper we have therefore studied in detail the signal coming from the pair production of the mirror quarks, ${\hat u}~{\rm and}~{\hat d}$ and their subsequent decays in our\begin{small}
•
\end{small} LRMM and compared it 
with the dominant SM background processes.} 

% Thus we get resonances in the jet + Z and jet +W channels. Such resonances are not present in the SM, and will
% be rather easy to observe above the background with a large $p_T$ cut on the jet, and the Z boson. This will be a 
% clear signal for new physics. In addition to the resonant signal, in this paper,  we have also studied  the final state 
% signals arising from  pair production of ${\hat u}~{\rm and}~{\hat d}$ quarks and their subsequent decays. This  
% gives rise to {\it two SM bosons in association with two light quarks} at the LHC.
% In this paper, we have studied {\it two $Z$-boson plus two light quark jets} and {\it one $Z$-boson and one $W$-boson plus two light quark jets} %signals at the LHC as a signature of LRMM.
 
The chapter is organized as follows. In Section \ref{sec:model}, we discuss our model and the formalism. Section \ref{sec:pheno} is devoted for the phenomenological implications of the model. Finally, a summary 
of our work, and the conclusions are given in Section \ref{sec:summary}.

\section{Left-Right symmetric mirror model (LRMM) and the formalism}\label{sec:model}

Our LR symmetric mirror model is based on the gauge symmetry $G_{LR}=SU(3)_C\otimes SU(2)_L\otimes SU(2)_R \otimes U(1)_{Y^{\prime}}$ supplemented by a discrete $Z_2$ symmetry.  Left-right symmetry, as in the usual left-right model, provides a natural explanation why the parity is violated at low energy. Inclusion of mirror particles gives an alternate realization of the  {LR} symmetry in the 
fermion sector.
%One of the motivation for proposing L-R model with singlet fermions is to solve strong CP problem. It was shown in Ref.~\cite{strongCP} that the %complete invariance of such a model under parity can guarantee a vanishing strong CP phase from the QCD $\theta$-vacuum and thus, solves strong CP %problem.
% In the present version of LRMM, the SM gauge group ($G_{SM}$) is extended to $G_{LR}=SU(3)_C\otimes SU(2)_L\otimes SU(2)_R \otimes U(1)_{Y^{\prime}}$ %with a discrete $Z_2$ symmetry and the matter content  is also enlarged by including new particles with mirror properties. For instance, this model %includes mirror fermions with opposite chirality relative to the SM fermions.
The fermion representations in our model for leptons and quarks in the first family is given by
%sector under the gauge group $G_{LR}$ are presented in the following:
\begin{eqnarray}
l_{L}^{0}={\begin{pmatrix} \nu^0 \\ e^0 \end{pmatrix}}_L\sim (1,2,1,-1) &,& e^0_R \sim (1,1,1,-2)~~,~~ \nu_R^0 \sim(1,1,1,0);\nonumber\\
\hat {l}_{R}^{0}={\begin{pmatrix} \hat{\nu}^0 \\ \hat{e}^0 \end{pmatrix}}_R\sim (1,1,2,-1) &,& \hat {e}^0_L \sim (1,1,1,-2)~~,~~ \hat {\nu}_L^0 \sim(1,1,1,0);\nonumber\\ 
Q_{L}^{0}={\begin{pmatrix} u^0 \\ d^0 \end{pmatrix}}_L\sim (3,2,1,\frac{1}{3}) &,& u^0_R \sim (3,1,1,\frac{4}{3})~~~,~~ d_R^0 \sim(3,1,1,-\frac{2}{3});\nonumber\\
\hat {Q}_{R}^{0}={\begin{pmatrix} \hat{u}^0 \\ \hat{d}^0 \end{pmatrix}}_R\sim (3,1,2,\frac{1}{3}) &,& \hat {u}^0_L \sim (3,1,1,\frac{4}{3})~~~,~~ \hat {d}_L^0 \sim(1,1,1,-\frac{2}{3});
\label{frep} 
\end{eqnarray}
where the bracketed entries correspond to the transformation properties under the symmetries of the group $G_{LR}$. Note that since the model is left-right symmetric, for every fermion representation of $SU(2)_L$, there is a there is a multiplet corresponding to the same representation of $SU(2)_R$.The superscripts ($^0$) denote gauge eigenstates and the hat symbol ($\hat~$) is associated with the mirror fermions. The charge generator is given by: $Q=T_{3L}+T_{3R}+Y^{\prime}/2$. In the usual LR symmetric model, $SU(2)_L\otimes SU(2)_R \otimes U(1)$, the $U(1)$ symmetry is $U(1)_{B-L}$. This is easily embedded into $SU(4)_C\otimes SU(2)_L\otimes SU(2)_R$ or $SO(10)$ GUT. The $U(1)_{Y^\prime}$ in our model is not $U(1)_{B-L}$. This can be seen from the $Y^\prime$ quantum numbers of the fermions in Eq.~\ref{frep}. Thus, $U(1)_{Y^\prime}$, in this model, can not be embedded in the usual $SO(10)$ GUT.

Under the $Z_2$ symmetry, the SM fermions as well as the right handed singlet neutrino (denoted by without hat) are even, whereas, the mirror fermions including the left handed singlet mirror neutrino (denoted by hat)  are odd.  This structure of $Z_2$ symmetry for the SM and corresponding mirror fermion is required to forbid the large (in general of the order of symmetry breaking scale) singlet mass terms between the SM and mirror singlets.The fermion representations for the second and third family are identical to the first family. 
%Note that the $Z_2$ is needed so that we do not have mass mixing of the charged fermions between the ordinary fermions and the mirror fermions. This avoids the ordinary charged fermions from getting masses in the first stage of symmetry breaking  {which happens at a high} scale.

Note that in the traditional  {LR} model, the fermion sector is completely symmetric for the ordinary SM fermions. For example, we have $(u,d)_L$ and $(u,d)_R$, and similarly for every  {fermion family}. Another version, proposed in  \cite{LY} is to introduce new fermions to make it  {LR} symmetric, {\it i.e.}  for every $(u,d)_L$, we have new fermions,  $(\hat{u},\hat{d})_R$. Hence it is the left-right mirror model (LRMM). It is this realization that we pursue here.
It was shown in Ref.~\cite{strongCP1,strongCP2} that the complete invariance of such a model under parity can guarantee a vanishing strong CP phase from the QCD $\theta$-vacuum and thus, solves strong CP problem.

\subsection{Symmetry breaking and the scalar sector}
In the framework of LRMM, spontaneous symmetry breaking is achieved via the following steps:
 {
\begin{equation}
%SU(3)_C\otimes SU(2)_L\otimes SU(2)_R \otimes U(1)_{Y^{\prime}} \to SU(3)_C\otimes SU(2)_L\otimes U(1)_{Y} \to SU(3)_C\otimes U(1)_{Q},
SU(2)_L\otimes SU(2)_R \otimes U(1)_{Y^{\prime}} \to SU(2)_L\otimes U(1)_{Y} \to U(1)_{Q},
\end{equation}} 
where, $Y/2=T_{3R}+Y^\prime/2$. In order to realize the above SSB, two Higgs doublets 
%(both even under the $Z_2$ symmetry) 
are required {\it i.e.}, the SM Higgs doublet ($\Phi$) and its mirror partner ($\hat {\Phi}$). In order to have the Yukawa interactions between doublet and singlet fermions for SM and mirror sector, both the Higgs doublets has to be even under the $Z_2$ symmetry. The gauge quantum numbers and  VEV's of these Higgs doublets are summarized below:
\begin{eqnarray}
\Phi \sim (1,2,1,1) &,& \hat {\Phi} \sim (1,1,2,1);\nonumber\\
\left < \Phi \right> = \frac{1}{\sqrt 2}{\begin{pmatrix} 0 \\ v \end{pmatrix}} &,& \hat {\left < {\Phi} \right>} = \frac{1}{\sqrt 2}{\begin{pmatrix} 0 \\ \hat {v} \end{pmatrix}} \label{vev}.
\end{eqnarray}
In addition to these two Higgs doublets, we have introduced a singlet (under both $SU(2)_L$ and $SU(2)_R$) real scalar which is odd under the $Z_2$ symmetry: $\chi\sim (1,1,1,0)$. The VEV of $\chi$: $\left< \chi \right>=v_{\chi}$, breaks the $Z_2$ symmetry spontaneously. This enables us to generate mixing between the SM fermions and the mirror fermions. This  {mixing with the SM fermions allows the mirror fermions to decay to 
lighter SM particles after they are pair produced at colliders  such as the LHC,} giving rise to interesting final state signals.  It is important to mention that spontaneous breaking of $Z_2$ discrete symmetry gives rise to domain walls problem in the theory. However, this problem is easily solved by breaking the $Z_2$ symmetry softly by introducing a $\mu_3 \chi^3$ in the potential. With this soft breaking the world will have no domain walls and choosing $\mu_3$ much smaller compared to $\mu_\chi$, there will be no significant effect on the collider phenomenology.

In order to generate the above structure of VEV's for $\Phi$ and $\hat{\Phi}$, the LR symmetry has to be broken, otherwise, we will end up with $v=\hat {v}$. The most general scalar potential that develops this pattern of VEV's is given by,
\begin{eqnarray}
V=&-&\left( \mu^2 \Phi^\dagger \Phi + \hat {\mu^2} \hat {\Phi}^\dagger \hat {\Phi}\right ) + \frac{\lambda}{2}\left[ \left(\Phi^\dagger \Phi \right)^2 + \left(\hat {\Phi}^\dagger \hat {\Phi} \right)^2\right ] + \lambda_1 \left(\Phi^\dagger \Phi \right) \left(\hat {\Phi}^\dagger \hat {\Phi} \right)\nonumber\\
&-&\frac{1}{2}\mu_\chi^2\chi^2+ \frac{1}{3}\mu_3 \chi^3+\frac{1}{4}\lambda_\chi \chi^4+ \lambda_{\phi\chi}\chi^2\left(\Phi^\dagger\Phi+\hat \Phi^\dagger \hat \Phi\right)
\end{eqnarray}
It is important to  {note} that in the above potential, the terms with $\mu,~\hat{\mu}$ break the parity symmetry softly, {\it i.e.}, only through the dimension-two mass terms of the scalar potential. Note that after the two stages of symmetry breaking, we are left with three neutral scalars, SM like Higgs, $h$,  Mirror Higgs, $\hat{h}$, and a singlet Higgs $\chi$. We consider a solution of the Higgs potential such that $v<< v_{\chi} << \hat{v}$, and so the mixing among these Higgses are negligible.

\subsection{Gauge bosons masses and mixings}
The gauge bosons masses and mixings are obtained from the kinetic terms of the scalars in the Lagrangian:
\begin{equation}
{\cal L} \supset \left({\cal D}_\mu\Phi\right)^\dagger\left({\cal D}^\mu\Phi\right)+\left(\hat {\cal D}_\mu\hat {\Phi}\right)^\dagger\left(\hat {\cal D}^\mu\hat{\Phi}\right),\label{ktl}
\end{equation}
where, ${\cal D}~{\rm and}~\hat {\cal D}$ are the covariant derivatives associated with the SM and mirror sector respectively.
 {\begin{equation}
%{\cal D}_\mu (\hat {\cal D}_\mu)=\partial_\mu + ig_s\frac{\lambda_a}{2}G^a_\mu + ig \frac{\tau_a}{2}W^a_\mu (\hat {W}^a_\mu)+ig^\prime \frac{Y^\prime}{2} B_\mu,
{\cal D}_\mu (\hat {\cal D}_\mu)=\partial_\mu + ig \frac{\tau_a}{2}W^a_\mu (\hat {W}^a_\mu)+ig^\prime \frac{Y^\prime}{2} B_\mu,
\end{equation}}
where, $\lambda_a$'s and $\tau_a$'s are the Gell Mann and Pauli matrices respectively. The gauge bosons and gauge couplings related to the gauge group 
%$SU(3)_C\otimes SU(2)_L\otimes SU(2)_R \otimes U(1)_{Y^{\prime}}$ 
 {$SU(2)_L\otimes SU(2)_R \otimes U(1)_{Y^{\prime}}$}
are respectively %$G_\mu^a$, 
 {$W^a_\mu,~\hat {W}^a_\mu$, $B_\mu$}  and  {$g,~g,~g^\prime$}.  {Note} that to ensure parity symmetry, we have chosen identical gauge coupling for $SU(2)_L$ and $SU(2)_R$.

Substituting the VEV's of Eq.~\ref{vev} in the kinetic terms for the scalars in Eq.~\ref{ktl}, we obtain the masses and mixings of the seven electroweak gauge bosons of this model. The light gauge bosons are denoted as: $W^\pm$, $Z$ and $\gamma$, which are identified with the SM ones, whereas the mirror gauge bosons are denoted by $\hat {W}^\pm$ and $\hat {Z}$. The mass matrix for the charged
gauge bosons is diagonal, with masses:
\begin{equation}
M_{W^\pm}~=~\frac{1}{2}gv~~~~~,~~~~~M_{\hat W^\pm}~=~\frac{1}{2}g\hat v.
\end{equation}    
The mass matrix for the neutral gauge boson sector is not diagonal and in the basis ($W^3,~\hat W^3,~B$), the neutral gauge boson mass matrix is given by,
\begin{equation}
M=\frac{1}{4}{\begin{pmatrix} g^2v^2 & 0 & -gg^\prime v^2 \\ 0 & g^2\hat v^2 & -gg^\prime \hat v^2 \\ -gg^\prime v^2 &  -gg^\prime \hat v^2 & g^{\prime 2}(v^2+\hat v^2) \end{pmatrix}}.\label{nmm}
\end{equation}
This mass matrix can be diagonalized by means of an orthogonal transformation $R$ which connects the weak eigenstates: ($W^3,~\hat W^3,~B$) to the physical mass eigenstates: ($Z,~\hat Z,~\gamma$);
\begin{equation}
{\begin{pmatrix} W^3 \\ \hat W^3 \\ B \end{pmatrix}}=R {\begin{pmatrix} Z \\\hat Z \\ \gamma \end{pmatrix}}.
\end{equation}
We have obtained the eigenvalues and eigenvectors of the matrix in Eq.~\ref{nmm}. The eigenvalues correspond to the masses of the physical states. One eigenstate ($\gamma$) has zero eigenvalue which is identified with the SM photon and the masses of other eigenstates are given by,
\begin{eqnarray}
M_Z^2&=& \frac{1}{4}v^2g^2\frac{g^2+2g^{\prime 2}}{g^2+g^{\prime 2}}\left[1-\frac{g^{\prime 4}}{\left(g^2+g^{\prime 2}\right)^{ {2}}}\epsilon\right],\nonumber\\
M_{\hat Z}^2&=& \frac{1}{4}\hat v^2\left(g^2+g^{\prime 2}\right)\left[1+\frac{g^{\prime 4}}{\left(g^2+g^{\prime 2}\right)^{ {2}}}\epsilon\right],\label{gbmass}
\end{eqnarray}
where, $\epsilon = v^2/\hat v^2$. Since  {we assume that} $\hat v >> v$, the ${\cal O}(\epsilon^2)$ terms in Eq.~\ref{gbmass}  {can be} neglected. The mixing matrix $R$ in the neutral gauge boson sector can be analytically expressed in terms of two mixing angle: $\theta_W$ and $\hat \theta_W$. The angles are defined in the following:
\begin{equation}
{\rm cos}^2\theta_W=\left(\frac{M_W^2}{M_Z^2}\right)_{\epsilon=0}=\frac{g^2+g^{\prime2}}{g^2+2g^{\prime 2}}~~,~~{\rm cos}^2\hat \theta_W=\left(\frac{M_{\hat W}^2}{M_{\hat Z}^2}\right)_{\epsilon=0}=\frac{g^2}{g^2+g^{\prime 2}}.
\end{equation}
The analytic expression for the mixing matrix upto ${\cal O}(\epsilon)$ is given by,
\begin{equation}
R={\begin{pmatrix}
-{\rm cos}\theta_W & -{\rm cos}\hat \theta_W{\rm sin}^2\hat \theta_W \epsilon & {\rm sin}\theta_W \\
{\rm sin}\theta_W {\rm sin} \hat\theta_W \left[ 1+\frac{{\rm cos}^2\hat \theta_W}{{\rm cos}^2\theta_W} \epsilon\right] & -{\rm cos}\hat \theta_W \left[1-{\rm sin}^4\hat\theta_W \epsilon \right] & {\rm sin}\theta_W \\
{\rm sin}\theta_W {\rm cos} \hat\theta_W \left[ 1-\frac{{\rm sin}^2\hat \theta_W}{{\rm cos}\theta_W} \epsilon\right] & {\rm sin}\hat \theta_W \left[1+{\rm sin}^2\hat\theta_W {\rm cos}^2\hat\theta_W \epsilon \right] & {\rm cos}\theta_W {\rm cos}\hat \theta_W
\end{pmatrix}}
\end{equation}  
It is important to  {note} that in the limit $\epsilon=0$, one recovers the SM gauge boson couplings. The couplings of our theory are related to the electric charge ($e$) by,
\begin{equation}
g=\frac{e}{{\rm sin}\theta_W},~~g^\prime=\frac{e}{{\rm cos}\theta_W {\rm cos}\hat \theta_W},~~
{\rm which~ implies},~~
\frac{1}{e^2}=\frac{2}{g^2}+\frac{1}{g^{\prime 2}}.
\end{equation}

Note that there are only two independent gauge couplings in the theory which we express in terms of 
$e$ and ${\rm cos}\theta_W$  {and therefore} $\hat{\theta}_W$ is not an independent angle, but 
is related to $\theta_W$  {as ${\rm sin} \hat \theta_W={\rm tan} \theta_W$}. 

\subsection{Fermion mass and mixing}
{\bf Charged fermion sector:}\\
The charged fermion mass Lagrangian includes Yukawa terms for the SM fermions and its mirror partners. Mass terms between the singlet SM fermions and mirror fermions are forbidden by the $Z_2$ symmetry. However, the Yukawa interactions between the singlet SM fermions and mirror fermions with the singlet scalar $\chi$ are allowed. The Lagrangian invariant under our gauge symmetry as well as the $Z_2$ symmetry for the down type quark and its mirror partner is given by,
\begin{eqnarray}
{\cal L} &\supset& y_d \left(\bar Q^0_L \Phi d^0_R + \bar {\hat Q}^0_R \hat \Phi \hat d^0_L\right) + h_d ~\chi \bar d_R \hat d_L + {\rm h.c.}\nonumber\\
&\supset& {\begin{pmatrix} \bar d_L^0 & \bar {\hat d}^0_L \end{pmatrix}}{\begin{pmatrix}\frac{y_d v}{\sqrt 2} & 0 \\ M^*_{d\hat d}  & \frac{y^*_d \hat v}{\sqrt 2}\end{pmatrix}}{ {\begin{pmatrix} d^0_R \\  {\hat d}^0_R \end{pmatrix}}}~+~{\rm h.c.},
\end{eqnarray}  
where, $y_d$ and $h_d$ are the Yukawa couplings for the SM $d$-quark and $M_{d \hat d}=h_d v_\chi$. It is important to mention that $h_d$, in general, is a $3\times 3$ matrix for 3 families and gives rise to flavor mixing. Flavor mixing in the leptonic sector results into lepton flavor violating (LFV) processes like $\mu \to e \gamma$, $\tau \to \mu \gamma$, $\mu \to eee$ e.t.c. which are highly constrained from BaBar \cite{babar} abd Belle \cite{belle} experiments. For example, in the present model, dominant contribution to the flavor violating $\mu$ or $\tau$ decay arises from the diagram with singlet scalar ($\chi$) and mirror lepton propagating in the loop. LFV processes in the context of models with TeV scale mirror fermions have already been studied in Ref. \cite{hung}.

To ensure LR symmetry, we have used  {the} same Yukawa coupling for the ordinary and  the mirror sector. Notice that the Yukawa terms involving $\chi$ introduce mixing between SM and mirror fermions. The charged fermion mass matrix can be diagonalized via bi-unitary transformation by introducing two mixing angles. The charged fermion mass (physical) eigenstates are related to the gauge eigenstates by the following relation:
\begin{equation}
{\begin{pmatrix} f^0 \\  {\hat f}^0 \end{pmatrix}}_{L,R}~=~{\begin{pmatrix}{\rm cos}\theta^f & {\rm sin}\theta^f \\-{\rm sin}\theta^f & {\rm cos}\theta^f \end{pmatrix}}_{L,R}{\begin{pmatrix}  f \\  {\hat f} \end{pmatrix}}_{L,R}
\end{equation}
where, $f_{L,R}$ can be identified with the L and R-handed component of the SM fermions and $\hat f_{L,R}$ corresponds to the heavy mirror fermions. The masses and mixing angles are given by:
\begin{eqnarray}
m_{f}~=~\frac{y_f v}{\sqrt 2}&,& m_{\hat f}~=~\sqrt{\frac{y^2_{ {f}}\hat v^2+ {2}M_{f\hat f}^2}{2}};\nonumber\\
{\rm tan}2\theta^f_{R}~=~\frac{2{\sqrt{2}}y_f M_{f\hat f}\hat v}{y_f^2\left(v^2-\hat v^2\right)+ {2}M_{f\hat f}^2}&,& {\rm tan}2\theta^f_{L}~=~\frac{2{\sqrt{2}}y_f M_{f\hat f}v}{y_f^2\left(v^2-\hat v^2\right)- {2}M_{f\hat f}^2}.
\label{tan}
\end{eqnarray}

\medskip
\noindent{\bf Neutrino Sector}:\\
The SM and singlet neutrinos (both in the ordinary and the mirror sector) are even under the $Z_2$ symmetry. Therefore, the mass terms between $SU(2)_L$ and $SU(2)_R$ singlet neutrinos are allowed.
The Lagrangian allowed by our gauge symmetry and  {respecting the} discrete $Z_2$ symmetry is given by
 {\begin{eqnarray*}
{\cal L} &\supset& f_{\nu} \left(\bar l^0_L \Phi \nu^0_R + \bar {\hat l}^0_R \hat \Phi \hat \nu^0_L\right) +M  \nu^{0T}_R C^{-1}  \nu^0_R +  h_\nu \chi \bar{\nu}^0_R \hat \nu^0_L
+ M \hat \nu^{0T}_L C^{-1} \hat \nu^0_L + {\rm h.c.}
%&\supset& {\begin{pmatrix} \bar d_L^0 & \bar {\hat d}^0_L \end{pmatrix}}{\begin{pmatrix}\frac{y_d v}{\sqrt 2} & 0 \\ M^*_{d\hat d}  & \frac{y^*_d \hat %v}{\sqrt 2}\end{pmatrix}}{\begin{pmatrix} \bar d^0_R \\ \bar {\hat d}^0_R \end{pmatrix}}~+~{\rm h.c.},
\end{eqnarray*}}
%where l's are lepton doublets.  
where $f_{\nu}$ is the neutrino Yukawa coupling, and M is the singlet neutrino mass of order $\hat{v}$.
The neutrino mass matrix with both Dirac  {mass ($m=f_\nu v/\sqrt{2}$, $m^\prime=f_\nu \hat v/\sqrt{2}$  and $M_{\nu \hat \nu}=h_\nu v_\chi$}) and Majorana mass  {($M$)} %\footnote{{\red Note that, it is not %necessary to have same Majorana mass terms for the SM and mirror singlet neutrinos. Here, $M \sim %\hat v$ represent order of magnitude of the Majorana mass terms.}} 
terms in $(\nu^0_L,\nu^0_R,\hat \nu^0_R, \hat \nu^0_L)$ basis is given by,
\begin{equation}
{\begin{pmatrix}
0 & m & 0 & 0 \\
m & M & 0 & M_{\nu \hat \nu} \\
0 & 0 & 0 & m' \\
0 & M_{\nu \hat \nu} & m'& M\\
\end{pmatrix}}.
\end{equation}  
%where, $f$ is the  neutrino Yukawa coupling and $M$ is the singlet neutrino mass which should be of order $\hat{v}$.  Assuming $M \sim \hat v$, 
Assuming $M_{\nu \hat \nu}\sim M$, the order of magnitude for the eigenvalues of the neutrino mass matrix are given 
\begin{equation}
-m^2/M,~m^\prime/ \sqrt{2}, -m^\prime/\sqrt{2},~2M.
\end{equation}
Thus to generate a light neutrino mass  $\simeq 10^{-11}$ GeV with a  {Yukawa coupling strength} of $f_\nu \sim 10^{-4(6)}$  (which is  somewhat similar to the Yukawa coupling of the electron), we need  $\hat{v}\sim 10^{7(3)}$ GeV. This $\hat{v}\sim 10^{7(3)}$ scale and $M_{f\hat f}$ (see Eq.~\ref{tan}) determines the masses of the mirror fermions. For the first family, the mirror fermion masses then  {come out to be} in the few hundred GeV to TeV range. 
 {Note that to fit the neutrino mass and mixing angles to experimental data would require a more detailed analysis of the neutrino sector which we leave for future studies. Another realization with a mirror like symmetry 
to generate neutrino masses was considered in Ref~\cite{Hung:2006ap}. }

\begin{table}[t!]
\begin{center}
\begin{tabular}{||c|c||c|c||}
\hline \hline
$f$ & $f^\prime$ & $A_{ff^\prime}^{W}$ & $A_{ff^\prime}^{\hat W}$ \\\hline\hline
$d$ & $u$ & ${\rm cos}^2 \theta_L$ & ${\rm sin}^2\theta_R$ \\
$d$ & $\hat u$ & ${\rm cos}\theta_L{\rm sin}\theta_L$ & -${\rm cos}\theta_R {\rm sin}\theta_R$ \\
$\hat d$ & $u$ & ${\rm cos}\theta_L{\rm sin}\theta_L$ & -${\rm cos}\theta_R {\rm sin}\theta_R$ \\
$\hat d$ & $\hat u$ & ${\rm sin}^2 \theta_L$ & ${\rm cos}^2\theta_R$ \\\hline\hline
\end{tabular}
\end{center}
\caption{\small{Analytical expressions for $A_{ff^\prime}^{W}$ and $A_{ff^\prime}^{\hat W}$. Note that we have assumed $V_{ud}=1$. We have also assumed fermion mixing angles ($\theta_L$ and $\theta_R$) are same for up and down flavor.}}
\label{tab:cc}
\end{table}
\section{Phenomenology}\label{sec:pheno}
In this section, we discuss the  {collider} phenomenology of the LRMM. Before going into the details of the collider signatures of LRMM, we  {first need to} study the properties of mirror fermions and bosons. From the point of view of collider phenomenology, we are interested in the interactions between SM particles and mirror particles  {which give the production and decay properties of the mirror particles}. The Lagrangian for the charge currents with $W^\pm$ and $\hat W^\pm$ boson contributions are given by, 
\begin{equation}
{\cal L}_{CC}=-\frac{g}{2\sqrt 2}\bar f \gamma^\mu \left[A_{ff^\prime}^{W}(1-\gamma^5)W_\mu^{-}+A_{ff^\prime}^{\hat{W}}(1+\gamma^5)\hat{W}_\mu^{-}\right]f^\prime,
\label{eq:cc}
\end{equation}
where the coefficients $A_{ff^\prime}^{W}~{\rm and}~A_{ff^\prime}^{\hat{W}}$ depend on the charged fermion mixing angles: $\theta_L~{\rm and}~\theta_R$. The analytical expressions for these coefficients are presented in Table~\ref{tab:cc}\footnote{Fermion mixing angles ($\theta_L~{\rm and}~\theta_R$) depend on the Yukawa coupling of the corresponding fermion. Therefore, the mixing angles are different for up and down flavor. However, we have used the same symbol for the mixing angles of up and down quarks.} for up and down flavored SM and mirror fermions. The neutral current interactions of fermions with neutral gauge bosons ($\gamma,~Z$ and $\hat Z$-bosons) are described by the following Lagrangian. 
\begin{eqnarray}
{\cal L}_{NC} = &-& e Q_f \bar f \gamma^\mu A_\mu f \nonumber\\
&-&\frac{1}{6}\frac{g}{{\rm cos}^3\theta_W}\bar f \gamma^\mu \left[A_{ff^\prime}^Z \frac{1-\gamma^5}{2}+B_{ff^\prime}^Z \frac{1+\gamma^5}{2}\right] Z_\mu f^\prime \nonumber\\
&-&\frac{1}{6}\frac{g}{{\rm cos}^3\theta_W \sqrt{{\rm cos}2\theta_W}}\bar f \gamma^\mu \left[A_{ff^\prime}^{\hat Z} \frac{1-\gamma^5}{2}+B_{ff^\prime}^{\hat Z} \frac{1+\gamma^5}{2}\right] {\hat Z}_\mu f^\prime ,
\label{eq:NC}
\end{eqnarray}
where, $e$ is electron charge and $Q_f$ is the charge of fermion $f$. For up and down flavored SM and mirror fermions, analytical expressions upto $O(\epsilon)$ for the coefficients 
\begin{table}[th!]
\small{
\begin{center}
\begin{tabular}{||c|c||c|c||}
\hline \hline
$f$ & $f^\prime$ & $A_{ff^\prime}^{Z}$ & $B_{ff^\prime}^{Z}$ \\\hline\hline
$d$ & $d$ & $3{\rm cos}^2\theta_L{\rm cos}^2\theta_W-2{\rm cos}^2\theta_W{\rm sin}^2\theta_W$ & $-2 {\rm cos}^2{\theta_W}{\rm sin}^2\theta_W-3 {\rm sin}^2{\theta_R}    {\rm sin}^2{\theta_W}    \sqrt{{\rm cos}2\theta_W}\epsilon$\\
& & $-(1-3{\rm sin}^2\theta_L){\rm sin}^2\theta_W\epsilon$ & $+(2-3 {\rm sin}^2{\theta_R})   {\rm sin}^3{\theta_W}    \epsilon $\\\hline
$d$ & $\hat d$ & $3{\rm cos}^2{\theta_W}    {\rm sin}\theta_L   {\rm cos}{\theta_L}- 3{\rm sin}\theta_L   {\rm sin}^3{\theta_W}    {\rm cos}{\theta_L}   \epsilon$ & $3{\rm sin}{\theta_R}   {\rm sin}^2{\theta_W}   {\rm cos}{\theta_R}   \sqrt{{\rm cos}2\theta_W}\epsilon$\\
& & & $+3{\rm sin}{\theta_R}   {\rm sin}^3{\theta_W}    {\rm cos}{\theta_R}   \epsilon$ \\\hline
$\hat d$ & $\hat d$ & $3 {\rm cos}^2{\theta_W}    {\rm sin}^2\theta_L-2 {\rm cos}^2{\theta_W}    {\rm sin}^2{\theta_W}$ & $-3 {\rm cos}^2{\theta_R}    {\rm sin}^2{\theta_W}    \sqrt{{\rm cos}2\theta_W}\epsilon$ \\
&& $+ (2-3 {\rm sin}^2\theta_L)   {\rm sin}^3{\theta_W}    \epsilon$ & $- (1-3 {\rm sin}^2{\theta_R})   {\rm sin}^3{\theta_W}    \epsilon$ \\\hline
$u$ & $u$ & $-3 {\rm cos}^2{\theta_L}    {\rm cos}^2{\theta_W}+4 {\rm cos}^2{\theta_W}    {\rm sin}^2{\theta_W}$ & $+4 {\rm cos}^2{\theta_W}    {\rm sin}^2{\theta_W}+3 {\rm sin}^2{\theta_R}    {\rm sin}^2{\theta_W}    \sqrt{{\rm cos}2\theta_W}\epsilon$\\
& & $- (1+3 {\rm sin}^2\theta_L)   {\rm sin}^3{\theta_W}    \epsilon$ & $- (4-3 {\rm sin}^2{\theta_R})   {\rm sin}^3{\theta_W}    \epsilon$\\\hline
$u$ & $\hat u$ & $-3{\rm cos}^2{\theta_W}    {\rm sin}\theta_L   {\rm cos}{\theta_L}+3 {\rm sin}\theta_L   {\rm sin}^3{\theta_W}    {\rm cos}{\theta_L}   \epsilon$ & $-3 {\rm sin}{\theta_R}   {\rm sin}^2{\theta_W}    {\rm cos}{\theta_R}   \sqrt{{\rm cos}2\theta_W}\epsilon$\\
& & & $-3 {\rm sin}{\theta_R}   {\rm sin}^3{\theta_W}    {\rm cos}{\theta_R}   \epsilon$ \\\hline
$\hat u$ & $\hat u$ & $-3 {\rm cos}^2{\theta_W}    {\rm sin}^2\theta_L+4 {\rm cos}^2{\theta_W}    {\rm sin}^2{\theta_W}$ & $4 {\rm cos}^2{\theta_W}    {\rm sin}^2{\theta_W}+3 {\rm cos}^2{\theta_R}    {\rm sin}^2{\theta_W}    \sqrt{{\rm cos}2\theta_W}\epsilon$\\
& & $- (4-3 {\rm sin}^2\theta_L)   {\rm sin}^3{\theta_W}    \epsilon$ & $- (1+3 {\rm sin}^2{\theta_R})   {\rm sin}^3{\theta_W}    \epsilon$\\\hline\hline
\end{tabular}
\end{center}
\caption{\small{Analytical expressions for $A_{ff^\prime}^{Z}$ and $B_{ff^\prime}^{Z}$.}}
\label{tab:NCZ} }
\end{table}
$A_{ff^\prime}^Z,~B_{ff^\prime}^Z$ 
%and  $A_{ff^\prime}^{\hat Z},~B_{ff^\prime}^{\hat Z}$ 
are presented in Table~\ref{tab:NCZ}.
% and~\ref{tab:NCZp} respectively. 
%
The interactions of fermions with the SM Higgs and mirror Higgs are described in Eq.~\ref{eq:higgs}.
\begin{eqnarray}
{\cal L}_{S}=&&\frac{y_f}{\sqrt 2}\bar f \left[A_{ff^\prime}^{H}\frac{1-\gamma^5}{2}+B_{ff^\prime}^{H}\frac{1+\gamma^5}{2}\right]H f^{\prime}\nonumber\\
&&\frac{y_f}{\sqrt 2}\bar f \left[A_{ff^\prime}^{\hat H}\frac{1-\gamma^5}{2}+B_{ff^\prime}^{\hat H}\frac{1+\gamma^5}{2}\right]\hat H f^{\prime},
\label{eq:higgs}
\end{eqnarray}
where, $y_f$ is the Yukawa coupling of fermion $f$. The expressions for the coefficients $A_{ff^\prime}^{H},~B_{ff^\prime}^{H},~A_{ff^\prime}^{\hat H}~{\rm and}~B_{ff^\prime}^{\hat H}$ can be found in Table~\ref{tab:higgs}. It is important to note that in the limit $\epsilon=0$ and ${\rm cos}\theta_{L,R}=1$, the SM fermions decouple from the mirror fermions and we recover the SM 
%gauge boson 
couplings.

The decays of the TeV scale mirror fermions into $\hat W$, $\hat Z$ or $\hat H$ are kinematically forbidden since the mass of these mirror bosons are proportional to $\hat v \sim 10^7$ GeV. Because of the mixing of the mirror fermions with the ordinary fermions, the mirror fermions can decay into a SM fermion, and a $Z,~ W$ or a Higgs boson. The  {expressions for the partial} decay widths are:
\begin{eqnarray}
\Gamma({\hat f} \to f Z)&=&\frac{g^2}{36{\rm cos}^6\theta_w}\frac{\left({A_{ff^\prime}^{Z}}\right)^2+\left({B_{ff^\prime}^{Z}}\right)^2}{64\pi}\frac{M_{\hat f}^3}{M_Z^2}\left(1-\frac{M_Z^2}{M_{\hat f}^2}\right)^2\left( 1+2\frac{M_Z^2}{M_{\hat f}^2}\right),\nonumber\\
\Gamma({\hat f} \to f^\prime W)&=&\frac{g^2}{8}\frac{\left({A_{ff^\prime}^{W}}\right)^2+\left({B_{ff^\prime}^{W}}\right)^2}{16\pi}\frac{M_{\hat f}^3}{M_W^2}\left(1-\frac{M_W^2}{M_{\hat f}^2}\right)^2\left( 1+2\frac{M_W^2}{M_{\hat f}^2}\right),\nonumber\\
\Gamma({\hat f} \to f H)&=&\frac{y_f^2}{2}\frac{\left({A_{ff^\prime}^{H}}\right)^2+\left({B_{ff^\prime}^{H}}\right)^2}{64\pi}M_{\hat f}\left(1-\frac{M_H^2}{M_{\hat f}^2}\right)^2,
\end{eqnarray}  
where, $M_Z,~M_W,~M_H~{\rm and}~M_{\hat f}$ are the masses of $Z,~W$, Higgs and mirror fermion 
\begin{table}[h!]
\begin{center}
\begin{tabular}{||c|c||c|c||c|c||}
\hline \hline
$f$ & $f^\prime$ & $A_{ff^\prime}^{H}$ & $B_{ff^\prime}^{H}$ & $A_{ff^\prime}^{\hat H}$ & $B_{ff^\prime}^{\hat H}$  \\\hline\hline
$f$ & $f$ & ${\rm cos}\theta_L{\rm cos}\theta_R$ & ${\rm cos}\theta_L{\rm cos}\theta_R$ & ${\rm sin}\theta_L{\rm sin}\theta_R$ & ${\rm sin}\theta_L{\rm sin}\theta_R$\\\hline
$f$ & $\hat f$ & ${\rm sin}\theta_L{\rm cos}\theta_R$ & ${\rm cos}\theta_L{\rm sin}\theta_R$ & -${\rm cos}\theta_L{\rm sin}\theta_R$ & -${\rm sin}\theta_L{\rm cos}\theta_R$\\\hline
$\hat f$ & $\hat f$ & ${\rm sin}\theta_L{\rm sin}\theta_R$ & ${\rm sin}\theta_L{\rm sin}\theta_R$& ${\rm cos}\theta_L{\rm cos}\theta_R$ & ${\rm cos}\theta_L{\rm cos}\theta_R$\\\hline\hline
\end{tabular}
\end{center}
\caption{\small{Analytical expressions for $A_{ff^\prime}^{H}$, $B_{ff^\prime}^{H}$, $A_{ff^\prime}^{\hat H}$ and $B_{ff^\prime}^{\hat H}$.}}
\label{tab:higgs}
\end{table}
respectively. Apart from the known SM parameters and mirror fermion masses, the decay widths of mirror fermions depend on $\epsilon$, $\theta_L~{\rm and}~\theta_R$. For $\hat v \sim 10^7$ GeV, the value of $\epsilon$ is about $10^{-10}$. Therefore, the terms proportional to $\epsilon$ in the decay widths can be safely neglected. The mirror fermions decay widths depend primarily on the fermion mixing angles. According to Eq.~\ref{tan}, the fermion mixing angles are determined in terms of two parameters, namely, $\hat v$ and $M_{f\hat f}$.  
%------------------------------------------------------------------
\begin{figure}
\begin{center}
\epsfig{file=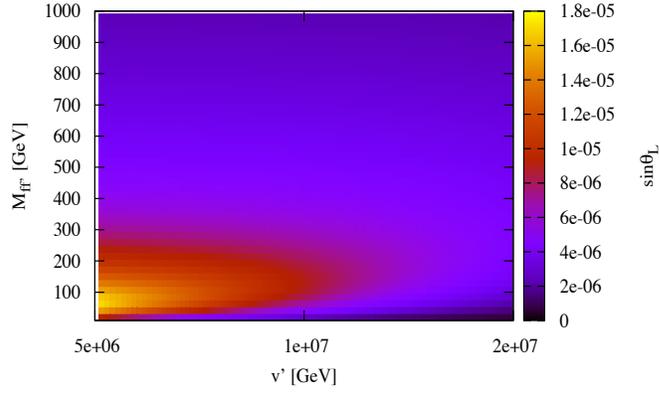,width=12cm,height=10cm}
\epsfig{file=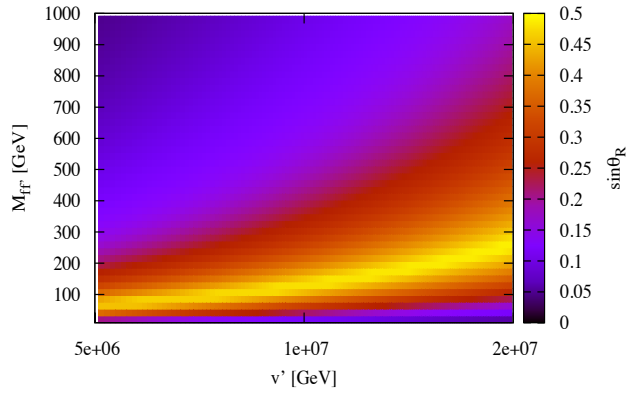,width=12cm,height=10cm}
\end{center}
\caption{\small{Fermion mixing angles, ${\rm sin}\theta_L$ (left panel) and ${\rm sin}\theta_R$ (right panel), for the up quark are presented by color gradient on the LRMM parameter space defined by $\hat v$ (along x-axis) and $M_{f\hat f}$ (along y-axis). The up quark Yukawa coupling, $y_u=1.3\times 10^{-5}$, and the SM VEV, $v=250$ GeV are assumed in these plots. }}
\label{mixing}
\end{figure}
%-------------------------------------------------------------------
Assuming the up quark Yukawa coupling, $y_u=1.3\times 10^{-5}$ and the SM VEV, $v=250$ GeV, in Fig.~\ref{mixing}, we show the mixing angles, ${\rm sin}\theta_L$ (left panel) and ${\rm sin}\theta_R$ (right panel), by color gradient, in the $\hat v$-$M_{f\hat f}$ plane. Eq.~\ref{tan}, shows that ${\rm tan}2\theta_L$ is suppressed by the SM quark mass ($\sim y_f v$) in the numerator and mirror 
quark mass ($\sim \sqrt{y_f^2 \hat v^2+2M_{f\hat f}^2}$) in the denominator. Therefore, for a MeV scale SM quark and TeV scale mirror partner, the value of ${\rm sin}\theta_L$ is about $10^{-6}$ which can be seen in Fig.~\ref{mixing} (left panel). Whereas, Fig.~\ref{mixing} (right panel) shows that,  ${\rm sin}\theta_R$ can be large depending on the values of $\hat v$ and $M_{f\hat f}$. 
%------------------------------------------------------------------
\begin{figure}
\begin{center}
\epsfig{file=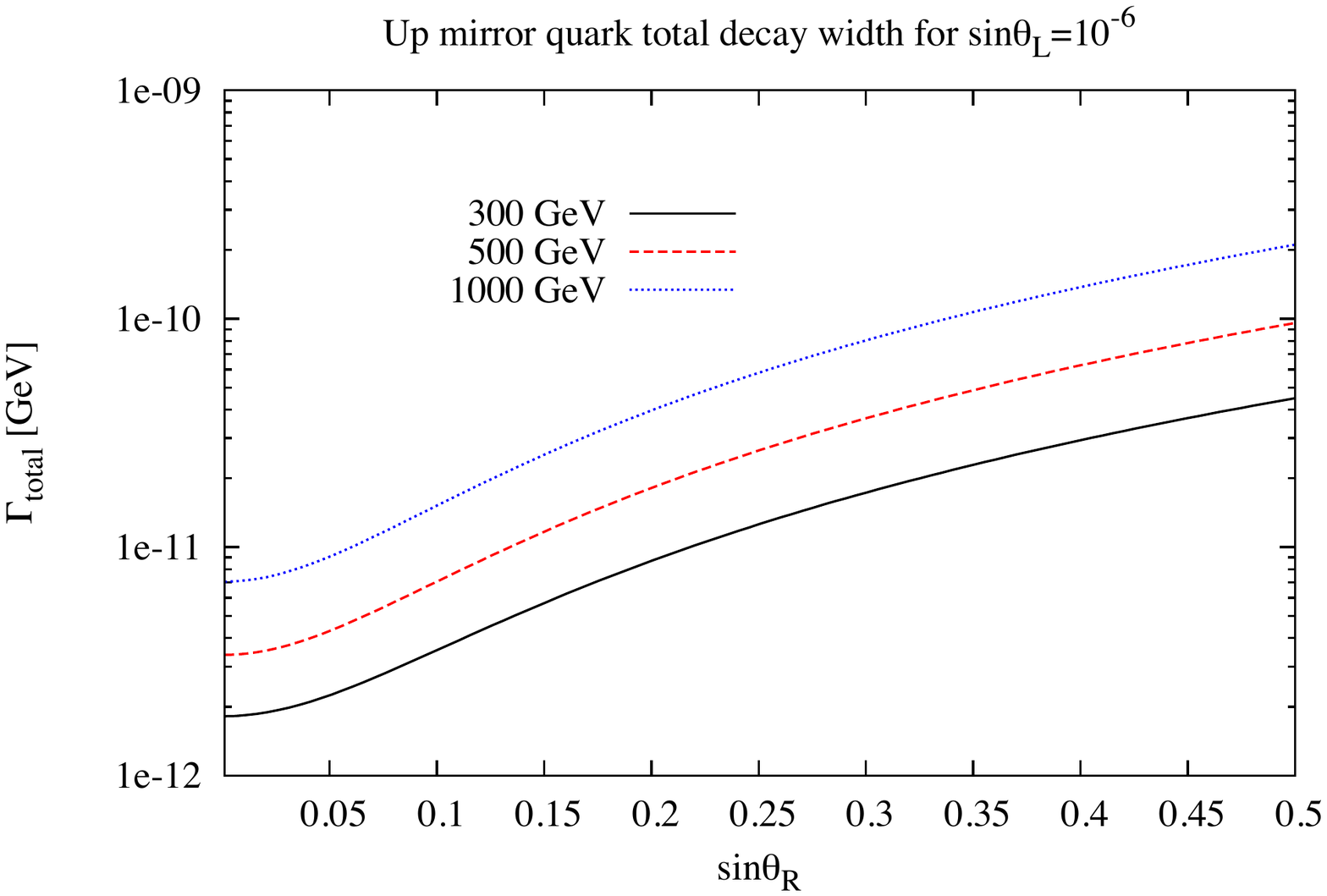,width=6in,height=5in}
\end{center}
\caption{\small{Total decay width of up-type mirror quark for three different values of 
$M_{\hat u}=300,~500~{\rm and}~1000$ GeV as a function of ${\rm sin}\theta_R$. We have 
assumed lowest possible value for ${\rm sin}\theta_L=10^{-6}$ in this plot.}}
\label{fig:width}
\end{figure}
%-------------------------------------------------------------------

The neutral (see Eq.~\ref{eq:NC}) and charge (see Eq.~\ref{eq:cc}) current interactions of mirror quarks with SM quarks and $Z/W$ bosons are suppressed by ${\rm sin}\theta_L$. Moreover, the interactions of mirror quarks with the SM quarks and Higgs boson are suppressed by the Yukawa couplings. Therefore, before going into the details of collider analysis, it is important to ensure that light mirror quarks decay inside the detectors of the LHC experiment. In Fig.~\ref{fig:width}, we plot the total decay width of up-type mirror 
quark as a function of ${\rm sin}\theta_R$ for three different values of the mirror quark mass, {\it viz.},
$M_{\hat u}=300,~500~{\rm and}~1000$ GeV. 
We have considered the lowest possible value of ${\rm sin}\theta_L=10^{-6}$ in Fig.~\ref{fig:width}. According to Fig.~\ref{fig:width}, the total decay width of up-type mirror quark is always greater than $10^{-12}$ GeV, $\Gamma_{total}>10^{-12}$ GeV, which  corresponds to a mean distance of $c\tau < 10^{-3}$ cm (without including Lorentz boost) traversed by a mirror quark inside a detector before its decay. These numbers assure us that 
the mirror quarks will always decay inside the detector for a wide range of model parameters.
%------------------------------------------------------------------
\begin{figure}
\begin{center}
\epsfig{file=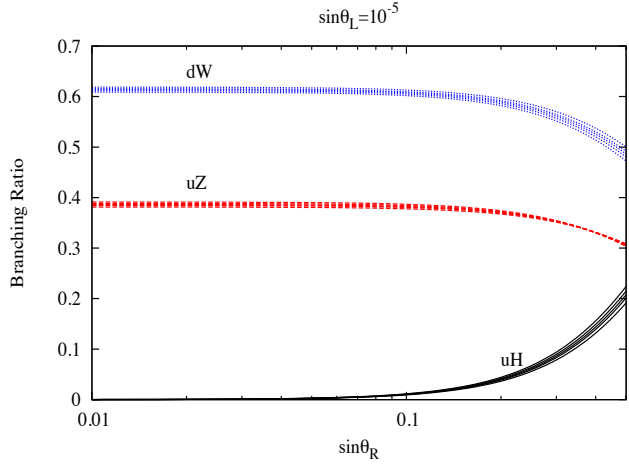,width=12cm,height=10cm}
\epsfig{file=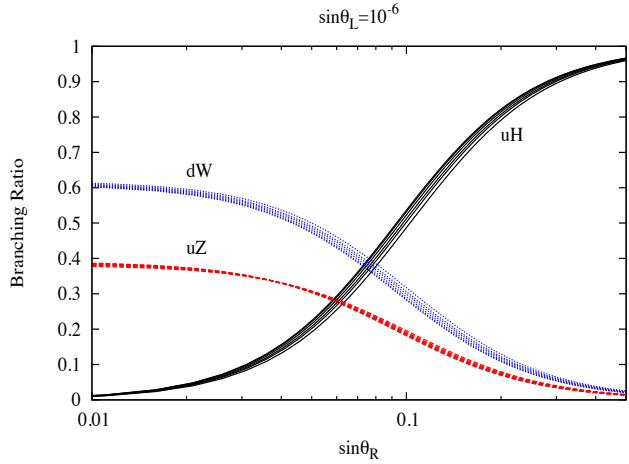,width=12cm,height=10cm}
\end{center}
\caption{\small{ Illustrating the up-type mirror quark branching ratios in $dW$, $uZ$ and $uH$ channel as a function of ${\rm sin}\theta_R$ for two different values of ${\rm sin}\theta_L=10^{-5}$ (left panel) and $10^{-6}$ (right panel). We have varied $\hat u$  mass over a range between 300 GeV to 1 TeV which gives rise to the bands instead of  lines.}}
\label{BR}
\end{figure}
%-------------------------------------------------------------------

In Fig.~\ref{BR}, we plot the branching ratios for the up-type mirror quark into $dW$, $uZ$ and $uH$ channel as a function of ${\rm sin}\theta_R$. We have assumed two different values of ${\rm sin}\theta_L=10^{-5}$ (left panel) and $10^{-6}$ (right panel). We have varied the mirror quark mass over 300 GeV to 1 TeV which gives rise to the bands in Fig.~\ref{BR}. Fig.~\ref{BR} (left panel) shows that for ${\rm sin}\theta_L=10^{-5}$, the decay of $\hat u$ into SM vector bosons dominates over the decay into Higgs boson. Whereas, for ${\rm sin}\theta_L=10^{-6}$ (right panel), the decay into vector bosons dominates only in the low ${\rm sin}\theta_R$ region (${\rm sin}\theta_R<0.08$).  

%%%%%%%
\subsection{Signature of mirror fermions at the LHC}
In this section, we will first discuss the production of TeV scale mirror quarks, namely ${\hat u} ~{\rm and}~{\hat d}$ quarks, at the LHC. As a consequence of the $Z_2$ symmetry, the couplings between a mirror quark and the SM particles are forbidden. Therefore, in presence of this $Z_2$ symmetry, the single production of the mirror fermions is not possible at the collider. As discussed in the previous section, spontaneous breaking of the $Z_2$ symmetry introduces mixing between the mirror and SM quarks and thus, gives rise to interactions between mirror and SM quarks with a $Z$, $W$ or Higgs boson. However, the single production rates of TeV scale mirror quarks via the $Z_2$ symmetry violating couplings are suppressed by the quark mixing angles. Therefore, in this work, we have considered the pair production of mirror quarks at the LHC. 

%------------------------------------------------------------------
\begin{figure}
\begin{center}
\epsfig{file=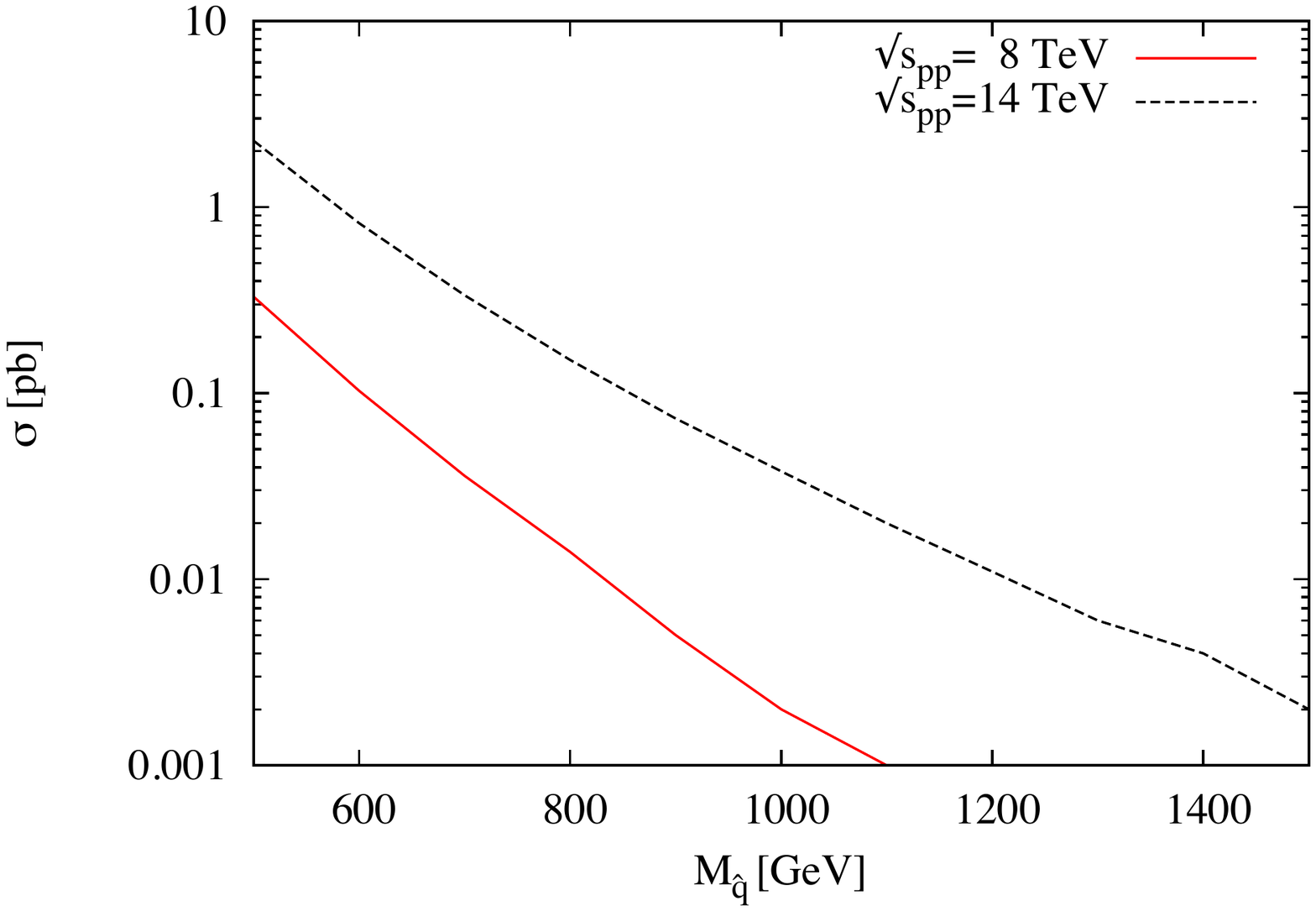,width=5in,height=5in}
\end{center}
\caption{\small{Pair production cross-sections of mirror quarks as a function of their
masses in proton proton collisions at center-of-mass energies 8 TeV and 14 TeV respectively.}}
\label{cross}
\end{figure}
\begin{figure}
\begin{center}
\epsfig{file=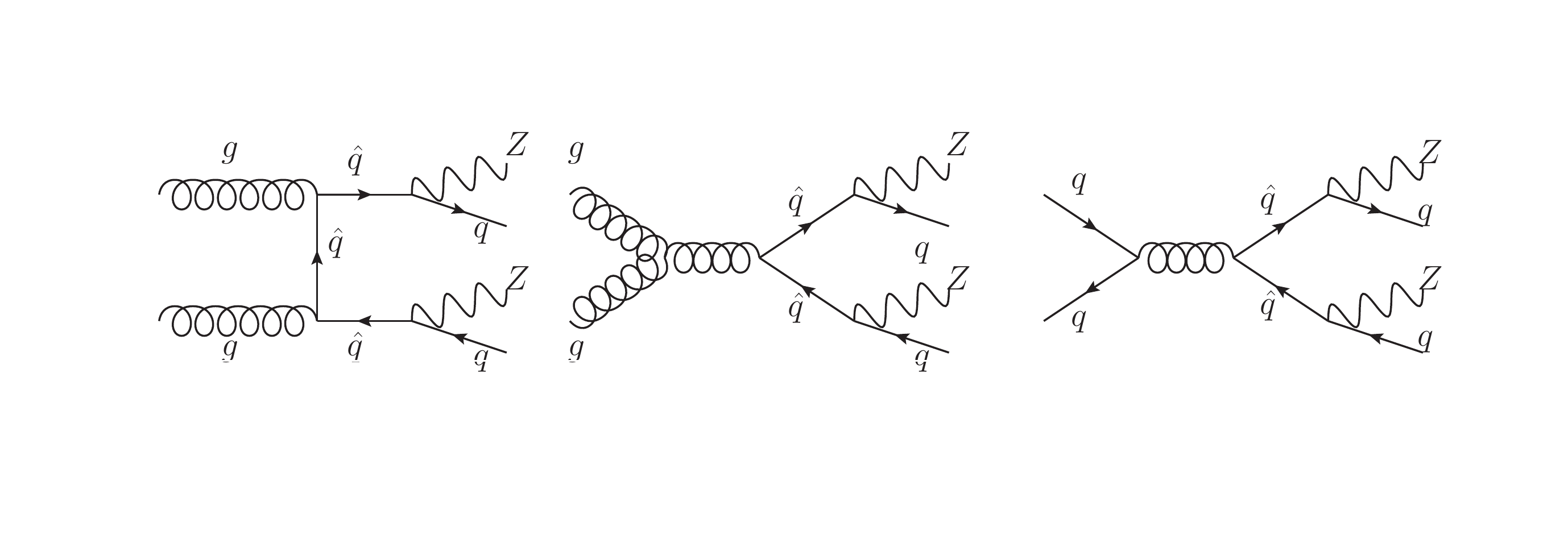,width=15cm,height=6cm}
\end{center}
\caption{\small{Feynman diagrams for the ${\hat q} \bar {\hat q}$ production and their subsequent decay to $q Z$.} }
\label{FD}
\end{figure}

%-------------------------------------------------------------------
As the mirror quarks carry $SU(3)_C$ quantum numbers, they couple directly to the gluons. 
%All the mirror quarks have tree level couplings with respective mirror anti-quarks and a SM gluon. 
The pair production of TeV scale mirror quarks, namely ${\hat u} \bar {\hat u}$ and ${\hat d} \bar {\hat d}$ production, in a proton-proton collision therefore is analogous to that of the pair production of SM heavy quarks, the analytic expressions for which can be found in Ref.~\cite{top}. 
Both gluon-gluon ($gg$) and  
%(through $t(u)$-channel mirror quark exchange and $s$-channel SM gluon exchange) as well as from 
quark-antiquark ($q \bar q$) initial states 
%(through $s$-channel SM gluon exchange diagram) 
contribute to the pair production (${\hat q} \bar {\hat q}$) of mirror quarks (see Fig.~\ref{FD}). For numerical evaluation of the cross-sections, we have used a tree-level Monte-Carlo program incorporating CTEQ6L \cite{cteq6l} parton distribution functions. Both the renormalization and the
factorization scales have been set equal to the subprocess center-of-mass energy $\sqrt {\hat s}$. The ensuing leading-order (LO) ${\hat q} \bar {\hat q}$ production cross-sections are presented in Fig.~\ref{cross} as a function of mirror quark mass ($M_{\hat q}$) for two different values of the proton-proton center-of-mass energy {\it viz.,} $\sqrt{s}_{pp}=8$ TeV and $14$ TeV. While the NLO and NLL corrections can be well estimated by a proper rescaling of the corresponding results for $t\bar t$ production, we deliberately resist from doing so. With the K-factor expected to be large \cite{ttk}, our results would thus be a conservative one. The pair production cross section is found to be a few hundred femtobarns (fb) for mirror quark mass 
of close to a TeV. As discussed before, these mirror quarks once produced will decay within the detector.  
 %After a very short discussion about the pair production cross-sections of TeV scale mirror quarks,
 We now analyze the possible signatures of mirror quarks at the LHC following its decay properties.  Mirror 
quarks can decay into a $Z$-boson, a $W$-boson or Higgs boson in association with a SM quark: ${\hat q} \to q Z,~q^\prime W~{\rm and}~qH$. Thus the pair production of mirror quarks, at the LHC, gives rise to a pair of heavy SM bosons ($Z$-boson, $W$-boson or Higgs boson) in association with multiple jets in the final state. 
%------------------------------------------------------------------
%-------------------------------------------------------------------
In this work, we have focused on the signal with the vector bosons in the final states. 
%We do not consider the final states with Higgs bosons. Therefore, our collider analysis is only applicable to those part of 
We choose the LRMM parameter space where the decay of mirror quarks into vector bosons dominates over its decay into Higgs boson. Fig.~\ref{BR} shows that for negligible $\hat q \to qH$ branching ratio, the mirror quarks decay into $qW$ and $qZ$ pairs with about 61\% and 39\% branching probability respectively. In the rest of our analysis, we have used the above mentioned values for the decay probability to compute the signal cross-sections. Pair production and the decay of mirror quarks in to $qW$ and $qZ$ channels gives rise to the following signatures:
\begin{itemize}
\item {\it {2 jets}+2 $Z$ } final state arises when both mirror quarks decay into $qZ$ pairs.
 $$
pp \to {\hat q} \bar {\hat q} \to (qZ)(\bar q Z)
$$
The production and decay of mirror quarks in this channel are schematically shown in Fig.~\ref{FD}. 
\item {\it {2 jets}+$Z$+$W$ } final state results when one mirror quark decays into $qZ$ channel and other one decays into $qW$ channel.
$$
pp \to {\hat q} \bar {\hat q} \to (qZ)(\bar q^\prime W)
$$
\item If both mirror quarks decay into $qW$ channel then pair production of mirror quarks gives rise to {\it {2 jets}+2 $W$} final state. 
\end{itemize}  
We consider the reconstruction of mirror quark mass from the invariant mass distribution of $qZ$ pairs which is possible for the first two signal topologies only. Therefore, we have only considered  {\it {2 jets}+2 $Z$} and {\it {2 jets}+$Z$+$W$} final states for further 
analysis. Note that in the leptonic channel the $Z$ reconstruction would be very clean while for the {\it {2 jets}+$Z$+$W$}, even the $W$ can be 
reconstructed well as there is only a single neutrino in the final state. The $W$'s can be reconstructed in the all hadronic mode but with significant
challenge in efficiencies in a hadronic machine such as the LHC. So we have chosen to neglect the {\it {2 jets}+2 $W$} final state in our analysis.

\subsubsection{{\it {2 jets}+2 $Z$-bosons} signature}
In this section, we have investigated {\it 2 jets + 2 $Z$ } final state as a signature of mirror quarks in the framework of LRMM. We have used a parton level Monte-Carlo simulation to evaluate the cross-sections and different kinematic distributions for the signal. We have assumed that $Z$-bosons decaying into leptons (electrons and muons) can be identified at the LHC with good efficiency. Therefore, in our parton level analysis, we consider $Z$-boson as a standard object\footnote{All the cross-sections (signal as well as background) presented in the next part of this article are multiplied by the leptonic branching fraction (6.7\% in electron and muon channel) of the $Z$-boson.} without simulating its decay to leptons.  {We must however point out that the total number of signal events are crucial 
in identifying the $Z$ boson in the leptonic channel because of the small branching probability of the $Z$ decaying to charged leptons.}

The dominant SM background to the signal comes from the pair production of $Z$-bosons in association with two jets. Before going into the details of signal and background, it is important to list a set of basic requirements for jets to be visible at the detector. To parametrize detector acceptance and enhance signal to background ratio, we have imposed kinematic cuts ({\it Acc. Cuts}), listed in Table~\ref{cuts}, on the jets (denoted by $j_1$ and $j_2$) after ordering the jets according to their transverse momentum ($p_T$) hardness ($p_T^{j_1}>p_T^{j_2}$). It should also be realized that any detector has only a finite resolution. For a realistic detector, this applies to both energy/transverse momentum measurements as well as determination of the angle of motion. For our purpose, the latter can be safely neglected\footnote {The angular resolution is, generically, far superior to the energy/momentum resolutions and too fine to be of any consequence at the level of sophistication of this analysis.} and we simulate the former by smearing the jet energy with Gaussian functions defined by an energy-dependent width, $\sigma_E$:
\begin{equation}
\frac{\sigma_E}{E}=\frac{0.80}{\sqrt E}\oplus 0.05,
\label{gau}
\end{equation}
where, $\oplus$ denotes a sum in quadrature.

\begin{table}[h]
\begin{center}
\begin{tabular}{c|c|c}
\hline \hline
Kinematic Variable & Minimum value & Maximum value \\\hline\hline
$p_T^{j_1,j_2}$ & 100 GeV & - \\
$\eta^{j_1,j_2}$ & -2.5 & 2.5 \\
$\Delta R(j_1,j_2)$  & 0.7 & - \\\hline\hline
\end{tabular}
\end{center}
\caption{\small{Acceptance cuts on the kinematical variables. $p_T^{j_1,j_2}$ is the transverse momentum and $\eta^{j_1,j_2}$ is the rapidity of the jets. $\Delta R(j_1,j_2)=\sqrt{(\Delta \eta)^2+(\Delta \phi)^2}$ is the distance among the jets in the $\eta-\phi$ plane, with $\phi$ being the azimuthal angle.}}

\label{cuts}
\end{table}

The signal jets arise from the decay of a significantly heavy mirror quark to a SM $Z$ and jet. Due to the large phase space available for the decay of the mirror quarks, the resulting jets will be predominantly hard. Therefore, the large jet $p_T$ cuts, listed in Table~\ref{cuts}, are mainly aimed to reduce the SM background contributions. With the set of acceptance cuts (see Table~\ref{cuts}) and detector resolution defined in the previous paragraph, we compute the signal and background cross-sections at the LHC operating with $\sqrt s$ = 8 TeV and 14 TeV respectively and display them in Table~\ref{cs}. Table~\ref{cs} shows that signal cross-sections are larger than the background for lower values of mirror quark masses. However, if we increase $M_{\hat q}$, signal cross-sections fall sharply as the pair production cross section for the mirror quarks fall with increasing mass.
\begin{table}[h]
\begin{center}
\begin{tabular}{||c|c|c||c|c||c|c|c||c|c||}
\hline \hline
\multicolumn{5}{||c||}{$\sqrt s$= 8 TeV} & \multicolumn{5}{|c||}{$\sqrt s$= 14 TeV}\\
\multicolumn{5}{||c||}{Cross-sections in fb} & \multicolumn{5}{|c||}{Cross-sections in fb}\\\hline\hline
\multicolumn{3}{||c||}{Signal} & \multicolumn{2}{|c||}{Background} & \multicolumn{3}{|c||}{Signal} &  \multicolumn{2}{|c||}{Background}\\\hline
%\cline{1-2}\cline{4-5}
$M_{\hat q}$ [GeV] & {\it A.C.} & {\it S.C.} & {\it A.C.} & {\it S.C.} & $M_{\hat q}$ [GeV]&  {\it A.C.} & {\it S.C.} & {\it A.C.} & {\it S.C.}\\\hline\hline
300 & 1.65 & 1.07 &       & 0.08 &   400 & 2.93 & 1.5  &      & 0.22 \\
350 & 0.92 & 0.52 & 0.35  & 0.07 &   500 & 1.04 & 0.48 & 1.36 & 0.14 \\
400 & 0.5  & 0.26 &       & 0.05 &   600 & 0.40 & 0.18 &      & 0.09\\\hline\hline
\end{tabular}
\end{center}
\caption{\small{Signal and SM background cross-section after the acceptance cuts ({\it A.C.}) and selection cuts ({\it S.C.}) for two different values of proton-proton center-of-mass energies. Signal cross-sections ($\sigma_{Signal}$) are presented for three different values of mirror quark masses ($M_{\hat q}$).}}
\label{cs}
\end{table}

%------------------------------------------------------------------
\begin{figure}
\begin{center}
\epsfig{file=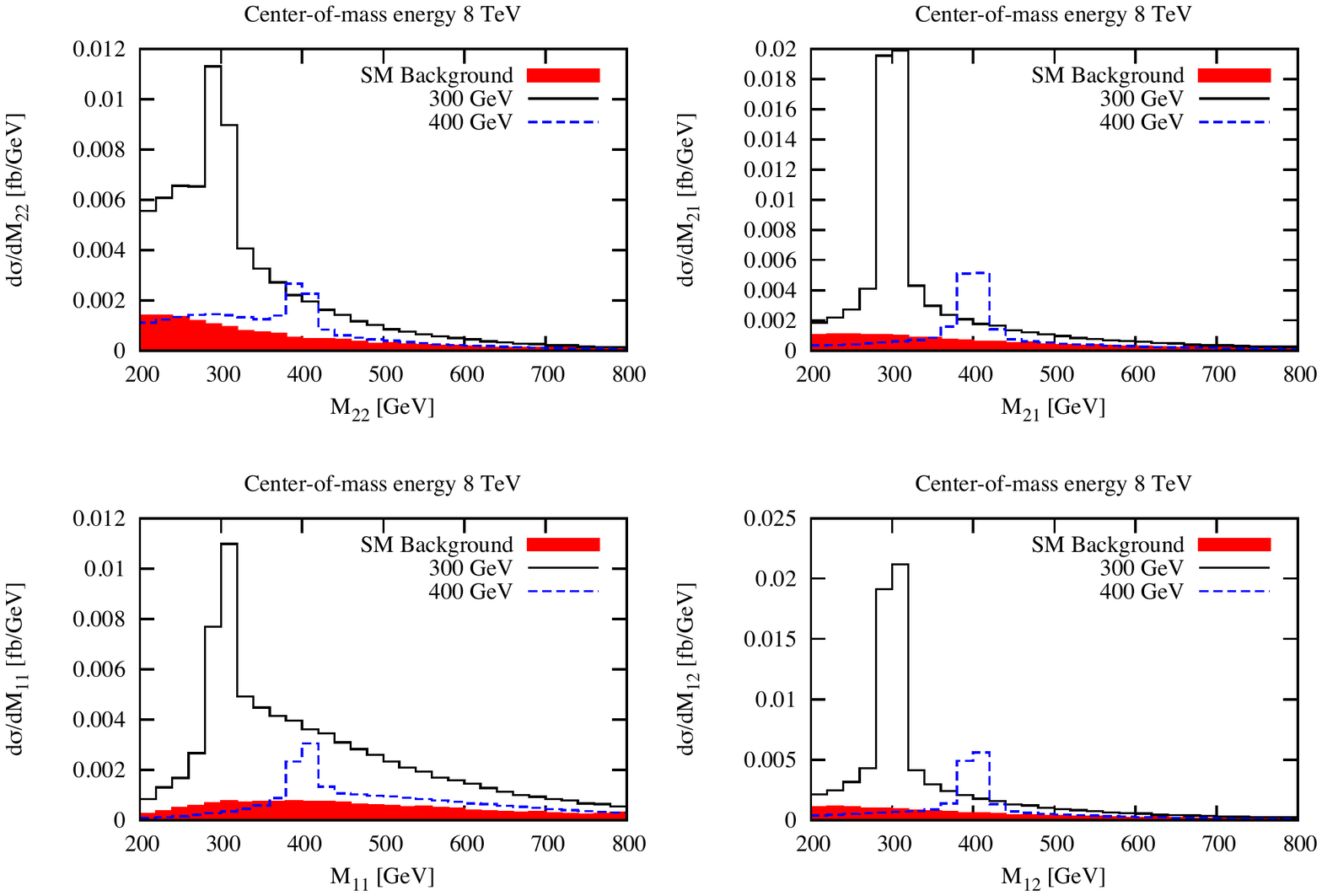,width=13cm,height=10cm}
\epsfig{file=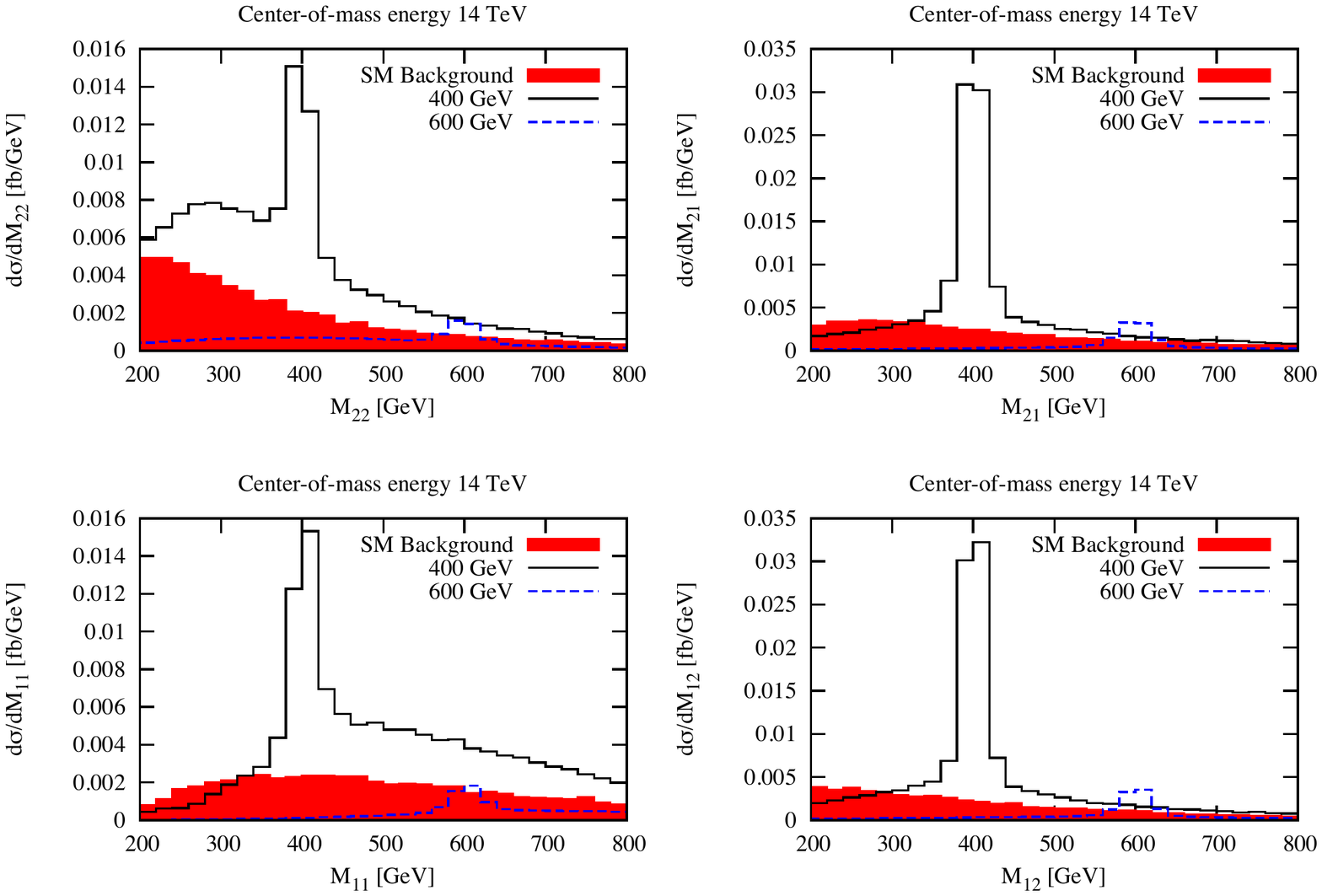,width=13cm,height=10cm}
\end{center}
\caption{\small{Jet-$Z$ invariant mass distributions after ordering the jets ($p_T^{j_1}>p_T^{j_2}$) and $Z$'s ($p_T^{Z_1}>p_T^{Z_2}$) according to their $p_T$ hardness for the LHC with center-of-mass energy 8 TeV (top panel) and 14 TeV (bottom panel).}}
\label{Invmass}
\end{figure}
%-------------------------------------------------------------------

Since the mirror quarks decay into a jet and $Z$-boson, the signal is characterized by a peak at $M_{\hat q}$ in the invariant mass distributions of jet-$Z$ pairs. The signal consists of two jets and two $Z$-bosons. In absence of any knowledge about the right jet-$Z$ pair arising from a particular $\hat q$ decay, we have ordered the jets and $Z$'s according to their $p_T$ hardness ($p_T^{j_1}>p_T^{j_2}$ and $p_T^{Z_1}>p_T^{Z_2}$) and constructed invariant mass distributions in the jet-$Z$ pairs as follows: $M_{11}=$ Invariant mass of $j_1~{\rm and}~Z_1$; $M_{12}=$ Invariant mass of $j_1~{\rm and}~Z_2$; $M_{21}=$ Invariant mass of $j_2~{\rm and}~Z_1$ and $M_{22}=$ Invariant mass of $j_2~{\rm and}~Z_2$. The four invariant mass distributions (for both signal and the SM background) are presented in Fig.~\ref{Invmass} for the LHC with center-of-mass energy 8 TeV (left panel) and 14 TeV (right panel). In Fig.~\ref{Invmass}, we have presented the signal invariant mass distributions for two different values of $M_{\hat q}$. We have included the leptonic branching ratio (6.7\% into electron and muon channel) of $Z$-boson into the cross-section in the Fig.~\ref{Invmass}. Fig.~\ref{Invmass} shows that the signal peaks are clearly visible over the SM background. Moreover, it is important to notice that signal peaks are more prominent in $M_{12}$ and $M_{21}$ distributions compared to $M_{11}$ and $M_{22}$ distributions. Due to the momentum conservation in the transverse direction at the LHC, both the mirror quarks are produced with equal and opposite transverse momentum\footnote{We do not consider initial/final state radiation (ISR/FSR) in our analysis. In presence of ISR/FSR, the transverse momentum of the mirror quarks might not be exactly equal and opposite.}. Therefore, if the decay of a particular mirror quark gives rise to the hardest jet then it is more likely that the $Z$-boson arising in the same decay will be the softest one. $M_{12}(M_{21})$ is the invariant mass of hardest-softest (softest-hardest) jet-$Z$ pairs which come from the decay of a particular $\hat q$ in most of the events. As a result, we observe more prominent peaks in the signal $M_{12} (M_{21})$ distribution compared to the $M_{11} (M_{22})$ distribution. In our analysis, we have utilized this feature of the signal for the further enhancement of signal to background ratio. Our final event selection criteria ({\it S.C.}) is summarized in the following:
\begin{itemize}
\item To ensure the observability of a peak for a given luminosity in the signal $M_{12}$ distribution, we have imposed the following criteria: (i) There are atleast 5 signal events in the peak bin. (ii) The number of signal events in the peak bin is greater than the $3\sigma$ fluctuation of SM background events in the same bin. 
\item If the signal peak in $M_{12}$ distribution is detectable then we selected events in the bins corresponding to the peak in the $M_{12}$ distribution and its four (two on the left hand side and two on the right hand side) adjacent bins as signal events. We have used a bin size of 20 GeV.
\item The total number of SM background events is given by the sum of events of the above mentioned five bins in the background $M_{12}$ distribution.  
\end{itemize}
After imposing the final event selection criteria, the signal and background cross-sections for different $M_{\hat q}$ and $\sqrt s$ are presented in Table~\ref{cs}. Table~\ref{cs} shows that selection cuts significantly suppress the SM background cross-section, whereas, signal cross-sections are reduced only by a factor $\sim$ 2.

After discussing the characteristics features of the signal and the SM background, we are now equipped enough to discuss the discovery reach of this scenario at the LHC with center-of-mass energy 8 TeV and 14 TeV. We define the signal to be observable over the background with confidence level (CL) $X$ for a integrated luminosity ${\cal L}$ if, $X$-CL upper limit on the background is smaller than the $X$-CL lower limit on the signal plus background \cite{tata}:
\begin{equation}
{\cal L}(\sigma_S+\sigma_B)-N\sqrt{{\cal L}(\sigma_S+\sigma_B)} > {\cal L}\sigma_B+N\sqrt{{\cal L}\sigma_B},
\end{equation}
or, equivalently,
\begin{equation}
\sigma_S > \frac{N^2}{{\cal L}}\left[ 1+2\frac{\sqrt {{\cal L} \sigma_B}}{N}\right],
\label{discovery}
\end{equation} 
where, $\sigma_S~{\rm and}~\sigma_B$ are the signal and background cross-sections, respectively and $N=2.5$ for $X=99.4\%$ \cite{handbook} CL discovery. 
The signal and background cross-sections in Table~\ref{cs} shows that at the LHC with center-of-mass energy 8 TeV (14 TeV), 350 GeV (550 GeV) mirror quark mass can be probed with integrated luminosity 25 fb$^{-1}$ (72 fb$^{-1}$). In Fig.~\ref{lumi}, we have presented the required luminosity for 99.4\% CL discovery as a function of $M_{\hat q}$ for the LHC with center-of-mass energy 8 TeV and 14 TeV.

%------------------------------------------------------------------
\begin{figure}
\begin{center}
\epsfig{file=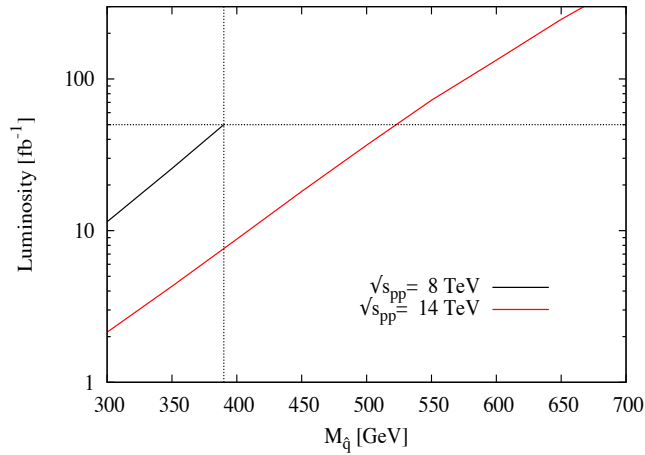,width=12cm,height=10cm}
\end{center}
\caption{\small{Required luminosity for $5\sigma$ discovery is plotted as a function of $M_{\hat q}$ for the LHC with center-of-mass energy 8 TeV and 14 TeV.} }
\label{lumi}
\end{figure}
%-------------------------------------------------------------------

\subsubsection{{\it {Two jets}+$Z$-boson+$W$-boson} signature}
Another interesting final state results from the pair production of mirror quarks which then decay to give {\it {2 jets}+$Z$+$W$} signal. This happens when one mirror quark decays into $qZ$ while the other one decays into $qW$. As before we have considered the $Z$ boson as a standard object without simulating its decay to leptons (electrons and muons). 
%However, in the hadron collider environment, reconstruction of $W$-bosons in the leptonic decay channel ($W^\pm \to l^\pm \nu$) is difficult because of the unknown longitudinal boost of the center-of-mass frame. Therefore, $W$-bosons can not be treated as standard object like $Z$-bosons.
Even the $W$ boson can be reconstructed to a certain efficiency in the leptonic channel, where the neutrino $p_z$ is determined by using the $W$ mass constraints. This is possible because of a single neutrino in the final state. However, we have chosen to ignore the $W$ as a standard object since the $qZ$ resonance will be much more 
well defined and with less ambiguity.
In our parton level Monte-Carlo analysis, we have simulated the decay of $W$ bosons into leptons (electron and muons only) and neutrinos. Electrons and muons show charge tracks in the tracker and are detected at the electromagnetic calorimeter and muon detector respectively. However, neutrinos remain invisible in the detector and give rise to a imbalance in the visible transverse momentum vector which is known as missing transverse momentum (${p}_T$). Therefore, the resulting signature in this case will be {\it {2 jets}+1 charged lepton + $Z$ + ${p}_T$}. 
\begin{table}[t!]
\begin{center}
\begin{tabular}{c|c|c}
\hline \hline
Kinematic Variable & Minimum value & Maximum value \\\hline\hline
$p_T^{l}$ & 25 GeV & - \\
$\eta^{l}$ & -2.5 & 2.5 \\
$\Delta R(l,j_{1,2})$  & 0.4 & - \\\hline\hline
\end{tabular}
\end{center}
\caption{\small{Acceptance cuts on the kinematical variables. $p_T^{l}$ is the transverse momentum and $\eta^{l}$ is the rapidity of the lepton. $\Delta R(l,j_{1,2})$ is the distance among the jet-lepton pairs in the $\eta-\phi$ plane, with $\phi$ being the azimuthal angle.}}
\label{cut_L}
\end{table}
%-----------------------------

The dominant SM background to the signal arises from the production of $ZW$ pairs in association with two jets. Both signal and background jets energy are smeared by a Gaussian function defined in Eq.~\ref{gau}. To ensure the visibility of the jets at the detector, acceptance cuts listed in Table~\ref{cuts} are applied on the jets. The 
%-----------------------------------------------
 \begin{table}[h!]
\begin{center}
\begin{tabular}{||c|c|c|c||c|c|c||c|c|c|c||c|c|c||}
\hline \hline
\multicolumn{7}{||c||}{$\sqrt s$= 8 TeV} & \multicolumn{7}{|c||}{$\sqrt s$= 14 TeV}\\
\multicolumn{7}{||c||}{Cross-sections in fb} & \multicolumn{7}{|c||}{Cross-sections in fb}\\\hline\hline
\multicolumn{4}{||c||}{Signal} & \multicolumn{3}{|c||}{Background} & \multicolumn{4}{|c||}{Signal} &  \multicolumn{3}{|c||}{Background}\\\hline
%\cline{1-2}\cline{4-5}
$M_{\hat q}$ & {\it A.C.} & {\it Cut} & {\it Cut} & {\it A.C.} & {\it Cut} & {\it Cut}  & $M_{\hat q}$ &  {\it A.C.} &  {\it Cut} & {\it Cut}  & {\it A.C.} & {\it Cut} & {\it Cut}\\
GeV & & {\it I}& {\it II} & & {\it I}& {\it II} &GeV & & {\it I}& {\it II} & & {\it I}& {\it II}\\\hline\hline
300 & 14.4 & 7.28 & 3.13 &       &      & 0.74 & 400 & 26.3 & 18.1 & 6.46 & & & 2.13\\
350 & 8.01  & 4.85 & 1.92 & 6.69 & 2.81 & 0.63 & 500 & 9.36  &  7.33 & 2.39 & 26.4 & 11.9 & 1.39\\
400 & 4.35 & 2.98 & 1.11 & & & 0.51 & 600 & 3.6 & 3.07 & 0.95 & & & 0.90 \\\hline\hline
\end{tabular}
\end{center}
\caption{\small{Signal and SM background cross-section after the acceptance cuts ({\it A.C.}), {\it Cut I} and {\it Cut II} for two different values of proton-proton center-of-mass energies. Signal cross-sections ($\sigma_{Signal}$) are presented for three different values of mirror quark masses ($M_{\hat q}$).}}
\label{cs_ZW}
\end{table}
%--------------------------------------
acceptance cuts for the lepton are listed in Table~\ref{cut_L}. We do not apply any cuts on the missing transverse momentum. With these set of cuts ({\it A.C.}) on jets (see Table~\ref{cuts}) and lepton (see Table~\ref{cut_L}), we have computed the signal and background cross-sections for the LHC with 8 TeV and 14 TeV center-of-mass energy and presented in Table~\ref{cs_ZW}. Table~\ref{cs_ZW} shows that for relatively large mirror quark masses, signal cross-sections are much smaller than the SM background cross-section. For example, at the LHC with 14 TeV center-of-mass energy, the signal to background ratio is 0.14 after acceptance cuts for $m_{\hat q}=600$ GeV. 

\begin{figure}
\begin{center}
\epsfig{file=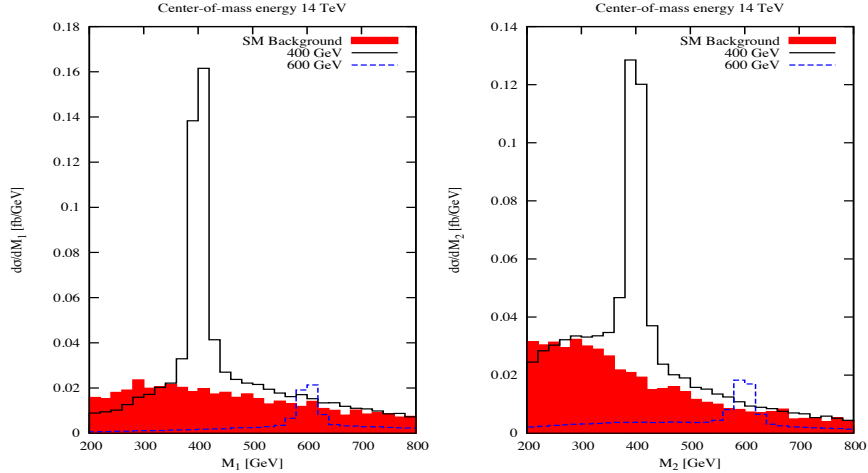,width=16cm,height=10cm}
\end{center}
\caption{\small{Jet-$Z$ invariant mass distributions after {\it Cut I} for the signal ($m_{\hat q}=400~{\rm and}~600$ GeV) and the SM background at the LHC with 14 TeV center-of-mass energy.}}
\label{inv_WZ}
\end{figure}

The signal contains a lepton and  ${p}_T$ arises from the decay of a $W$-boson. The SM background  lepton and  ${p}_T$ also results from the $W$-boson decay. However, the signal $W$-boson will be boosted in most of the events since it arises from the decay of a TeV scale mirror quark. We have tried to exploit this feature of the signal for the further enhancement of signal to background ratio. We have examined the following kinematic distributions:
\begin{itemize}
\item In Fig.~\ref{ptlep}, we have presented normalized lepton $p_T$ (left panel) and missing $p_T$ (right panel) distributions for the signal ($m_{\hat q}=400~{\rm and}~600$ GeV) and the SM background at the LHC with $\sqrt s=14$ TeV. The boost of the signal $W$-boson results into a long tail in the signal lepton and missing $p_T$ distributions. Fig.~\ref{ptlep} shows that harder cuts on the lepton and/or missing $p_T$ will suppress the SM background significantly. However, these cuts will also reduce signal cross-sections considerably. For example,  
a kinematic requirement of $p_T> 75$ GeV on the charged lepton at 14 TeV LHC will reduce 45\% of the SM background and 25\% of the signal for $m_{\hat q}=600$ GeV. As a result, we do not use any further cuts on 
lepton and/or missing $p_T$.

%------------------------------------------------------------------
\begin{figure}
\begin{center}
\epsfig{file=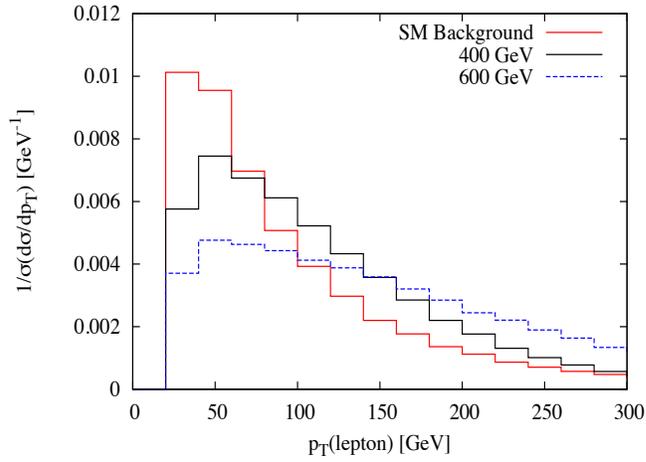,width=12cm,height=10cm}
\epsfig{file=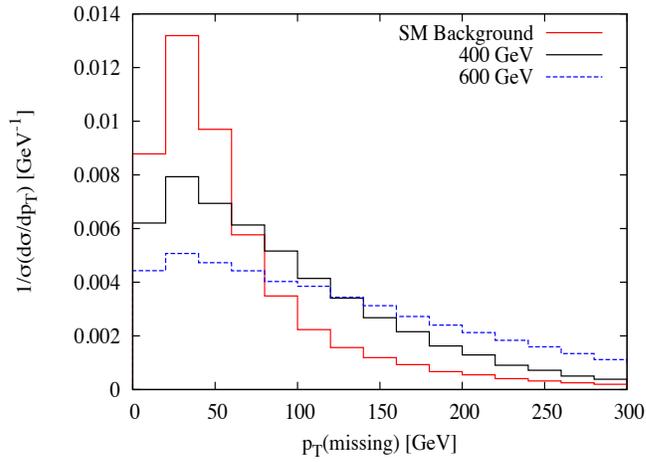,width=12cm,height=10cm}
\end{center}
\caption{\small{Normalized lepton $p_T$ (top panel) and missing $p_T$ (bottom panel) distributions for the signal ($m_{\hat q}=400~{\rm and}~600$ GeV) and the SM background after the acceptance cuts at the LHC with $\sqrt s=14$ TeV.} }
\label{ptlep}
\end{figure}
%-------------------------------------------------------------------
\item Since the signal $W$-boson is boosted, we expect that the signal lepton and neutrino will be collimated. Therefore, it is viable to study the azimuthal angle ($\Delta \phi$) between lepton transverse momentum vector ($\vec p_T^{l}$) and missing transverse momentum vector ($\vec{{p}}_T$). In Fig.~\ref{angle}, we have presented normalized $\Delta \phi(\vec p_T^{l},\vec{{p}}_T)$ distributions for the signal ($m_{\hat q}=400~{\rm and}~600$ GeV) and the SM background  at the LHC with $\sqrt s=14$ TeV. Since the background $W$-bosons are predominantly produced with small transverse momentum, background $\Delta \phi(\vec p_T^{l},\vec{{p}}_T)$ distribution is almost flat (see Fig.~\ref{angle}). Whereas, the signal $\Delta \phi(\vec p_T^{l},\vec{{p}}_T)$ distributions peaks in the small $\Delta \phi(\vec p_T^{l},\vec{{p}}_T)$ region. As a result, we have imposed an upper bound of 1 on the azimuthal angle between lepton $p_T$ vector and missing $p_T$ vector: $\Delta \phi(\vec p_T^{l},\vec{{p}}_T) < 1$. We collectively call acceptance cuts and $\Delta \phi(\vec p_T^{l},\vec{{p}}_T) < 1$ cut as {\it Cut I}. The signal and background cross-sections after {\it Cut I} are presented in Table~\ref{cs_ZW}. For 14 TeV center-of-mass energy, $\Delta \phi(\vec p_T^{l},\vec{{p}}_T) < 1$ cut reduces 55\% of the SM background and 14\% of the signal for $m_{\hat q}=600$ GeV and thus, enhances the signal to background ratio by a factor about 2. 
%------------------------------------------------------------------
\begin{figure}
\begin{center}
\epsfig{file=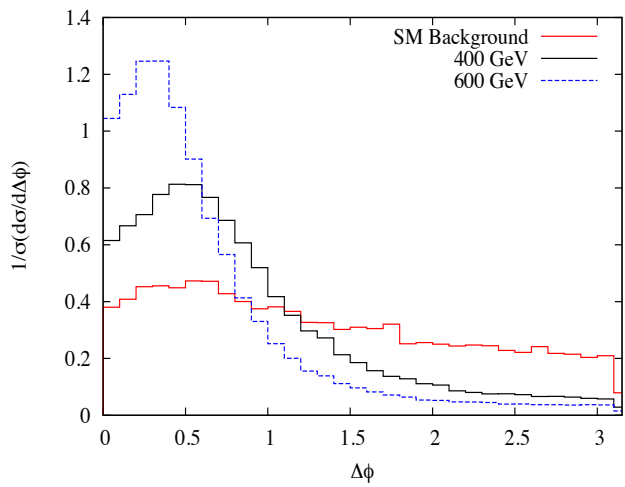,width=12cm,height=10cm}
\end{center}
\caption{\small{Normalized azimuthal angle $\Delta \phi(\vec p_T^{l},\vec{{p}}_T)$  distributions between lepton $p_T$ vector and missing $p_T$ vector after the acceptance cuts at the LHC with $\sqrt s=14$ TeV. Signal distributions are presented for two different values of mirror quark mass ($m_{\hat q}=400~{\rm and}~600$ GeV).}}
\label{angle}
\end{figure}
%-------------------------------------------------------------------

\item After $\hat q \bar{\hat q}$ production, one mirror quark decays into $qZ$ pair. Therefore, signal jet-$Z$ invariant mass distribution is characterized by a peak at $m_{\hat q}$. After ordering the jets according to their $p_T$ hardness ($p_T^{j_1}>p_T^{j_2}$), we have constructed two invariant mass: (i) $M_1$: invariant mass of $j_1$-$Z$ pair and (ii) $M_2$: invariant mass of $j_2$-$Z$ pair. The signal and background invariant mass distributions are presented in Fig.~\ref{inv_WZ} for the LHC with $\sqrt s=14$ TeV. For the further enhancement of signal to background ratio, we have imposed cuts on $M_2$ in a way similar to that discussed in the previous section. This cut and {\it Cut I} are collectively called as {\it Cut II} in Table~\ref{cs_ZW}. Table~\ref{cs_ZW} shows that for $m_{\hat q}=600$ GeV, $j_2$-$Z$ invariant mass cut suppress the SM background by a factor about 13, whereas, the signal is reduced by a factor of 3 only.   
\end{itemize}    

%------------------------------------------------------------------
%-------------------------------------------------------------------

\begin{figure}
\begin{center}
\epsfig{file=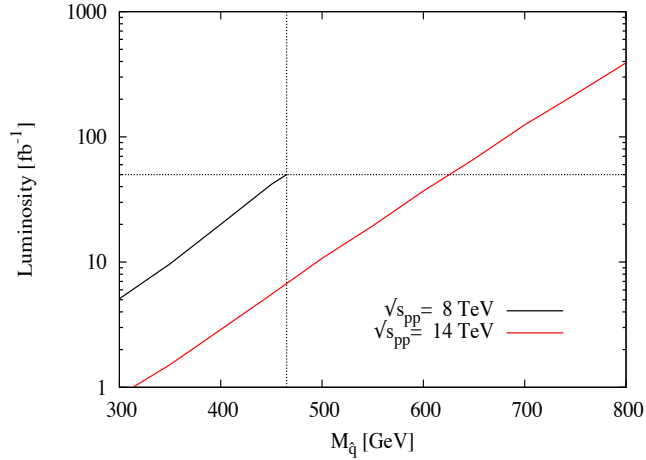,width=12cm,height=10cm}
\end{center}
\caption{\small{Required luminosity for $5\sigma$ discovery in {\it {two-jets}+one lepton + one $Z$-boson + $p_T\!\!\!\!\!/~$} channel is plotted as a function of $M_{\hat q}$ for the LHC with center-of-mass energy 8 TeV and 14 TeV.}}
\label{lumi_WZ}
\end{figure}

To estimate the required integrated luminosity for the discovery of the mirror quarks in {\it {two jets}+one charged lepton + $Z$ + ${p}_T$} channel, we have used Eq.~\ref{discovery}. The signal and background cross-sections after {\it Cut II} in Table~\ref{cs_ZW} shows that at the LHC with center-of-mass energy 8 TeV (14 TeV), 400 GeV (600 GeV) mirror quark mass can be probed with integrated luminosity 20 fb$^{-1}$ (37 fb$^{-1}$). In Fig.~\ref{lumi_WZ}, we have presented the required luminosity for 99.4\% CL discovery in {\it {two jets}+one charged lepton + $Z$ + ${p}_T$} channel as a function of $M_{\hat q}$ for the LHC with center-of-mass energy 8 TeV and 14 TeV. Inspired by this work, ATLAS collaboration has looked for a heavy quark decaying into a W boson and a light quark and has set a mass limit of 690 GeV at center of mass energy 8 TeV at 95$\%$ CL.

%------------------------------------------------------------------
%-------------------------------------------------------------------

\section{Conclusions}\label{sec:summary}

In this work, we have a realistic left-right symmetric model with mirror fermions and mirror Higgs, and the possibility of discovering the low lying mirror fermions at the LHC. The model is $SU(3)_C \otimes SU(2)_L\otimes SU(2)_R \otimes U(1)_Y^\prime$ supplemented by a discrete $Z_2$. For each chiral multiplet of the SM fermions, we have corresponding mirror fermions of opposite chirality. The symmetry is broken to the usual SM symmetry by a mirror Higgs doublet.  The mixing between the SM fermions and the mirror fermions is achieved by using a Higgs multiplet which is a singlet under the gauge symmetry, but odd under the $Z_2$ symmetry. The model has singlet right handed neutrinos, and the corresponding mirror neutrinos which are even under $Z_2$. These are used to generate tiny neutrino masses $\simeq 10^{-11}$ GeV with a primary symmetry breaking scale of $\simeq 10^7$ GeV (which is the VEV of the mirror Higgs doublet). In this model, only the mirror fermion of the 1st family ($\hat{e}, \hat{u}, \hat{d}$) are light with well-defined relative spectrum. All the other mirror fermions are much heavier, and well above the LHC reach. Since the model is completely left-right symmetric in the fermion sector, it is naturally anomaly free. Parity conservation, and the nature of the fermion mass matrices also provides a solution for the strong CP in the model.

The light mirror fermions, $ \hat{u}, \hat{d}$, with masses around few hundred GeV to about a TeV, can be pair produced at the LHC via their QCD color interactions. They dominantly decay to a $Z$ boson plus the corresponding ordinary fermion ($\hat{u} \rightarrow{u + Z}, \hat{d}\rightarrow {d + Z}$), or to a W boson and the corresponding ordinary fermions ($\hat{u} \rightarrow{d + W}, \hat{d}\rightarrow {u + W}$). (The decays ($\hat{u} \rightarrow{u + H}, \hat{d}\rightarrow {d + H}$) are highly suppressed for most of the parameter space). Thus the most striking signal of the model is the existence of 
of resonances in the jet plus $Z$ channel. Since both the jet and the $Z$ is coming from the decay of a very heavy particle, both will have very high $p_T$. We have shown that putting a high $p_T$ cut on the jet, and reconstructing the $Z$ in the $e^+ e^-$ or $\mu^+ \mu^-$ channels, these resonances 
$\hat{u}, \hat{d}$ can be reconstructed upto  a mass of $\simeq 350$ GeV at the 8 TeV LHC, and upto  a mass of $\simeq 550$ GeV at the 14 TeV LHC. We are not aware of any other model which predicts such a resonance. We have also studied, in some detail, the final states arising from the pair productions of these light mirror fermions at the LHC. These final states are $(u Z) (\bar{u} Z), (d Z) (\bar{d} Z), (u Z) (\bar{d} W), (d Z) (\bar{u} W)$, and the subsequent decays of $W$ and $Z$ into the leptonic channels. The signals are much more observable in the (jet jet $Z Z$) channel than the (jet jet $Z W$) channel because of the missing neutrino in the latter. (The {resonance in the} signals involving the two W's will be difficult to observe). We have studied these final states and the corresponding backgrounds, and find that the reaches for the light mirror quarks can be$\simeq 450$ GeV at the 8 TeV LHC with  luminosity of $ ~30~ fb^{-1}$, and upto $ ~ 750 $ GeV  at the $14$ TeV LHC with $~300 ~fb^{-1}$ luminosity.

Our model predicts a definite pattern of spectrum for the light mirror fermions, $\hat{e}, \hat{u}, \hat{d}$. Thus with $m_{\hat{u}}< m_{\hat{d}}$, if a resonance $\hat{u}$ is observed, we expect a nearby $\hat{d}$ within few hundred GeV. This makes the prediction of the model somewhat unique. Also the $\hat{e}$ will have even lower mass, and can be looked for in the proposed future $e^+ e^{-}$ collider.

In this work, we have studied the collider phenomenology of TeV scale mirror fermions in the framework of a particular variant of LRMM in which mirror fermions dominantly decays into the SM fermion and $W/Z$-boson.  However, our collider analysis is general enough to be applicable to a class of models with TeV scale fermions decaying into a SM fermion and $W/Z$-boson.

\chapter{Prospects and discovery potential of Non-universal SUGRA at the LHC}\label{chap:chap4}
\section{Introduction}

Supersymmetry (SUSY) \cite{SUSY,KaneKingRev} has been under scanner 
since last forty years or more. On-going Large Hadron Collider (LHC) has put strong 
bounds on the squark and gluino masses of minimal supersymmetric Standard Model (MSSM); 
particularly on minimal supergravity (mSUGRA) or constrained minimal 
supersymmetric Standard Model (CMSSM) \cite{msugra}, not seeing 
any of those supersymmetric particles. Still, SUSY search in different forms is the most 
studied subject of particle physics research due to its unparalleled theoretical appeal
and phenomenological implications.     

Out of different SUSY-breaking schemes, mSUGRA has been most popular 
due to its economy of parameters; the universal gaugino mass
($M_{1/2}$), the universal scalar mass ($m_0$), the universal
trilinear coupling ($A_0$) all at the GUT scale, $\tan\beta$, the
ratio of the vacuum expectation values (vev) of the two Higgses and
the sign of SUSY-conserving Higgsino mass parameter $\mu$. However,
this framework has been highly constrained by direct and indirect search experiments 
\cite{msgbound,b-jets-search,bsg-recent,mupmum} and 
non-universality in scalar \cite{3rd1,3rd2,3rd3,ucddOct2008,berez,Nath:1997qm,Cerdeno:2004zj,ellis-all,baer-all-non,so102,Datta:1999uh,BM-SB-AD2}
 and gaugino masses \cite{nonunigaugino,BM-AD-SB,sb-sn} are getting more and more importance 
 to keep low-scale SUSY alive.

Recent discovery of Higgs boson with $m_H\simeq$ 125 GeV at LHC 
by the ATLAS and CMS Collaborations \cite{higgs} has put a severe 
constraint on SUSY parameter space. SUSY Higgs gets significant correction 
from the top squark (stop) loop, which increases with increasing 
stop mixing and/or stop mass scale. Therefore, in order to get a Higgs boson 
around 125 GeV, significant stop mixing or a large stop mass scale is required. Large stop mixing 
results into large mass splitting in the stop sector and consequently gives rise to a lighter stop ($\tilde t_1$) 
in the mass spectrum. Hence, Higgs boson mass at 125 GeV results in a 
SUSY mass spectrum with light third family scalars.

 Light third family scalars, but relatively heavy first two families\footnote{Such scenarios 
 have already been considered for studies in different contexts \cite{3rd1,3rd2,3rd3}.} favor 
 SUSY discovery at future LHC runs given gluino ($\tilde g$) dominantly decays 
 into top-stop pairs ($\tilde g \to t \tilde t_1$) and subsequently stop decays into top-neutralino or b-chargino 
 where $t\rightarrow bW^\pm$ gives rise to multiple b-jets, leptons and large missing energy ($E_{T}\!\!\!\!/$). 
 Final states with multiple b jets and charged leptons, together with large missing energy,
 cut down the SM background much more than the usual SUSY signals with multijets plus large missing energy, and both 
 ATLAS and CMS experiments has achieved  b-tagging efficiency $50\%$ or more and have put bounds on 
 SUSY from the available data \cite{b-jets-search}. 
 
 Another important aspect of SUSY is the dark matter (DM); $R$-parity conservation yields a natural candidate namely, 
 the lightest supersymmetric particle (LSP). DM relic density limits from 
 WMAP \cite{WMAPdata} and PLANCK \cite{PLANCK} can be easily satisfied in the 
 non-universal gaugino and/or scalar mass scenarios where CMSSM is tightly constrained. 
 For example, if wino mass is smaller than bino mass at GUT scale ($M_2 \le M_1$), we obtain wino dominated LSP 
 yielding correct abundance in a larger parameter space. Similarly, non-universality in the scalar sector
  may results in a higgsino like LSP (from non-universality in the Higgs sector) or 
  stau-LSP co annihilation (from non-universality in the soft SUSY breaking stau mass). 
 We have systematically studied such non-universal gaugino  and/or scalar mass scenarios
 and proposed benchmark points for collider studies at LHC with $E_{CM}$= 14 TeV. 
 
 %We show that 
 %SUSY can leave imprints through $4b+{E_{T}}\!\!\!\!/$, $4b+\ell+{E_{T}}\!\!\!\!/$ and $2b+2\ell+{E_{T}}\!\!\!\!/$ 
 %channels with proper cuts. Lepton-rich final states such as Same-sign dilepton, trilepton may also 
 %be observable with suitable cuts and high luminosity. 

%Efforts have already been made to study the allowed parameter space in mSUGRA [] and 
%non-universal scenarios [] after LHC data, but our analysis may pin-point the particular 
%regions to look at following their discovery potential at LHC.  

Vast amount of work has already been done in mSUGRA to discover SUSY at the LHC. However, because of the 
observed Higgs mass, and the dark matter constraint, the only region left in mSUGRA and accessible at the LHC 
is the stop co-annihilation region (where the lighter top squark $\tilde{t_1}$ and the lightest neutralino $\tilde \chi_{1}^0$ annihilate to satisfy the
dark matter constraint. However, in this parameter space, $\tilde{t_1}$ mass is very close to the  $\tilde \chi_{1}^0$ mass
giving rise to very little high $p_T$ multijet activity from its decay \cite{degenerate-stop}. Significant number of works have also been done 
by increasing the number of parameters, with non-universal gaugino masses and non-universality in the scalar masses satisfying all
the existing constraints \cite{recentworks}. However, we pin point that to survive Higgs mass and dark matter 
constraint in the framework of gravity mediated supersymmetry breaking, a larger region of parameter 
space is available with specific non-universal gaugino and scalar mass patterns with a generic signature in 
bottom rich, and bottom quark plus charged lepton rich final states with large missing energy, 
which with suitable cuts can be observed over the SM background at the 14 TeV LHC. 
We claim that these will be the most favorable final states at the 14 TeV LHC to discover SUSY or to put 
strongest bounds on them.

The chapter is organized as follows. In Section 2, we discuss the model
under consideration and the selected benchmark points. We also review dark matter 
constraints on SUSY parameter space to motivate our benchmark points. In Section 3, we
discuss the  final states in which SUSY signals can be observed over the SM background, including the details of the
collider simulation strategy and the numerical results at the 14 TeV LHC. We conclude in Section 4.

\section {Model, Constraints and Benchmark Points}

\subsection{Constraints on SUSY models:} 
 
%We choose benchmark points that satisfy all existing constraints including Higgs mass and Dark matter.  

%The benchmark points are chosen as follows. BP1 describes a sample point in chargino co-annihilation region, 
%while BP2 has a higgsino LSP, BP3 depicts stau co-annihilation and MSG represents a sample point in stop 
%co-annihilation region from mSUGRA. The values of the high-scale parameters and low scale masses are indicated.
 %The values of  $\tan \beta$ has been chosen
%such that it satisfies the experimental constraint branching ratio for the 
Following measurements play a key role to constrain SUSY parameter space. We discuss their effect 
and motivate how that leads eventually to the benchmark points chosen in this article for SUSY 
searches at LHC.

\begin{itemize}
\item
The main constraint on the SUSY parameter space after LHC 7/8 TeV data is that the 
CP even Higgs mass to be within  \cite{higgs}:

\begin{equation}
123 \le m_h \le 127.
\label{Higgs}
\end{equation}

\item
The branching ratio for $b \longrightarrow s \gamma$ \cite{bsg-recent} which
at the $3 \sigma$ level is
\begin{equation}
2.13 \times 10^{-4} < Br (b \rightarrow s \gamma) < 4.97 \times 10^{-4}.
\label{bsgammalimits}
\end{equation}

\item
We also take into account the constraint coming from $B_s  \longrightarrow \mu^+ \mu^-$ branching ratio 
which by LHCb observation \cite{mupmum} at 95\% CL is given as 
\begin{equation}
2 \times 10^{-9} < Br (B_s \rightarrow \mu^+ \mu^-) <  4.7 \times 10^{-9}.
\label{bstomumu}
\end{equation}

\item
Parameters are fine-tuned in a way that it gives a correct cold dark matter
relic abundance according to WMAP data \cite{WMAPdata},
which at $3 \sigma$ is
\begin{equation}
0.091 < \Omega_{CDM}h^2 < 0.128 \ ,
\label{relicdensity}
\end{equation}
where $\Omega_{CDM}$ is the dark matter
relic density in units of the critical
density and $h=0.71\pm0.026$ is the reduced Hubble constant
(namely, in units of
$100 \ \rm km \ \rm s^{-1}\ \rm Mpc^{-1}$).
\end{itemize}

To note here, the PLANCK 
constraints $0.112 \leq \Omega_{\rm DM} h^2 \leq 0.128$ \cite{PLANCK}
is more stringent, and cuts a significant amount of dark matter allowed 
SUSY parameter space. We choose our benchmark points satisfying 
PLANCK on top of WMAP.

In the following subsection, we discuss mainly the dark matter 
and Higgs mass constraints on SUSY parameter space as they have 
been the key to choose our benchmark points. 

\subsection{Dark matter and Higgs mass on SUSY: Benchmark Points}  

One of the main motivations for postulating $R$-parity conserving SUSY is the presence 
of a stable weakly interacting massive particle (WIMP) which can be a good cold dark matter.
Lightest neutralino $\tilde{\chi}_1^0$ is most often the lightest supersymmetric particle (LSP) 
and a good candidate for cold dark matter. In some regions of the parameter space, it has 
the annihilation cross-section to Standard Model (SM) particles yielding correct relic 
abundance satisfying WMAP/PLANCK \cite{WMAPdata,PLANCK}. 

In mSUGRA, $\tilde{\chi}_1^0$ is bino dominated in a large  
part of the parameter space. For a bino DM, WIMP miracle occurs when they annihilate to 
leptons via $t$-channel exchange of sleptons with mass in the 30-80 GeV range \cite{WIMP-miracle}. 
However, slepton masses that light was already discarded by direct slepton searches at LEP2 \cite{slepton-mass}. 
Therefore, after LEP2, some distinct parts of mSUGRA parameter space that satisfies relic abundance are as follows:
\begin{itemize}
\item {\bf The $h$-resonance region }\cite{h-resonance} is characterized by $2m_{\tilde{\chi_1^0}} \sim m_h$ which 
occurs at low $m_{1/2}$. In this region, $\tilde{\chi}_1^0$ annihilation cross-section enhances due to the presence of a $s$-channel $h$-resonance.
\item  {\bf $A$-funnel region}  \cite{A-funnel} is where $2m_{\tilde{\chi_1^0}} \sim m_A$; $A$ is the CP-odd Higgs boson. 
This region is characterized by large ${\rm tan}\beta \sim 50$.
\item {\bf Hyperbolic branch/focus point (HB/FP) region} \cite{HB/FP} is the parameter space where large 
$m_0$ region corresponds to small $\mu$ and thus Higgsino dominates $\tilde{\chi}_1^0$ and annihilates to $WW$, 
$ZZ$ and $Ah$ significantly. 
\item  {\bf Stau co-annihilation region}\cite{stau-co} arises if neutralino-LSP is nearly degenerate with the stau 
($m_{\tilde{\chi_1^0}} \simeq m_{\tilde \tau_1}$). In mSUGRA, this occurs at low $m_0$ and high $M_{1/2}$.
\item {\bf Stop co-annihilation} \cite{stop-co} occurs in mSUGRA with some particular values of $A_0$, 
where lighter stop ($\tilde t_1$) becomes nearly degenerate with the LSP.   
\end{itemize}

After LHC data with the discovery of Higgs and exclusion limits 
on the squark/gluino masses, many of the above DM regions in mSUGRA 
are highly constrained. With 20.3 fb$^{-1}$ integrated luminosity and $E_{CM}$= 8 TeV, 
ATLAS \cite{ATLAS} and CMS \cite{CMS} collaborations have excluded equal squark and 
gluino mass below 1.7 TeV completely ruling out  $h$-resonance region whereas, 
the $A$-funnel, stau and stop co-annihilation regions are partly excluded. 
Observation of Higgs mass at about 125 GeV indicates towards large $m_0$  ($m_0> 0.8$ TeV) 
and large $A_0$ ( $|A_0| > 1.8m_0$ for $m_0<5$ TeV) \cite{higgs-m0}. 
For $m_0>0.8$ TeV, stau co-annihilation is only viable at very large $M_{1/2}$ values 
which makes the SUSY discovery at the collider very challenging. 
The HB/FP region remains unscathed by the LHC squark/gluino searches 
as  it requires low $\mu$ at very large $m_0 \sim 3-10$ TeV for $A_0=0$. 
However, Higgs mass at 125 GeV (requires large $|A_0|$) push the 
region to much higher $m_0\sim 10-50$ TeV values. 
A small part of stop co-annihilation is the only region of mSUGRA 
parameter space alive, having some possibilities of seeing at 14 TeV LHC.

Non-universality in the gaugino and/or scalar sector on the other hand, 
can provide a lot more breathing space. The implications of direct search bound 
from LHC on neutralino dark matter have been studied extensively. See for example, \cite{DM-LHC1,DM-LHC2,AD-recent}. 
In our analysis, we choose four benchmark 
points (BP) which are motivated from different LSP annihilation and co-annihilation 
mechanism and consistent with all experimental limits. 

\begin{figure}[htbp]
\begin{center}
%\vspace*{-2.0cm}
\centerline{\psfig{file=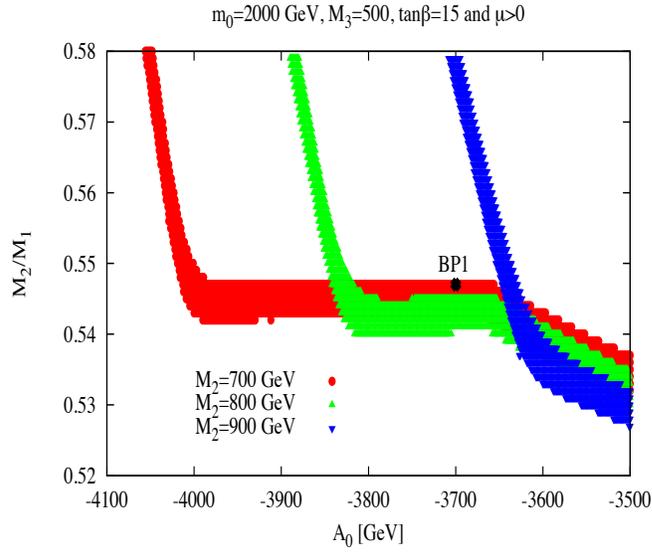,width=12 cm,height=12cm}}
\caption{A sample parameter space scan for gaugino mass non-universality with $M_3<M_2<M_1$ in $A_0$ vs $M_2/M_1$ 
plane to satisfy DM abundance. 
$M_2$= {(700, 800, 900)} GeV, yields three discrete consistent regions in red, blue and green respectively with $M_2/M_1$ varying along 
y-axis with $A_0$ varying along x-axis. We choose $M_3$= 500 GeV, $m_0$= 2000 GeV, $\tan\beta$= 15. BP1 represent a benchmark point of this sort.} 
\label{fig:BP1}
\end{center}
\end{figure}

\noindent {\bf i) BP1}:  If $M_2<M_1$ at GUT scale and EW scale, and $M_2< \mu$ at low scale, the LSP $\tilde{\chi}_1^0$ 
is wino dominated and then lightest chargino is almost degenerate with LSP. 
Chargino co-annihilation crucially controls relic abundance in such a region of parameter 
space, apart from larger wino component itself increases annihilation cross-section. A large part of purely 
wino DM hence provides under-abundance \cite{wino-DM}.  However, we scan the wino dominated 
parameter space where it is consistent with relic abundance from WMAP.
As an example, we have scanned the parameter space over $M_1,~M_2$ and $A_0$ for 
$m_0=2000$ GeV, $M_3=500$ GeV ${\rm tan}\beta=15$ and 
$\mu>0$. The allowed values of $M_2/M_1$ as a function of $A_0$ are plotted in 
Fig.~\ref{fig:BP1} for three different values $M_2=$ 700, 800 and 900 GeV in 
red, blue and green respectively. When we vary $M_2$ continuously, they merge 
into a continuous region. It is important to note in Fig.~\ref{fig:BP1} 
the vertical high $A_0$ region is dominated by stop co annihilation as the stop becomes 
lighter with increasing $A_0$ and a small change in $A_0$ results in a big change in $M_2/M_1$ to 
keep relic abundance within proper limit. The horizontal part of red, blue and green region 
with smaller $A_0$ on the other hand, represent wino dominated dark matter with nearly 
degenerate chargino and co-annihilation to yield proper abundance. 
For example, with $M_2=700$ GeV, $|A_0|>4000$ GeV is dominated by stop co-annihilation 
and $|A_0|<4000$ GeV characterizes wino DM. Our first benchmark point BP1 is a representative 
of this particular non-universal gaugino mass scenario $M_3<M_2<M_1$ with wino dominated DM. 
The benchmark points are explicitly written in Table \ref{tab:benchmark}. While 
gaugino mass non-universality has been used to obtain BP1, scalar masses are 
kept universal. 

Also note that gaugino non-universality with $M_3<M_2<M_1$ is obtained within the framework of
SUSY-GUT in $SU(5)$ or $SO(10)$ \cite{nonunigaugino, BM-AD-SB} with dimension five operator in the
extension of the gauge kinetic function $f_{ab}(\Phi^{j})$
%\bea
 %Re f_{a b}(\phi)F_{\mu \nu}^{\a}F^{\beta \mu \nu}= \frac {\eta (\Phi^s)}
%{M}Tr(F_{\mu \nu}\Phi^N F^{\mu \nu})
%\eea
\noindent
where non-singlet chiral superfields $\Phi^N$ belongs to the symmetric product of the adjoint representation of
the underlying gauge group as\\
\begin{eqnarray}
SU(5):    &  (24\times 24)_{symm} = 1+24+75+200 \\ \nonumber
SO(10):   &   (45\times 45)_{symm}=1+54+210+770
\end{eqnarray}

\begin{table}[htb]
\begin{center}
\begin{tabular}{|c|c|}
\hline
\hline
 Representation & $M_{3}:M_{2}:M_{1}$ at $M_{GUT}$ \\
\hline
{\bf 75} of $SU(5)$ & 1:3:(-5) \\
\hline
{\bf 200} of $SU(5)$ & 1:2:10 \\
\hline
{\bf 770} of $SO(10)$: {$H \rightarrow SU(4) \times SU(2) \times SU(2)$} &
1:(2.5):(1.9) \\
\hline
\hline
\end {tabular}
\end{center}
\label{tab:ratio}
\caption {Non-universal gaugino mass ratios for different non-singlet
representations belonging to $SU(5)$ or $SO(10)$ GUT-group that gives rise
to the hierarchy of $M_3 < M_1, M_2$ at the GUT scale.}
\end{table}

Gaugino masses become non-universal if these non-singlet Higgses are responsible for
the GUT-breaking. $75$ and $200$ belonging to
$SU(5)$ or $770$\footnote{For breaking through $770$, we quote
the result, when it breaks through the Pati-Salam gauge group
$G_{422D}$ ($SU(4)_C \times SU(2)_L\times SU(2)_R$ with even
D-parity and assumed to break at the GUT scale itself.} of $SO(10)$
yield a  hierarchy of $M_3 < M_1, M_2$ shown in Table 1. The specific 
non-universal ratio(s) used in the scan can be motivated from GUT breaking
with a linear combination of aforementioned non-singlet representations.

 \noindent {\bf ii) BP2}: Our second benchmark point BP2 is motivated from the Hyperbolic branch/Focus Point 
 region of DM. As has already been mentioned, for mSUGRA, very large values 
 $m_0\sim 10-50$ TeV is required to make $\mu$ small such that LSP becomes 
 predominantly a Higgsino, that paves the way for correct relic abundance through annihilation to 
 $WW$, $ZZ$ and $Ah$ final states. However, introduction of non-universality in the 
 scalar sector, in particular in the Higgs parameters $m_{H_u}$ and $m_{H_d}$ at GUT scale, gives rise to 
 small $\mu$, even without going to such high scalar masses, making it accessible to collider events at LHC.
 Again, following our strategy to minimize the number of parameters to choose BP2, we kept all gaugino and 
 other scalar masses universal at the high scale.

 \noindent {\bf iii) BP3}: Our third benchmark point BP3 represents stau co-annihilation region 
 exploiting non-universality in the scalar sector. We have used squark-slepton non-universality as well as 
 non-universality in the family to make the third family slepton masses lighter than other scalars at the high scale.
 Although such scalar non-universality is mostly phenomenological, 
 having impacts on CP and FCNC issues, it can be motivated from string-inspired models 
 with flavor dependent couplings to the modular fields \cite{3rd1,3rd2}.
 In Table \ref{tab:benchmark} we show all the inputs at high scale as well as the low-scale SUSY masses.

\noindent {\bf iv) MSG}: The mSUGRA benchmark point represent stop co-annihilation region of DM parameter space. 
In mSUGRA, stop co-annihilation occurs at distinct non-zero values of $|A_0|$ in a narrow range, for 
particular values of $m_0,~M_{1/2},~{\rm tan}\beta~{\rm and~Sign}{(\mu)}$.
Higgs mass of 125 GeV can also be obtained in the whole $m_0-M_{1/2}$ plane with 
$m_0> 0.8$ TeV for large $A_0$. Hence, a tiny region of $m_0,~M_{1/2}~{\rm and}~A_0$ 
parameter space simultaneously satisfy right Higgs mass and dark matter constraints. 

However, the situation changes dramatically if we introduce non-universality in gaugino sector, if we assume 
$M_3<M_2=M_1$, effectively adding one more parameter to mSUGRA. Then Higgs mass of 125 GeV can be 
satisfied in a larger range of $A_0$ values; while for a given $A_0$, dark matter density can be satisfied by varying 
$M_{1,2}$ appropriately through stop co-annihilation.

 \begin{figure}
\begin{center}
%\vspace*{-2.0cm}
\centerline{\psfig{file=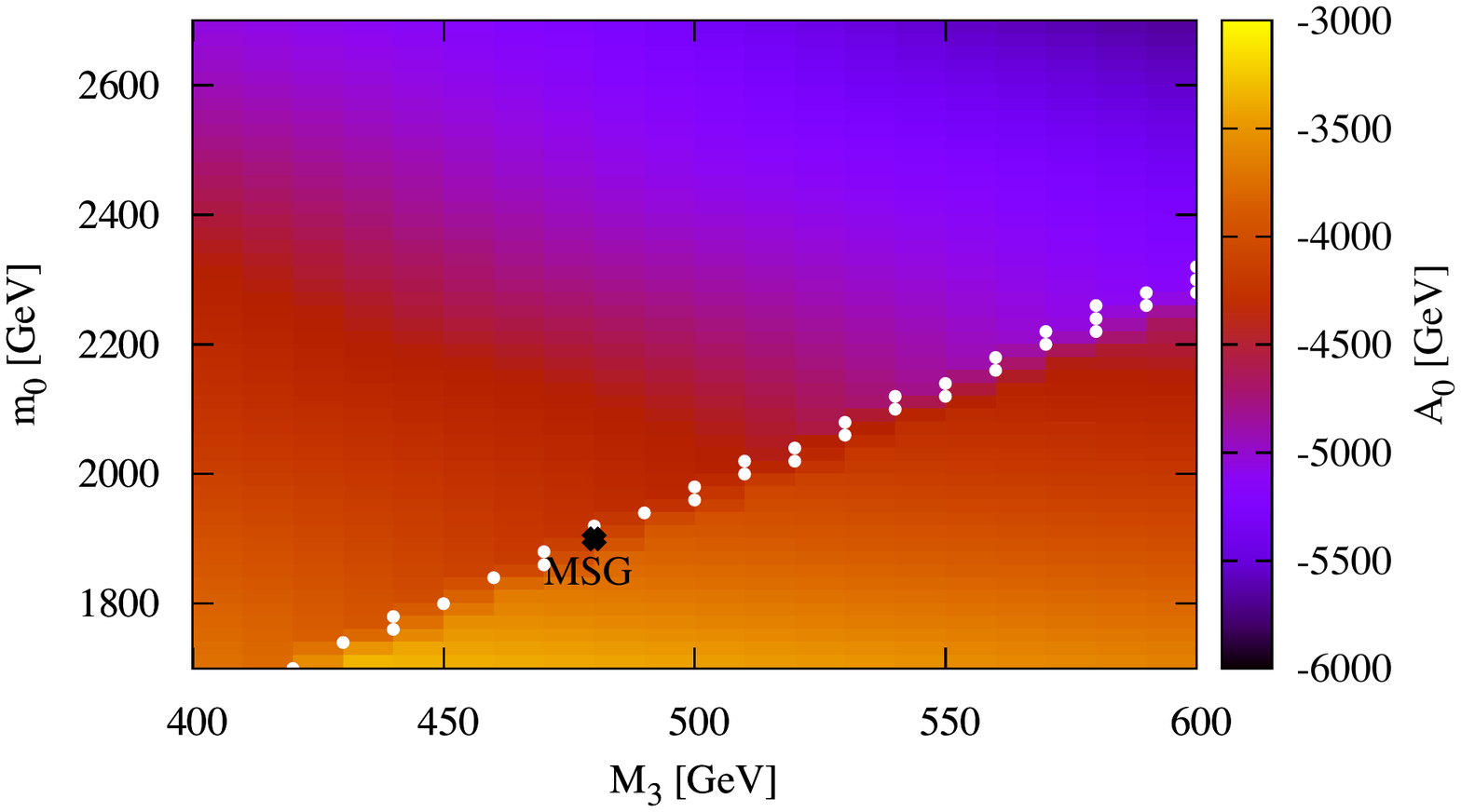,width=8 cm,height=10cm}
\hskip 20pt \psfig{file=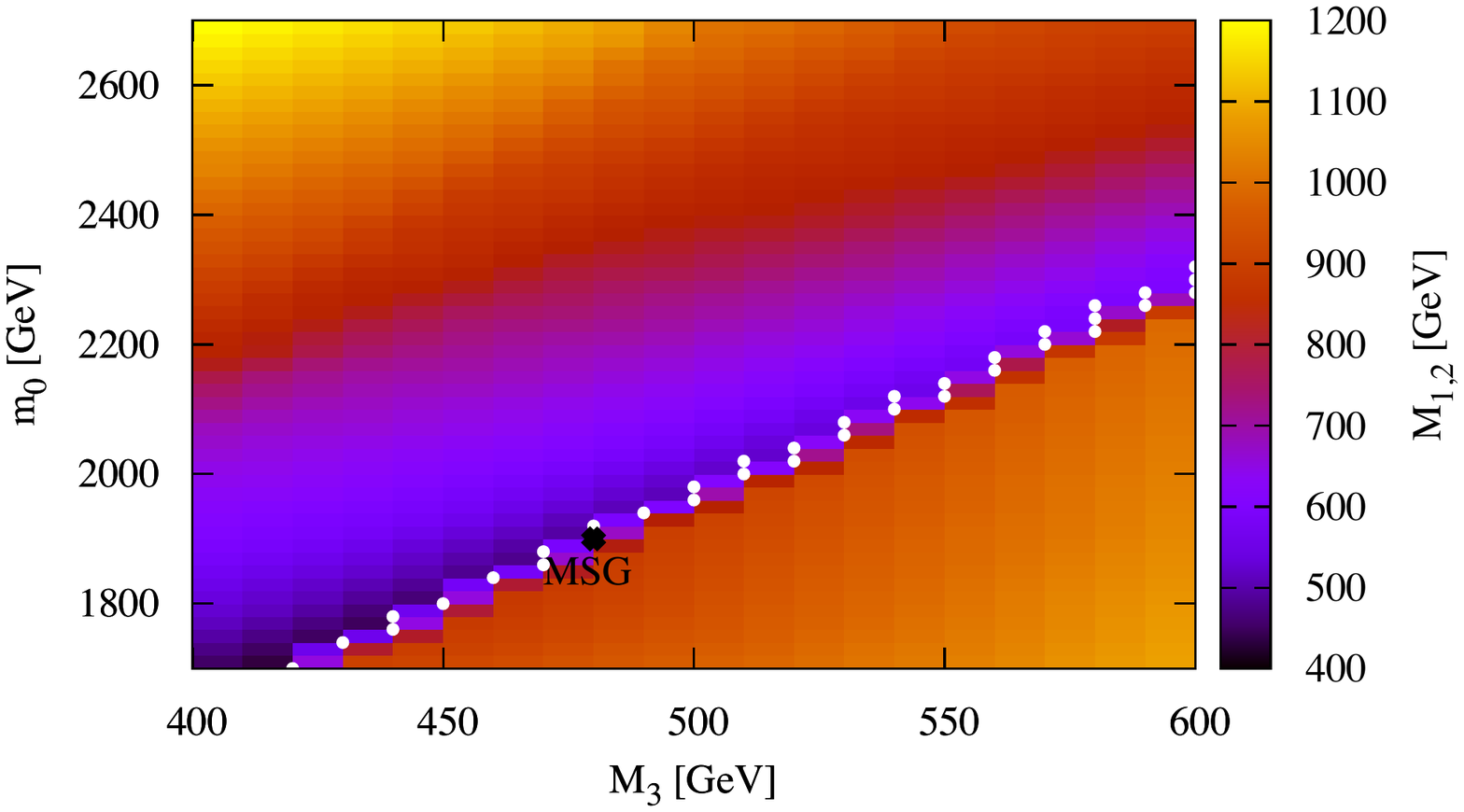,width=8 cm,height=10cm}}
\caption{Four-dimensional parameter space scan with $m_0,~M_3,~M_{1,2}~{\rm and}~A_0$; 
for ${\rm tan}\beta=15$ and positive $\mu$ to obtain correct dark matter relic abundance, Higgs mass and 
other low energy constraints. LHS: Three-dimensional subset of the scan with $M_3$ (along x-axis), 
$m_0$ (along y-axis), $A_0$ (color gradient); RHS:  $M_3$ (along x-axis), $m_0$ (along y-axis) and
$M_{1,2}$ (color gradient). mSUGRA points are represented in white dots and our benchmark MSG is one of them.} 
\label{fig:MSG}
\end{center}
\end{figure}

In Fig.~\ref{fig:MSG}, 
we have presented a sample scan of such a four-dimensional parameter space $m_0,~M_3,~M_{1,2}~{\rm and}~A_0$, 
for ${\rm tan}\beta=15$ and positive $\mu$. Left panel shows a three-dimensional subset of the scan with $M_3$ (along x-axis), 
$m_0$ (along y-axis), $A_0$ (color gradient) and on the right panel we have $M_3$ (along x-axis), $m_0$ (along y-axis) and
$M_{1,2}$ (color gradient). For a given $M_3$ and $m_0$, there is a range of $A_0$ and $M_{1,2}$ which gives rise to right 
relic abundance and Higgs mass. For simplicity, in Fig.~\ref{fig:MSG}, we consider the minimum possible values of 
$A_0$ and $M_{1,2}$ which are consistent with experimental constraints. As a result, the whole parameter space shown in the 
figure is allowed by dark matter and Higgs mass constraint. White dots in Fig.~\ref{fig:MSG} corresponds to $M_3=M_{1,2}$, 
i.e. mSUGRA points as a subspace of such gaugino non-universality. Our benchmark point MSG is represented by one of these white dots. 
We didn't chose a non-universal benchmark point from this region as the collider signature is expected to be the same as the chosen MSG point.

For renormalization group equation RGE, we use the code {\tt SuSpect v2.3} \cite{suspect} 
with $m_t = 173.2$ GeV, $m_b = 4.2$ GeV, $m_\tau = 1.777$ GeV and stick to 
two-loop RGE with radiative corrections to the gauginos and squarks. We use full one loop and
dominant two loop corrections for the Higgs mass. We ensure radiative electroweak symmetry breaking 
to evaluate the Higgsino parameter $\mu$ at the low scale out of high-scale inputs $m_{H_u}^2$ and $m_{H_d}^2$
and the electroweak symmetry breaking scale has been set
at $\sqrt{m_{\tilde{t_{L}}}m_{\tilde{t_{R}}}}$, the default value in the code
{\tt SuSpect}. The low scale value of the strong
coupling constant has been chosen at
${\alpha_3 (M_{Z})}^{\overline{MS}}= 0.1172$. We compute the cold dark matter relic density with the code micrOMEGAs3.1 \cite {micromegas}.

\begin{table}
\begin{center}
\begin{tabular}{|c|c|c|c|c|}
\hline
%\hline
parameter & BP1 &  BP2 &  BP3 &  MSG \\
\hline
%\hline
$\small{\tan\beta} $ &\small15.00  &\small 15.00 & \small 15.00 & \small 15.00\\
$\small{(M_3,M_2,M_1)}$ &\small(500,700,1282)  &\small(500,500,500) & \small(500,500,500) & \small(480,480,480) \\
$\small{(m_{\widetilde f},m_{\widetilde \tau})}$ & \small(2000,2000) &\small(2500,2500) & \small(2000,518) & \small(1900,1900)\\
$ \small{(m_{H_u},m_{H_d})}$ & \small(2000,2000) & \small(3047,4000) & \small(2000,2000) & \small(1900,1900)\\
$\small{A_0}$ &\small -3700  &\small -3500 & \small-3500 & \small-4239 \\
$\small{sgn(\mu)}$ &\small+  &\small+ &\small+ & \small+\\
\hline
%\hline
%$\mu$ & 0 & 0 & 0 & 0 \\
%$m_{h}$ & 124.1 & 123.3 & 123.2 & 123.8\\
$m_{\widetilde g}$ & 1251  & 1277 & 1265.2 & 1201.3 \\
%$m_{\widetilde q_{1,2}}$ & 2200  & 3100 & 2900 & 3100  \\
%$m_{\widetilde l_{1,2}}$ & 3000  & 3000  & 2700 & 3000 \\
$m_{\widetilde u_{L}}$ & 2234  & 2667  & 2217.6 & 2108 \\
$m_{\widetilde t_{1}}$ & 761  & 785.6  & 865 &  243\\
$m_{\widetilde t_{2}}$ & 1656.5  & 1950.2  & 1670 & 1487\\
$m_{\widetilde b_{1}}$ & 1635  & 1940.5  & 1651 & 1442 \\
$m_{\widetilde b_{2}}$ & 2117  & 2558.3  & 2124.6 & 1988 \\
$m_{\widetilde e_{L}}$ & 2054  & 2473  & 2019 &  1918.3\\
$m_{\widetilde \tau_{1}}$ & 1962  & 2420.3  & 219.7 &  1797\\
$m_{\widetilde \tau_{2}}$ & 2013.8 & 2467.2 & 492.2 & 1870 \\
$m_{\widetilde \chi_{1}^{\pm}}$ & 588.3  & 262.6  & 417.6 & 404.6 \\
$m_{\widetilde \chi_{2}^{\pm}}$ & 1584.4  & 447.5  & 1523 & 1742\\
$m_{\widetilde \chi_{4}^{0}}$ & 1584.3  & 447.7  & 1522.4 & 1741 \\
$m_{\widetilde \chi_{3}^{0}}$ & 1581.3  & 285.3  & 1520.3 & 1739.4 \\
$m_{\widetilde \chi_{2}^{0}}$ & 588.4  &  275.3  & 417.6 & 404.6 \\
$m_{\widetilde \chi_{1}^{0}}$ & 561.7  & 201.7  & 211.4 & 208.3\\
\hline
%\hline
$m_{h}$ & 124.1 & 123.4 & 123.2 & 123.8\\
$\Omega_{\tilde\chi_1}h^2$& 0.118 & 0.127 & 0.116 & 0.112\\
$BF(b\to s\gamma)$ & $2.98\times 10^{-4}$ & $2.83\times 10^{-4}$
& $3.00\times 10^{-4}$ & $3.25\times 10^{-4}$\\
$Br (B_s \rightarrow \mu^+ \mu^-)$ & $3.10\times 10^{-9}$ & $3.07\times 10^{-9}$
& $3.09\times 10^{-9}$ & $3.13\times 10^{-9}$\\
\hline
\hline
\end{tabular}
\end{center}
\caption {\small Benchmark points BP1, BP2, BP3 and MSG. Model inputs, low scale
predictions and constraints are mentioned.}
\label{benchmark-points}
%\end{table}
\label{tab:benchmark}
\end{table}

\section {Collider Simulation and Results}

Non-universal SUGRA points advocated in the earlier section 
can be seen at the future run of LHC in bottom rich and 
leptonic final states. This also serves as a major distinguishing feature from mSUGRA points
surviving Higgs mass and dark matter constraints. 

We first discuss the strategy for the simulation including the
final state observables and the cuts employed therein 
and then we discuss the numerical results in next subsection.

\subsection{Strategy for Simulation} 

The spectrum generated by {\tt SuSpect} as described in the earlier
section, at the benchmark points are fed into the event generator
{\tt Pythia} 6.4.16 \cite{Pythia} by {\tt SLHA} interface \cite{sLHA}
for the simulation of $pp$ collision with center of mass energy 14 TeV for 
LHC.

The default parton distribution functions {\tt CTEQ5L} \cite{CTEQ}, 
QCD scale $\sqrt{\hat{s}}$ in {\tt Pythia} 
has been used. All possible SUSY processes  (mainly 2$\rightarrow$2) and decay chains consistent
with conserved $R$-parity have been kept open with initial and final state radiation on. 
We take hadronization into account using the fragmentation functions
inbuilt in {\tt Pythia}.

The main 'physics objects' that are reconstructed in a collider, are:
\begin{itemize}
\item Isolated leptons identified from electrons and muons
\item Hadronic Jets formed after identifying isolated leptons
\item Unclustered Energy made of calorimeter clusters with $p_T~>$ 0.5 GeV
(ATLAS) and $|\eta|<5$, not associated to any of the above types of
high-$E_T$ objects (jets or isolated leptons).
\end{itemize}
We try to mimic the experimental reconstruction for these objects 
in Pythia as follows.
%\bei
Isolated leptons: Isolated leptons are identified as electrons and muons with $p_T>$ 10 GeV
and $|\eta|<$2.5. An isolated lepton is separated from another lepton by 
${\bigtriangleup R}_{\ell\ell}~ \geq $0.2, from jet 
(jets with $E_T >$ 20 GeV) with ${\bigtriangleup R}_{{\ell}j}~ \geq 0.4$, while 
the energy deposit $\sum {E_{T}}$ due to low-$E_T$ hadron activity around a
lepton within $\bigtriangleup R~ \leq 0.2$ of the lepton axis should be
$\leq$ 10 GeV. $\bigtriangleup R = \sqrt {{\bigtriangleup \eta}^2
+ {\bigtriangleup \phi}^2}$ is the separation in pseudo rapidity and
azimuthal angle plane. The smearing functions of isolated electrons, photons
and muons are described below.

%\bei
 Jets:
%\eei 
Jets are formed with all the final
state particles after removing the isolated leptons from the list
with {\tt PYCELL}, an inbuilt cluster routine in {\tt Pythia}. The
detector is assumed to stretch within the pseudorapidity range
$|\eta|$ from -5 to +5 and is segmented in 100 pseudorapidity
($\eta$) bins and 64 azimuthal ($\phi$) bins. The minimum $E_T$ of
each cell is considered as 0.5 GeV, while the minimum $E_T$ for a
cell to act as a jet initiator is taken as 2 GeV. All the partons
within $\bigtriangleup R$=0.4 from the jet initiator cell is
considered for the jet formation and the minimum $\sum_{parton}
{E_{T}}^{jet}$ for a collected cell to be considered as a jet is
taken to be 20 GeV. We have used the smearing function and
parameters for jets that are used in {\tt PYCELL} in {\tt Pythia}.

%\bei
 b-jets:
%\eei
We identify partonic $b$ jets by simple $b$-tagging algorithm with
efficiency of $\epsilon_b = 0.5$ for $p_T >$ 40 GeV and $|\eta| <$ 2.5 
\cite{b-tagging-ref}.

%\bei
Unclustered Objects:
%\eei
 All the other final state
particles, which are not isolated leptons and separated from jets by
$\bigtriangleup R \ge$0.4 are considered as unclustered objects.
This clearly means all the particles (electron/photon/muon) with
$0.5< E_T< 10$GeV and $|\eta|< 5$ (for muon-like track $|\eta|<
2.5$) and jets with $0.5< E_T< 20$GeV and $|\eta|< 5$, which are
detected at the detector, are considered as unclustered objects.

\begin{itemize}

\item Electron/Photon Energy Resolution 

%\be
$\sigma(E)/E=a/\sqrt{E}\oplus b\oplus c/E$  \footnote{$\oplus$ indicates addition in quadrature}
%\ee

Where,\\
$~~~~~~~~$ $a$ = 0.03 [GeV$^{1/2}$], $~~$   $b$ = 0.005 \& $~$   $c$ = 0.2 [GeV] 
$~~~$  for $|\eta|< 1.5$ \\     
$~~~~~~~~~~~~$= 0.055 $~~~~~~~~~~~~~~~~$ = 0.005  $~~~~~~~$ = 0.6 $~~~~~~~~~$  
for $1.5< |\eta|< 5$ \\

\item { Muon $P_T$ Resolution :}

\begin{eqnarray}
%\nonumber
\sigma(P_T)/P_T&=&a  ~~~~~~~~~~~~~~~~~~~~~~~{\rm if} ~ P_T< 100 GeV\\
%\nonumber
&=&a+b\log(P_T/\xi) ~~~~~ {\rm if}~ P_T> 100 GeV
\end{eqnarray}

Where,\\
$~~~~~~~~$ $a$= 0.008 $~~$\&    $b$= 0.037 $~~~~~$  for $|\eta|< 1.5$ \\     
$~~~~~~~~~~$= 0.02 $~~~~~~$    = 0.05  $~~~~~~~~~$ $1.5< |\eta|<2.5$\\
and $\xi=100$ GeV.

\item {Jet Energy Resolution :}

%\be
$\sigma(E_T)/E_T=a/\sqrt{E_T}$
%\ee

Where,\\
$~~~~~~~~$ a= 0.55 [GeV$^{1/2}$], default value used in {\tt PYCELL}.

\item {Unclustered Energy Resolution :}

%\be
$\sigma(E_T)=\sqrt{\Sigma_{i}E^{(Unc.O)i}_T}$

Where, $\approx0.55$. One should keep in mind that the x and y component of 
$E^{Unc. O}_T$ need to be smeared independently with same smearing 
parameter.

\end{itemize}

We sum vectorially the x and y components of the momenta separately for all visible
objects to form visible transverse momentum $(p_T)_{vis}$,
%\bea
$(p_T)_{vis}=\sqrt{(\sum p_x)^2+(\sum p_y)^2}$
%\eea
where, $\sum p_x =\sum (p_x)_{iso~\ell}+\sum (p_x)_{jet}+\sum (p_x)_{Unc.O}$
and similarly for $\sum p_y$.
We identify $(p_T)_{vis}$ as missing energy
$E_{T}\!\!\!\!/$:
%\bea
$E_{T}\!\!\!\!/ = (p_T)_{vis}$
%\eea

We also define Effective mass $H_T$ as the scalar sum of transverse momenta of 
visible objects like lepton and jets with missing energy 

%\bea
$H_T=\sum {p_T}^{\ell_i}+ {p_T}^{jets}+{E_{T}}\!\!\!\!/ $
%\eea

Effective mass cuts have really been useful to reduce SM background for 
the signals as we will see shortly.

We studied the benchmark points in multi-lepton final states as well as in 
b-rich final states at $E_{CM}$= 14 TeV at LHC with varying cuts. The 
channels we study are:

\begin{itemize}

\item Four b-jet with inclusive lepton and jets $(4b)$ :
$4b~+ X~ + {E_{T}}\!\!\!\!/$ ; Here $X$ implies any number of inclusive jets or leptons without any specific veto on that. 
 Basic cuts applied here are ${p_T}^b>$ 40 GeV,  ${E_{T}}\!\!\!\!/ >$100 GeV.

\item Four b-jet with single lepton $(4b\ell)$ :
 $4b~+ \ell~ + X~ + {E_{T}}\!\!\!\!/$ ; Here $X$ implies any number of inclusive jets without any specific veto on that. The lepton can have any charge $\pm$.
 Basic cuts applied here are ${p_T}^b>$ 40 GeV, ${p_T}^{\ell}>$ 20 GeV, $|\eta| <$ 2.5, ${E_{T}}\!\!\!\!/ >$100 GeV.

\item Two b-jets with di-lepton $(2b2\ell)$:
$2b~ + 2\ell~ + X~ + {E_{T}}\!\!\!\!/$ ; Here $X$ implies any number of inclusive jets without any specific veto on that. Leptons can have any charge $\pm$ 
(including same and opposite sign). Basic cuts applied here are ${p_T}^b>$ 40 GeV, ${p_T}^{\ell}>$ 20 GeV, $|\eta| <$ 2.5, ${E_{T}}\!\!\!\!/ >$100 GeV.

\item Same sign dilepton with inclusive jets $(\ell^{\pm}\ell^{\pm})$: 
$\ell^{\pm}\ell^{\pm}~+X~ + {E_{T}}\!\!\!\!/$ ;
The basic cuts applied are ${E_{T}}\!\!\!\!/ >$ 30 GeV, ${p_T}^{\ell_1}>$ 40 GeV and ${p_T}^{\ell_2} >$ 30 GeV with $|\eta| <$ 2.5.

\item Trilepton with inclusive jets ($\ell^{\pm}\ell^{\pm}\ell^{\pm}$): 
$\ell^{\pm}\ell^{\pm}\ell^{\pm}~+X~ + {E_{T}}\!\!\!\!/$ ;
Basic cuts ${E_{T}}\!\!\!\!/ >$ 30 GeV, ${p_T}^{\ell_1} >$ 30 GeV, ${p_T}^{\ell_2} >$ 30 GeV and ${p_T}^{\ell_3} >$ 20 GeV with $|\eta| <$ 2.5.

\item Four-lepton with inclusive jets ($\ell^{\pm}\ell^{\pm}\ell^{\pm}$): 
$\ell^{\pm}\ell^{\pm}\ell^{\pm}\ell^{\pm}~+X~ + {E_{T}}\!\!\!\!/$ ;
For basic cuts no missing energy cut is employed while, lepton transverse momentum cuts are as follows: ${p_T}^{\ell_1}>$ 20 GeV, ${p_T}^{\ell_2} >$ 20 GeV and ${p_T}^{\ell_3} >$ 20 GeV and ${p_T}^{\ell_4} >$ 20 GeV with $|\eta|<$ 2.5.
\end{itemize}

$\ell$ stands for final state isolated electrons and or muons as discussed above and 
${E_{T}}\!\!\!\!/$ depicts the missing energy. Opposite-sign dilepton was not considered mainly because of the huge SM background from $t\bar{t}$ process. 

Apart from the basic cuts including a Z-veto of $|M_Z-M_{\ell^+\ell^-}|\ge$15 GeV on same flavor opposite sign dilepton arising in $2b2l$, trilepton and 
four lepton final states, we apply sum of lepton $p_T$ cut ($\sum {p_T}^{\ell_i}$) and combination of lepton  $p_T$ cut with MET, called modified effective mass cut 
$H_{T1}=\sum {p_T}^{\ell_i}+{E_{T}}\!\!\!\!/ $ to the leptonic final states, and harder $H_T$ cuts on b-rich final states and we refer to them as follows: 

\begin{itemize}
\item $C1$: $\sum {p_T}^{\ell_i}>$ 200 GeV
\item $C2$: $\sum {p_T}^{\ell_i}>$ 400 GeV
\item $C3$: $H_{T1} >$ 400 GeV
\item $C4$: $H_{T1} >$ 500 GeV
\item $C1'$: $\sum {p_T}^{\ell_i}>$ 100 GeV
\item $C2'$: $\sum {p_T}^{\ell_i}>$ 200 GeV
\item $C3'$: $H_{T1} >$ 150 GeV
\item $C4'$: $H_{T1} >$ 250 GeV
\item$C5$: $H_T >$ 1000 GeV, ${E_{T}}\!\!\!\!/ >$ 200 GeV, ${p_T}^b>$ 60 GeV.
\end{itemize}

 We have generated dominant SM events from $t\bar t$ in {\tt Pythia} for the same final 
states with same cuts and multiplied the corresponding events in different channels 
by proper $K$-factor ($1.59$) to obtain the usually noted next to leading order 
(NLO) and next to leading log re summed (NLL) cross-section at LHC \cite{ttbar}. 
$b\bar{b}b\bar{b}$ ,$b\bar{b}b\bar{b}W/Z$ and $t\bar{t}b\bar{b}$ background have 
been calculated in {\tt Madgraph5}\cite{Madgraph}.
The cuts are motivated such that we reduce the background to a great extent as shown 
in next subsection. Note that softer cuts $C1',C2',C3',C4'$ have been used 
for four lepton channel where the SM background is much smaller. 

\subsection{Numerical results}
\begin{table}[!ht]
\begin{center}
\begin{tabular}{|c|r|r|r|r|r|r|r|r|r|r}
\hline
Model Points & \multicolumn{1}{c|}{Total} &
  \multicolumn{1}{c|}{$\tilde g \tilde g$} &
\multicolumn{1}{c|}{$\tilde t_1 {\tilde t_1}^*$} 
& \multicolumn{1}{c|}{$\tilde{\chi}_i^0 \tilde{\chi}_j^0$} 
& \multicolumn{1}{c|}{$\tilde{\chi}_i^{\pm} \tilde{\chi}_j^{\mp}$} 
& \multicolumn{1}{c|}{$\tilde{\chi}_i^{0} \tilde{\chi}_j^{\pm}$} 
& \multicolumn{1}{c|}{$\tilde g \rightarrow \tilde t_1 \bar{t} $}
& \multicolumn{1}{c|}{$\tilde t_1 \rightarrow t \chi_1^0$} & 
\multicolumn{1}{c|}{$\tilde t_1 \rightarrow b \chi_1^{+}$}\\
\hline
\hline
 {\bf BP1} & 107.6 & 29.20 & 32.06& 0.11 & 7.18 & 14.6 & 99 $\%$ & 76.3 $\%$ & 23.7 $\%$\\
\hline
 {\bf BP2} & 607 & 26.3 & 15.1 & 64.7 & 126.9 & 354.8 & 99 $\%$ &10.3$\%$ & 45.2$\%$\\
\hline
 {\bf BP3} & 188 & 26.5 & 13.8 & 0.33 & 37.1 & 74.2 & 99 $\%$& 86.5$\%$ & 9.4$\%$  \\
\hline
 {\bf MSG} &18208 & 39  &  18010  &0.1 &28 &84 & 99$\%$ & 0$\%$ & 0$\%$\\
\hline
\hline
\end {tabular}
\end{center}
\caption {Total supersymmetric particle production cross-sections
(in fb) as well as some leading contributions from  $\tilde g \tilde g$ and $\tilde t_1 {\tilde t_1}^*$ and electroweak 
neutralino-chargino productions for each of the benchmark points with $E_{CM}$= 14 TeV. 
We also quote the significant decay branching fractions (in percentage). }
\label{production}
\end{table}

The main SUSY production cross-sections for the benchmark points have been noted in Table \ref{production} with the total 
cross-section for all 2$\rightarrow$2 SUSY processes.  All the non-universal benchmark points have 
similar gluino production and third family stop production, while the mSUGRA point has a huge stop production 
due to very light stop mass and the total cross-section for this point is also dominated by that.
Although other benchmark points have sufficiently large branching fraction of stop going to 
bottom chargino or stop neutralino, MSG has nothing in these channels as the stop is almost degenerate 
with the lightest neutralino, it only decays to $c \tilde{\chi}_1^{0}$ in loop.  For MSG, $\tilde{\chi}_2^{0}$ decays to 
$\tilde{\chi}_1^{0}h$ 95$\%$ and first chargino dominantly decays to $\tilde{t}_1\bar{b}$. Hence 3b channel can be a better 
channel to look for such MSG points. As mSUGRA is only alive in such a region of parameter space for the sake of dark matter, 
all MSG points will be similar in this aspect. We also note that for  BP1: $\tilde{\chi}_1^{\pm}$ decays into $\ell + \nu_{\ell}+ \tilde{\chi}_1^{0}$ 
through off-shell sleptons in $33\%$ while $\tilde{\chi}_2^{0}$ decays to leptonic final state is only $\simeq 1\%$.
BP2 has dominant production in electroweak gauginos. Associated production of the gluinos with neutralinos are also quite heavy. 
Here $\tilde t_1 \rightarrow t \tilde{\chi}_{2,3}^0$ branchings are also of the same order of $\tilde t_1 \rightarrow t \tilde{\chi}_{1}^0$. 
Although $\tilde{\chi}_2^{0}$ decays to  leptonic final state is 1$\%$, 
$\tilde{\chi}_1^{\pm}$ decays into $\ell + \nu_{\ell}+ \tilde{\chi}_1^{0}$ in 33$\%$. Huge electroweak production will significantly 
contribute to leptonic final states for BP2. For BP3, chargino and neutralino decays to tau-rich final state as a 
result of lighter stau. Hence, in addition to the standard leptons, channels with tau-tagging can be a better channel to look for 
this benchmark point.

\begin{figure}[htbp]
\begin{center}
%\vspace*{-2.0cm}
\centerline{\psfig{file=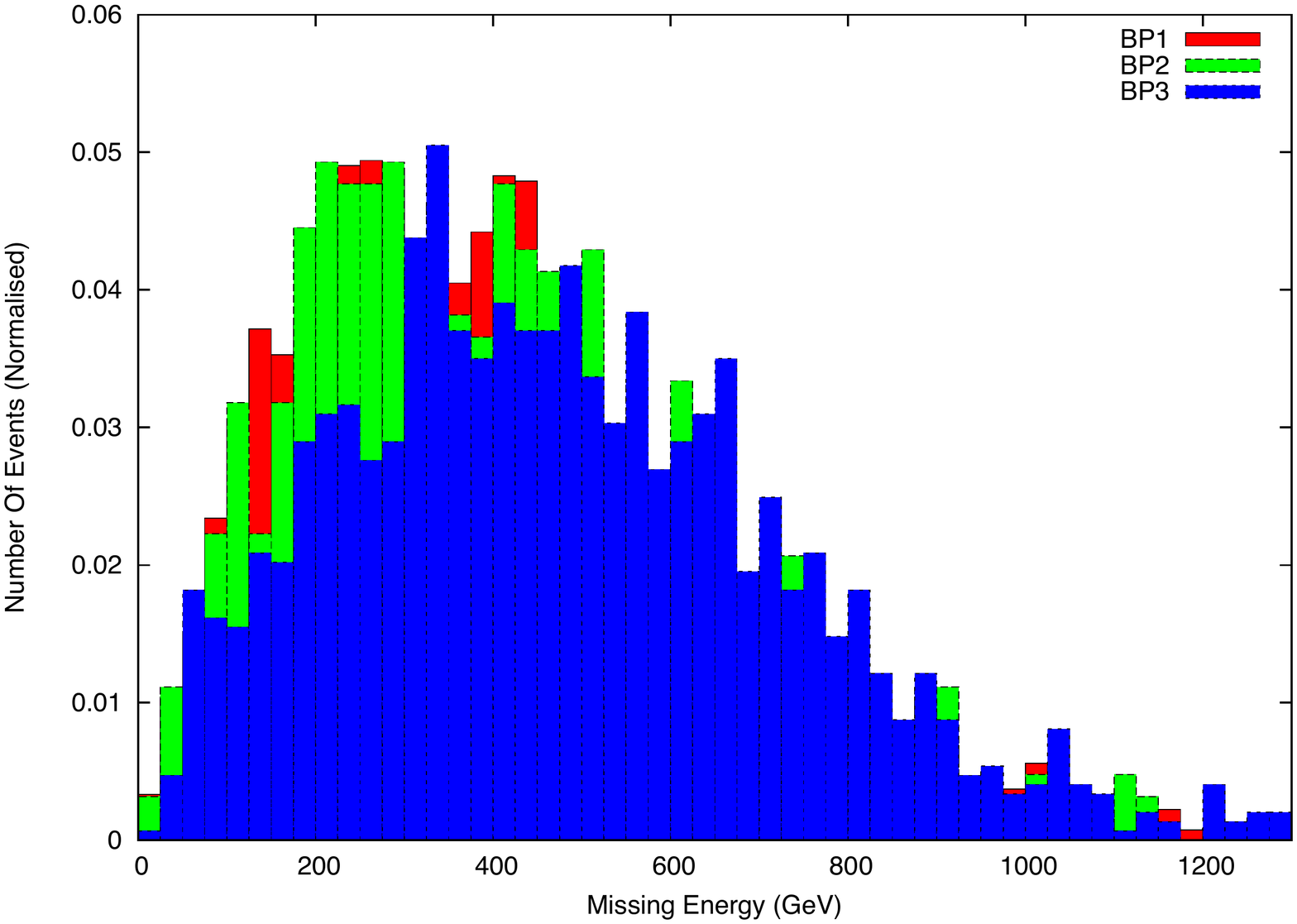,width=6.5 cm,height=7.5cm}
\hskip 20pt \psfig{file=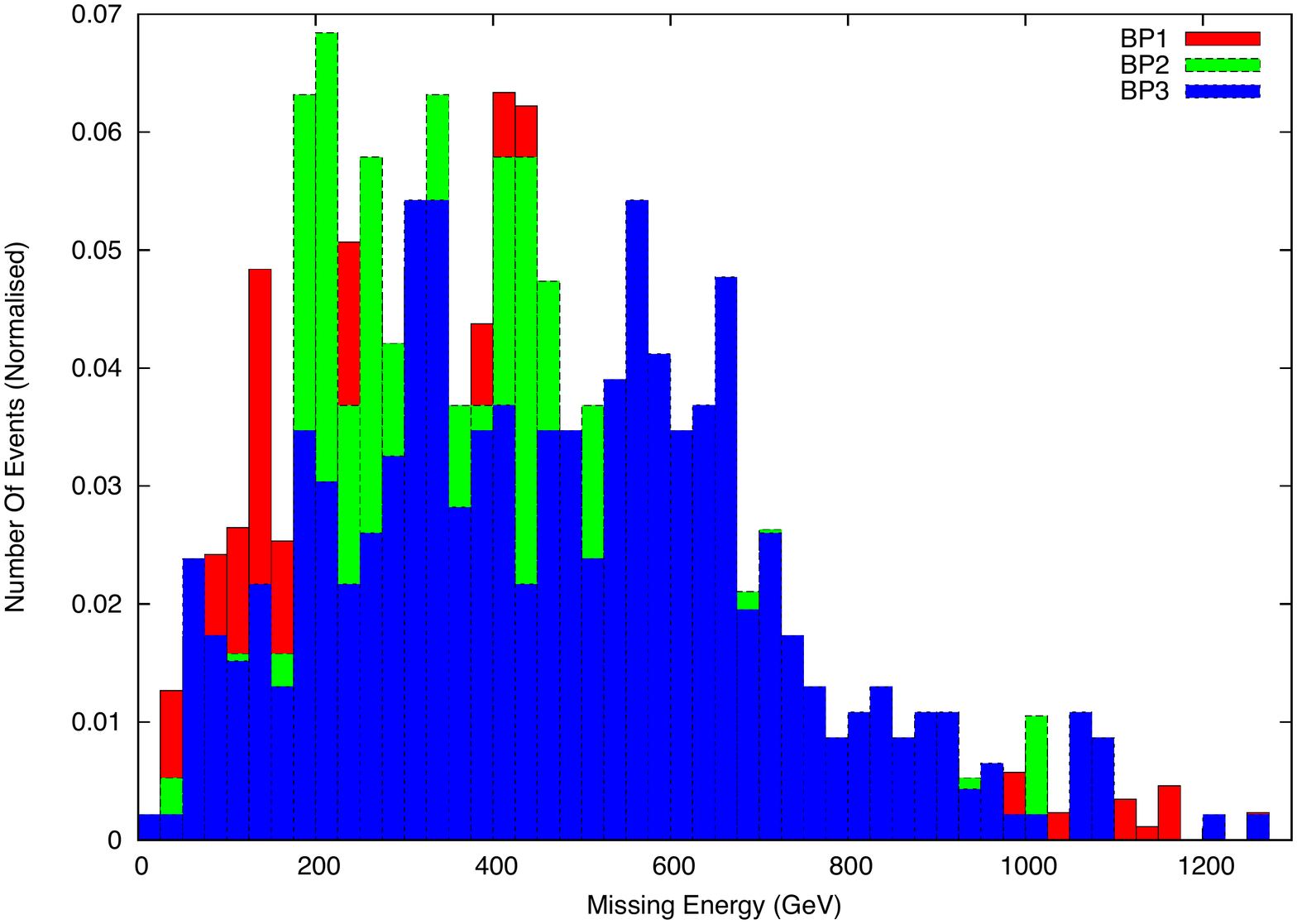,width=6.5 cm,height=7.5cm}}
\vskip 10pt
\centerline{\psfig{file=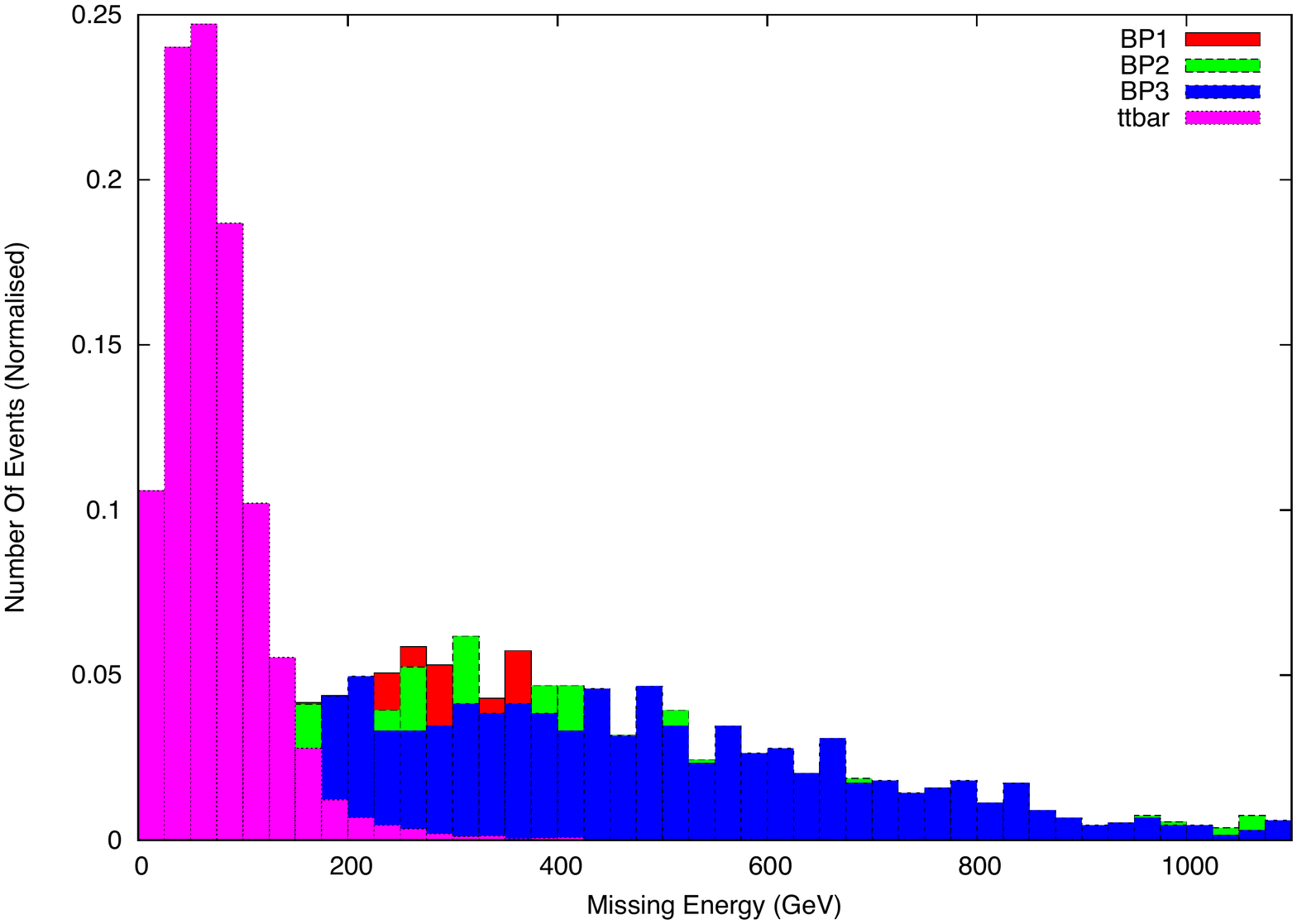,width=6.5 cm,height=7.5cm}}
\caption{Missing energy distribution in bottom rich final states at the benchmark points.
Top left: $4b$ channel, Top right: $4b\ell$ channel; bottom: $2b2\ell$ channel.
{\tt CTEQ5L} pdfset was used. Factorization and 
Renormalization scale has been set to $\mu_F=\mu_R=\sqrt{\hat s}$, 
sub-process center of mass energy.}
\label{fig:MET1}
\end{center}
\end{figure}

\begin{figure}[htbp]
\begin{center}
%\vspace*{-2.0cm}
\centerline{\psfig{file=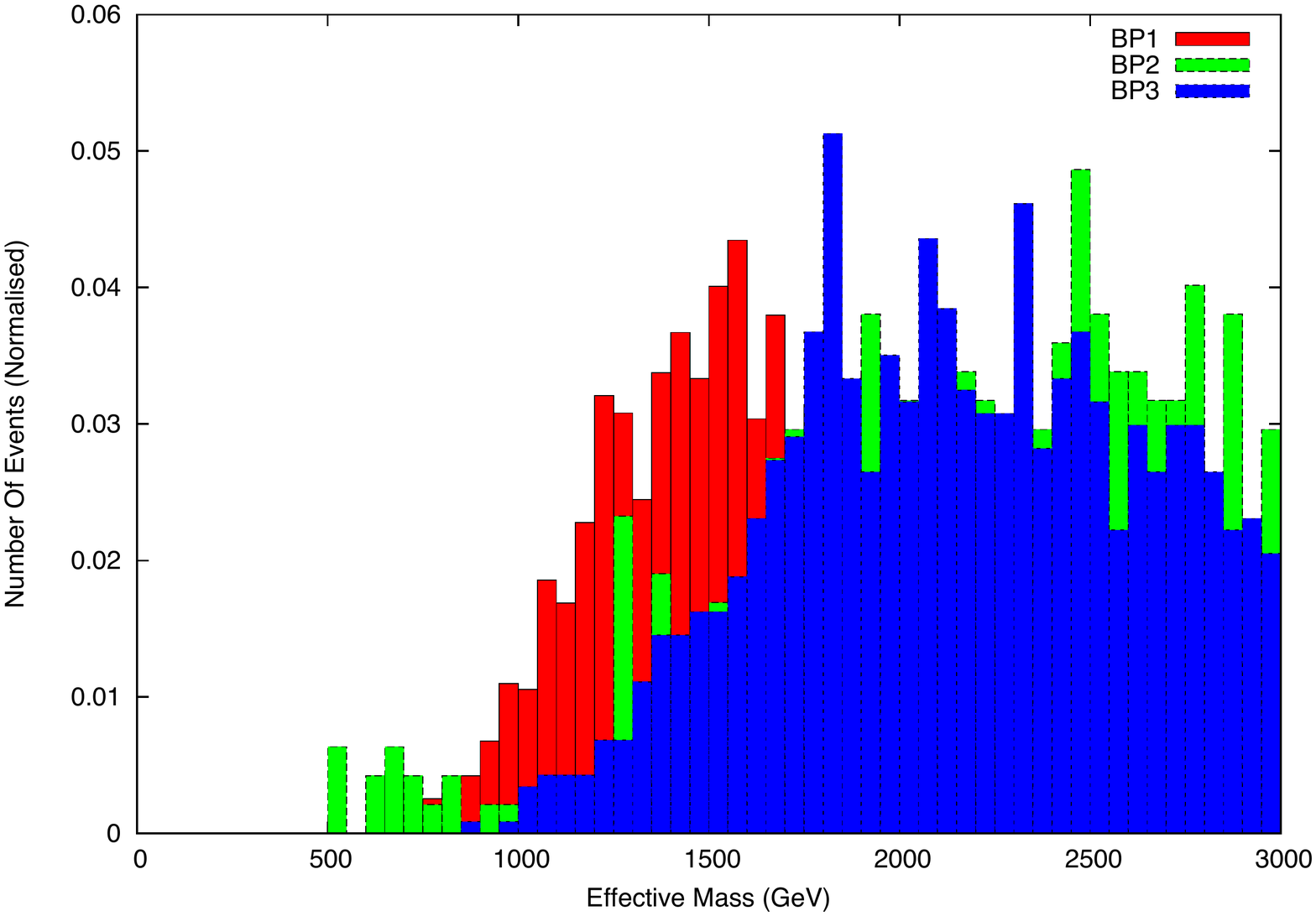,width=6.5 cm,height=7.5cm}
\hskip 20pt \psfig{file=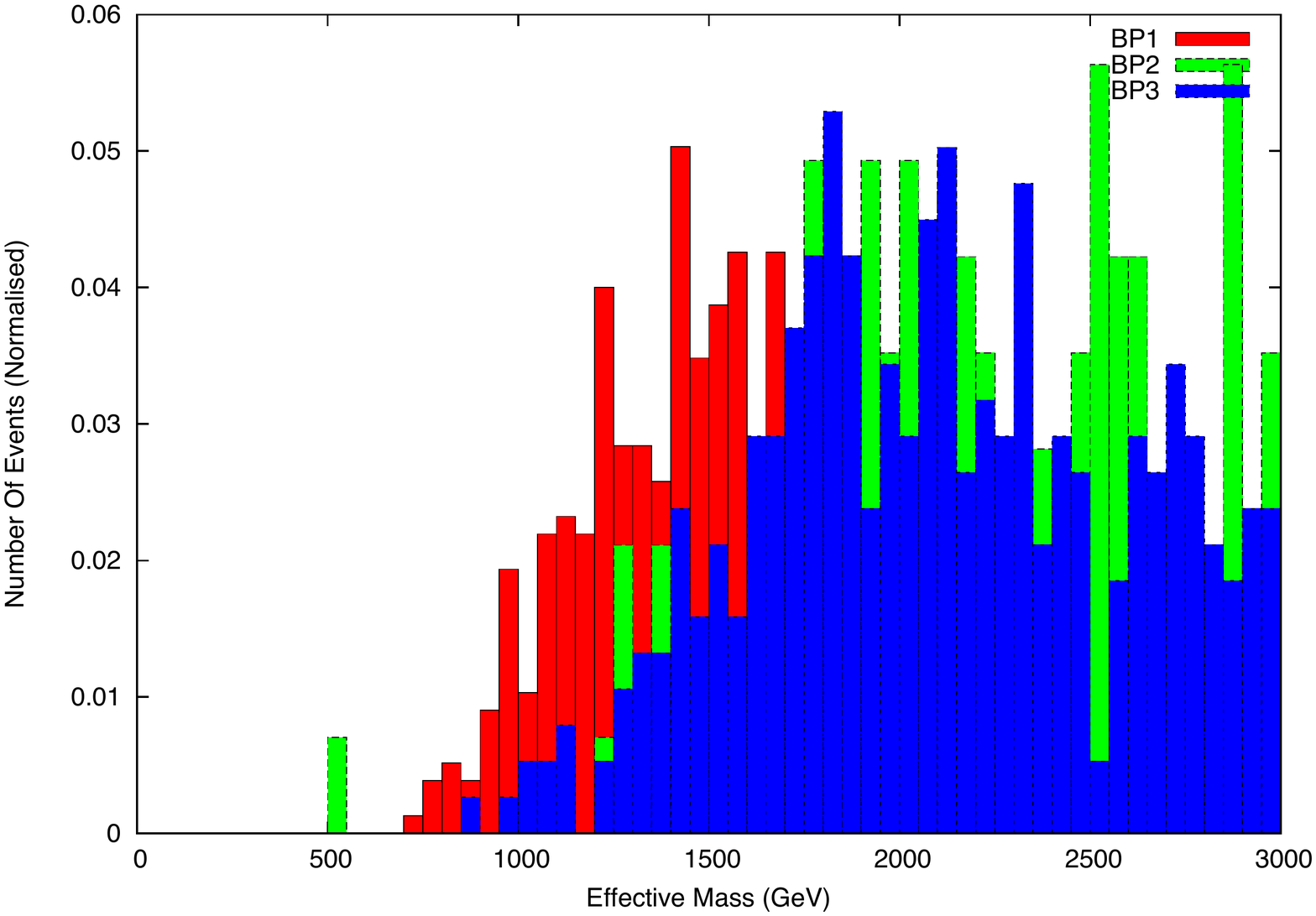,width=6.5 cm,height=7.5cm}}
\vskip 10pt
\centerline{\psfig{file=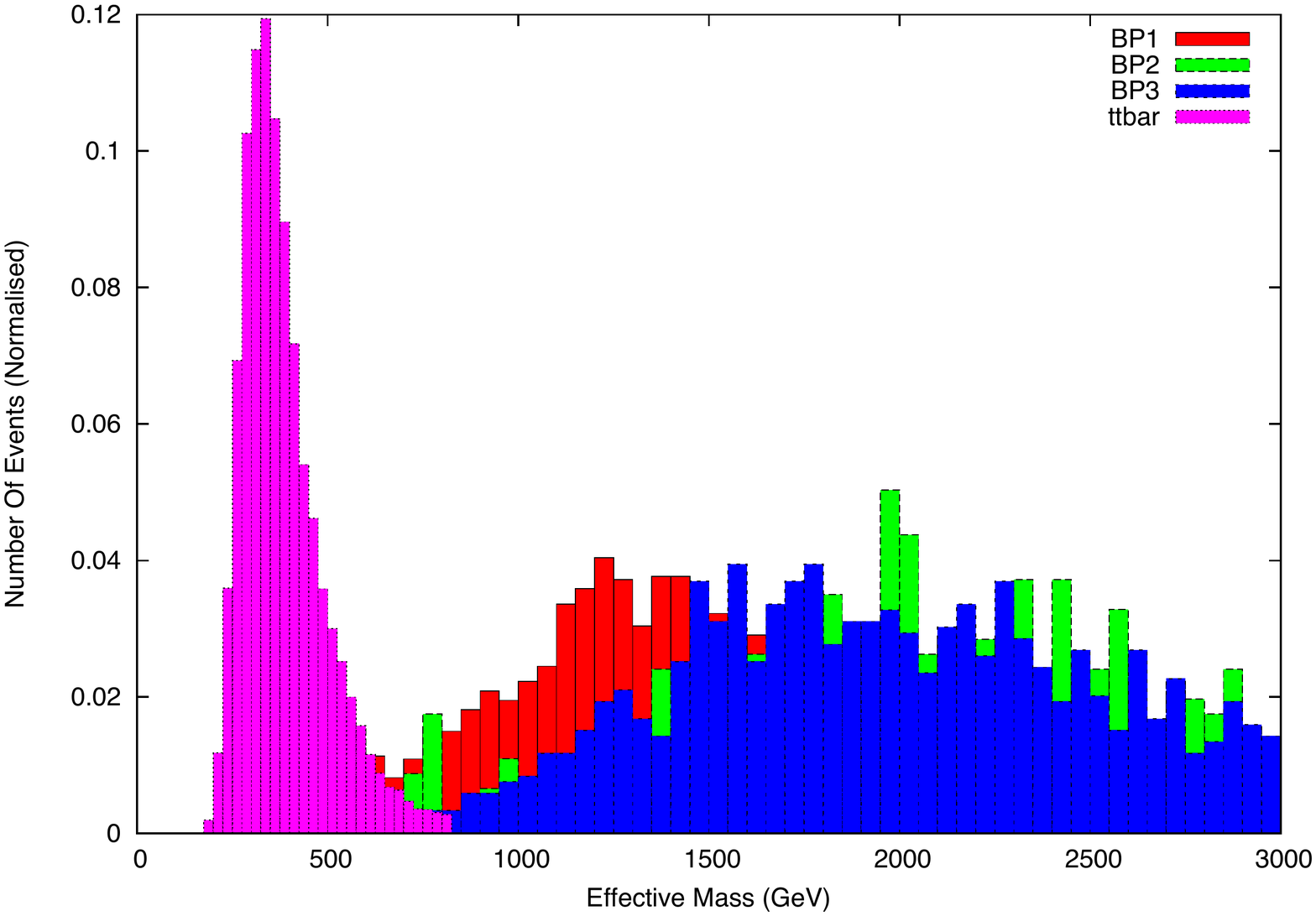,width=6.5 cm,height=7.5cm}}
\caption{ Effective mass distribution in bottom rich final states at the benchmark points. 
Top left: $4b$ channel, Top right: $4b\ell$ channel; bottom: $2b2\ell$ channel.
{\tt CTEQ5L} pdfset was used. Factorization and 
Renormalization scale has been set to $\mu_F=\mu_R=\sqrt{\hat s}$, 
sub-process center of mass energy.}
\label{fig:EFT1}
\end{center}
\end{figure} 

\begin{table}
\begin{center}
\begin{tabular}{|c|c|c|c|c|c|c|c|c|}
\hline
Benchmark Points & $\sigma_{4b}$ & $\sigma_{4bl}$ &$\sigma_{2b2l}$ &$\sigma_{4b}$($C5$) & $\sigma_{4bl}$($C5$) &$\sigma_{2b2l}$($C5$)\\
\hline
\hline
BP1 & 1.35 & 0.44 & 1.15 & 0.60 & 0.18 & 0.84\\
\hline
BP2 & 1.56 & 0.50 & 1.24 & 1.53 & 0.49 & 1.11\\
\hline
BP3 & 1.34 & 0.41 & 1.17 & 0.76 & 0.22 & 0.91 \\
\hline
MSG & 0.004 & 0.004  &  0.1 & $\le$0.001 & $\le$0.001  &  0.01 \\
\hline
\hline
$t\bar{t}$ & $\le$0.01 & $\le$0.01 & 973.1 & $\le$0.01 & $\le$0.01 & $\le$0.01 \\
\hline
$b\bar{b}b\bar{b},b\bar{b}b\bar{b} + W/Z $& 0.106 & $\le$0.01 & $\le$0.01 & $\le$0.01 & $\le$0.01 & $\le$0.01\\
\hline
$t\bar{t}b\bar{b}$ &0.8825 & 0.634 & 1.03 & 0.005 & $\le$0.01  & $\le$0.01 \\
\hline
\hline
\end {tabular}
\end{center}
\vspace{0.2cm}
%\begin{center}
\caption{ Event-rates (fb) in bottom rich final states at the chosen
benchmark points for $E_{CM}$= 14 TeV with basic cuts and cuts $C5$. {\tt CTEQ5L} pdfset was
used. Factorization and Renormalization scale has been set to 
$\mu_F=\mu_R=\sqrt{\hat s}$, subprocess center of mass energy. Contributions from dominant SM 
backgrounds are also noted. } 
\label{b-events}
\end{table}

Missing energy distribution of the benchmark points in bottom rich final states 
are shown in figure \ref{fig:MET1}. Missing Energy has been normalized to 1. $4b$ and $4b\ell$ 
final states doesn't have a significant background, hence only signal events are shown. It occurs 
that the benchmark points have a similar missing energy pattern, while for $2b2\ell$, the $t\bar{t}$ background has a 
sharper peak at low missing energy as can be expected. Similarly effective mass  $H_T$ distribution in bottom-rich final states 
is shown in figure \ref{fig:EFT1}. 
There is no significant difference between the benchmark points in terms of this distribution either. 
We can see for $4b\ell$ channel (Fig \ref{fig:MET1}, top right), the peaks of the distributions are a bit separated. 
For $2b2\ell$, background $t\bar{t}$ peaks at a much lower value while the signal events have a peak $\ge$ 1000 GeV. 
This gives us the opportunity to put a very hard effective mass $H_T$ cut, which reduces the background to almost zero, 
while retaining the signal. Hard effective mass cut also helps to remove other hadronic and QCD backgrounds as 
shown in Table \ref{b-events}.   

In summary, from Table \ref{b-events}, BP1, BP2 and BP3 have very good prospects of being 
discovered at LHC in $4b$, $4b\ell$ and $2b2\ell$ final states while the corresponding MSG 
point doesn't contribute at all in such final states. The main reason of this is clear from 
Table \ref{production}. Although $\tilde t_1 {\tilde t_1}^*$ 
production is huge for MSG, stop being almost degenerate with LSP, 
it can not decay to $t \tilde{\chi_1^0}$ or $b \tilde{\chi_1^{+}}$ and hence
it doesn't produce any $b$-jets. We might however, see 3$b$ events from 
electroweak production. 

\begin{figure}[htbp]
\begin{center}
%\vspace*{-2.0cm}
\centerline{\psfig{file=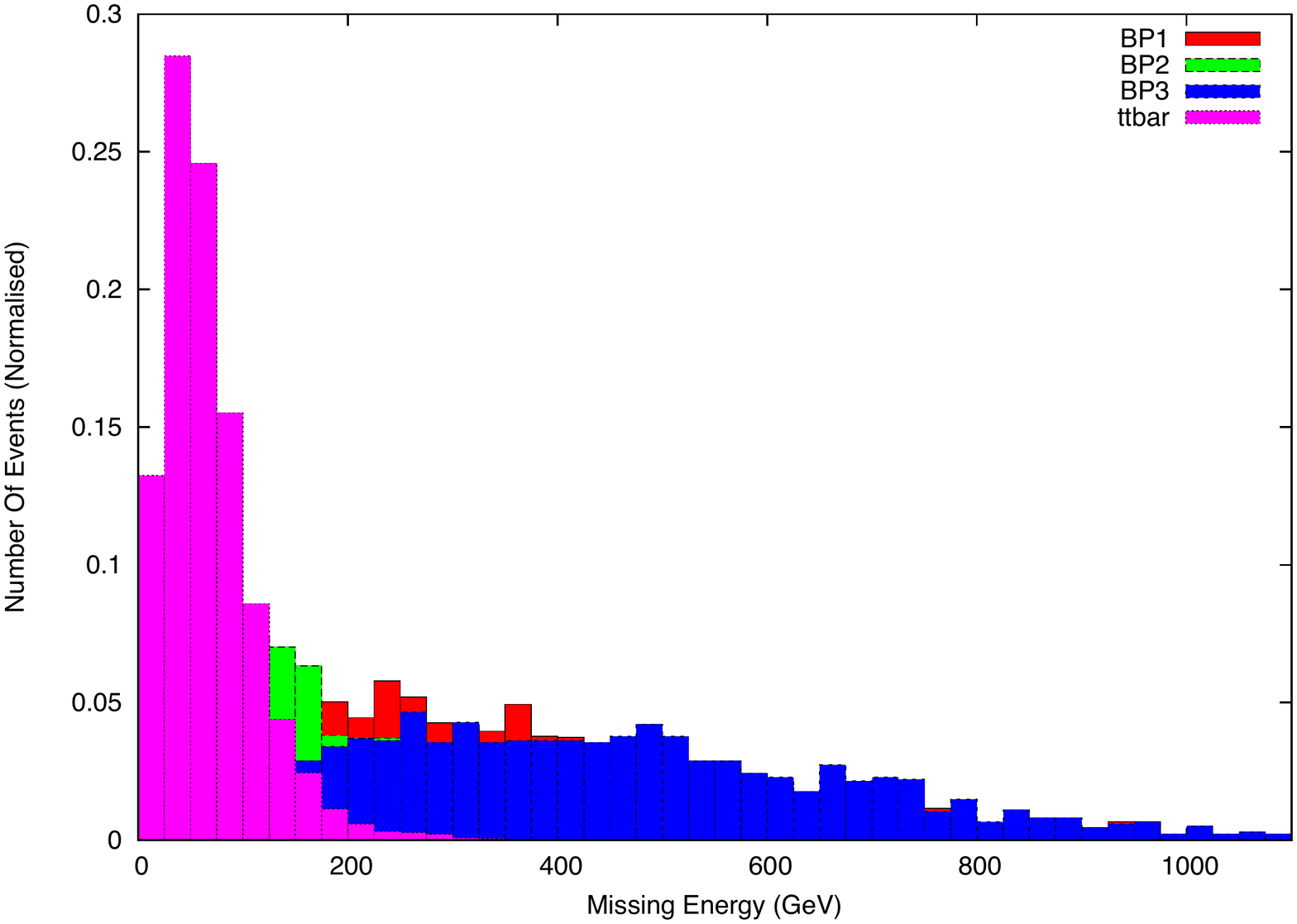,width=6.5 cm,height=7.5cm}
\hskip 20pt \psfig{file=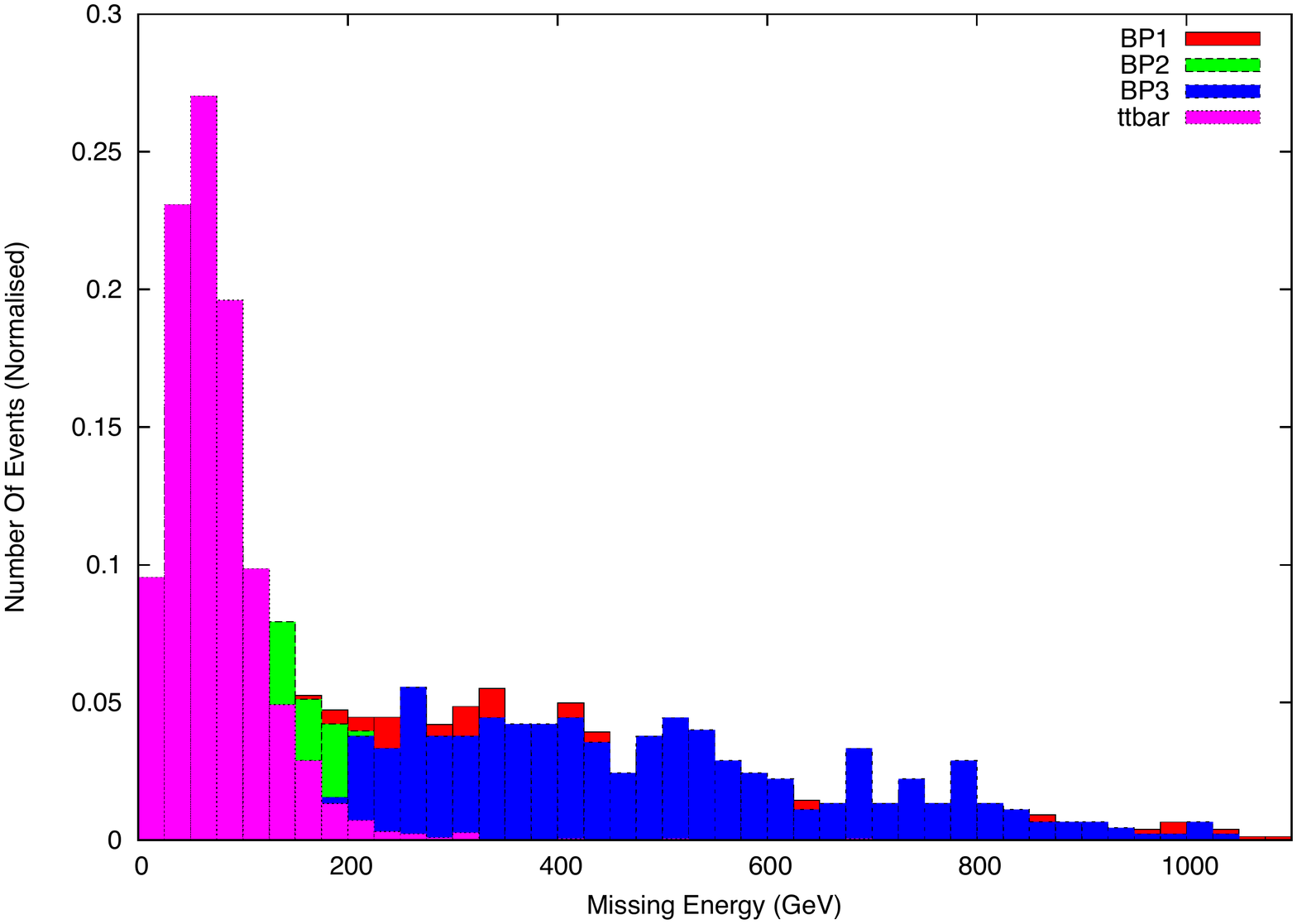,width=6.5 cm,height=7.5cm}}
\caption{Missing energy distribution in  $\ell^{\pm}\ell^{\pm}$ (left) and 
$\ell^{\pm}\ell^{\pm}\ell^{\pm}$ (right) final states at the benchmark points.}
\label{fig:MET2}
\end{center}
\end{figure} 

\begin{figure}[htbp]
\begin{center}
%\vspace*{-2.0cm}
\centerline{\psfig{file=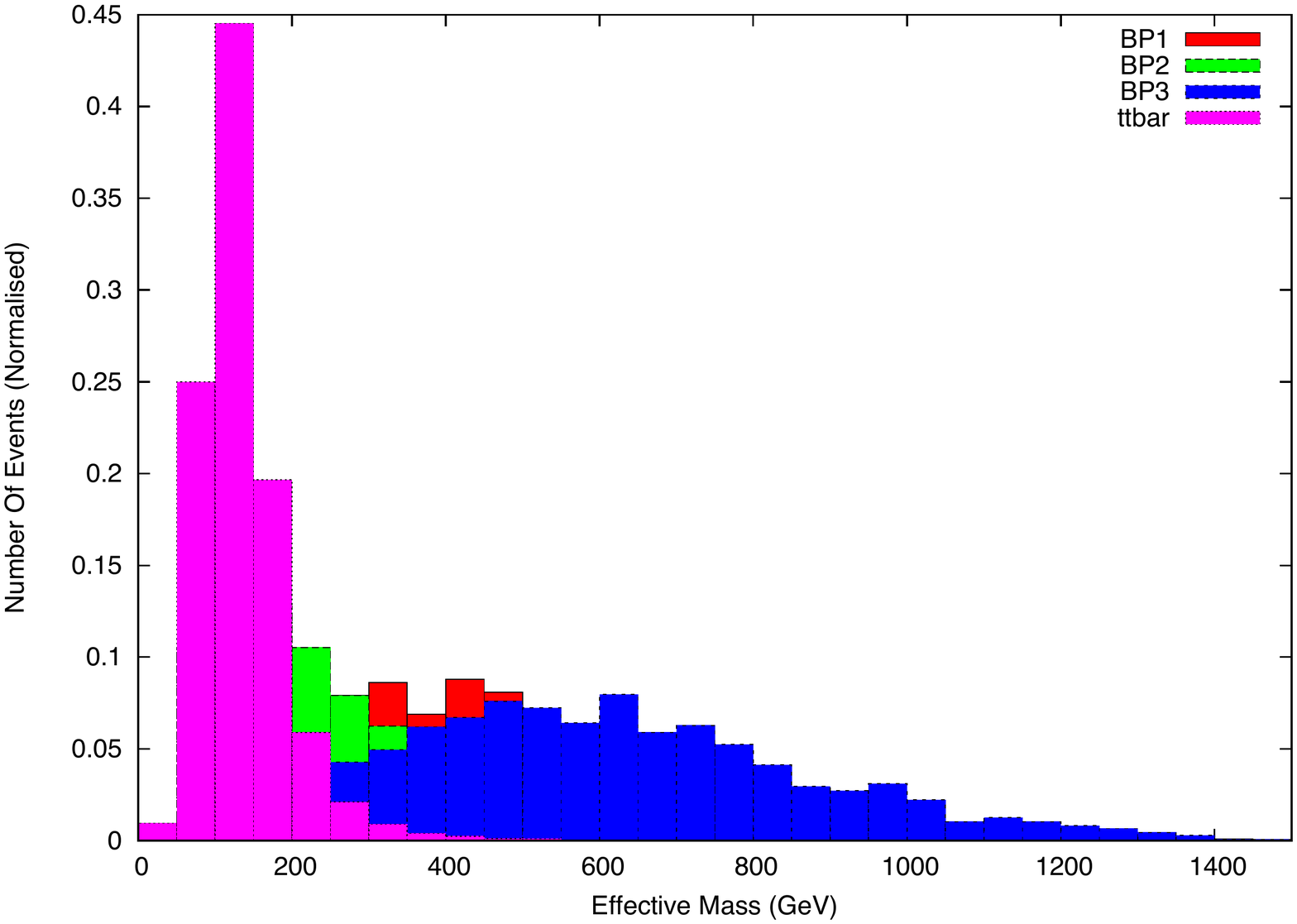,width=6.5 cm,height=7.5cm}
\hskip 20pt \psfig{file=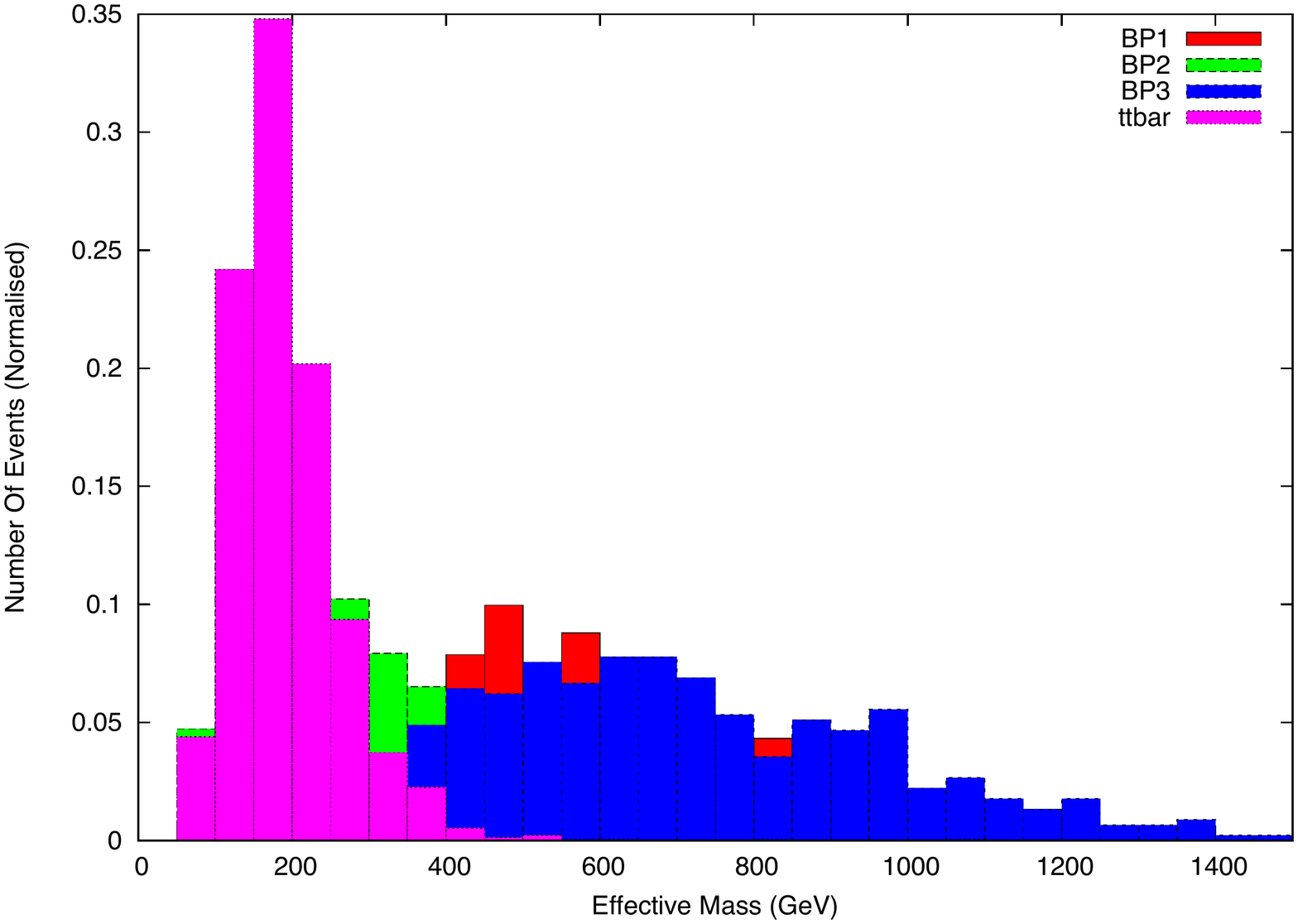,width=6.5 cm,height=7.5cm}}
\caption{Effective mass $H_{T_1}$ distribution in  $\ell^{\pm}\ell^{\pm}$ (left) and 
$\ell^{\pm}\ell^{\pm}\ell^{\pm}$ (right) final states at the benchmark points.} 
\label{fig:EFT2}
\end{center}
\end{figure}

The SM backgrounds are negligible in bottom rich channels excepting $2b2\ell$, which suffers from  
a sufficiently large background from $t\bar{t}$ production. But, a heavy Effective mass cut ($H_T$) eliminates this to 
a large extent, while retaining the signals. The Effective mass distribution in Fig. \ref{fig:EFT1} bears the testimony 
to the fact. We also note that SM background events were simulated with very high 
number of events, such that each event carries a small weight, 0.01 fb of cross-section; hence, null 
events in simulation corresponds to cross-section less than that.

\begin{table}
\begin{center}
\begin{tabular}[ht]{|c|c|c|c|c|c|}
\hline
\hline
Channels and Event rates ($fb$) & BP1 & BP2 & BP3 & MSG & $t\bar{t}$\\
\hline
\hline
 $\ell^{\pm}\ell^{\pm}$ (Basic) & 0.48  & 1.03 & 0.65 &0.2 & 40.32\\
 $\ell^{\pm}\ell^{\pm}$+$C1$& 0.16  & 0.30 & 0.30 & 0.1 & 1.08\\
 $\ell^{\pm}\ell^{\pm}$+$C2$& 0.03 & 0.03 & 0.05& $\le$0.001 & $\le$0.01\\
 $\ell^{\pm}\ell^{\pm}$+$C3$& 0.35 & 0.53 & 0.54 & 0.1 & 0.54\\
$\ell^{\pm}\ell^{\pm}$+$C4$& 0.26 & 0.40 &  0.44 & $\le$0.001 & 0.17\\
\hline
$\ell^{\pm}\ell^{\pm}\ell^{\pm}$ (Basic) & 0.18 &  0.96 & 0.24 & $\le$0.001 & 33.96\\
 $\ell^{\pm}\ell^{\pm}\ell^{\pm}$+$C1$& 0.11  & 0.40 & 0.18 & $\le$0.001 & 3.62\\
$\ell^{\pm}\ell^{\pm}\ell^{\pm}$+$C2$& 0.02 & 0.06 & 0.04 & $\le$0.001 & 0.17\\
$\ell^{\pm}\ell^{\pm}\ell^{\pm}$+$C3$& 0.15 & 0.38 & 0.22 & $\le$0.001 & 0.54\\
$\ell^{\pm}\ell^{\pm}\ell^{\pm}$+$C4$& 0.11 & 0.31 &  0.19 & $\le$0.001 & $\le$0.01\\
\hline
$\ell^{\pm}\ell^{\pm}\ell^{\pm}\ell^{\pm}$ (Basic) & 0.018  & 0.21 & 0.019& $\le$0.001 & 0.17\\
$\ell^{\pm}\ell^{\pm}\ell^{\pm}\ell^{\pm}$+$C1'$& 0.018  & 0.20 & 0.019& $\le$0.001 & 0.17\\
$\ell^{\pm}\ell^{\pm}\ell^{\pm}\ell^{\pm}$+$C2'$& 0.013 & 0.11 & 0.017& $\le$0.001 & $\le$0.01\\
$\ell^{\pm}\ell^{\pm}\ell^{\pm}\ell^{\pm}$+$C3'$& 0.017 & 0.21 & 0.019 & $\le$0.001 & 0.17\\
$\ell^{\pm}\ell^{\pm}\ell^{\pm}\ell^{\pm}$+$C4'$& 0.016 & 0.16 &  0.019 & $\le$0.001 & $\le$0.01\\
\hline
\end{tabular}
\end{center}
\caption {Event-rates (fb) in leptonic final states at the chosen
benchmark points for $E_{CM}$= 14 TeV with basic cuts and cuts $C1$, $C2$, $C3$, $C4$ as described. 
The main background $t\bar{t}$ is also noted. {\tt CTEQ5L} pdfset was used. Factorization and Renormalization scale has been set to
$\mu_F=\mu_R=\sqrt{\hat s}$, subprocess center of mass energy. Note that trilepton and four-lepton final states include $Z-$veto.}
\label{lep-events}
\end{table}

Missing energy and effective mass distribution for Same-sign dilepton and trilepton 
events are shown in fig \ref{fig:MET2} and \ref{fig:EFT2} respectively. Again all the benchmark 
points show very similar distribution, while the $t \bar{t}$ can be reduced with a heavy $H_{T_1}$ cut.
All the leptonic event numbers for the benchmark points are shown in Table \ref{lep-events}. 

Table \ref{lep-events} tells us, that trilepton events are 
still good for all the benchmark points while 4-lepton channel is good for BP2 and BP3 only. 
We also need to note that the background for 4-lepton channel is negligible 
(hadronically quiet part comes from $4W$ or $ZZZ$ production). After the cuts they vanish almost completely. 
Similarly $ZW$, which contributes to trilepton reduces to a great extent after the Z-veto. 
Hence, we didn't quote those background events here. We also see that $C2$ and $C4$ cut
reduce the $t\bar{t}$ background significantly. 
$C2$ kills the signal events to a great extent too, hence, $C4$ is a better choice to 
reduce background and retain signal. Hence, these leptonic final states are also good channels to study 
such benchmark points. The reason of BP2 having larger leptonic events, 
comes also from huge electroweak gaugino productions as pointed in Table \ref{production}. 
Hence, a significant part of these leptonic final states should contain hadronically quiet lepton events. 
The minimal supergravity benchmark point doesn't contribute at all to the leptonic final states, the reason 
being simply understood as not having lighter stops to decay through top or sleptons leading  to leptons. 
Hence, such mSUGRA points can only be studied in hadronic channels or perhaps $3b$ final states as mentioned earlier. 
After mSUGRA being alive only in stop co-annihilation region, this seems to be a generic feature 
for all mSUGRA parameter space points to obey Higgs mass and dark matter constraint.  
This in turn, can help distinguishing such non-universal frameworks from mSUGRA in LHC signature space.
\newpage

\section {Summary and Conclusions}

It is remarkable that a Higgs boson has been discovered with a mass $\simeq 125$ GeV. In pure SM, theoretically
there is no reason why its mass should be at the EW scale, or even it is, why it is not much higher or lower
than 125 GeV. (In fact, in pure SM,  best fit to the EW data prefers a much lower mass). This gives us hope
that some symmetry principle is there beyond the pure SM, and supersymmetry being the most natural candidate,
because it solves the hierarchy problem, as well as it constraints the Higgs mass to be less than $\sim 135$ GeV.
In addition, supersymmetry has a natural candidate for the dark matter. However, the minimal version of the 
most desirable version of MSSM, mSUGRA, is in very tight corner to satisfy all the existing experimental
constraints, as well as being within the reach of LHC. We find that mSUGRA is still viable in the stop co-annihilation region
in which the classic SUSY signal (multijet plus missing $E_T$) is essentially unobservable beyond the SM background
at the LHC. (The other allowed region such as hyperbolic/ focus point has SUSY particle masses well beyond the
reach of LHC). However, if we relax little bit from mSUGRA with non-universal gaugino and /or scalar masses, 
the situation becomes much more favorable to discover SUSY at the LHC.

In this work, we have shown that SUSY with non-universalities in gaugino or scalar  masses within high scale SUGRA
set up can still be accessible at LHC with $E_{CM}=$ 14 TeV. In particular, 
we show the consistency of the parameter space in different dark matter annihilation regions. 
Wino dominated LSP with chargino co-annihilation can be achieved with gaugino mass 
non-universality with $M_3<M_2<M_1$. Hyperbolic Branch/Focus point 
region with Higgsino dominated LSP can be obtained easily with Higgs non-universality as BP2. 
Such parameter space automatically occurs with lighter gauginos and hence they may dominate 
the production and leptonic final states at LHC. Stau co annihilation can occur with scalar non-universality 
while stop co annihilation can arise simply with high-scale gaugino non-universality with $M_3<M_2=M_1$. 
mSUGRA, though viable in only stop co-annihilation region, do not yield lepton or b-rich final states due to 
lack of phase space for the stop to decay leptonically. There exist a reasonable region of parameter space
 in the non-universal scenario which not only satisfy all the existing constraints, but also can unravel SUSY in bottom and 
lepton rich final states with third family squarks being lighter than the first two automatically. We have made detailed studied
of three benchmark points in these allowed parameter spaces, and find that SUSY signal in the bottom or bottom plus 
lepton-rich final state  stands over the SM  background with suitable cuts. We have also investigated pure leptonic final states
with suitable cuts, and find some of these final state have viable prospects. 
Finally we also emphasize that  with good luminosity in the upcoming 14 TeV LHC runs, these allowed parameter space can be 
ruled out easily, or we we will discover SUSY.

%\section*{\large \bf References}

\chapter{Parallel Universe, Dark Matter and Invisible Higgs Decays}\label{chap:chap5}
\section{Introduction}

Symmetry seems to play an important role in the classification and interactions of the elementary particles. The Standard Model (SM) based on the gauge symmetry $SU(3)_C \times SU(2)_L \times U_Y(1)$ has been extremely successful in describing all experimental results so far to a precision less than one percent.
The final ingredient of the SM, namely the Higgs boson, has finally been observed at the LHC \cite{ATLAS}. However, SM is unable to explain why the charges of the elementary particle are quantized because of the presence of $U(1)_Y$. This was remedied by enlarging the $SU(3)_C$ symmetry to $SU(4)_C$ with the lepton number as the fourth color,(or grand unifying all three interaction in SM in $SU(5)$ \cite{GG}
or $SO(10)$ \cite{gfm}).

 SM also has no candidate for the dark matter whose existence is now well established experimentally \cite{dm}. Many extensions of the SM models, such as models with weakly interacting massive particles (WIMP) can explain the dark matter \cite{dm}. The most poplar examples are the lightest stable particles in supersymmetry \cite{dm}, or the lightest Kaluza-Klein particle in extra dimensions \cite{kkdm}. Of course, axion \cite{ww} is also a good candidate for dark matter. Several experiments are ongoing to detect signals of dark matter in the laboratory. However, it is possible that the dark matter is just the analogue of ordinary matter belonging to a parallel universe. Such a parallel universe  naturally appears in the superstring theory with the $E_8 \times E'_8$ gauge symmetry before compactification \cite{chsw}. Parallel universe in which the gauge symmetry is just the replication of our ordinary universe, i,e the gauge symmetry in the parallel universe being $SU(3)'\times SU(2)'\times U(1)'$ has also been considered \cite{fv}. If the particles analogous to the proton and neutron in the parallel universe is about five times heavier than the proton and neutron of our universe, then that will naturally explain why the dark matter of the universe is about five times the ordinary matter. This can be easily arranged by assuming strong coupling constant square$/4\pi, \alpha'_s$ is about five times larger than the QCD $\alpha_s$. Thus, in this work, we assume that the two universe where the electroweak sector is exactly symmetric, whereas the corresponding couplings in the strong sector are different, explaining why the dark matter is larger than the ordinary matter. Also, we assume that both universes are described by non-abelian gauge symmetry so that the kinetic mixing  between the photon ($\gamma$) and the parallel photon ($\gamma '$) is forbidden.
We also assume that post-inflationary reheating in the two worlds are different, and the the parallel universe is colder than our universe \cite {mohapatra}. This makes it possible to maintain the successful prediction of the big bang nucleosynthesis, though the number of degrees of freedom is increased from the usual SM of 10.75 at the time of nucleosynthesis  due the extra light degrees of freedom (due to the $\gamma ', e'$ and three $\nu '$s). 

In this work, we explore the LHC and ILC implications of this scenario due to the mixing among the Higgs bosons in the two electroweak sectors. Such a mixing, which is allowed by the gauge symmetry, will mix the lightest Higgs bosons of our universe ($h_1$) and the lightest Higgs boson of the parallel universe ($h_2$), which we will call the dark Higgs. One of the corresponding mass eigenstates, $h_{SM}$ we identify with the observed Higgs boson with mass of $125$ GeV. The other mass eigenstate, which we denote by $h_{DS}$, the dark Higgs, will also have a mass in the electroweak scale. Due to the mixing effects, both Higgs will decay to the kinematically allowed  modes in our universe and as well as to the modes of the dark universe. One particularly interesting scenario is when the two Higgs bosons are very close in mass, say within 4 GeV so that the LHC can not resolve it \cite {ATLASHgaga}. However, this scenario will lead to the invisible decay modes\cite{ATLASINV}. The existence of such invisible decay modes can be established at the LHC when sufficient data accumulates. (The current upper limit on the invisible decay branching ratio of the observed Higgs at the LHC is $0.65$). At the proposed future International Linear Collider (ILC) \cite{ILC}, the existence of such invisible modes can be easily established, and the model can be tested in much more detail.\\
The chapter is organized as follows. In Section 2, we discuss the model
and formalism. We also discuss the symmetry breaking. In Section 3, we
discuss the phenomenological implications at the LHC, including the interactions and decays of light higgses, details of the data used in collider analysis and bounds on mixing angle. In section 4, we discuss in detail possible phenomenological implications of the model at the ILC. At the end, we conclude in Section 5.

\section{Model and the Formalism}

The gauge symmetry we propose for our work is $ SU(4)_C \times SU(2)_L \times SU(2)_R $ for our universe, and $ SU(4)'_C \times SU(2)'_L \times SU(2)'_R $ for the parallel universe. Note that we choose this non-abelian symmetry not only to explain charge quantization (as in Pati-Salam model \cite{Pati:1974yy}), but also to avoid the kinetic mixing of $\gamma$ and $\gamma '$ as would be allowed in the Standard Model. All the elementary particles belong to the representations of this symmetry group and their interactions are governed by this symmetry. The 21 gauge bosons belong to the adjoint representations $(15,1,1)$, $(1,3,1)$, $(1,1,3)$. $(15,1,1)$ contain the 8 usual colored gluons, 6 lepto-quark gauge bosons $(X, \bar{X})$, and one $(B-L)$ gauge boson \cite{mm}. $(1,3,1)$ contain the 3 left handed weak gauge bosons, while $(1,1,3)$ contain the 3 right handed weak gauge bosons. The parallel universe contains the corresponding parallel gauge bosons. However, so far as the gauge interactions are concerned, we do not assume that the coupling for $SU(4)$ and $SU(4)'$ interactions are the same, but strong coupling in the parallel universe is larger in order to account for the $p'$ (proton of the parallel universe) mass to be about five times larger than the proton. For the electroweak sector, we assume the exact symmetry between our universe and the parallel universe.

 The fermions belong to the fundamental representations $(4, 2,1) + (4,1,2)$. The  4 represent three color of quarks and the lepton number as the 4th color, $(2,1)$ and $(1,2)$ represent the left and right handed doublets. The forty eight Weyl fermions belonging to three generations may be represented by the matrix

\begin{equation}
{\begin{pmatrix}

{\begin{pmatrix} u \\ d \end{pmatrix}}_1 &
 {\begin{pmatrix} u \\ d \end{pmatrix}}_2 & {\begin{pmatrix} u \\ d \end{pmatrix}}_3
& {\begin{pmatrix} \nu_e \\ e \end{pmatrix}}_4\\

{\begin{pmatrix} c \\ s \end{pmatrix}}_1 &
 {\begin{pmatrix} c \\ s \end{pmatrix}}_2 & {\begin{pmatrix} c \\ s \end{pmatrix}}_3
& {\begin{pmatrix} \nu_{\mu} \\ \mu \end{pmatrix}}_4\\

{\begin{pmatrix} t \\ b \end{pmatrix}}_1 &
 {\begin{pmatrix} t \\ b \end{pmatrix}}_2 & {\begin{pmatrix} t \\ b \end{pmatrix}}_3
& {\begin{pmatrix} \nu_{\tau} \\ \tau \end{pmatrix}}_4\\
\end{pmatrix}}_{L, R}.
\end{equation} 

We have similar fermion representations for the parallel universe, denoted by primes.

 The model has 3 gauge  coupling constants: $g_4$ for $SU(4)$ color which we will identify with the strong coupling constant of our universe,  $g'_4$ for $SU(4)'$ color of the parallel universe, and $g$ for $SU(2)_L$ and $SU(2)_R$, and corresponding electroweak couplings for the parallel universe ($g_L = g_R = g'_L = g'_R = g$ (we assume that the gauge couplings of the electroweak sectors of the two universe are the same).

\subsection{Symmetry breaking}
 $SU(4)$ color symmetry is spontaneously broken to $SU(3)_C \times U(1)_{B-L}$ in the usual Pati-Salam way using  the Higgs fields $(15, 1,1)$ at a scale $V_c$. The most stringent limit on the scale of this symmetry breaking comes from the upper limit of the rare decay mode $K_L \rightarrow \mu e$ \cite{KL}. 
$SU(2)_L \times SU(2)_R \times U(1)_{B-L}$ can be broken to the SM using the Higgs representations  $(1,2,1)$ and $1,1,2)$ at a scale $V_{LR}$. Alternatively, one can use the Higgs multiplets $(1.3,1)$ and $(1,1,3)$ if we want to generate the light neutrino masses at the observed scale.  Finally the remaining symmetry is broken to the $U(1)_{EM}$ using the Higgs bi-doublet $(1,2,2)$ as in the left-right model. The $(15, 2, 2)$ Higgs multiplet  could also be added to eliminate unwanted mass relations among the charged fermions. Similar Higgs representations are used to break the symmetry in the parallel universe to $U'(1)_{EM}$. A study of the Higgs potential shows that there exist a parameter space  where only one neutral Higgs in the bi-doublet remains light, and becomes very similar to the SM Higgs in our universe \cite{Senjanovic}. All other Higgs fields become very heavy compared to the EW scale. Similar is true in the parallel universe. The symmetry of the Higgs fields in   the EW sector between our universe and the parallel universe will make the two  electroweak VEV's the same. Thus the mixing terms between the two bi-doublets (one in our universe and one in the parallel universe) then leads to mixing between the two remaining SM like Higgs fields. The resulting mass terms for the remaining two light Higgs fields  can be written as 
$m_{VS}^2 h_{1}^2 + m_{DS}^2 h_{2}^2 + 2 \lambda v_{VS} v_{DS}h_1 h_2$, (where $v_{VS}$ and $v_{DS}$ are the electroweak symmetric breaking scale in the visible sector and dark sector respectively)  from which the two mass eigenstates and the mixing can be calculated.

\begin{figure}
\begin{center}
\includegraphics[width=8 cm,height=8cm]{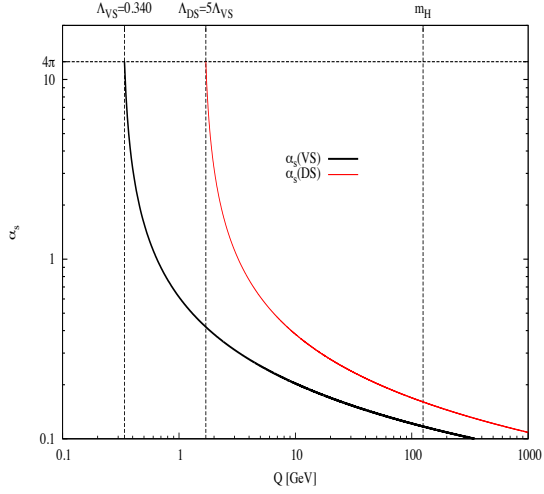}
\end{center}
\caption{Running of the strong coupling constant in the visible sector ($\alpha_S^{VS}(Q)$) and dark sector ($\alpha_S^{DS}(Q)$)}
\label{running}
\end{figure}

The implications for this is when the two light Higgses are very close in mass (within about 4 GeV, which LHC can not resolve) leads to the invisible decay of the observed Higgs boson.
The main motivation for postulating this kind of parallel universe scenario is to explain the dark matter density which is five time larger than the ordinary matter density. The particles analogous to the proton and neutron in the dark sector, namely, the dark proton and dark neutron, are stable due to the conservation of the dark baryon number. Moreover, the gauge symmetry and the particle content  of the model does not allow any gauge or Yukawa interactions of dark protons and dark neutrons with the visible sector particles. Therefore, in the framework of this model, the  dark protons and dark neutrons are  the candidates for the dark matter. If the dark protons and dark neutrons in the parallel universe are about five times heavier than the protons and neutrons of our universe, then that will naturally explain why the dark matter of the universe is about five times the ordinary matter. The Lagrangian quark masses for the up and down quarks are only of order of 10 MeV or less. Therefore,$99\%$ of the mass of the proton or neutron arises from the strong interaction of the constituent quarks and the gluons. The mass scale associated with these interactions is set by the value of the three-quark QCD scale $\Lambda_{QCD}$. Therefore, one can easily achieve a dark proton or neutron mass which is five times larger than the visible proton or neutron mass by assuming the QCD scale in the dark sector ($\Lambda_{DS}$) to be five times larger than the QCD scale in the visible sector ($\Lambda_{VS}=340$ MeV). Different QCD scales give rise to different running of the strong coupling constant in the visible sector ($\alpha_S^{VS}(Q)$) and dark sector ($\alpha_S^{DS}(Q)$). In Fig.~\ref{running}, we have presented the running of the strong coupling constant in the visible sector and dark sector. Fig.~\ref{running} shows that $\alpha_S^{DS}(Q=m_H=125~{\rm GeV})$= $ 1.4~ \alpha_S^{VS}(Q=m_H=125~{\rm GeV})$.

 Below we discuss the  phenomenological implications for this scenario at the LHC, and briefly at the proposed ILC \cite{ILC}.

%\section{Phenomenological Implications at LHC}

\section{Phenomenological Implications at the LHC}

%\subsection{Study of parameter space}
In the framework of this model, interaction between fermions and/or gauge bosons of dark sector and visible sector (the SM particles) are forbidden by the gauge symmetry. However, quartic Higgs interactions of the form $\lambda (H_{VS}^{\dag}H_{VS})(H_{DS}^{\dag}H_{DS})$ (where $H_{VS}$ and $H_{DS}$ symbols denote the Higgs fields in the visible sector and dark sector respectively) are allowed by the gauge symmetry and gives rise to mixing between the Higgses of dark and visible sector. The mixing between the lightest Higgses of dark sector and visible sector gives rise to interesting phenomenological implications at the collider experiments. In this section, we will discuss the phenomenological implications of the lightest dark and visible neutral Higgs mixing ($h_1$ and $h_2$). As discussed in the previous section, the bi-linear terms involving the lightest visible sector (denoted by $h_{1}$) and dark sector (denoted by $h_2$) Higgses in the scalar potential are given by,
\begin{equation}
{\cal L}_{Scalar} \supset m_{VS}^2h_1^2 + m_{DS}^2h_2^2 + 2 \lambda v_{VS} v_{DS} h_1 h_2
\end{equation}     
where, $v_{VS}$ and $v_{DS}$ are the electroweak symmetric breaking scale in the visible sector and dark sector respectively. In our analysis, we have assumed  the both $v_{VS}$ and $v_{DS}$ are equal to the SM electroweak symmetry breaking scale $v_{SM} \sim 250$ GeV. $m_{VS}$, $m_{DS}$ and $\lambda$ are the free parameters in the theory and the masses ($m_{h_1^{(p)}}$ and $m_{h_2^{(p)}}$) and mixing between physical light Higgs states (denoted by $h_1^{(p)}$ and $h_{2}^{(p)}$) are determined by these parameters:   
\begin{eqnarray}
h_{1}^{(p)}&=&{\rm cos}\theta ~h_1 + {\rm sin}\theta ~h_2, \nonumber\\
h_{2}^{(p)}&=&-{\rm sin}\theta ~h_1 + {\rm cos}\theta ~h_2,
\end{eqnarray}
where the masses and the mixing angle of these physical states are given by,
\begin{eqnarray}
m_{h_1^{(p)},h_2^{(p)}}^2&=&\frac{1}{2}[(m_{VS}^2+m_{DS}^2)\mp\sqrt{(m_{VS}^2-m_{DS}^2)^2+4\lambda^2v_{VS}^2v_{DS}^2}]\nonumber\\
{\rm tan}2\theta &=&\frac{2\lambda ~v_{VS}~v_{DS}}{m_{DS}^2-m_{VS}^2}.
\end{eqnarray}
 In the framework of this model, we have two light physical neutral Higgs ($h_1^{(p)}$ and $h_{2}^{(p)}$) states. Out of these two Higgs states, we define the SM like Higgs $h_{SM}$ is the state which is dominantly $h_1$-like, i.e., if ${\rm cos}\theta > {\rm sin}\theta$ then $h_{SM}=h_{1}^{(p)}$ and vice versa. The other Higgs is denoted as dark Higgs ($h_{DS}$).  Since ATLAS and CMS collaborations have already detected a SM like Higgs boson with mass about 125 GeV, we only studied the scenario where the mass of $h_{SM}$ is between 123 to 127 GeV. Before going into the details of collider implication of visible sector and dark sector Higgs mixing, it is important to understand the correlation between the mixing and mass of the dark Higgs ($m_{h_{DS}}$). To understand the correlation, for few fixed values of $\lambda$, we have scanned the $m_{VS}-m_{DS}$ parameter space.
 We have only considered the points which gives rise to a $h_{SM}$ in the mass range between 123 GeV to 127 GeV. For these points, the resulting dark Higgs masses ($m_{h_{DS}}$) and mixing ($\theta$) are plotted in Fig.~\ref{mixing}.
The scatter plot in Fig.~\ref{mixing} shows that large mixing in the visible and dark sector is possible only when the dark Higgs mass is near 125 GeV i.e., near the mass of SM like Higgs boson. It is important to note that the LHC is a proton-proton collider, i.e., LHC collides the visible sector particles only. Therefore, the production cross-section of dark Higgs at the LHC is proportional to the square of the visible sector Higgs component in $h_{DS}$. Therefore, in order to detect the signature of dark Higgs at the collider experiments, we must have significant mixing between the visible and dark sector Higgses. And Fig.~\ref{mixing} shows that significant mixing arises only when dark Higgs and SM like Higgs are nearly degenerate in mass. Therefore, in this article, we studied the phenomenology of two nearly degenerate Higgs bosons with mass about 125 GeV.

\begin{figure}
\begin{center}
\includegraphics[width=10 cm,height=8cm]{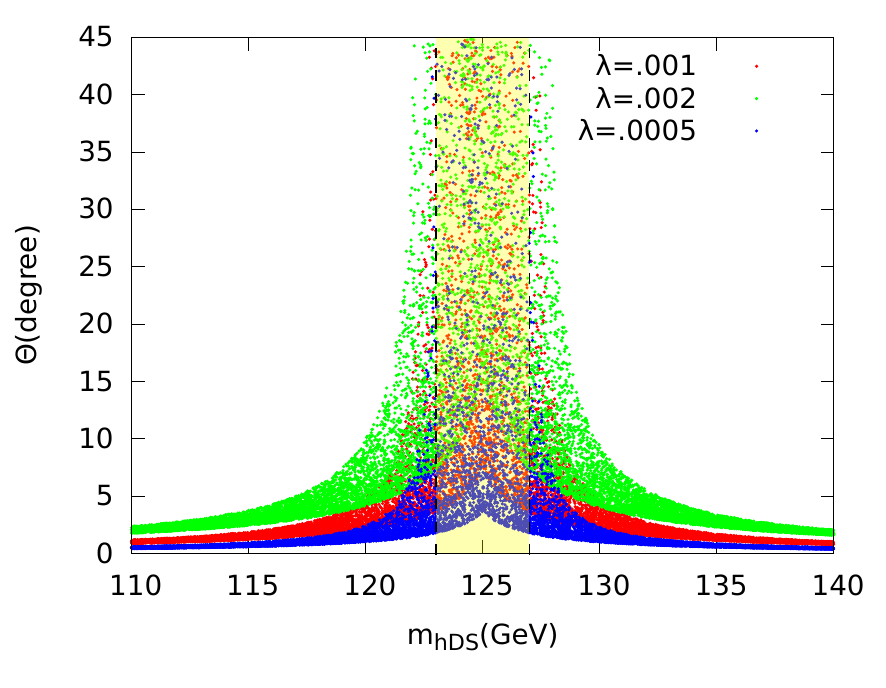}
\end{center}
\caption{Scatter plot of dark Higgs mass vs mixing angle for different values of $\lambda$. The SM-like Higgs mass is kept fixed in the range between 123 to 127 GeV denoted by the shaded region in the plot.}
\label{mixing}
\end{figure}    
\subsection{Interactions and Decays of light Higgses}

In the present model, two light Higgs physical states ($h_1^{(p)}$ and $h_2^{(p)}$) result from the mixing of  visible sector and dark sector light Higgs weak eigenstate $h_1$ and $h_2$ respectively. Visible sector light Higgs weak eigenstates, $h_1$ interacts only with the visible sector fermions ($f$) via Yukawa interactions and gauge bosons ($V$) via gauge interactions. Whereas the dark sector light Higgs weak eigenstate interacts only with the dark fermions $f_D$ and dark gauge bosons $V_D$. However, as a result of mixing, the physical light Higgses interact with both the visible particles and dark particles and thus, they can be produced at the Large Hadron Collider(LHC) experiment. The coupling of the physical states $h_1^{(p)}$ and $h_2^{(p)}$ with the visible as well as dark fermions and gauge bosons can be written as a product of corresponding SM coupling and sine or cosine of the mixing angle. As a result the production cross sections of $h_1^{(p)}$ and $h_2^{(p)}$ and decay widths into visible as well as dark particles can be computed in terms of the SM Higgs production cross-sections/decay widths and the mixing angle. For example, total $h_1^{(p)}$ production cross section at the LHC is given by $\sigma_{SM}{\rm cos}^2\theta$, where $\sigma_{SM}$ is the production cross-section of the SM Higgs with equal mass. Similarly, the decay widths of $h_{1}^{(p)} (h_{2}^{(p)})$ into visible and dark sector fermions are given by $\Gamma_{SM}^{H\to f\bar f} {\rm cos}^2\theta$ ($\Gamma_{SM}^{H\to f\bar f} {\rm sin}^2\theta$) and $\Gamma_{SM}^{H\to f\bar f} {\rm sin}^2\theta$ ($\Gamma_{SM}^{H\to f\bar f} {\rm sin}^2\theta$) respectively, where $\Gamma_{SM}^{H\to f\bar f}$ is the decay width of the SM Higgs into fermions. It is important to note that since the QCD coupling in the dark sector is about $5$ times
larger than the QCD coupling in the visible sector, the Higgs coupling with dark gluon in this model is enhanced by a factor about $5$.\\
In this analysis we are considering both the higgs states in the mass range between $123-127$ GeV. Here we present the expressions for $\mu= \sigma/\sigma_{SM}$ and total  $\sigma\times BR_{invisible}$ for present model,
\begin{eqnarray}
\mu &=&\frac{(\sigma_{h1}{cos}^4\theta BR_{h1}/(1 + 24 BR_{h1}^{gg}{sin}^2\theta)) + (\sigma_{h2}{sin}^4\theta BR_{h2}/(1 + 24 BR_{h2}^{gg} {cos}^2\theta)) }{\sigma_{SM}*BR}\nonumber\\
\sigma \times BR_{inv} &=&\frac{\sigma_{h1}{cos}^2\theta {sin}^2\theta (BR_{h1}^{inv} + 25BR_{h1}^{gg})}{1+ 24BR_{h1}^{gg} {sin}^2\theta} + \frac{\sigma_{h2}{cos}^2\theta {sin}^2\theta (BR_{h2}^{inv} + 25BR_{h2}^{gg})}{1+ 24BR_{h2}^{gg} {cos}^2\theta}\
\end{eqnarray}
where $\sigma_{h1}$ corresponds to Standard Model Higgs production cross-section at mass of $h_{1}^{(p)}$ and $\sigma_{h2}$ corresponds to Standard Model production cross-section at mass of $h_{2}^{(p)}$ (see Table \ref{production cross section}) and $BR_{h1}$ and $BR_{h2}$ corresponds to Branching ratios of Higgs boson at mass $h_{1}^{(p)}$ and $h_{2}^{(p)}$ respectively(see Table \ref{Decay Branching Ratio1}). For calculating the $\mu$ values in present model we have used Branching Ratios of $H \rightarrow WW \rightarrow l \nu l \nu$ and $H \rightarrow \gamma \gamma$ channels(see Table \ref{Decay Branching Ratio}).
\subsection{Data used in Collider Analysis}

In this section, we discuss the collider phenomenology of invisible Higgs Decays. Before going into
the details of the collider prediction, we first need to study the constraints on the parameter space
coming from present Standard Model predictions and experimental data. The Higgs mass eigenstates of $h_{SM}$ and $h_{DS}$ will be produced in Colliders through the top loop as top quark has Standard Model couplings to the $h_{SM}$ mass eigen state.  
The Higgs, which comprises of both $h_1$ and $h_2$ eigen states, will then decay in both the Standard Model decay modes along with Dark sector decay modes. We will perceive these dark sector decay modes as enhancement in the invisible Branching Fraction of the Higgs.\

 We first discuss the different constraints on the mixing angle $\theta$ between the two eigenstates coming from experimental data of $H \rightarrow WW \rightarrow l \nu l \nu$ and $H \rightarrow \gamma \gamma$ channels. Along with these experimental data in Higgs decays in different modes, we have also taken into account constraints on the mixing angle parameter space coming from the ATLAS search for the invisible decays of a $125$ GeV Higgs Boson produced in association with a Z boson \cite{ATLASINV}. \ 
\begin{table}
\begin{center}
\begin{tabular}{|c|c|c|c|c|}
\hline
Mass of Higgs(GeV) & $\sigma_{ggf}$ & $\sigma_{ttH}$ &$\sigma_{VBF}$ &$\sigma_{Vh}$ \\
\hline
\hline
123 & 20.15 & 1.608 & 1.15 & 0.1366 \\
\hline
124 & 19.83 & 1.595 & 1.12 & 0.1334 \\
\hline
125 & 19.52 & 1.578 & 1.09 & 0.1302 \\
\hline
126 & 19.22 & 1.568 & 1.06 & 0.1271 \\
\hline
127 & 18.92 & 1.552  & 1.03 & 0.1241  \\
\hline
\end {tabular}
\end{center}
\vspace{0.2cm}
%\begin{center}
\caption{ Standard Model production cross section (pb) in different channels for $E_{CM}$ = 8 TeV.} 
\label{production cross section}
\end{table}

\begin{table}
\begin{center}
\begin{tabular}{|c|c|c|c|c|c|}
\hline
Mass of Higgs(GeV) & BR($H {\rightarrow}WW$) & BR($H {\rightarrow}ZZ$) & BR($H {\rightarrow}{\gamma\gamma}$) & BR($H {\rightarrow} gg $)& BR($H {\rightarrow}ff$) \\
\hline
\hline
123 & 0.183 & $2.18\times 10^{-2}$ & $2.27 \times 10^{-3}$  & $8.71\times 10^{-2}$  & 0.687 \\
\hline
124 & 0.199 & $2.41\times 10^{-2}$ & $2.27\times 10^{-3}$ & $8.65\times 10^{-2}$ &  0.687 \\
\hline
125 & 0.215 & $2.64\times 10^{-2}$ & $2.28\times 10^{-3}$ & $8.57\times 10^{-2}$ & 0.670 \\
\hline
126 & 0.231 & $2.89\times 10^{-2}$ & $2.28\times 10^{-3}$ & $8.48\times 10^{-2}$ & 0.651 \\
\hline
127 & 0.248 & $3.15\times 10^{-2}$  & $2.27\times 10^{-3}$ & $8.37\times 10^{-2}$  & 0.633 \\
\hline
\end {tabular}
\end{center}
\vspace{0.2cm}
%\begin{center}
\caption{ Standard Model Decay Branching Ratio in different channels.} 
\label{Decay Branching Ratio1}
\end{table}

\begin{table}
\begin{center}
\begin{tabular}{|c|c|c|}
\hline
Channels for Higgs Decay & $\mu$ value by ATLAS & $\mu$ value by CMS \\
\hline
\hline
$H \rightarrow WW \rightarrow l \nu l \nu$  & $1.01 \pm 0.31$ & $0.76 \pm 0.21$ \\
\hline
 $H \rightarrow  \gamma \gamma$ & $1.65 \pm 0.24(stat) ^ {+0.25}_{-0.18}(syst) $  & $0.78 \pm 0.27$ \\
\hline
\end {tabular}
\end{center}
\vspace{0.2cm}
%\begin{center}
\caption{ Experimental values of best fit signal strength $\mu = \sigma/\sigma_{SM}$ at $E_{CM}$ = 8 TeV.} 
\label{Decay Branching Ratio}
\end{table}

The Standard Model production cross-sections in different channels (such as gluon-gluon fusion, ttH, vector boson fusion and vector boson (both W boson and Z boson) in association with a Higgs boson) at $E_{CM}$ = 8 TeV and Decay Branching ratios in different channels (such as $H {\rightarrow}WW$, $H {\rightarrow}ZZ$,$H {\rightarrow}{\gamma\gamma}$,$H {\rightarrow} gg$,$H {\rightarrow}ff$)  has been given by ATLAS collaboration in reference [17] [18]. We have used these cross-sections and branching ratios in different channels in our analysis.
The relevant cross-sections and branching ratios used for our analysis are presented in Table \ref{production cross section} and Table \ref{Decay Branching Ratio1} respectively. We have taken the mass range between $123 - 127$ GeV which is the interesting parameter space for our analysis.\

In Table \ref{Decay Branching Ratio} we present the results of the different experimental searches in the $H \rightarrow WW \rightarrow l \nu l \nu$ channel by ATLAS collaborations \cite{ATLASWW} and CMS collaboration \cite{CMSWW} and in $H \rightarrow  \gamma \gamma$ channel by ATLAS collaborations\cite{ATLASGG} CMS collaborations\cite{CMSGG} .

\subsection{Bounds on Mixing Angle}
\begin{figure}
\begin{center}
\vspace*{-2.0cm}
\includegraphics[width=10.5cm,height=8.5cm]{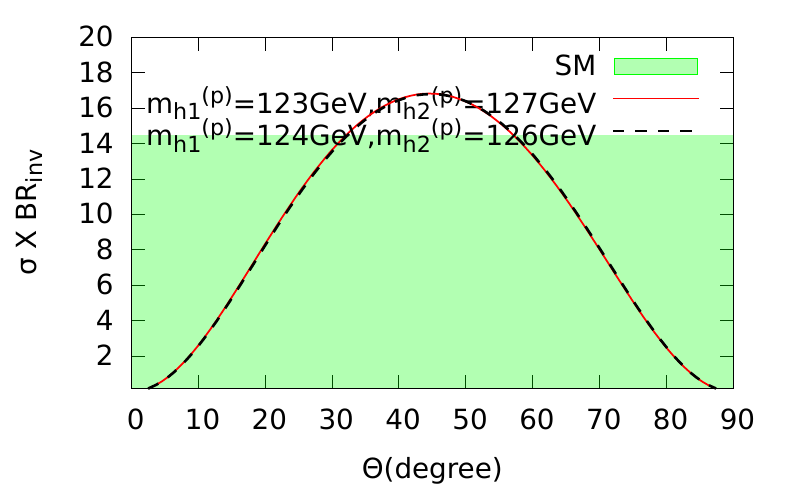}
\caption{Decay rate in invisible channels in present model as a function of mixing angle $\theta$. The shaded regions correspond to SM allowed values for $\sigma \times BR_{inv}$.} 
\label{fig:decayrate}
\end{center}
\end{figure}

In this section we use the data that we presented in the previous section to constrain the mixing angle parameter space.
In Fig \ref{fig:decayrate}, we present the total invisible decay rate i.e $\sigma \times BR$ in the invisible channel vs the mixing angle $\theta$ for
  $ m_{h1}^{(p)} = 123$ GeV  and $ m_{h2}^{(p)} = 127$ GeV ($m_{h1}^{(p)} = 124 GeV$ and $m_{h2}^{(p)} = 126 GeV)$. ATLAS collaboration has searched for the invisible decay of higgs boson in Z H production channel at $E_{CM} = 8 TeV$. In absence of any significant deviation of data from the Standard Model background prediction, ATLAS collaboration has set an upper limit of $65\%$ on the invisible decay branching of a SM higgs boson of mass $125$ GeV \cite{ATLASINV}. Assuming $\sigma_{total}$ = $22.32$ pb Higgs cross-section at $125$ GeV (see Table \ref{production cross section}),  $65\%$ upper limit on invisible decay branching ratio corresponds to $14.5$ pb upper limit on the invisible Higgs decay rate. This limit is shown in the shaded green region in Fig \ref{fig:decayrate}. It can be seen from the plot that present model is consistent with ATLAS experimental data for $\theta < 33^{o} $ and $\theta > 58^{o}$ in the parameter space region. 

\begin{figure}
\begin{center}
\vspace*{-2.0cm}
\includegraphics[width=10.5cm,height=8.5cm]{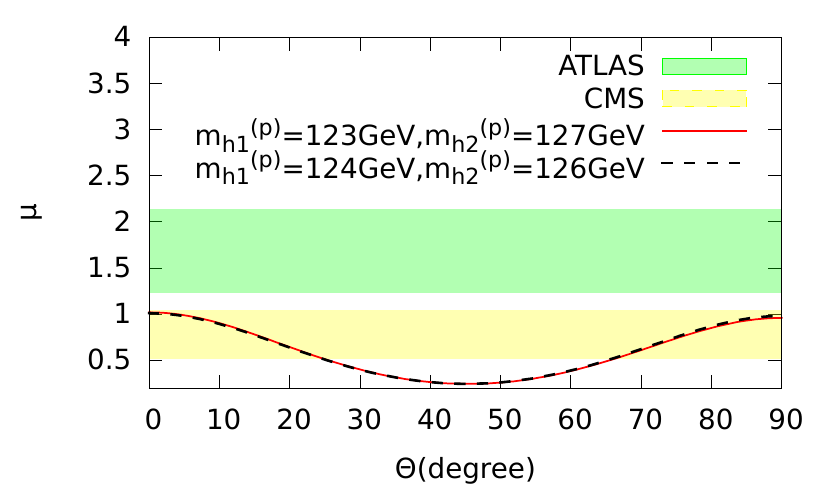}
\caption{Higgs decaying into diphoton rate in present model as a function of mixing angle $\theta$. The shaded regions again correspond to ATLAS and CMS allowed $\mu = \sigma/\sigma_{SM}$ values.} 
\label{fig:H_gaga}
\end{center}
\end{figure} 

\begin{figure}
\begin{center}
\vspace*{-2.0cm}
\includegraphics[width=10.5cm,height=8.5cm]{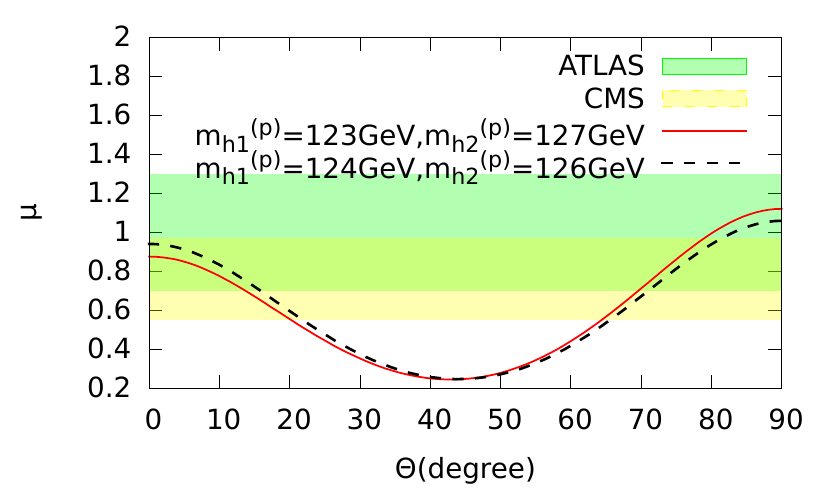}
\caption{$H \rightarrow WW \rightarrow l \nu l \nu$ rate in present model as a function of mixing angle $\theta$. The shaded regions correspond to ATLAS and CMS allowed $\mu = \sigma/\sigma_{SM}$ values.} 
\label{fig:H_WW}
\end{center}
\end{figure}

In Fig. \ref{fig:H_gaga} we have presented a plot of $\mu = \sigma/\sigma_{SM}$ in the $H \rightarrow \gamma \gamma$ channel as a function of the  mixing angle $\theta$. The plot shows prediction in present model for $m_{h1}^{(p)} = 123$ GeV  and $ m_{h2}^{(p)} = 127$ GeV $(m_{h1}^{(p)} = 124$ GeV and $m_{h2}^{(p)} = 126$ GeV) mass values.The yellow shaded region corresponds for allowed region by CMS collaboration and green shaded region is allowed region for ATLAS collaboration in this channel. It can be seen from the plot that CMS allowed region is consistent for all $\theta$'s for the present model,but present model is not consistent with ATLAS allowed region for any values of $\theta$. We point out that  $H \rightarrow  \gamma \gamma$ data for ATLAS, is well above the SM expectation. If the present model is realized by  Nature, with the accumulation of more data with higher luminosities at the Large Hadron Collider(LHC) the $H \rightarrow  \gamma \gamma$ branching ratio measured by ATLAS experiment should should come down significantly from present experimental value of $1.65 \pm 0.24(stat) ^ {+0.25}_{-0.18}(syst) $. Our model is consistent with the lower $\mu$ value of $0.78 \pm 0.27$  for $H \rightarrow  \gamma \gamma$ as measured by the CMS experiment for the whole parameter of the parameter space.

In Fig.~\ref{fig:H_WW} we present a plot of $\mu = \sigma/\sigma_{SM}$ in the $H \rightarrow WW \rightarrow l \nu l \nu$ channel with mixing angle $\theta$. Two curves for $ m_{h1}^{(p)} = 123$ GeV  and $ m_{h2}^{(p)} = 127$ GeV $(m_{h1}^{(p)} = 124$ GeV and $m_{h2}^{(p)} = 126$ GeV) present the prediction for present model. The yellow shaded region corresponds for allowed region by CMS collaboration and green shaded region is for allowed region by ATLAS collaboration in this channel. It can be seen from the plot that ATLAS allowed region is consistent with present model for $\theta < 13 (16)^{o} $ and $\theta > 70 (71)^{o} $ region in the parameter space. It can also be seen that present model is also consistent with CMS allowed region for $\theta < 20 (23)^{o}$ and $\theta > 65 (66)^{o}$ parameter space. It is  interesting to note that the prediction curves for the present model with mass values of $ m_{h1}^{(p)} = 123$ GeV  and $ m_{h2}^{(p)} = 127$ GeV $(m_{h1}^{(p)} = 124$ GeV and $m_{h2}^{(p)} = 126$ GeV) are not symmetric. It can be understood by taking into the fact that in low $\theta$ region $ m_{h1}^{(p)}$ is SM like. As $ m_{h1}^{(p)}$ is lower than $ m_{h2}^{(p)}$ for both curves, the cross-section $\times$ Branching ratio is smaller in lower $\theta$ region. Whereas for high $\theta$ region $ m_{h2}^{(p)}$ is SM like and as it is heavier than $ m_{h1}^{(p)}$ for both curves the cross section $\times$ Branching Ratio is higher in this region,which makes the curves non-symmetric.\\
This present analysis in the $H \rightarrow WW \rightarrow l \nu l \nu$ channel gives the most stringent constraint of $\theta < 13 (16)^{o} $ and $\theta > 70 (71)^{o} $ on the parameter space for the mixing angle $\theta$ taking into account all the constraints coming from analysis in $\sigma \times BR_{invisible}$, $H \rightarrow \gamma \gamma$ and $H \rightarrow WW \rightarrow l \nu l \nu$ channels. From this analysis in different channels it is certain that there is still plenty of parameter space available for the present model taking into account all the known experimental constraints at the LHC.\\
We would also like to comment that in a linear collider like the proposed International Linear Collider(ILC) this analysis can be done without any ambiguity about the resolution of the two Higgs in the close range of $4GeV$. In a $e+e-$ collider the Higgs will be produced in association with a Z boson and from the mass recoil of the Z boson the peak resolution of the Higgs boson can be measured in the limit of $40$ MeV \cite{ILC}. So from linear colliders we will be able to tell for sure if there are two Higgs bosons in the comparable mass range between ($123-127$GeV), which is not possible in this precision from Hadron Collider like LHC.

\section{Phenomenological Implications at the ILC}

After discussing the mixings and the decays of the physical Higgs states $h_1^{(p)}$ and $h_2^{(p)}$, we are now equipped
enough to discuss the collider phenomenology in the context of an electron-positron collider. In this section, we are particularly interested in the scenario in which two light  physical Higgs states are quasi-degenerate. Due to the large background, it will be challenging to distinguish such a scenario at the hadron collider experiments. The advantages of an electron-positron collider compared to a hadron collider are the cleanliness of the environment, the precision of the measurements and the large number of Higgs bosons production. Therefore, it could be possible for an  electron-positron collider to probe a scenario with two quasi-degenerate Higgs bosons. It has recently been shown by the {\it   International Large Detector (ILD) Concept Group} in Ref.~\cite{ILD_resolution} that the proposed electron-positron collider will be able to determine the Higgs boson mass with a statistical precision of 40 MeV. Motivated by the results of Ref.~\cite{ILD_resolution}, in our analysis, we have considered two different mass splittings between the Higgs bosons:
\begin{itemize}
\item {\bf Scenario I:} We have considered the mass splitting between the two Higgs bosons to be about 40 MeV. Therefore, electron-positron collider can not resolve two Higgs bosons mass peaks in this case. The LHC experiment has already observed a Higgs boson with mass about 125 GeV. Therefore, we have assumed one Higgs boson mass to be 124.98 GeV and the other Higgs boson mass is 125.02 GeV.
\item {\bf Scenario II:} In this case, we assume relatively large mass splitting (about 500 MeV) between the two Higgs bosons so that the electron-positron collider can resolve the Higgs bosons mass peaks. The numerical values of the Higgs boson masses are chosen to be  124.75 GeV and 125.25 GeV.  
\end{itemize}

%%%%%%%%%%%%%%%%%%%%%%%%%%%%%%
\begin{figure}
\begin{center}
\includegraphics[width=12 cm,height=6cm]{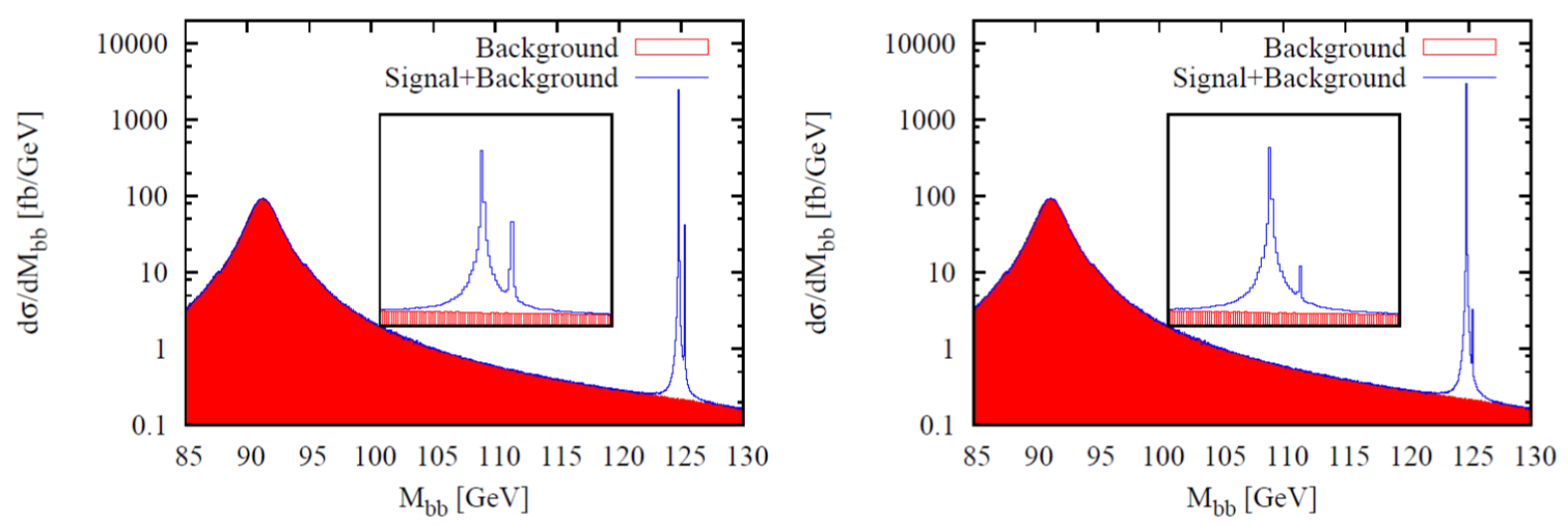}
\end{center}
\caption{Invariant mass distributions of $b\bar b$ pairs for two different values of the mixing angle, $\theta=20^0$ (left panel) and $10^0$ (right panel). The masses of the two physical Higgs states are given by $m_{h_{SM}}=124.75$ GeV and $m_{h_{DH}}=125.25$ GeV. $123~{\rm GeV}<m_{bb}<127~{\rm GeV}$ regions are magnified in the insets.}
\label{inv_mass_large}
\end{figure}    
%%%%%%%%%%%%%%%%%%%%%%%%%%%%%%

At the electron-positron collider the main production mechanism of the Higgs boson is the Higgs-strahlung process i.e., the production of the Higgs boson in association with a $Z$-boson: $e^+e^- \to ZH$. The Higgs-strahlung process is an s-channel process so that its production cross section is maximal just above the threshold of the process. As a result, for the proposed electron-positron collider to study the observed Higgs boson properties in detail, we should start with the $e^+e^-$ collision initial center-of-mass energy of 250 GeV at which the Higgs-strahlung cross-section reaches its  maximum value for a $125$ GeV  Higgs boson mass. In our analysis, we have also considered 250 GeV center-of-mass energy for the electron-positron collider. In the framework of the present model, both physical Higgs states will be produced in association with a $Z$-boson. The production cross-sections of $h_1^{(p)}$ and $h_2^{(p)}$ are given in terms of the SM Higgs production cross-section ($\sigma_{SM}(ZH)$) and the mixing angle: $\sigma(e^+e^- \to Z h_1^{(p)})={\rm cos}^2\theta \sigma_{SM}(ZH)$ and $\sigma(e^+e^- \to Z h_2^{(p)})={\rm sin}^2\theta \sigma_{SM}(ZH)$ . From the Higgs-strahlung process, the Higgs signature could be detected and hence, Higgs mass could be measured by the direct Higgs decays and the recoiling to the $Z$-boson. In our analysis, we have studied both the direct Higgs decays as well as  the recoiling to the $Z$-boson. The recoil mass to the $Z$-boson is the invariant mass of the decay products against which the $Z$-boson recoils assuming the collision occurs at the nominal center of mass energy $\sqrt s$ and is defined as,
$$
M_{recoil}^2=(\sqrt s-E_Z)^2-|\vec p_Z|^2=s+M_Z^2-2E_Z\sqrt s,
$$
where $M_Z$ denote the mass of the $Z$-boson as reconstructed from the decay products of the $Z$-boson and $E_Z$ is the corresponding energy. Recoil mass to the $Z$-boson do not depend on the decay products of the Higgs boson and hence, provides an unique opportunity for the reconstruction of the Higgs mass from its invisible decays. In the framework of the present model, both  $h_1^{(p)}$ and $h_2^{(p)}$ decay to the dark sector particles which remain invisible in the detector. Therefore, in order to detect the decays $h_1^{(p)}$ and $h_2^{(p)}$ into a pair of dark particles, we have used the recoil mass to the $Z$-boson.

%%%%%%%%%%%%%%%%%%%%%%%%%%%%%%
\begin{figure}
\begin{center}
\includegraphics[width=13cm,height=7cm]{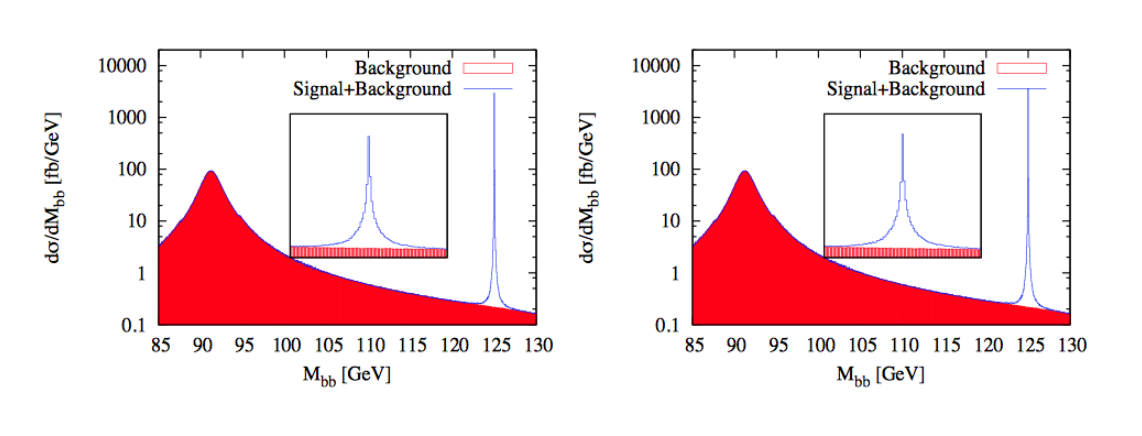}
\end{center}
\caption{Same as Fig.~\ref{inv_mass_large} for $m_{h_{SM}}=124.98$ GeV and $m_{h_{DH}}=125.02$ GeV.}
\label{inv_mass_small}
\end{figure}    
%%%%%%%%%%%%%%%%%%%%%%%%%%%%%%

We have used CalcHEP package \cite{calchep} for the simulation of the signal and the background. The recoil mass to the $Z$-boson  crucially depends on the initial state radiation (ISR) and beamstrahlung. CalcHEP implements the Jadach, Skrzypek and Ward expressions of Refs.~\cite{ISR} for the simulation of ISR. Whereas for the Beamstrahlung, we have used the parameterizations specified for the ILD project \cite{ILD}: {\bf beam size (x + y)} = 645:7 nm, {\bf bunch length} = 300 $\mu$m and {\bf bunch population} = $2\times 10^{10}$.

%====================================================================%
\begin{table}[h]

\begin{center}

\begin{tabular}{c|c|c}
\hline\hline

Kinematic Variable & Minimum value & Maximum value \\\hline\hline
$\Delta R(l^{+}l^{-})$ & 0.3  & -    \\
$\Delta R(bb)$ & 0.7  & -    \\
$\Delta R(l^\pm b)$ & 0.7  & -    \\
$p_{T}^{l^{+},l^{-}}$  & 10 GeV   & -    \\
$\eta_{l^{+},l^{-}}$ & -2.5  & 2.5 \\
$p_{T}^{b}$  & 20 GeV   & -    \\
$\eta_{b}$ & -2.5  & 2.5 \\
$M(l^{+}l^{-})$    & 80 GeV      & 100 GeV \\\hline\hline

\end{tabular}

\end{center}

\caption{Acceptance cuts,used in this calculation, on the kinematical variables for {\it
2-lepton + 2-b} signal.}

\label{cuts}
\end{table}

%====================================================================%
%%%%%%%%%%%%%%%%%%%%%%%%%%%%%%
\begin{figure}
\begin{center}
\includegraphics[width=12cm,height=6cm]{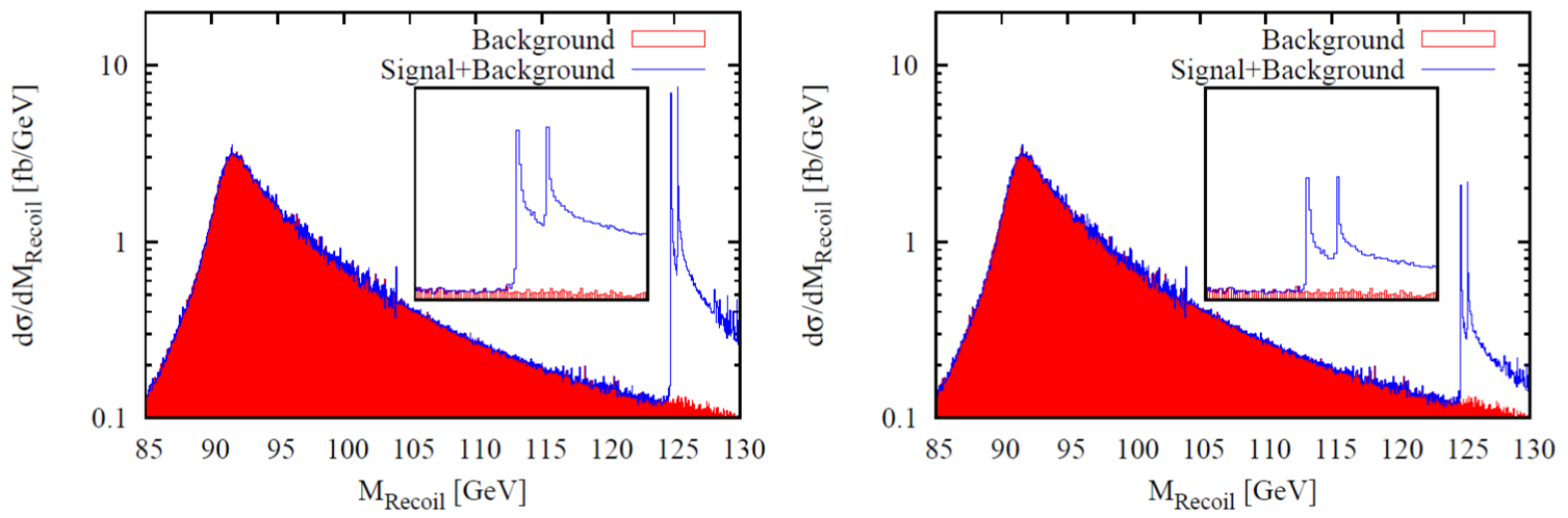}
\end{center}
\caption{Recoil mass distribution for invisible Higgs decays for two different values of the mixing angle, $\theta=20^o$ (left panel) and $10^o$ (right panel). The masses of the two physical Higgs states are given by $m_{h_{SM}}=124.75$ GeV and $m_{h_{DH}}=125.25$ GeV. $123~{\rm GeV}<m_{bb}<127~{\rm GeV}$ regions are magnified in the insets. }
\label{re_mass}
\end{figure}    
%%%%%%%%%%%%%%%%%%%%%%%%%%%%%%

Higgs boson dominantly decays into a pair of bottom-quarks. For the reconstruction of the Higgs bosons from its decay products, we consider the decays of Higgs boson into a pair of $b$-quarks. For the associated $Z$-boson, we consider its leptonic (electron and muon) decay modes only. Therefore, the signal is characterized by two opposite sign same flavor leptons and two bottom quarks. The dominant background arises from the production of a pair of $Z$-bosons when one $Z$ decays in to a pair of b-quarks and another $Z$ decays leptonically. To parameterize detector acceptance and enhance signal to background ratio, we have imposed kinematic cuts, listed in Table~\ref{cuts}.  In Fig.~\ref{inv_mass_large}, we have presented the invariant mass distributions of $b\bar b$ pairs for two different values of the mixing angle, $\theta=20^o$ (left panel) and $10^o$ (right panel) for {\bf Scenario II}. We have used a bin size of 40 MeV. {\bf Scenario II} corresponds to relatively large mass splitting between the two Higgs bosons. As a result, in Fig.~\ref{inv_mass_large}, two characteristic mass peaks in the $b\bar b$ invariant mass distributions are clearly visible. Fig.~\ref{inv_mass_small} corresponds to the $b\bar b$ invariant mass distributions for the {\bf Scenario I}.

We have also studied the recoil mass to the $Z$-boson when the two Higgs bosons decay invisibly to a pair of dark sector particles. The $Z$-boson is assumed to decay into a pair of leptons (electrons and muons only). Although the branching ratio of the $Z \to l^+l^-$, where $l$ refers to $e$ or $\mu$, is only about 3.4\%, which is about 20 times smaller than that of the $Z \to q\bar q$, the high momentum resolution of leptons could overcome the shortage in statistics to gain even higher precision on the Higgs mass measurement. The signal in this case is characterized by two opposite sign same flavor leptons and missing energy. The dominant background in this case is again $ZZ$ production followed by leptonic decay of one $Z$-boson and invisible decay of the other $Z$-boson. The cuts listed in Table~\ref{cuts} are applied for the leptons. Fig.~\ref{re_mass} gives the background and signal+background recoil mass distributions for {\bf Scenario II} for two different values of the mixing angle, $\theta=20^o$ (left panel) and $10^o$ (right panel). The Higgs recoil mass distributions are crucially affected by the beamstrahlung and the initial state radiation which are responsible for the long tail of the distributions as visible in Fig.~\ref{re_mass}. However, 
the two Higgs bosons mass peaks are clearly visible in Fig.~\ref{re_mass}. 
%%%%%%%%%%%%%%%%%%%%%%%%%%%%%%
\begin{figure}
\begin{center}
\includegraphics[width=8 cm,height=6cm]{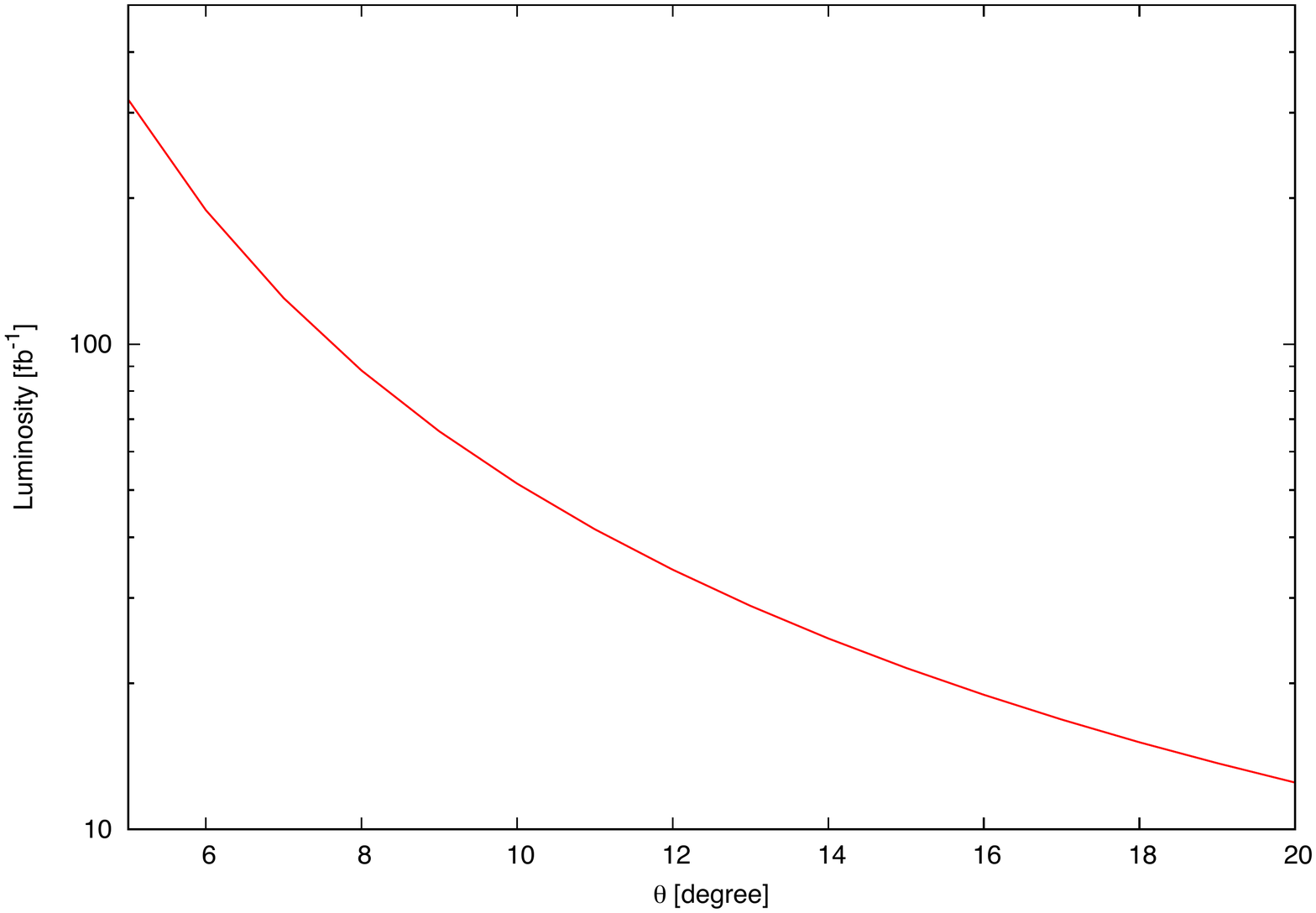}
\end{center}
\caption{ Required luminosity for 5$\sigma$ discovery as a function of the mixing angle $\theta$.}
\label{lumi}
\end{figure}    
%%%%%%%%%%%%%%%%%%%%%%%%%%%%%%

Invisible Higgs decay is a characteristic signature of this model. The invisible decay rate directly depends on the mixing angle and hence, could be used to probe the mixing angle.  In view of Fig.~\ref{re_mass}, we impose further cuts on the $M_{Recoil}$: $123~{\rm GeV} \le M_{Recoil} \le 127~{\rm GeV}$, to enhance the signal to background ratio. In order to quantify the ability of extracting signal event,  $N_S=\sigma_S{\cal L}$, for a given integrated luminosity ${\cal L}$ over the SM background events, $N_B=\sigma_B{\cal L}$, we define the significance $S=N_S/{\sqrt {N_B+N_S}}$. In Fig.~\ref{lumi}, we have presented the required luminosity for 5$\sigma$ (i.e., $s=5$) discovery as a function of the mixing angle $\theta$ for an electron-positron collider with center-of-mass energy of 250 GeV. Fig.~\ref{lumi} shows that a 250 GeV electron-positron collider with 300 fb$^{-1}$ integrated luminosity will be able to probe $\theta$ upto $5^o$.

\section{Summary and Conclusions}

Motivated by the fact that the dark matter is about five  times the ordinary matter, we have proposed that the dark matter can just be like the ordinary matter in a parallel universe with the QCD scale in the dark sector ($\Lambda_{DS}$) is about five times larger than the QCD scale in the visible sector ($\Lambda_{VS}$). The parallel universe needs to be much colder than our universe to keep the successful prediction for the big bang nucleosynthesis. We have used the non-abelian Pati-Salam gauge symmetry for both universe to have the charge quantization, as well as, to avoid any kinetic mixing between the photon of our universe and the parallel universe. However, the two universes will be connected via the electroweak Higgs bosons of the two universes. If the electroweak sector of the two universes are symmetric, the lightest Higgs bosons of the two universes will mix. In particular, if these two Higgses mix significantly, and their masses are close (say within 4 GeV), LHC will not be able to resolve if it is observing one Higgs or two Higgses. However, each Higgs will decay to the particles of our universe as well as to the corresponding particles of the the parallel universe. This leads to the invisible decays  of the observed Higgs boson (or bosons). We have used all the available experimental data at the LHC to set constraint on this mixing angle, and find that in can be as large as $16^{o}$. If the mixing angle is not very small, LHC will be able to infer the existence of such invisible decays when sufficient data accumulates. (The current limit on the invisible branching ratio from the LHC data is
$<65\%$). We also find that the cross section times the branching ratio for Higgs to $\gamma \gamma$ channel is fully consistent with our model as measured by the CMS collaboration, but not by the ATLAS collaboration. The results by the ATLAS collaboration for this channel has to come down if our model is realized by nature. We get very interesting phenomenology in the special case when the two light Higgs bosons of the two Universes are almost degenerate in mass. Specifically, if their mass difference is  $\sim 100 $MeV, the LHC would not be able to resolve them as two separate mass peaks in its entire run time. We consider proposed ILC where because of its clean environment, the precise measurements and large number of Higgs boson production, Higgs mass splittings upto $\sim 100$ MeV may be  possible with high luminosity. We investigate two scenarios where the mass difference between the two Higgs bosons are 40 MeV and 500 MeV for two mixing angles of $20^o$ and $10^o$. We find that with a $250 $ GeV ILC, for 500 MeV mass splitting we can see two clear mass peaks when the Higges decay to $b \bar{b}$ or invisibly. But for the more ambitious $\sim 40$ MeV mass splitting, it is not possible to resolve them. We also  study the sensitivity to the mixing angle to the recoil mass to the $Z$-boson when the both the Higgs bosons decay invisibly to a pair of dark sector particles. We show that with a 250 GeV ILC  with 300 fb$^{-1}$ integrated luminosity it will be possible to probe the mixing angle $\theta$ between the two Higgs bosons upto $5^o$.

% ***************************************************************

\chapter{Predictive models of the Dirac Neutrinos}\label{chap:chap6}

\section{Introduction}
In the past 20 years, there has been a great deal of progress  in neutrino physics from the atmospheric neutrino experiments (Super-K \cite{Wendell:2013kxa}, K2K \cite{Mariani:2008zz}, MINOS \cite{Barr:2013wta}), solar neutrino experiments ( SNO \cite{Aharmim:2011vm}, Super-K \cite{Koshio:2013dta} , KamLAND \cite{Mitsui:2011zz}) as well as  reactor/accelerator neutrino experiments (Daya Bay \cite{An:2012eh}, RENO \cite{Ahn:2012nd}, Double Chooz \cite{Abe:2011fz}, T2K\cite{Abe:2011sj}, NO$\nu$a \cite{Patterson:2012zs}). These experiments have pinned down three mixing angles - $\theta_{12}$, $\theta_{23}$, $\theta_{13}$ and two mass squared differences $\Delta m^2_{ij} = m^2_{i}-m^2_{j}$ with reasonable accuracy
\cite{deGouvea:2013onf}. However there are several important  parameters yet to be measured. These include the value of the CP phase $\delta$ which will determine the magnitude of CP violation in the leptonic sector and the sign of $\Delta m^2_{32}$ which will determine whether the neutrino mass hierarchy is normal or inverted. We also don't know yet if the neutrinos are Majorana or Dirac particles.

On the theory side, the most popular mechanism for neutrino mass generation is the see-saw \cite{Gell-Mann}. This requires heavy right handed neutrinos, and this comes naturally in the $SO(10)$ grand unified theory
(GUT) \cite{Georgi} in the $16$ dimensional fermion representation. The tiny neutrino masses require the scale of these right handed neutrinos close the GUT scale.  The light neutrinos generated via the sea-saw mechanism are Majorana particles. However, the neutrinos can also be Dirac particles just like ordinary quarks and lepton.This can be achieved by adding right handed neutrinos to the Standard Model. The neutrinos can get tiny Dirac masses via the usual Yukawa couplings with the SM Higgs. In this case, we have to assume that the corresponding Yukawa couplings are very tiny, $\sim 10^{-12}$. Interesting works in Dirac neutrinos can be found in these references\cite{Aranda:2013gga}. 
%Alternatively, we can introduce a 2nd Higgs doublet responsible for the generation of neutrino masses by extending the SM with a discrete  $Z_2$ symmetry. 
Alternatively, we can introduce a 2nd Higgs doublet and a discrete  $Z_2$ symmetry so that the  neutrino masses are generated only from the 2nd Higgs doublet. 
The neutrino masses are generated from the spontaneous breaking of this discrete symmetry from a tiny vev of this 2nd Higgs doublet in the eV or keV range, and then the associated Yukawa couplings need not be so tiny \cite{Gabriel:2006ns}. At this stage of neutrino physics, we can not  determine which of these two possibilities are realized by nature.

In the first work, we show that  with the three known mixing angles and two known mass difference squares, we find an interesting pattern in the neutrino mass matrix if the neutrinos are Dirac particles. With three reasonable assumptions : (i) lepton number conservation, (ii) hermiticity of the neutrino mass matrix, and (iii) $\nu_{\mu}$ - $\nu_{\tau}$ exchange symmetry, we can construct the neutrino mass matrix completely. The resulting mass matrix satisfies all the constraints implied by the above three assumptions, and gives an inverted hierarchy (IH) (very close to the degenerate) pattern. We can now predict the absolute values of the masses of the three neutrinos, as well as the value of the CP violating phase $\delta$. We also predict the absence of neutrinoless double $\beta \beta$  decay.     

In the second work, we have considered a general symmetry involving the interchange of the right handed muon neutrino ($\nu_{\mu R}$) and tau neutrino ($\nu_{\tau R}$). The three RH charged leptons and neutrinos are singlet under $SU(2)_L$ and thus they do not form a multiplate. Therefore, we can invoke any symmetry in the RH neutrino sector without imposing that symmetry in the charged lepton sector. If any symmetry exists in the Dirac neutrino mass matrix under interchange of $\nu_{\mu R}$-$\nu_{\tau R}$ then this will be symmetry of the whole Lagrangian. We have constructed the different Dirac neutrino mass matrices assuming different kinds of symmetries in the $\nu_{\mu R}$ and $\nu_{\tau R}$ sector and tried to fit the experimentally observed quantities. Finally, we end up with a four parameter Dirac neutrino mass matrix which is based on the assumption of the Hermiticity\footnote{It is important to note that the assumption of Hermiticity is somewhat ad hoc i.e., Hermiticity of neutrino mass matrix is not an outcome of symmetry argument. However, we have shown in the following that with this assumption, the existing neutrino data can completely determine the mass matrix for the Dirac neutrinos with particular predictions for the neutrino masses and the CP violating phase which can be tested at the ongoing and future neutrino experiments. Therefore, in our analysis, the assumption of hermiticity of neutrino mass matrix is a purely phenomenological assumption. However, in the future, there might be some compelling theoretical framework which requires the hermiticity of neutrino mass matrix.} of the Dirac neutrino mass matrix and a particular symmetry between  $\nu_{\mu R}$ and $\nu_{\tau R}$. We have also shown that assuming IH in the neutrino sector, this four parameter neutrino mass matrix is consistent with the observed values of the three mixing angles and two squared-mass differences listed in Table \ref{tab:expdata}, and also makes definite
predictions for the values of the three neutrino masses and the leptonic CP violating phase.\\
The chapter is organized as follows. In Section 2, we discuss the details of the Model with five parameters in the neutrino mass matrix. In Section 3, we
discuss the phenomenological implications and results of this framework. In section 4, we discuss in detail the Model with only four parameters in the neutrino mass matrix. We present the summary and conclusions in Section 5.

%\section{The Models and the neutrino mass matrix}
\section{The Model with five parameters in the neutrino mass matrix}
Our first model is based on the Standard Model (SM) Gauge symmetry, $SU(3)_C \times SU(2)_L \times U(1)_Y$, supplemented by a discrete $ Z_2$ symmetry. \cite{Gabriel:2006ns}.  In addition to the SM  particles,  we have three SM singlet right handed neutrinos, $N_{Ri}$, i = 1,2,3, one for each family of fermions.  We also have one additional  Higgs doublet $\phi$, in addition to the usual SM Higgs doublet $\chi$.
 All the SM particles are even under $Z_2$, while the $N_{Ri}$ and  the  $\phi$ are odd under $Z_2$. Thus while the SM quarks and leptons obtain their masses from the usual Yukawa couplings with $\chi$ with vev of $\sim 250$ GeV, the neutrinos get masses only from its Yukawa coupling with $\phi$ for which we assume the vev is $\sim$ keV to satisfy the cosmological constraints which we will discuss later briefly. Note that even with as large as a keV vev for $\phi$, the corresponding Yukawa coupling is of order $10 ^{-4}$ which is not too different from the light quarks and leptons Yukawa coupling in the SM.
 %t As a consequence to the even $Z_2$ assignment, the SM fermions(except %the left-handed neutrinos) and the Higgs doublet $\chi$ couple to each %other and due to the odd $Z_2$ assignment,the SM left-handed neutrinos, %RH neutrinos couple with the other Higgs doublet $\phi$. The gauge %symmetry $SU(2)\times U(1)$
%is broken spontaneously at the EW scale by the VEV of $\chi$, while the %discrete
%symmetry $Z_2$ is broken by a VEV of $\phi$, and we take VEV of $\phi %\sim 10^{-2}$ eV to generate tiny neutrino masses. 
The Yukawa interactions of the Higgs fields $\chi$ and $\phi$ and the leptons can be written as,
\begin{equation}
L_Y = y_l \bar{\Psi}_L^l l_R \chi + y_{\nu l} \bar{\Psi}_L^l N_R \tilde{\Phi} + h.c.,
\end{equation}
where $\bar{\Psi}_L^l = (\bar{\nu_l}, \bar{l})_L$ is the usual lepton doublet and $l_R$ is the charged lepton singlet, and we have omitted the family indices. The
first term gives rise to the masses of the charged leptons, while the second term gives tiny 
neutrino masses. The interactions with the quarks are the same as in the Standard Model
with $\chi$ playing the role of the SM Higgs doublet. 
%Here, a SM left-handed neutrino $\nu_L$ combining with right handed %neutrino, $N_R$ makes a Dirac neutrino with a mass of the order of %$10^{-2}$, which is the scale of $Z_2$ symmetry breaking. 
Note that in our model, the tiny neutrino masses are generated from the spontaneous breaking of the discrete $Z_2$ symmetry with its tiny vev of 
$\sim$ keV. The left handed doublet neutrino combine with its corresponding 
right handed singlet neutrino to produce a massive Dirac neutrino.
Since we assume lepton number conservation, the Majorana mass terms for the right handed neutrinos, having the form, $M \nu_R^TC^{-1}\nu_R$ are not allowed. 
%In our model, the origin of the tiny neutrino masses is due to the %spontaneous breaking of the discrete symmetry, which is a very different %scenario than the very popular and well cultivated traditional see-saw %mechanism for $\nu$ mass generation \cite{Mohapatra:1979ia}. In the see-saw model, light neutrinos are Majorana particles, and thus neutrinoless %double beta
%decay is allowed. The current limit on the double beta decay is $m_{ee} %\sim 0.3 eV$ . This limit is
%expected to go down to about $m_{ee} \sim 0.01 eV$ in future experiments.%\cite{Aalseth:2002rf} If no neutrinoless
%double beta decay is observed to that limit, that will cast serious %doubts on the see-saw
%model. In our model, of course, it is not allowed at any level. As a %result, on the other hand, observing neutrinoless
%double beta decay in future will certainly rule out this model.

The model has a very light neutral scalar $\sigma$ with mass of the order of this $Z_2$ symmetry breaking scale. 
Detailed phenomenology of this light scalar $\sigma$
 in context of $e+e-$ collider has been done previously \cite {Gabriel:2006ns} and also some phenomenological works have been done on the chromophobic charged Higgs of this model at the LHC whose signal are very different from the charged Higgs in the usual two Higgs doublet model \cite{Gabriel:2008es}. There are bounds on $v_{\phi}$ from cosmology, big bang nucleosynthesis, because of the presence of extra degree of freedom compared to the SM; puts a lower limit on $v_{\phi}\geq 2$ eV \cite{logan}, while the bound from supernova neutrino observation is $v_{\phi}\geq 1$ keV \cite{Raffelt:2011nc}.
 
 In this work, we study the neutrino sector of the model using the input of all the experimental information regarding the neutrino mass difference squares and the three mixing angles. Our additional theoretical inputs are  that the neutrino mass matrix is hermitian and also has  $\nu_{\mu} - \nu_{\tau}$ exchange symmetry. We find that in order for our model to be consistent with the current available experimental data, the neutrino mass hierarchy has to be inverted type (with neutrino mass values close to degenerate case). We also predict the values of all three neutrino masses, as well as the CP violating phase $\delta$.
 
 % in particular predictions for the hierarchical structure and the value %of the neutrino masses, their mixings, and phase, are testing the %neutrinos of this model in the parameter space of all known experimental %data available and as a result we get some very interesting answers about %$\nu$ masses and CP-violating phase $\delta$.
%\section*{The Neutrino mass matrix}
%%%%%%%%%%%%%%
With the three assumptions stated in the introduction, namely, lepton number conservation, Hermiticity of the neutrino mass matrix, and the $\nu_{\mu} - \nu_{\tau}$ exchange symmetry, the neutrino mass matrix can be written as

\begin{equation} 
 M_{\nu} = \left(\begin{matrix}
a & b & b \\
b^* & c & d\\
b^* & d & c
\end{matrix}
\right).  
%\nonumber
\label{eqn1}
\end{equation}
The parameters a, c and d are real, while the parameter b is complex.
Thus the model has a total of five real parameters.
The important question at this point is whether the experimental data is consistent with this form. Choosing a basis in which the Yukawa couplings for the charged leptons are diagonal, the PMNS matrix in our model is simply given by $U_{\nu}$, where $U_{\nu}$ is the matrix which diagonalizes the neutrino mass matrix. Since the neutrino mass matrix is hermitian, it can then be obtained from
\begin{equation} 
M_{\nu} = U_{\nu} \bold {M}_{\nu}^{diag} U_{\nu}^{\dagger}
\label{eqn0}
\end{equation}

where 

\begin{equation} 
 \bold {M}_{\nu}^{diag}= \left(\begin{matrix}
m_1 & 0 & 0 \\
0 & m_2 & 0\\
0 & 0 & m_3
\end{matrix}
\right). 
%\nonumber
\end{equation}

The matrix $U_{\nu}$ is the PMNS matrix for our model (since $U_l$ is the identity matrix from our choice of basis), and  is conventionally written as: 
\begin{equation} 
U_{\nu}
% = U_{l}^{\dagger}U_{\nu} 
= \left(\begin{matrix}
c_{12}c_{13} & s_{12}c_{13} & s_{13}e^{-i\delta} \\
-s_{12}c_{23}-c_{12}s_{23}s_{13}e^{i\delta} & c_{12}c_{23}-s_{12}s_{23}s_{13}e^{i\delta} & s_{23}c_{13} \\
s_{12}s_{23}-c_{12}c_{23}s_{13}e^{i\delta} & -c_{12}s_{23}-s_{12}c_{23}s_{13}e^{i\delta} & c_{23}c_{13}
\end{matrix}
\right),  
%\nonumber
\label{eqn2}
\end{equation}
where, $c_{ij}={\rm Cos}\theta_{ij}$ and $s_{ij}={\rm Sin}\theta_{ij}$.

\section{Phenomenological Implications and Results}
The values of three mixing angles and the two neutrino mass squared differences are now determined from the various solar, reactor and accelerator neutrino experiments with reasonable accuracy (the sign of $\Delta m^2_{32}$ is still unknown).
%known today from various experiments. In one hand, using the $\nu_\mu$ %disappearance channel with atmospheric and accelerator neutrinos, %Super-%Kamiokande,~\cite{Wendell:2013kxa}, K2K~\cite{Mariani:2008zz}, MINOS~\cite{Barr:2013wta},  T2K~\cite{Abe:2011sj}, and %IceCube~\cite{icecube} experiments have
%determined the angle $\theta_{23}$ and the mass difference $|\Delta %m^2_{32}|$ , while with reactor 
%antineutrinos and $\nu_e$ disappearance channel using solar neutrinos, %the KamLAND~\cite{Mitsui:2011zz} and SNO~\cite{Aharmim:2011vm} %experiments have measured $\theta_{12}$ and $\Delta m^2_{21}$ with %$\bar{\nu}_e$ disappearance channel. Recently, the Daya %Bay~\cite{An:2012eh}, Double Chooz~\cite{Abe:2011fz}, and RENO~\cite{Ahn:2012nd}
%successfully measured $\theta_{13}$ and are on their ways to measure $|%\Delta m^2_{31}|$ with $\bar{\nu}_e$ disappearance
%using reactor antineutrinos.
The current knowledge of the mixing angles and mass squared differences are given by~\cite{PDG} Table \ref{tab:expdata}.
\begin{table}[htb]
%\label{tab:expdata}
\begin{center}
\begin{tabular}{|c|c|}
\hline
\hline
Parameter & best-fit ($\pm\sigma$) \\
\hline
$\Delta m^2_{21}[10^{-5} eV^2]$ & $7.53_{-0.22}^{+0.26}$ \\
\hline
$\Delta m^2[10^{-3} eV^2]$ & $2.43_{-0.10}^{+0.06}$ \\
\hline
$\sin^2\theta_{12}$ & $0.307_{-0.016}^{+0.018}$\\
\hline
$\sin^2\theta_{23}$ & $0.392_{-0.022}^{+0.039}$\\
\hline
$\sin^2\theta_{13}$ & $0.0244_{-0.0025}^{+0.0023}$\\
\hline
\end {tabular}
\end{center}
\caption {The best-fit values and $1\sigma$ allowed ranges of the 3-neutrino oscillation parameters. The definition of $\Delta m^2$ used is $\Delta m^2 = m_{3}^2 - (m_{2}^2 + m_{1}^2)/2$. Thus $\Delta m^2 = \Delta m_{31}^2 - m_{21}^2/2$ if $m_1 < m_2 < m_3$ and $\Delta m^2 = \Delta m_{32}^2 + m_{21}^2/2$ for $m_3 < m_1 < m_2$.}
\label{tab:expdata}
\end{table}
\\
It is not at all sure that the data will satisfy our model given by Eqn. (\ref{eqn1}),
either for the direct hierarchy or the indirect hierarchy.
We first try the indirect hierarchy. In this case, the diagonal neutrino mass matrix, using the experimental mass difference squares, can be written as

\begin{equation} 
 \bold {M}_{\nu}^{diag}= \left(\begin{matrix}
\sqrt{m_3^2 + 0.002315} & 0 & 0 \\
0 & \sqrt{m_3^2 + 0.00239} & 0\\
0 & 0 & m_3
\end{matrix}
\right),  
%\nonumber
\end{equation}
where we have used the definition of $\Delta m^2$ in the inverse hierarchy mode as referred in Table \ref{tab:expdata}.

Taking these experimental values in the best-fit($\pm\sigma$) region from Table \ref{tab:expdata},  for the PMNS mixing matrix, we get from Eqn (\ref{eqn2})
% the neutrino mass matrix can be calculated via $ M_{\nu} = U_{\nu} \bold %{M}_{\nu}^{diag} U_{\nu}^{\dagger}$
%where,
\begin{equation} 
 U_{\nu} = \left(\begin{matrix}
0.822 & 0.547 & 0.156 \exp(-i\delta) \\
-0.432-0.081 \exp(i\delta) & 0.649-0.054 \exp(i\delta) & 0.618\\
0.347-0.101 \exp(i\delta) & -0.521-0.067 \exp(i\delta) & 0.771
\end{matrix}
\right).  
%\nonumber
\label{eqn3}
\end{equation}
%and
%\begin{equation} 
% \bold {M}_{\nu}^{diag}= \left(\begin{matrix}
%\sqrt{m_3^2 + 0.002315} & 0 & 0 \\
%0 & \sqrt{m_3^2 + 0.00239} & 0\\
%0 & 0 & m_3
%\end{matrix}
%\right), 
% \nonumber
%\end{equation} 
%and $\bold {M}_{\nu}^{diag}$ is taken in the Inverse hierarchical(IH) %mode with the definition of $\Delta m^2$ in the inverse hierarchy mode %being used as referred in Table \ref{tab:expdata}.\\
We plug  these expressions  for $\bold {M}_{\nu}^{diag}$ and  $U_{\nu}$ in $ M_{\nu} = U_{\nu} \bold {M}_{\nu}^{diag} U_{\nu}^{\dagger}$ and 
demand that  the resulting mass matrix satisfy the form of our model predicted Eqn. (\ref{eqn1}). First, using
%the first assumption of $\mu$-$\tau$ symmetry 
$M_{\mu \mu}$ = $M_{\tau \tau}$ as in Eqn. (\ref{eqn1}),  we obtain  the following  $2^{nd}$ order equation for  $\cos\delta$ 
\begin{eqnarray}
(-123.27m_3^4 - 0.15m_3^2 + 0.0026)\cos^2\delta +(6.66m_3^4 - 6.7m_3^2 - 0.006)\cos\delta 
\nonumber\\ 
+ 29.654m_3^4 - 3.19m_3^2  
 + 0.0031  = &0& ,
\label{equation}
\end{eqnarray}
where, we have used some approximations while simplifying the equation analytically, which would not affect our result, if it is done numerically. Further, Eqn. (\ref{equation}) is satisfied only for certain range of values of $m_3$  demanding that $ -1 < \cos\delta < 1 $.
For that range of $m_3$, now we demand  that 
$M_{e\mu}$ = $M_{e\tau}$ to be satisfied. This takes into account separately satisfying the equality of the real and imaginary parts of $M_{e\mu}$ and $M_{e\tau}$ elements. It is intriguing that a solution exists, and gives the values of 
$m_3 = 7.8 \times 10^{-2}$ eV and  $\delta = 109.63^o$.

%the values of $m_3$(where $m_3 < m_1 < m_2$) and $\delta$ as 
%$m_3 = 7.5 \times 10^{-2} $ eV and $\delta = 110.09^o$.
%So the three $\nu$ mass pattern, which satisfy all the experimental data %is given by,
Thus the prediction for the three neutrino masses and the CP violating phase in our model are,
\begin{eqnarray}
m_1 = \sqrt{m_3^2 + 0.002315} = 9.16 \times 10^{-2} eV ,\\\nonumber
m_2 = \sqrt{m_3^2 + 0.00239} = 9.21 \times 10^{-2} eV ,\\\nonumber
m_3 = 7.8 \times 10^{-2} eV ,\\\nonumber
\delta = 109.63^o .
\end{eqnarray}
%where $m_3 \leq m_1 \simeq m_2$ \nonumber i.e three neutrino masses obey Thus our model gives a  Quasi-degenerative Inverse Hierarchical pattern
with $\delta$ being close to the maximum CP violating phase.

%Thus the $\nu$ mass matrix in this particular case with $m_3 = 7.5 \times %10^{-2} $ eV and $\delta = 110.09 ^o$ looks like,
As a double check of our calculation, we have calculated the neutrino mass matrix numerically using the above obtained values of $m_1, m_2 , m_3$ and $\delta$ as given by mass matrix  Eqn.(\ref{eqn0}).  The resulting numerical neutrino mass matrix we obtain is given by,  
%Eq. (8), and 
\begin{equation} 
 M_{\nu} = \left(\begin{matrix}
0.091 & 0.00048 + 0.001i & 0.00044 + 0.0015i \\
0.00048 - 0.001i & 0.086 & -0.0066 \\
0.00044 - 0.0015i & -0.0066 & 0.084
\end{matrix}
\right).  
%\nonumber
\end{equation}

We see that with this verification, the mass matrix predicted by our model in Eqn.(\ref{eqn1}), is well satisfied.

%for which $M_{e\mu} \sim M_{e\tau}$ and $M_{\mu \mu} \sim M_{\tau \tau}$ %is satisfied. Due to $\mu$-$\tau$ symmetry and Hermiticity, the resulting %mass matrix $ M_{\nu}$ seems to have the following pattern,
%\begin{equation} 
% M_{\nu} = \left(\begin{matrix}
%a & b & b \\
%b^* & c & d\\
%b^* & d & c
%\end{matrix}
%\right),  \nonumber
%\end{equation}
%where $b^*(d^*)$ is the complex conjugate of b(d) and $M_{e\mu}$ = %$M_{e\tau}$ = b as well $M_{\mu \mu}$ = $M_{\tau \tau}$ = c, due to the %$\mu$-$\tau$ symmetry. Thus we have 5 parameters, with which we can %explain all the experimental data.

We note that we also investigated the normal hierarchy case for our model satisfying hermiticity and $\nu_{\mu} - \nu_{\tau}$ exchange symmetry. We found no solution for $\cos\delta$ for that case. Thus normal hierarchy for the neutrino masses  can not be accommodated in our model.

% We also find that for the Normal Hierarchy (NH) case, all the %experimental data could not be fit along with the assumption of %Hermiticity and $\mu$-$\tau$ symmetry in the neutrino mass matrix. We %conclude that the normal mass hierarchy (NH)($m_1 < m_2 < m_3$) is ruled %out by the present $\nu$ observational data with $\mu$-$\tau$ symmetry
%and Hermiticity assumptions; but with Quasi-degenerate/inverted mass %hierarchy (IH), all the experimental observables alongwith the above %mentioned assumptions are satisfied.\\
  Our model predicts the electron type neutrino mass to be rather large ($9.16 \times 10^{-2}$ eV), and the  CP violating parameter $\delta$ close to the maximal value ($\delta \simeq 109 ^o$). Let us now discuss briefly how our model can be tested in the proposed future experiments of electron type neutrino mass measurement directly and also for the leptonic CP violation. 
The measurement of the electron anti-neutrino mass from tritium $\beta$ decay in Troitsk $\nu$-mass experiment set a limit of $m_\nu < 2.2 eV$ 
 \cite {Aseev:2012zz}. 
New experimental approaches such as the MARE \cite{Schaeffer:2011zz} will perform  measurements of the neutrino mass in the sub-eV region. So with a little more improvement, it may be possible to reach our predicted value of $\sim 0.1$ eV. 

The magnitude of the CP violation effect depends directly on the magnitude of the well known Jarlskog Invariant \cite{Jarlskog:1985cw}, which is a function of the three mixing angles and CP violating phase $\delta$ in standard parametrization of the mixing matrix:
\begin{equation}
J_{CP} = 1/8 \cos\theta_{13} \sin2\theta_{12}\sin2\theta_{23}\sin2\theta_{13}\sin\delta
%\nonumber
\end{equation}
Given the best fit values for the mixing angles in Table \ref{tab:expdata} and the value of CP violating phase $\delta = 110^o$ in our model, we find the value of Jarlskog Invariant,
\begin{equation}
J_{CP} = 0.032 , 
%\nonumber
\end{equation}
which corresponds to large CP violating effects. The study of $\nu_\mu \rightarrow \nu_e$ and $\bar{\nu}_\mu \rightarrow  \bar{\nu}_e$ transitions using accelerator based beams is sensitive to the CP violating phenomena arising from the CP violating phase $\delta$. We are particularly interested in the Long Baseline Neutrino Experiment (LBNE) \cite{LBNE}, which with its baseline of $1300$ Km and neutrino energy $E_\nu$ between $1-6$ GeV would be able to unambiguously shed light both on the 
mass hierarchy and the CP phase simultaneously. Evidence of the CP violation in the neutrino sector requires the explicit observation of asymmetry between $P(\nu_\mu \rightarrow \nu_e)$ and $P(\bar{\nu}_\mu \rightarrow  \bar{\nu}_e)$, which is defined as the CP asymmetry $\mathcal{A}_{CP}$,
\begin{equation}
\mathcal{A}_{CP} = \frac{P(\nu_\mu \rightarrow \nu_e) - P(\bar{\nu}_\mu \rightarrow  \bar{\nu}_e)}{P(\nu_\mu \rightarrow \nu_e) + P(\bar{\nu}_\mu \rightarrow  \bar{\nu}_e)}
%\nonumber
\end{equation}
In three-flavor model the asymmetry can be approximated to leading order in $\Delta m_{21}^2$ as, \cite{Marciano:2006uc}
\begin{equation}
\mathcal{A}_{CP} \sim \frac{\cos\theta_{23} \sin2\theta_{12} \sin\delta}{\sin\theta_{23}\sin\theta_{13}}(\frac{\Delta m_{21}^2 L}{4E_\nu}) + \text{matter effects} 
%\nonumber
\end{equation}  
For our model, taking LBNE Baseline value L = $1300$ Km and $E_\nu = 1$ GeV, we get the value of $\mathcal{A}_{CP} = 0.17$ + matter effects.
With this relatively large values of $\mathcal{A}_{CP}$,
% and $\delta = 110\degree$, 
LBNE10 in first phase with values of 700 KW wide-band muon neutrino and muon anti-neutrino beams and $100$ kt.yrs will be sensitive to our predicted value of CP violating phase $\delta$ with 3-Sigma significance \cite{Kronfeld:2013uoa}.\\ 
Finally, we compare our model for the sum of the three neutrino masses against the cosmological observation.
%\section*{Sum of neutrino masses}
The sum of neutrino masses $m_1 + m_2 + m_3 < (0.32 \pm 0.081)$ eV \cite {Battye:2013xqa} from (Planck + WMAP + CMB + BAO) for an active neutrino model with three degenerate neutrinos has become an important cosmological bound. For our model, we find $m_1 + m_2 + m_3 \simeq 0.26 $ eV, which is consistent with this bound.

\section{The Model with four parameters in the neutrino mass matrix}

The most general Dirac neutrino mass matrix contain 9 complex parameters and can be written as: 
\begin{equation} 
 \bold {M}_{\nu}= \left(\begin{matrix}
m_{e_L e_R} & m_{e_L \mu_R} & m_{e_L \tau_R} \\
m_{\mu_L e_R} & m_{\mu_L \mu_R} & m_{\mu_L \tau_R} \\
m_{\tau_L e_R} & m_{\tau_L \mu_R} & m_{e_L \tau_R} \\
\end{matrix}
\right). 
\label{general_matrix}
%\nonumber
\end{equation}
On this 18 parameter Dirac neutrino mass matrix, we have imposed the following conditions:
\begin{itemize}
\item We have assumed the hermiticity of the neutrino mass matrix. As a result of this assumption, the diagonal elements of Eq.~\ref{general_matrix} become real and off-diagonal elements become complex conjugate of each other: $m_{\mu_L e_R}=m_{e_L \mu_R}^*$, $m_{\tau_L e_R}=m_{e_L \tau_R}^*$ and $m_{\tau_L \mu_R}=m_{\mu_L \tau_R}^*$. Therefore, after demanding the hermiticity, we have a 9 parameter neutrino mass matrix. 
%
%\item We have demanded the invariance under interchange of $\nu_{\mu R} \leftrightarrow -\nu_{\tau R}$ followed by a complex conjugation of the couplings. Complex conjugation of the couplings is equivalent to making a CP transformation. In the rest of this article, the symmetry  under interchange of $\nu_{\mu R} \leftrightarrow -\nu_{\tau R}$ followed by a  CP transformation is denoted as {\it $\nu_{\mu R}$-$\nu_{\tau R}$ reflection} symmetry. As a result of the {\it $\nu_{\mu R}$-$\nu_{\tau R}$ reflection} symmetry, we can derive three more constraints on the elements of the hermitian Dirac neutrino mass matrix: $m_{e_L \tau_R}=-m_{e_L \mu_R}^*$, $m_{\mu_L \tau_R}=-m_{\mu_L \mu_R}^*$ and $m_{\tau_L \tau_R}=-m_{\tau_L \mu_R}^*$. These constraints reduces 5 more parameters and the 4 parameter hermitian Dirac neutrino mass matrix which respects {\it $\nu_{\mu R}$-$\nu_{\tau R}$ reflection} symmetry can be written as: 
%\begin{equation} 
% \bold {M}_{\nu}= \left(\begin{matrix}
%a & b & -b^* \\
%b^* & c & -c \\
%b & -c & c \\
%\end{matrix}
%\right), 
%\label{final_matrix}
%\nonumber
%\end{equation} 
\end{itemize}
%where, $a$ and $c$ are real and $b$ is complex parameter. 

The hermitian neutrino mass matrix is given in the flavor basis by
\begin{equation} 
M_{\nu} = U_{\nu} \bold {M}_{\nu}^{diag} U_{\nu}^{\dagger},
\label{eqn0}
\end{equation}
where, ${M}_{\nu}^{diag}$ is the diagonal neutrino mass matrix in the mass basis. Two squared-mass differences of the neutrinos are known from the experiments. Therefore, ${M}_{\nu}^{diag}$ can be constructed with only one mass as unknown. For IH, the diagonal neutrino mass matrix is given by,
\begin{equation} 
 \bold {M}_{\nu}^{diag}= \left(\begin{matrix}
\sqrt{m_3^2 + 0.002315} & 0 & 0 \\
0 & \sqrt{m_3^2 + 0.00239} & 0\\
0 & 0 & m_3
\end{matrix}
\right),  
%\nonumber
\end{equation}
where, $m_3$ is the unknown mass and we have used the central values of the squared-mass differences listed in Table~\ref{tab:expdata} for IH. In the mixing matrix $U$, there are three angles and one phase. The mixing angles are already measured (see Table~\ref{tab:expdata} for their central values and $3\sigma$ range) with good precision. In our analysis, we have considered the IH central values for the $s_{12}^2$ and $s_{13}^2$. However, we have considered $s_{23}^2=0.5$ which is not the central value but well within $3\sigma$ of the central value. 

If we assume one particular neutrino mass hierarchy, there are still two quantities unknown in for the Dirac neutrinos namely, the mass $m_3$ in the diagonal mass matrix and the CP violating phase ($\delta$) in the mixing matrix. In our analysis, we have scanned unknown parameters ($m_3$ and $\delta$) over a range of values and tried to find out a constrained phenomenological neutrino mass matrix which is consistent with the 5 experimental results (three mixing angles and two squared-mass differences). 
%in the framework of a 4 parameter neutrino mass matrix (which results as a consequence of the assumption of hermiticity and {\it $\nu_{\mu R}$-$\nu_{\tau R}$ reflection} symmetry) in Eq.~\ref{final_matrix}. 
Our phenomenological results are summarized in the following:
%----------------------------------------------------------------
\begin{figure}[t]
\begin{center}
\includegraphics[width=0.7\textwidth]{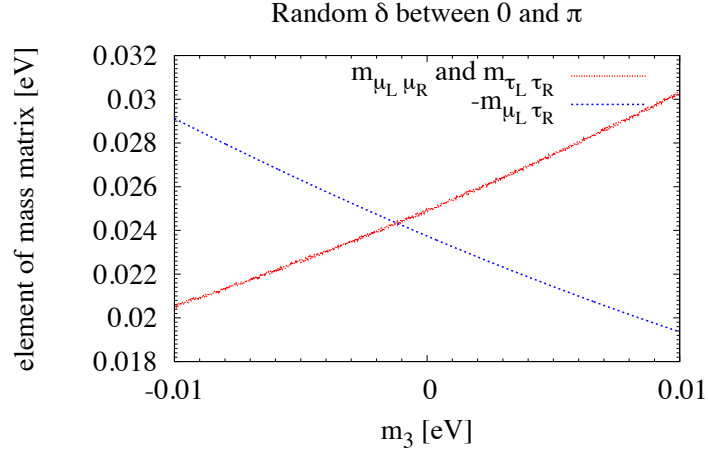}
\end{center}
\caption{The elements $m_{\mu_L \mu_R}$ and real part of -$m_{\mu_L \tau_R}$ of the Dirac neutrino mass matrix in Eq.~\ref{general_matrix} as a function of $m_3$. The other free parameter $\delta$ was randomly varied between 0 and $\pi$. We have used IH central values for the $\Delta m_{21}^2$, $\Delta m^2$, $s^2_{12}$ and $s^{2}_{13}$ from Table~\ref{tab:expdata} and for $s^2_{23}$, we choose $s^2_{23}=0.5$.} 
\label{xm}
\end{figure}
%--------------------------------------------------

\begin{figure}[t]
\begin{center}
\includegraphics[width=0.49\textwidth]{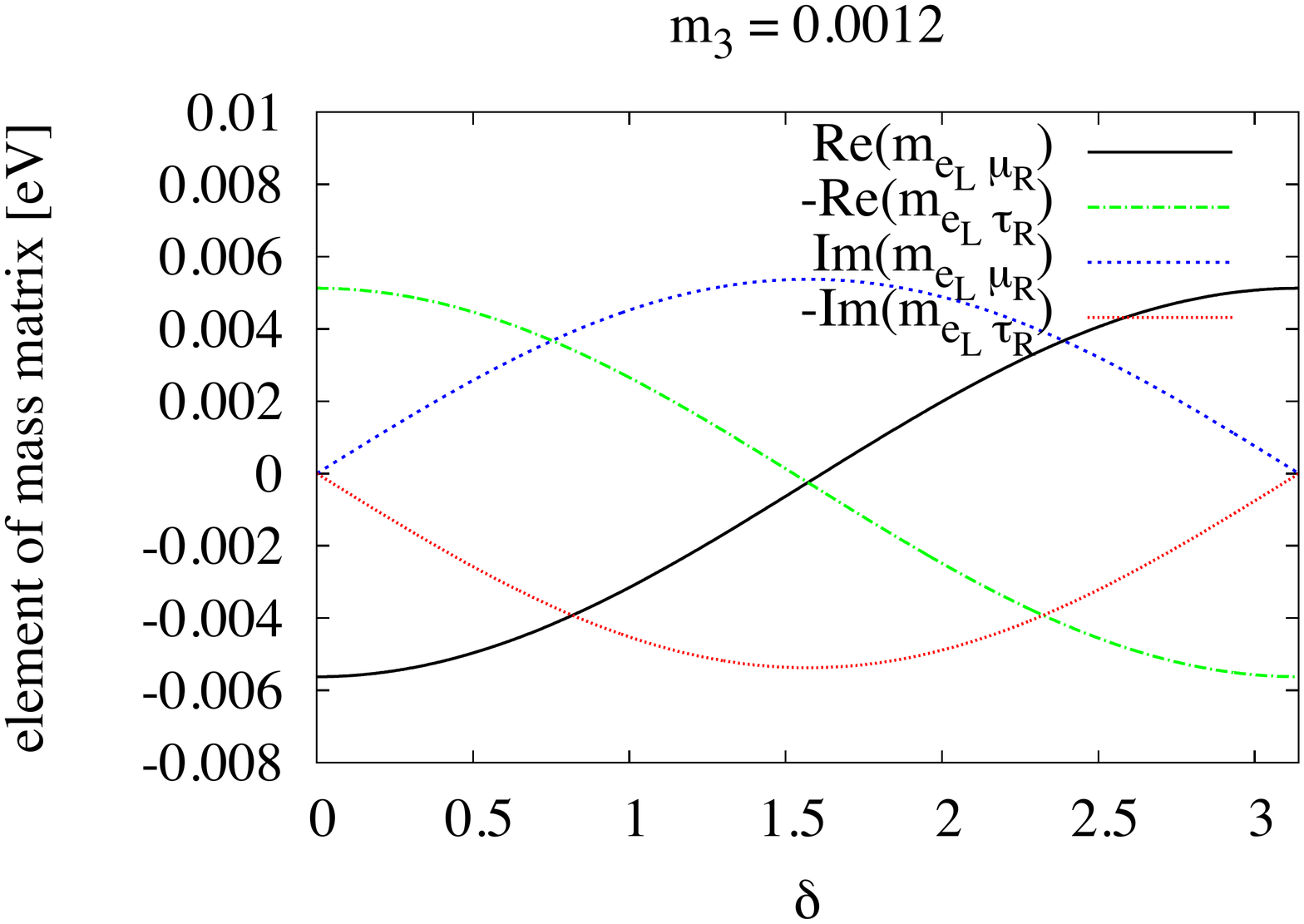}
\includegraphics[width=0.49\textwidth]{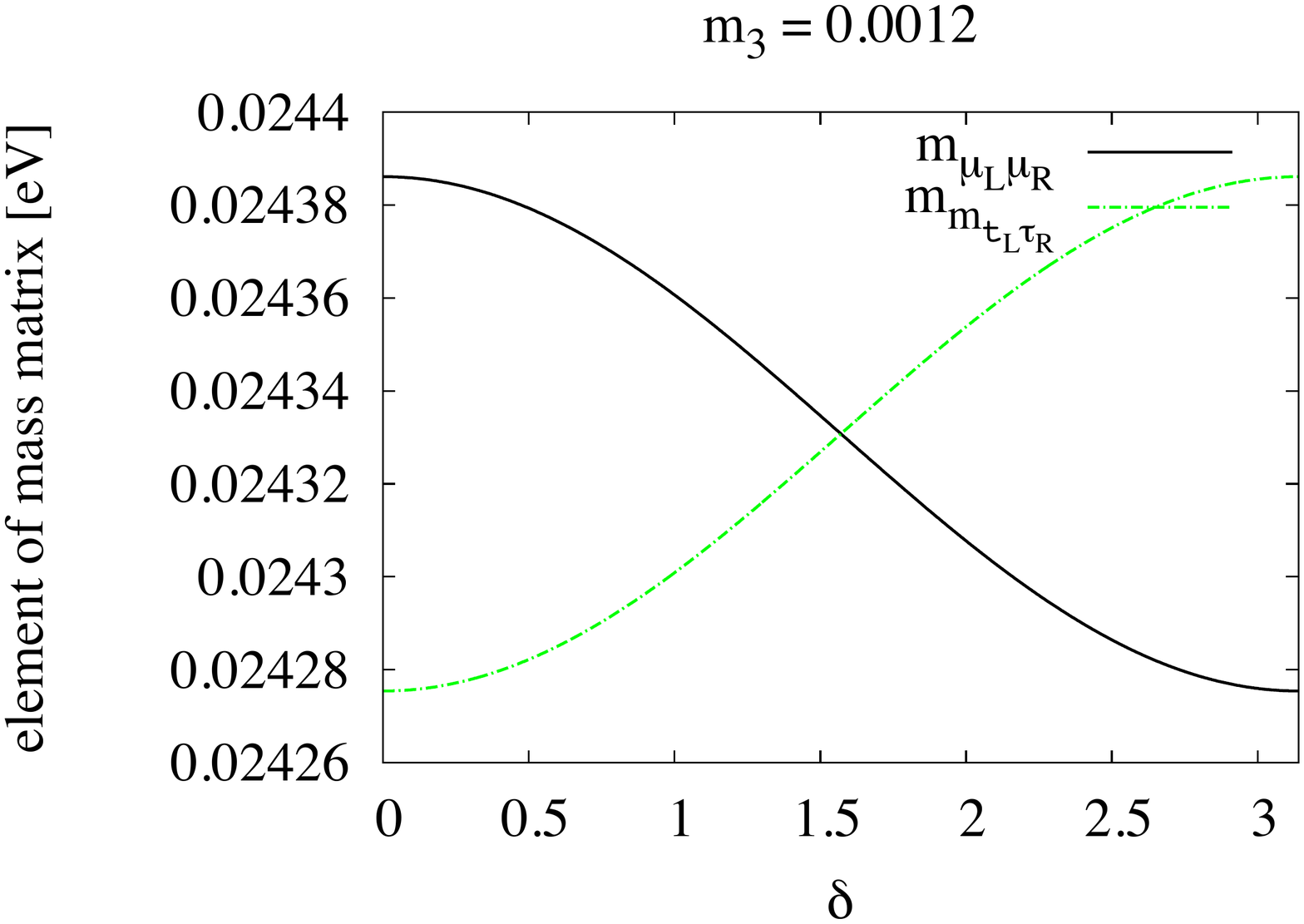}
\end{center}
\caption{Left panel: The real and imaginary part of the elements $m_{e_L \mu_R}$ and $m_{e_L \tau_R}$ of the Dirac neutrino mass matrix in Eq.~\ref{general_matrix} as a function of $\delta$ (in radian) for $m_3=-1.198 \times 10^{-3}$. Right panel: The diagonal elements $m_{\mu_L\mu_R}$ and $m_{\tau_L \tau_R}$ of the Dirac neutrino mass matrix in Eq.~\ref{general_matrix} as a function of $\delta$ for $m_3=-1.198 \times 10^{-3}$. We have used IH central values for the $\Delta m_{21}^2$, $\Delta m^2$, $s^2_{12}$ and $s^{2}_{13}$ from Table~\ref{tab:expdata} and for $s^2_{23}$, we choose $s^2_{23}=0.5$.} 
\label{del}
\end{figure}

\begin{itemize}
\item In Fig.~\ref{xm}, we have presented  $m_{\mu_L \mu_R}$ and real part of -$m_{\mu_L \tau_R}$ elements of the Dirac neutrino mass matrix in Eq.~\ref{general_matrix} as a function of $m_3$. The other free parameter $\delta$ was randomly varied between 0 and $\pi$. Fig.~\ref{xm} shows that two curves interests each other at $m_3=-1.198 \times 10^{-3}$eV. 
\item In Fig.~\ref{del}, we have presented real and imaginary parts of the elements $m_{e_L \mu_R}$ and $m_{e_L \tau_R}$ (left panel) and diagonal elements $m_{\mu_L\mu_R}$ and $m_{\tau_L \tau_R}$ (right panel) of the Dirac neutrino mass matrix in Eq.~\ref{general_matrix} as a function of $\delta$ for $m_3=-1.198 \times 10^{-3}$ eV. Fig.~\ref{del} shows that 
%the desired structure (see Eq.~\ref{final_matrix}) of the Dirac neutrino mass matrix i.e., hermitian and {\it $\nu_{\mu R}$-$\nu_{\tau R}$ reflection} symmetric, 
a constrained neutrino mass matrix is obtained for $\delta=\pi/2$ and $m_3=-1.198 \times 10^{-3}$ eV. The numerical form of the mass matrix in the flavor basis for $\delta=\pi/2$ and $m_3=-1.198 \times 10^{-3}$ eV is given by,
\end{itemize}
\begin{equation} 
\left(\begin{matrix}
4.72\times 10^{-2} & 2.49 \times 10^{-4}-5.37 \times 10^{-3}i & -2.49 \times 10^{-4}-5.37 \times 10^{-3}i \\
 2.49 \times 10^{-4}+5.37 \times 10^{-3}i & 2.43 \times 10^{-2}  & -2.43 \times 10^{-2} \\
%-  5.37 \times 10^{-5}i \\
-2.49 \times 10^{-4}+5.37 \times 10^{-3}i  & -2.43 \times 10^{-2} 
%+  5.37 \times 10^{-5}i 
&  2.43 \times 10^{-2} \\
\end{matrix}
\right), 
\label{num_matrix}
%\nonumber
\end{equation} 
 %----------------------------------------------------------------
%--------------------------------------------------
It is important to note that the mass matrix in Eq.~\ref{num_matrix} is a four parameter matrix can be written as,
\begin{equation}
 \bold {M}_{\nu}^{pheno}=
\left(\begin{matrix}
a & b e^{i\eta} & -b e^{-i\eta} \\
 b e^{-i\eta} & c & -c \\
%-  5.37 \times 10^{-5}i \\
-b e^{i\eta}  & -c &  c \\
\end{matrix}
\right), 
\label{param_matrix}
%\nonumber
\end{equation} 
 %----------------------------------------------------------------
with $a=4.72\times 10^{-2}$, $b=5.38\times 10^{-3}$, $c=2.43 \times 10^{-2}$ and $\eta=272.6^{0}$. We  now  search for  symmetry in the $\nu_{\mu R}$-$\nu_{\tau R}$ sector which is consistent with the structure of the phenomenological neutrino mass matrix in Eq.~\ref{param_matrix}. 

The most general transformation in the $\nu_{\mu R}$-$\nu_{\tau R}$ sector can be written as,
\begin{equation} 
\Psi_{R}=\left(\begin{matrix}
\nu_e\\
\nu_\mu\\
\nu_\tau\\
\end{matrix}
\right) \to
\left(\begin{matrix}
e^{i\phi_1} & 0 & 0 \\
0 & p e^{i\phi_2} & -q e^{-i\phi_3}\\
0 & q e^{i\phi_3} & p e^{-i\phi_2}\\
\end{matrix}
\right)~\Psi_{R} \to U_{R}\Psi_R, 
\label{r_trans}
%\nonumber
\end{equation} 
where, $p^2+q^2=1$ and $\phi_1,~\phi_2$ and $\phi_3$ are the arbitrary phases. As already discussed in the beginning of this paper, we do not want to introduce any symmetry in $\nu_{\mu L}$-$\nu_{\tau L}$ sector in order to make the symmetry as the symmetry of the Lagrangian. However, phase transformation for the left-handed neutrino fields are still allowed:
 \begin{equation} 
\Psi_{L}=\left(\begin{matrix}
\nu_e\\
\nu_\mu\\
\nu_\tau\\
\end{matrix}
\right) \to
\left(\begin{matrix}
e^{-i\theta_1} & 0 & 0 \\
0 & e^{-i\theta_2} & 0\\
0 & 0 & e^{-i\theta_3}\\
\end{matrix}
\right)~\Psi_{L}\to U_{L}\Psi_L, 
\label{l_trans}
%\nonumber
\end{equation}  
We have demanded the invariance under simultaneous transformations $\Psi_R \rightarrow  U_R \Psi_R$ and $\Psi_L \to U_L \Psi_L$ followed by a complex conjugation of the couplings. Complex conjugation of the couplings is equivalent to making a CP transformation. In the rest of this article, the symmetry  under above mentioned transformations followed by a  CP transformation is denoted as {\it $\nu_{\mu R}$-$\nu_{\tau R}$ reflection} symmetry. As a consequence of the {\it $\nu_{\mu R}$-$\nu_{\tau R}$ reflection} symmetry, we obtain the following matrix equation:
\begin{equation} 
\left [ U_L^\dagger  \bold {M}_{\nu}^{pheno} U_R \right]^*~=~\bold {M}_{\nu}^{pheno}.
\label{meq}
\end{equation}  
The most general solution of Eq.~\ref{meq} is given by
\begin{eqnarray}
\phi_1=n_1\pi - {\rm cos}^{-1}\left[(-1)^{n_2} p\right ]~&;&~\theta_1=n_1\pi + {\rm cos}^{-1}\left[(-1)^{n_2} p\right ];\nonumber\\
\phi_2=n_2\pi~&;&~\theta_2={\rm cos}^{-1}\left[(-1)^{n_2} p\right ];\nonumber\\
\phi_2=\left(n_3+\frac{1}{2}\right)\pi~&;&~\theta_2={\rm sin}^{-1}\left[(-1)^{n_3} q\right ]
\end{eqnarray}
and
\begin{equation}
\eta~=~\frac{n\pi}{2};
\end{equation}
where, $n,~n_1,~n_2$ and $n_3$ are arbitrary integers. The trivial solution ($n_1=0,~n_2=0$ and $n_3=0$) of Eq.~\ref{meq} physically corresponds to  a symmetry under interchange of $\nu_{\mu R} \leftrightarrow -i\nu_{\tau R}$ followed by a CP transformation with $\eta=0^0,~90^0,~180^0,~270^0,....$. However, the phenomenological neutrino mass matrix under consideration (Eq.~\ref{num_matrix} and Eq.~\ref{param_matrix}) corresponds to $\eta=272.6^0$. Therefore, tiny violation of the symmetry under interchange of $\nu_{\mu R} \leftrightarrow -i\nu_{\tau R}$ followed by a CP transformation is required to satisfy all the experimental results.

%As we get a value of this CP phase $\delta$ different from 0 or $180 %\degree$, it is a signal for CP violation in the lepton sector.The %experimental determination of leptonic CP phase is very important as it %could hold the key to the matter-antimatter asymmetry in the universe %through the mechanism of Leptogenesis. It has been recognized that the %most direct and promising way to probe the Dirac CP violation phase %$\delta$ is through observation in neutrino oscillation experiments. T2K %experiment has successfully observed the $\nu_e$ appearance %\cite{Ikeda:2013oya} and is expected to shed some light in the CP %violation phase in its current run-time. NO$\nu$A,\cite{Ayres:2004js} %which is expected to start taking data in near future will also play an %imp role in this context.

\section{Summary and Conclusions}

In this work, we have presented predictive models for Dirac neutrinos. 
The first model has three assumptions: (i) lepton number conservation, (ii) hermiticity of the neutrino mass matrix, and (iii) $\nu_{\mu}$ - $\nu_{\tau}$ exchange symmetry. The resulting neutrino mass matrix is of Dirac type, and has five real parameters, (three real and one complex). We have shown that the data on neutrino mass differences squares, and three mixing angles are consistent with this model yielding a solution for the neutrino masses with inverted mass hierarchy (close the degenerate pattern). The values predicted by the model for the three  neutrino masses are $9.16 \times 10^{-2}$ eV, $9.21 \times 10^{-2}$ eV and $7.80 \times 10^ {-2}$ eV. In addition, the model also predicts the CP violating phase $\delta$ to be $109.63 ^o$, thus predicting a rather large CP violation in the neutrino sector, and will be easily tested in the early runs of the LBNE. The mass of the electron type neutrino is also rather large, and has a good possibility for being accessible for measurement in the proposed tritium beta decay experiments. Neutrinos being Dirac, neutrinoless double beta decay is also forbidden in this model.  Thus, all of these predictions can be tested in the upcoming and future precision neutrino experiments.\\
In the second model, we have also considered Dirac neutrino mass matrix and investigated the possible symmetries in the $\nu_{\mu R}$-$\nu_{\tau R}$ sector. In order to ensure that the imposed condition is a symmetry of the Lagrangian (not only the symmetry of the neutrino mass matrix in the flavor basis), we have restricted the requirements only to the singlet right-handed muon and tau neutrinos. Assuming the hermiticity of the neutrino mass matrix, we have obtained a particular structure of the phenomenological Dirac neutrino mass matrix with only 4 parameters. This 4 parameter Dirac neutrino mass matrix can explain all five (two squared-mass differences and three mixing angles) experimental results in the neutrino sector with particular predictions for the absolute values of the neutrino masses ($m_1=4.81\times 10^{-2},~m_2=4.89\times 10^{-2}~{\rm and}~m_3=-1.198\times 10^{-3}$ eV) and CP violating phase $\delta=270^{0}$. We have shown that the 4 parameters phenomenological mass matrix corresponds to a symmetry under interchange of $\nu_{\mu R} \leftrightarrow -i\nu_{\tau R}$ followed by a CP transformation  with a tiny violation of this symmetry to accommodate a value of the phase $\delta=272.6^{0}$ as required by the mass matrix in Eq. (6). \\

%%%%%%%%%%%%%%%%%%%%%%%%%%%%%%%%%%%%%%%%%%%%%%%%%%%%%%%%%%%%%%%%%%%%%%

\chapter{Warm Dark Matter in two higgs doublet models}\label{chap:chap7}

\section{Introduction}

One of the simplest extensions of the Standard Model is the addition of a second Higgs doublet to its spectrum.  A second Higgs doublet appears naturally in a variety of well motivated scenarios that go beyond the Standard Model.  These include supersymmetric models \cite{susy},  left-right symmetric models \cite{lr}, axion models \cite{dfsz} and models of spontaneous CP violation \cite{lee}, to name a few. These models have the potential for rich phenomenology that may be subject to tests at colliders and in low energy experiments.  A notable feature of these models is the presence of additional scalar states, two neutral and one charged, which may be accessible experimentally at the LHC.  Naturally, two Higgs doublet models have been extensively studied in the literature \cite{review}.

In this chapter we focus on certain cosmological and astrophysical aspects of the two Higgs doublet models in a regime that has not been previously considered. It is well known that no particle in the Standard Model can fit the observed properties of the dark matter in the universe inferred from astrophysical and cosmological data.  New particles are postulated to fulfill this role.  Two Higgs doublet models do contain a candidate for dark matter in one of its neutral scalar bosons.  It is generally assumed that this particle, which is stable on cosmological time scales owing to an approximate (or exact) symmetry, is a {\it cold} dark matter candidate with masses in the several 100  GeV range\cite{ma,inert}.  These particles annihilates into lighter Standard Model particles in the early thermal history of the universe with cross sections of order picobarn. In this work we show that there is an alternative possibility where the extra neutral scalar boson of these models can have mass of the order of a keV and be identified as a {\it warm} dark matter candidate. This scenario is completely consistent with known observations and would have distinct signatures at colliders as well as in cosmology and astrophysics, which we outline here.

The $\Lambda$CDM cosmological paradigm, which assumes a significant cold dark matter component along with a dark energy component in the energy density of the universe, has been immensely successful in confronting cosmological and astrophysical data over a wide range of distance scales, of order Gpc to about 10 Mpc.  However, at distance scales below a Mpc, cold dark matter, which has negligible free--streaming velocity, appears to show some inconsistencies. There is a shortage in the number of galactic satellites observed compared to CDM $N$--body simulations;
density profiles of galactic dark matter haloes are too centrally concentrated in simulations
compared to data; and the central density profile of dwarf galaxies are observed to be shallower than predicted by CDM \cite{CDM}. These problems can be remedied if the dark matter is {\it warm}~\cite{WDM}, rather than cold.  Warm dark matter (WDM) has non-negligible free--streaming velocity, and is able to wipe out structures at distance scales below a Mpc, while behaving very much like CDM at larger distance scales.  This would alleviate the small scale problems of CDM, while preserving its
success at larger distance scales. The free streaming length of warm dark matter can be written down very roughly as \cite{kusenko}
\begin{equation}
R_{fs} \approx 1 \,{\rm Mpc} \left({{\rm keV} \over m_\sigma}\right) \left({\langle p_\sigma\rangle \over 3.15 T}\right)_{T \approx {\rm keV}}~,
\end{equation}
where $m_\sigma$ is the dark matter mass and $\langle p_\sigma \rangle$ its average momentum.  For a fully thermalized
WDM, $\langle p_\sigma \rangle = 3.15 T$.  In the WDM of two Higgs doublet model, as we shall see later, $\langle
p_\sigma \rangle/(3.15 T) \simeq 0.18$, so that an effective thermal mass of $\sigma$, about six times larger than
$m_\sigma$ can be defined corresponding to fully thermalized momentum distribution.  For $m_\sigma$ of order few keV, we see that the free--streaming length is of order
Mpc, as required for solving the CDM small scale problems.  Note that structures at larger scales would not be
significantly effected, and thus WDM scenario would preserve the success of CDM at large scales.

The WDM candidate of two Higgs doublet extensions of the Standard Model is a neutral scalar, $\sigma$,  which can have a mass
of order keV.  Such a particle, which remains in thermal equilibrium in the early universe
down to temperatures of order 150 MeV through weak interaction
processes (see below), would contribute too much to the energy density of the universe, by about a factor of 34
(for $m_\sigma = 1$ keV).  This unpleasant situation is remedied by the late decay of a particle that dumps entropy into other species and heats up the photons relative to $\sigma$.  A natural candidate for such a late decay is a right-handed neutrino $N$ that takes part in neutrino mass generation via the seesaw mechanism.  We find that for $M_N = (25 ~{\rm GeV}- 20~{\rm TeV})$,
and $\tau_N = (10^{-4} -1)$ sec. for the mass and lifetime of $N$, consistency with dark matter abundance can be realized.  Novel signals for collider experiments as well as for cosmology and astrophysics for this scenario are outlined.  In particular, by introducing a tiny breaking of a $Z_2$ symmetry that acts on the second Higgs doublet and makes the dark matter stable, the decay $\sigma \rightarrow \gamma \gamma$ can occur with a lifetime longer than the age of the universe.
This can explain the recently reported anomaly in the $X$-ray spectrum from extra-galactic sources, if $m_\sigma = 7.1$ keV
is adopted, which is compatible with other WDM requirements. This feature is somewhat analogous to the proposal of Ref. \cite{bm} where a SM singlet scalar which coupled
very feebly with the SM sector played the role of the 7.1 keV particle decaying into two photons. The present model with $\sigma$ belonging to a Higgs doublet has an entirely different cosmological history; in particular $\sigma$ interacts with the weak gauge bosons with a coupling strength of $g^2 \sim {\cal O} (1)$ and remains in thermal equilibrium in the early universe
down to $T \approx 150$ MeV, while the singlet scalar of Ref. \cite{bm} was never thermalized.

The rest of the chapter is organized as follows.  In Sec. 2 we describe the two Higgs doublet model for warm dark matter.
Here we also study the experimental constraints on the model parameter.  In Sec. 3 we derive the freeze-out temperature of
the WDM particle $\sigma$ and compute its relic abundance including the late decays of $N$.  Here we show the
full consistency of the framework.  In Sec. 4 we analyze some other implications of the model.  These include supernova energy loss, dark matter self interactions, 7.1 keV $X$-ray anomaly, and collider signals of the model.  Finally in Sec. 5 we conclude.

\section{Two Higgs Doublet Model for Warm Dark Matter}

The model we study is a specific realization of two Higgs doublet models that have been widely studied in the context
of dark matter \cite{ma,inert}.  The two Higgs doublet fields are denoted as $\phi_1$ and $\phi_2$.  A discrete
$Z_2$ symmetry acts on $\phi_2$ and not on any other field.  This $Z_2$ prevents any Yukawa couplings of $\phi_2$.
While $\phi_1$ acquires a vacuum expectation value $v \simeq 174$ GeV, $\langle\phi_2^0\rangle = 0$, so that the $Z_2$ symmetry remains unbroken.  The lightest member of the $\phi_2$ doublet will then be stable.  We shall identify one of the neutral
members of $\phi_2$ as the WDM $\sigma$ with a mass of order keV.

Neutrino masses are generated via the seesaw mechanism.  Three $Z_2$ even singlet neutrinos, $N_i$, are introduced.
The Yukawa Lagrangian of the model is
\begin{equation}
{\cal L}_{\rm Yuk} = {\cal L}_{\rm Yuk}^{\rm SM} + (Y_N)_{ij} \ell_i N_j \,\phi_1 + \frac{M_{N_i}}{2}N_i^T C N_i + h.c.
\end{equation}
Here ${\cal L}_{\rm Yuk}^{\rm SM}$ is the SM Yukawa coupling Lagrangian and involves only the $\phi_1$ field owing to the
$\phi_2 \rightarrow -\phi_2$ reflection ($Z_2$) symmetry.  The Higgs potential of the model is
\begin{eqnarray}
V &=& -m_1^2 |\phi_1|^2 + m_2^2 |\phi_2|^2 + \lambda_1 |\phi_1|^4 + \lambda_2 |\phi_2|^4 + \lambda_3 |\phi_1|^2 |\phi_2|^2
\nonumber \\
&+& \lambda_4 |\phi_1^\dagger \phi_2|^2 + \{\frac{\lambda_5}{2}(\phi_1^\dagger \phi_2)^2 + h.c.\}.
\end{eqnarray}
With $\langle\phi_1^0\rangle = v \simeq 174$ GeV and $\langle \phi_2^0 \rangle = 0$, the masses of the various fields
are obtained as
\begin{eqnarray}
m_h^2 &=& 4 \lambda_1 v^2,~~~ m_\sigma^2 = m_2^2 + (\lambda_3+ \lambda_4 +\lambda_5)\, v^2; \nonumber \\
m_A^2 &=& m_2^2 + (\lambda_3 + \lambda_4 - \lambda_5)\, v^2;~~~ m_{H^\pm}^2 = m_2^2 + \lambda_3 v^2~.
\label{masses}
\end{eqnarray}
Here $h$ is the SM Higgs boson with a mass of 126 GeV;  $\sigma$ and $A$ are the second scalar and
pseudoscalar fields, while $H^\pm$ are the charged scalars.  We wish to identify $\sigma$ as the keV warm dark matter
candidate.\footnote{Alternatively, $A$ can be identified as the WDM candidate. With some redefinitions of couplings, this
scenario would lead to identical phenomenology as in the case of $\sigma$ WDM.}
In order to go from 100 GeV to a few keV for the mass of $\sigma$, some fine-tuning
will have to be done. While this may be viewed as not natural, we note that in any
non-supersymmetric model of this type, such fine-tunings are needed to protect the
scalar masses from quadratic divergences.  This is true in the inert doublet model
with cold dark matter as well.
An immediate concern is whether the other
scalars can all be made heavy, of order 100 GeV or above, to be consistent with experimental data.  This can indeed be done,
as can be seen from Eq. (\ref{masses}). Note that $m_A^2 = m_\sigma^2 - 2 \lambda_5 v^2$ and $m_{H^\pm}^2 = m_\sigma^2
 - (\lambda_4+\lambda_5)v^2$, so that even for $m_\sigma \sim$ keV, $m_A$ and $m_{H^\pm}$ can be large.
 However, the masses of $A$ and $H^\pm$ cannot be taken to arbitrary large
values, since  $\lambda_i v^2$ are at most of order a few hundred (GeV)$^2$ for perturbative values of $\lambda_i$.
The boundedness conditions on the Higgs potential can all be satisfied \cite{review} with the choice of positive $\lambda_{1,2,3}$ and negative $\lambda_5$ and $(\lambda_{4}+\lambda_5)$.
The keV WDM version of the two Higgs doublet model would thus predict that the neutral scalar $A$ and the charged scalar $H^\pm$ have masses not more than a few hundred GeV.  The present limits on the masses of $A$ and $H^\pm$ are approximately
$m_A > 90$ GeV (from $Z$ boson decays into $\sigma + A$) and $m_{H^\pm} > 100$ GeV from LEP searches for charged scalars.
This would mean that $|\lambda_5| > 0.13$ and $|\lambda_4+\lambda_5| > 0.17$.

\subsection{Electroweak precision data and Higgs decay constraints}

The precision electroweak parameter $T$ receives an additional contribution from the second Higgs doublet, which is given by\cite{inert}
\begin{equation}
\Delta T = \frac{m^2_{H^\pm}}{32 \pi^2 \alpha v^2} \left[1 - \frac{m_A^2}{m^2_{H^\pm}-m_A^2} {\rm log}\frac{m^2_{H^\pm}}{m_A^2}
\right]
\end{equation}
where the mass of $\sigma$ has been neglected.  For $\{m_{H^\pm}, \,m_A\} = \{150, \, 200\}$ GeV, $\Delta T \simeq -0.095$
while for $\{m_{H^\pm}, \,m_A\} = \{200, \, 150\}$ GeV, $\Delta T \simeq +0.139$. Both these numbers are consistent with current precision electroweak data constraints, $T = 0.01 \pm 0.12$ \cite{pdg}.  Note, however, that the mass splitting between $H^\pm$ and $A$ cannot be too much, or else the limits on $T$ will be violated. For example, if $\{m_{H^\pm}, \,m_A\} = \{150, \, 300\}$ GeV,
$\Delta T \simeq -0.255$, which may be disfavored.

The parameter $S$ receives a new contribution from the second Higgs doublet, which is evaluated to be
\begin{equation}
\Delta S = \frac{1}{12 \pi}\left({\rm log} \frac{m_A^2} {\,m^2_{H^\pm}} - \frac{5}{6} \right)~.
\end{equation}
If $\{m_{H^\pm}, \,m_A\} = \{150, \, 200\}$ GeV, $\Delta S \simeq +0.025$, while for  $\{m_{H^\pm}, \,m_A\} = \{200, \, 150\}$ GeV, $\Delta S \simeq -0.007$.  These values are consistent with precision electroweak data which has $S = -0.03 \pm 0.10$
\cite{pdg}.

In this model the decay $h \rightarrow \sigma \sigma$ can occur proportional to the quartic coupling combination
$(\lambda_3 + \lambda_4 + \lambda_5)$.  The decay rate is given by
\begin{equation}
\Gamma (h \rightarrow \sigma \sigma) = \frac{|\lambda_3+\lambda_4 + \lambda_5|^2}{16 \pi} \frac{v^2}{m_h}~.
\end{equation}
Since the invisible decay of the SM Higgs should have a branching ratio less than 23\% \cite{gunion}, we obtain the limit
(using $\Gamma = 4.2 \pm 0.08$ MeV for the SM Higgs width)
\begin{equation}
|\lambda_3 + \lambda_4+\lambda_5| < 1.4 \times 10^{-2}~.
\label{fine-tune}
\end{equation}
The cubic scalar coupling $h \sigma \sigma$ can also arise through loops mediated by gauge bosons.
The dominant such contribution is from a $W^+ W^-$ loop, which has
an amplitude of order $g^4 v/(16 \pi^2)$, which should be compared with the tree-level
amplitude of $|\lambda_3 + \lambda_4 + \lambda_5| v$.  As long as $|\lambda_3 + \lambda_4 + \lambda_5|
> 10^{-3}$, the tree level contribution will dominate.
There is a mini fine-tuning needed to realize the constraint quoted in Eq. (\ref{fine-tune}),
since $|\lambda_5| > 0.13$ and $|\lambda_4+\lambda_5| > 0.17$ are required to meet
the constraints on the masses of $A$ and $H^\pm$.  This tuning is at the level of
10\%, which is quite stable under radiative corrections.
We thus see broad agreement with all experimental constraints in the two Higgs doublet models with a keV neutral scalar
identified as warm dark matter.

\subsection{Late decay of right-handed neutrino \boldmath{$N$}}

\label{latedecay}

Before proceeding to discuss the early universe cosmology within the two Higgs doublet model with warm dark matter, let us identify the parameter space of the model where the late decay of a particle occurs with a lifetime in the range of $(10^{-4}-1)$ sec. Such a decay is necessary in order to dilute the warm dark matter abundance within the model, which would otherwise be too large. A natural candidate for such late decays is one of the heavy right-handed neutrinos, $N$, that participates in the seesaw mechanism for small neutrino mass generation.  If its lifetime were longer than 1 sec. that
would affect adversely the highly successful big bang nucleosynthesis scheme. Lifetime shorter than $10^{-4}$ sec. would not
lead to efficient reheating of radiation in the present model, as that would also reheat the warm dark matter field.

It turns out that the masses and couplings of the late--decaying field $N$ are such that its contribution to the
light neutrino mass is negligibly small.  The smallest neutrino mass being essentially zero can be taken as one of the
predictions of the present model.  We can therefore focus on the mixing of this nearly decoupled $N$ field with light
neutrinos.  For simplicity we shall assume mixing of $N$ with one flavor of light neutrino, denoted simply as $\nu$.
The mass matrix of the $\nu-N$ system is then given by
\begin{eqnarray}
M_\nu = \left(\begin{matrix} 0 & Y_N v \\ Y_N v & M_N\end{matrix}    \right)~.
\label{seesaw}
\end{eqnarray}
A light--heavy neutrino mixing angle can be defined from Eq. (\ref{seesaw}):
\begin{equation}
\sin\theta_{\nu N} \simeq \frac{Y v}{M_N}~.
\label{angle}
\end{equation}
This mixing angle will determine the lifetime of $N$.

If kinematically allowed, $N$ would decay into $ h \nu,\, h \overline{\nu},\, W^+ e^-,\, W^- e^+,\, Z\nu$ and
$Z \overline{\nu}$.  These decays arise through the $\nu-N$ mixing.  The total two body decay rate of $N$ is given by
\begin{equation}
\Gamma(N \rightarrow h \nu,\, h \overline{\nu},\, W^+ e^-,\, W^- e^+,\, Z\nu,\;Z \overline{\nu}) = ~~~~~~~~~~~~~~~~~~~~~~~~~~~~~~~~~~~~~~~~~~~~`
\nonumber
\end{equation}
\begin{equation}
\frac{Y_N^2 M_N} {32 \pi} \left[\left(1-\frac{m_h^2}{M_N^2}\right)^2 + 2\left(1-\frac{m_W^2}{M_N^2}\right)^2\left(1+  \frac{2 m_W^2}{M_N^2}\right)
+ \left(1-\frac{m_Z^2}{M_N^2}\right)^2 \left(1+ \frac{2m_Z^2}{M_N^2}\right) \right] ~.
\label{2body}
\end{equation}

\noindent Here the first term inside the square bracket arises from the decays $N \rightarrow h \nu$ and $N \rightarrow h \overline{\nu}$, the second term from decays of $N$ into $W^\pm e^\mp$ and the last term from $N$ decays into
$Z \nu$ and $Z \overline{\nu}$.  We have made use of the expression for the mixing angle given in Eq. (\ref{angle}),
which is assumed to be small.

When the mass of $N$ is smaller than 80 GeV, these two body decays are kinematically not allowed.  In this case,
three body decays involving virtual $W$ and $Z$ will be dominant.  The total decay rate for $N$ decaying into three body final
states through the exchange of the $W$ boson is given by
\begin{equation}
\Gamma(N \rightarrow 3~{\rm body}) = \frac{G_F^2 M_N^5}{192 \pi^3} \sin^2\theta_{\nu N}\left(1 + \frac{3}{5} \frac{M_N^2}{m_W^2}\right) (2) \left[5 + 3 F\left(\frac{m_c^2}{M_N^2}\right) + F\left(\frac{m_\tau^2}{M_N^2}\right) \right]~.
\label{3body}
\end{equation}
This expression is analogous to the standard muon decay rate.  An overall factor of 2 appears here since $N$ being Majorana
fermion decays into conjugate channels. The factor 5 inside the square bracket accounts for the virtual $W^+$ boson decaying
into $e^+ \nu_e,\, \mu^+ \nu_\mu$ and $\overline{d} u$ for which the kinematic function $F(x) = \{1- 8 x + 8 x^3 - x^4-12 x^2\, {\rm ln} x\}$ is close to one \cite{kuno}.  For $M_N > 175$ GeV, an additional piece, $3F(m_t^2/M_N^2)$, should
be included inside the square bracket of Eq. (\ref{3body}). Analogous expressions for three body decay of $N$ via virtual $Z$ boson are found
to be numerically less important (about 10\% of the virtual $W$ contributions) and we ignore them here.  Virtual Higgs boson
exchange for three body $N$ decays are negligible owing to small Yukawa coupling suppressions.  We shall utilize expressions
(\ref{2body}) and (\ref{3body}) in the next section where the relic density of $\sigma$ WDM is computed.

\section{Relic Abundance of Warm Dark Matter \boldmath{$\sigma$} }

Here we present a calculation of the relic abundance of $\sigma$ which is taken to have a mass of order keV, and  which serves
as warm dark matter of the universe.  Since $\sigma$ has thermal abundance, it turns out that relic abundance
today is too large compared to observations.  This situation is remedied in the model by the late decay of
$N$, the right--handed neutrino present in the seesaw sector.  To see consistency of such a scheme, we should
follow carefully the thermal history of the WDM particle $\sigma$.

When the universe was hot, at temperatures above the $W$ boson mass, $\sigma$ was in thermal equilibrium via
its weak interactions through scattering processes such as $W^+ W^- \rightarrow \sigma \sigma$. As temperature
dropped below the $W$ boson mass, such processes became rare, since the number density of $W$ boson got depleted.
The cross section for the process $W^+ W^-\rightarrow \sigma+ \sigma$ is given by
\begin{equation}
\sigma(W^+ W^- \rightarrow \sigma+ \sigma) \simeq \left(\frac{g^4}{64 \pi}\right)\frac{1}{m_W^2}~.
\end{equation}
The interaction rate $\langle \sigma n v\rangle$ is then given by
\begin{equation}
\langle n \sigma v \rangle \approx \left(\frac{g^4}{64 \pi }\right)\frac{1}{m_W^2} T^3 \left(\frac{m_W}{T}\right)^{3} e^{-2m_W/T}
\end{equation}
where the Boltzmann suppression factor in number density of $W$s appears explicitly.  Demanding this interaction rate
to be below the Hubble expansion rate at temperature $T$, given by $H(T) = 1.66 g_*^{1/2} T^2/M_P$, with
$g_*$ being the effective degrees of freedom at $T$ and $M_P = 1.19 \times 10^{19}$ GeV, we obtain the
freeze--out temperature for this process to be $T_f \simeq 4.5$ GeV (with $g_* \approx 80$ used).

$\sigma$ may remain in thermal equilibrium through other processes.  The scattering $b \bar{b}  \rightarrow \sigma \sigma$
mediated by the Higgs boson $h$ of mass 126 GeV is worth considering.  ($b$ quark has the largest Yukawa coupling among
light fermions.)  The cross section for this process at energies below the $b$-quark mass is given by
\begin{equation}
\sigma (b \bar{b}  \rightarrow \sigma \sigma) \simeq \frac{|\lambda_3+\lambda_4+\lambda_5|^2 m_b^2}{4 \pi m_h^4}~.
\end{equation}
If $|\lambda_3+\lambda_4+\lambda_5| = 10^{-2}$, this process will freeze out at $T_f \approx 240$ MeV ($g_* = 70$ is
used in this estimate, along with Boltzmann suppression.)  For smaller values of $|\lambda_3+\lambda_4+\lambda_5|$, the freeze--out temperature will be higher.

The process $\mu^+ \mu^- \rightarrow \sigma \sigma$ mediated by the Higgs boson $h$ can potentially
keep $\sigma$ in thermal equilibrium down to lower temperatures,
since the $\mu^\pm$ abundance is not Boltzmann suppressed.  (Note however, that this process suffers from a stronger chiral suppression
compared to the process $b \bar{b} \rightarrow \sigma \sigma$.)  The cross section is given by
\begin{equation}
\sigma(\mu^+ \mu^- \rightarrow \sigma \sigma) = \frac{|\lambda_3+\lambda_4+\lambda_5|^2}{64\pi} \frac{m_\mu^2}{m_h^4}~.
\label{muonEq}
\end{equation}
The number density of $\mu^\pm$, which are in equilibrium, is given by $0.2 T^3$, from which we find that this process
would go out of thermal equilibrium at $T \approx 250$ MeV for $|\lambda_3+\lambda_4+\lambda_5|= 10^{-2}$.  This process
could freeze out at higher temperatures for smaller values of $|\lambda_3+\lambda_4+\lambda_5|$.

\begin{center}
\begin{figure}
\hspace*{0.8in}
\includegraphics[width = 2.5in]{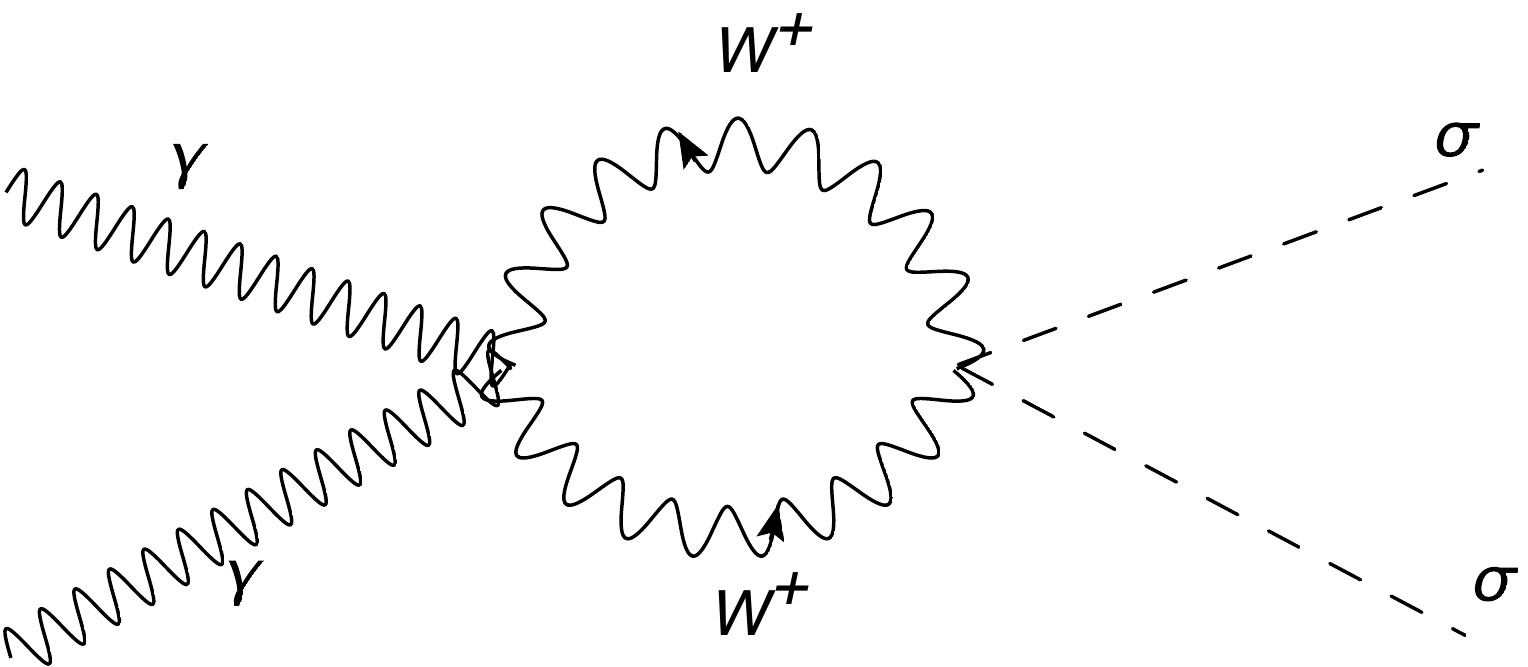}
\includegraphics[width = 2.5in]{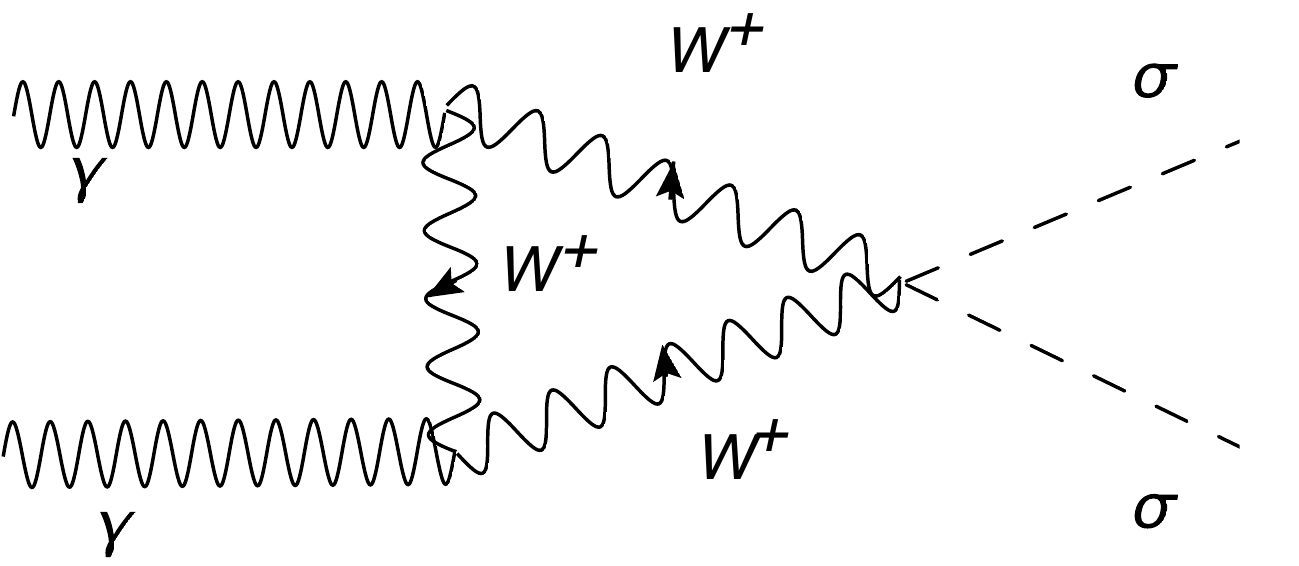}
\vspace*{0.05in}
\caption{Loop diagrams leading to $\gamma \gamma \rightarrow \sigma \sigma$.}
\label{loop}
\end{figure}
\end{center}

There is one process which remains in thermal equilibrium independent of the values of the Higgs quartic couplings.
This is the scattering $\gamma \gamma \rightarrow \sigma \sigma$ mediated by the $W^\pm$ gauge bosons shown in Fig. \ref{loop}.  The relevant couplings are all fixed, so that the cross section has no free parameters.  We find it to be
\begin{equation}
\sigma(\gamma \gamma \rightarrow \sigma \sigma) = \frac{E_\sigma^2 F_W^2}{64 \pi} \left[\frac{e^2 g^2}{32 \pi^2 m_W^2}\right]^2
\label{prod}
\end{equation}
where $F_W = 7$ is a loop function. Using $E_\sigma = 3.15 T$ and $n_\gamma = 0.2 T^3$, the interaction rate $\langle
\sigma n v\rangle$ can be computed.  Setting this rate to be equal to the Hubble expansion rate, we find that this process
freezes out at $T \approx 150$ MeV (with $g_* = 17.25$ appropriate for this temperature used).  Among all scattering
processes, this one keeps $\sigma$ to the lowest temperature, and thus the freeze-out of $\sigma$ occurs at
$T_{f, \sigma} \approx 150$ MeV with a corresponding $g_*^\sigma = 17.25$.

Having determined the freeze--out temperature of $\sigma$ to be $T_f^\sigma \approx 150$ MeV, we can now proceed to
compute the relic abundance of $\sigma$.  We define the abundance of $\sigma$ as
\begin{equation}
Y_\sigma = \frac{n_\sigma}{s}
\end{equation}
where $n_\sigma$ is the number density of $\sigma$ and $s$ is the entropy density.  These two quantities are given
for relativistic species to be
\begin{equation}
n_\sigma = \frac{g_\sigma \zeta(3)}{\pi^2} T^3,~~~s = \frac{2 \pi^4}{45}g_{\rm efff} T^3,
\end{equation}
where
\begin{equation}
g_{\rm efff} = \sum_{\rm bosons}g_b + \frac{7}{8} \sum_{\rm fermions} g_f~.
\end{equation}
Thus
\begin{equation}
Y_\sigma = \frac{45 \zeta(3)}{2\pi^4} \frac{g_\sigma}{g_{\rm eff}}~.
\end{equation}
Since $Y_\sigma$ is a thermally conserved quantity as the universe cools, we can obtain the abundance of $\sigma$ today as
\begin{equation}
\Omega_\sigma = Y_\sigma m_\sigma \frac{s_0}{\rho_c},
\end{equation}
where $s_0 = 2889.2/{\rm cm}^3$ is the present entropy density and $\rho_c = 1.05368 \times 10^{-5} h^2 \,{\rm GeV}/{\rm cm}^3$
is the critical density.  Using $g_\sigma = 1$ appropriate for a real scalar field and with $h = 0.7$ we thus obtain
\begin{equation}
\Omega_\sigma = 9.02 \left(\frac{17.25}{\rm g_{\rm eff}}\right) \left(\frac{m_\sigma}{1 \, {\rm keV}}\right)~.
\label{abund0}
\end{equation}
Here we have normalized $g_{\rm eff} = 17.25$, appropriate for the freeze--out temperature of $\sigma$.  We see
from Eq. (\ref{abund0}) that for a keV warm dark matter, $\Omega_\sigma$ is a factor of 34 larger than the observed
value of $0.265$.  For a clear discussion of the relic abundance in a different context see Ref. \cite{zhang}.

\subsection{Dilution of \boldmath{$\sigma$} abundance via late decay of \boldmath{$N$}}

The decay of $N$ involved in the seesaw mechanism, as discussed in Sec. \ref{latedecay}, can dilute the abundance
of $\sigma$ and make the scenario consistent.  We assume that at very high temperature $N$ was in thermal equilibrium.
This requires going beyond the model described in Sec. 2.
This could happen in a variety of ways.\footnote{Late decays of heavy particles have been used in order to dilute
dark matter abundance in other contexts \cite{lindner,zhang}.}  For example, one could have inflaton field $S$ couple to $N$ via a Yukawa coupling of type $SNN$
which would then produce enough $N$'s in the process of reheating after inflation~\cite{shafi}. Alternatively, the two Higgs extension of SM model could be an effective low energy theory which at high energies could have a local $B-L$ symmetry.
%which results in an effective 2HDM at low energies after symmetry breaking.
% under which the $\phi_{1,2}$ are neutral. We will assume in this case that the $B-L$ symmetry is spontaneously broken at very %high scale so that the low energy theory is the 2HDM model.
The $B-L$ gauge interactions would keep $N$ in thermal equilibrium down to temperatures a few times
below the gauge boson mass, at which point $N$ freezes out.  As the universe cools, the Hubble expansion rate also slows down.
The two body and three body decays of $N$, given in Eqs. (\ref{2body})-(\ref{3body}), will come into equilibrium at some
temperature at which time $N$ would begin to decay.  If this temperature $T_d$ is in the range of 150 MeV to 1 MeV, the decay
products (electron, muon, neutrinos, up quark and down quark) will gain entropy as do the photons which are in thermal
equilibrium with these species.\footnote{At $T= 150$ MeV, it is not completely clear if we
should include the light quark degrees of freedom or the hadronic degrees.  We have kept the $u$ and $d$ quarks in our
decay rate evaluations.}  Since $\sigma$ froze out at $T \approx 150$ MeV, and since $\sigma$ is not a decay product
of $N$, the decay of $N$ will cause the temperature of photons to increase relative to that of $\sigma$.  Thus a dilution
in the abundance of $\sigma$ is realized.  Note that the decay temperature $T_d$ should be above one MeV, so that big bang
nucleosynthesis is not affected.  The desired range for the lifetime of $N$ is thus $\tau_N = (10^{-4} - 1)$ sec.

The reheat temperature $T_r$ of the thermal plasma due to the decays of $N$ is given by \cite{turner}
\begin{equation}
T_r = 0.78 [g_*(T_r)]^{-1/4} \sqrt{\Gamma_N M_P}~.
\end{equation}
Energy conservation then implies the relation
\begin{equation}
M_N Y_N s_{\rm before} = \frac{3}{4} s_{\rm after} T_r~.
\end{equation}
If the final state particles are relativistic, as in our case, a dilution factor defined as
\begin{equation}
d=\frac{s_{\rm before}}{s_{\rm after}}
\end{equation}
takes the form
\begin{equation}
d = 0.58\, [g_*(T_r)]^{-1/4} \sqrt{\Gamma_N M_P}/(M_N Y_N)~.
\end{equation}
The abundance of $N$ is given by
\begin{equation}
Y_N = \frac{135}{4\pi^4} \frac{\zeta(3)}{g(T_{f,N})},
\end{equation}
where $g(T_{f,N})$ stands for the degrees of freedom at $N$ freeze--out.  Putting all these together we
obtain the final abundance of $\sigma$ as
\begin{equation}
\Omega_\sigma = (0.265) \left(\frac{m_\sigma}{1 \, {\rm keV}}\right) \left(\frac{7.87 \, {\rm GeV}}{M_N} \right)
\left(\frac{1\, {\rm sec.}}{\tau_N} \right)^{1/2} \left(\frac{g(T_{f,N})}{106.75}\right)\left(\frac{17.25}{g_f^\sigma} \right).
\label{abundance}
\end{equation}
Here we have normalized various parameters to their likely central values and used $g_*(T_r) = 10.75$.  The value of
$g(T_{f,N}) = 106.75$ counts all SM degrees and nothing else.

\begin{center}
\begin{figure}
\includegraphics[width=12cm]{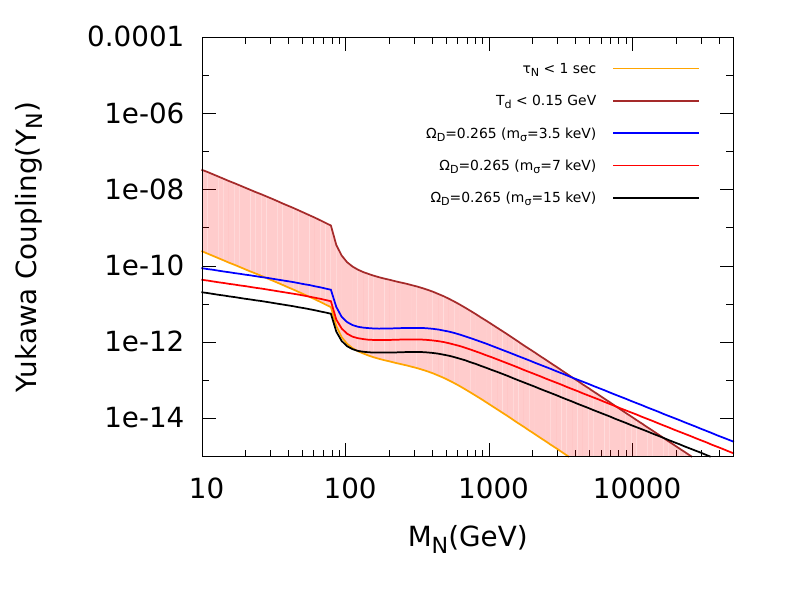}
\caption{Allowed parameter space of the model in the $M_N-Y_N$ plane. The shaded region corresponds to
the decay temperature $T_d$ of $N$ lying in the range 150 MeV -- 1 MeV.  The three solid curves generate
the correct dark matter density $\Omega_D$ for three different values of the WDM mass $m_\sigma = \{3.5,\, 7,\, 15\}$ keV.}
\label{plot}
\end{figure}
\end{center}

From Eq. (\ref{abundance}) we see that the correct relic abundance of $\sigma$ can be obtained for $M_N \sim 10$ GeV
and $\tau_N \sim 1$ sec.  In Fig. \ref{plot} we have plotted the dark matter abundance as a function of $M_N$ and its Yukawa coupling $Y_N$
for three different values of $m_\sigma$ (3.5, 7 and 15 keV).  Also shown in the figure are the allowed band for
$\tau_N$ to lie in the range of $(10^{-4} - 1)$ sec., or equivalently for $T_d = (150-1)$ MeV.  We see that there is a significant region allowed by the model
parameters.  We also note that the mass of $N$ should lie in the range $M_N = 25 \,{\rm GeV} - 20 \, {\rm TeV}$
for the correct abundance of dark matter.

A remark on the average momentum $\langle p_\sigma\rangle$ of the dark matter is in order.  The dilution factor
$d \simeq 1/34$ for $m_\sigma = 1$ keV.  The temperature of $\sigma$ is thus cooler by a factor of $1/(34)^{1/3} =
0.31$ relative to the photon.  The momentum of $\sigma$ gets redshifted by a factor $\xi^{-1/3} = 0.58$ where
$\xi = g_f^\sigma/g_{\rm today} = 17.25/3.36$.  The net effect is to make $\langle p_\sigma \rangle/(3.15 T) = 0.18$.

\section{Other Implications of the Model}

In this section we discuss briefly some of the other implications of the model.

\subsection{Supernova energy loss}

The process $\gamma \gamma \rightarrow \sigma \sigma$ can lead to the production of $\sigma$ inside supernova core.
Once produced these particles will freely escape, thus contributing to new channels of supernova energy loss.
Note that $\sigma$ does not have interactions with the light fermions. The cross section for $\sigma$ production is
given in Eq. (\ref{prod}).  Here we make a rough estimate of the energy lost via this process and ensure that this
is not the dominant cooling mechanism of supernovae. We follow the steps of Ref. \cite{bm} here.
The rate of energy loss is given by
\begin{equation}
Q = V_{\rm core} n_\gamma^2 \langle E \rangle \sigma(\gamma \gamma \rightarrow \sigma \sigma)
\end{equation}
where $V_{\rm core} = 4\pi R_{\rm core}^3/3$ is the core volume and we take $R_{\rm core} = 10$ km.
$n_\gamma \simeq 0.2 T_\gamma^3$ is the photon number density, and $\langle E \rangle = 3.15 T_\gamma$ is the
average energy of the photon.  Using Eq. (\ref{prod}) for the cross section we obtain $Q \sim 2.8 \times 10^{51}$ erg/sec,
when $T_\gamma = 30$ MeV is used.  Since the supernova explosion from 1987A lasted for about 10 seconds, the total
energy loss in $\sigma$ would be about $2.8 \times 10^{52}$ erg, which is to be compared with the total energy loss
of about $10^{53}$ erg.  This crude estimate suggests that energy loss in the new channel is not excessive.
We should note that the energy loss scales as the ninth power of core temperature, so for larger values of
$T_\gamma$, this process could be significant. A more detailed study of this problem would be desirable.

\subsection{Dark matter self interaction}

In our model dark matter self interaction, $\sigma \sigma \rightarrow \sigma \sigma$, occurs proportional to
$|\lambda_2|^2$.  There are
rather severe constraints on self-interaction of dark matter from dense cores of galaxies and
galaxy clusters where the velocity distribution can be isotropized. Constraints from such
halo shapes, as well as from dynamics of bullet cluster merger have been used to infer an
upper limit on the dark matter self-interaction cross section \cite{manoj}:
\begin{equation}
\frac{\sigma}{m_\sigma} < 1~ {\rm barn}/{\rm GeV}~.
\end{equation}
The self interaction cross section in the model is given by
\begin{equation}
\sigma (\sigma \sigma \rightarrow \sigma \sigma) = \frac{9 \hat{\lambda}_2^2}{8 \pi s}~
\end{equation}
where $\hat{\lambda_2} = \lambda_2 - |\lambda_3+\lambda_4+\lambda_5|^2 \,(v^2/m_h^2)$, with the
second term arising from integrating out the SM Higgs field $h$.
This leads to a limit on the coupling $\hat{\lambda}_2$ given by
\begin{equation}
\hat{\lambda}_2 < 5.4 \times 10^{-6} \left(\frac{m_\sigma}{10\,{\rm keV}} \right)^{3/2}~.
\end{equation}
The one loop corrections to $\sigma$ self interaction strength is of order $g^4/(16 \pi^2)\sim 10^{-3}$. So we use the tree level $\lambda_2$ to cancel this to make the effective self interaction strength of order $10^{-6}$ as needed.
Such a fine-tuning is unpleasant, but nevertheless can be done within the model consistently.
One can choose $|\lambda_3+\lambda_4+\lambda_5| \sim 10^{-3}$, so that the effective quartic coupling $\hat{\lambda_2}$
is positive.

\subsection{The extra-galactic $X$-ray anomaly}

Recently two independent groups have reported the observation of a peak in the extra-galactic
$X$-ray spectrum at 3.55 keV \cite{Xray1,Xray2}, which appear to be not understood in terms of known physics and
astrophysics. While these claims still have to be confirmed by other observations, it is tempting to speculate
that they arise from the decay of WDM into two photons.  If the $Z_2$ symmetry remains unbroken, $\sigma$ is
absolutely stable in our model and will not explain this anomaly.  
(The cross section for $\sigma \sigma \rightarrow \gamma \gamma$ is
orders of magnitude smaller than required to be relevant for the X-ray anomaly.)
However, extremely tiny breaking of this symmetry via a soft term of the type
$m^2_{12}\phi_1^\dagger \phi_2+h.c.$ can generate the reported signal.  Such a soft breaking term would induce a nonzero
vacuum expectation value for $\sigma$ which we denote as $u$.  Explicitly, the relevant potential for the $\sigma$ field will be
\begin{equation}
V(\sigma) = \frac{m^2_\sigma}{2} \sigma^2 + \sqrt{2} m^2_{12} v \sigma + ...
\end{equation}
which minimizes to
\begin{equation}
u = \left\langle \sigma \right \rangle = -\frac{\sqrt{2} m_{12}^2}{m^2_\sigma}\, v~.
\end{equation}
Such an induced VEV is quite stable, since $m_{12} \sim 0.003$ eV and $u \sim 0.03$ eV will be needed, which
are much smaller than $m_\sigma \sim 7$ keV.
In order to explain the $X$-ray anomaly, this VEV has to be in the range
$u = (0.03-0.09)$ eV.  This comes about from the decay rate, which is given by
\begin{equation}
\Gamma(\sigma \rightarrow \gamma \gamma) = \left(\frac{\alpha}{4\pi} \right)^2 F_W^2 \left(\frac{u^2}{v^2} \right)
\frac{G_F m_\sigma^3}{8 \sqrt{2} \pi}
\end{equation}
with $F_W = 7$, which is matched to a partial lifetime in the range $\Gamma^{-1} (\sigma \rightarrow \gamma \gamma) = (4 \times 10^{27} -
4 \times 10^{28})$ sec \cite{Xray1,Xray2}.  Once $\sigma$ develops a vacuum expectation value, it also mixes with
SM Higgs field $h$, but this effect is subleading for the decay $\sigma \rightarrow \gamma \gamma$.  Such mixing
was the main source of the two photon decay of WDM in the case of a singlet scalar WDM of Ref. \cite{bm}.

\subsection{Effective number of neutrinos for BBN}

Since the warm dark matter $\sigma$ is in thermal equilibrium down to temperatures of order 150 MeV, it can modify the
number of neutrino species that affect big bang nucleosynthesis.  A fully thermalized real scalar would count as
4/7 of a neutrino species, but the abundance of $\sigma$ is diluted via the late decay of $N$ in our case.  The dilution
factor is about 1/36 (compare Eq. (23) and (29)), which would mean that the effective number of neutrinos for BBN is
shifted from 3 only by a tiny amount of about 0.02.

\subsection{Collider signals}

The charged scalar $H^\pm$ of the model can be pair produced at the Large Hadron Collider via the Drell-Yan process.
$H^+$ will decay into a  $W^+$ and a $\sigma$. This signal has been analyzed in Ref. \cite{rai} within the context
of a similar model \cite{nandi}.  Sensitivity for these charged scalars would require 300 $fb^{-1}$ of luminosity
of LHC running at 14 TeV.

The pseudoscalar $A$ can be produced in pair with a $\sigma$ via $Z$ boson exchange. $A$ will decay into
a $\sigma$ and a $Z$.  The $Z$ can be tagged by its leptonic decay.  Thus the final states will have two leptons
and missing energy.  The Standard Model $ZZ$ background with the same final states would be much larger.
We can make use of the fact that in the signal events, the $Z$ boson which originates from the decay $A \rightarrow Z \sigma$
with a heavy $A$ and a massless $\sigma$ will be boosted in comparison with the background $Z$ events.
This  will  reflect in the $p_T$ distribution which would be
different for the signal events compared to the SM background $Z$'s.
Studies to look for this kind of signals in this particular framework are in order.

\section{Conclusions}
In this chapter we have proposed a novel warm dark matter candidate in the context of two Higgs doublet extensions of the
Standard Model.  We have shown that a neutral scalar boson of these models can have a mass in the keV range.  The abundance of
such a thermal dark matter is generally much higher than observations; we have proposed a way to dilute this by the late
decay of a heavy right-handed neutrino which takes part in the seesaw mechanism.  A consistent picture emerges where
the mass of $N$ is in the range 25 GeV to 20 TeV.  The model has several testable consequences at colliders as well
as in astrophysical settings.  The charged scalar and the pseudoscalar in the model cannot be much heavier than a few
hundred GeV.  It will be difficult to  see such a warm dark matter candidate in direct detection experiments. The cross
sections for the processes $\sigma e \rightarrow \sigma e$ and $\sigma N \rightarrow \sigma N$ are of order $10^{-49}$ cm$^2$
and $10^{-45}$ cm$^2$ respectively in the model (the expressions for these cross sections are analogous to Eq. (\ref{muonEq})).
Since the warm dark matter has a velocity of $10^{-3}$, the kinetic energy of $\sigma$ today is of order $10^{-2}$ eV, which
would mean that the recoil energy will be well below the detection threshold in ongoing direct detection experiments.
Supernova dynamics may be significantly modified by the production of $\sigma$ pairs in photon--photon
collisions.  The model can also explain the anomalous $X$-ray signal reported by different groups in the extra-galactic
spectrum.

 \chapter{Conclusions and Outlook}\label{chap:conclusions}

%\section{Conclusions and Outlook}
The individual chapters of this thesis contain their own conclusions, so here there is
only a brief overview, and a few comments about the future of the field. Some different 
approaches for new physics beyond the Standard Model have been examined which includes topSU(5) model, left-right mirror symmetry, supergravity, Standard model extensions with neutrino sector, dark matter sector and Higgs sectors. Although the phenomenology of each model is diverse and different in outcome, the particular common aspects addressed in each chapter were the detection of new particles in TeV scales. In all the cases it was shown that the next generation of particle colliders like the high luminosity Large Hadron Collider at 14 TeV and proposed International linear collider will be able to investigate these models.

There can be many possibilities of the nature of these new particles. The study in the chapter 2 on leptoquarks as well as diquark gauge bosons showed that, in the particular framework of TopSU(5) model considered, the discovery mass range extends upto 1.5 TeV at the LHC with center of mass energy of 14 TeV with a luminosity of $100 fb^{-1}$. On the other hand, in chapter 3 we have studied the collider phenomenology of TeV scale mirror fermions in the framework of a particular variant of Left-Right Mirror model in which mirror fermions dominantly decays into the SM fermion and $W/Z$-boson. We find that the reaches for the light mirror quarks can be upto $ ~ 750 $ GeV  at the $14$ TeV LHC with $~300 ~fb^{-1}$ luminosity.
 
One of the most exciting possibilities beyond the Standard Model physics is the well motivated and well studied case of Supersymmetry. In chapter 4, we have shown that SUSY with non-universalities in gaugino or scalar masses within high scale SUGRA set up can still be accessible at LHC with $E_{CM}=$ 14 TeV. In particular, 
we show the consistency of the parameter space in different dark matter annihilation regions. We find that 
there exists a reasonable region of parameter space in the non-universal scenario which not only satisfy all the existing constraints, but also can unravel SUSY in bottom and 
lepton rich final states with third family squarks being lighter than the first two automatically. With the next run of high luminosity LHC at 14 TeV, this allowed parameter space can be 
ruled out easily or we we will be able to discover SUSY in its glory.

In chapter 5, motivated by the fact that the dark matter is about five times the ordinary matter, we have proposed that the dark matter can just be like the ordinary matter in a parallel universe with the two sectors  connecting via the electroweak Higgs bosons of the respective universes. If the electroweak sector of the two universes are symmetric, the lightest Higgs bosons of the two universes will mix. In particular, if these two Higgses mix significantly, and their masses are close (say within 4 GeV), LHC will not be able to resolve if it is observing one Higgs or two Higgses. We take refuge in the proposed International Linear Collider (ILC) where because of its clean environment, the precise measurements and large number of Higgs boson production, Higgs mass splittings upto $\sim 100$ MeV may be possible with high luminosity. 
We show that with a 250 GeV ILC  with 300 fb$^{-1}$ integrated luminosity it will be possible to probe the mixing angle $\theta$ between the two Higgs bosons upto $5^o$.

Chapter 6 is dedicated to one of the most important and exciting sectors of beyond the standard model physics, the Neutrinos. We have shown that the experimental data on neutrino mass differences squares and three mixing angles are consistent with the proposed 4/5 parameter Dirac neutrino models, based on some reasonable assumptions yielding a solution for the neutrino masses with inverted mass hierarchy (close the degenerate pattern). In addition, the model also predicts a large CP violating phase $\delta$, thus predicting a rather large CP violation in the neutrino sector, and will be easily tested in the early runs of the The Deep Underground Neutrino Experiment (DUNE). Neutrinos being Dirac, neutrinoless double beta decay is also forbidden in this model. Thus, all of these predictions can be tested in the upcoming and future precision neutrino experiments.

In chapter 7 we present a novel warm dark matter candidate in the context of two Higgs doublet extensions of the Standard Model. We have shown that such a warm dark matter candidate in these models can have a mass in the keV range, which although will be difficult to see in direct detection experiments. The model has several testable consequences at colliders as well as in astrophysical settings as the charged scalar and the pseudoscalar in the model are in few hundred GeV mass range. The model also has the added feature of explaining the anomalous $X$-ray signal reported by different groups in the extra-galactic spectrum. 

Thus this thesis presents a well rounded effort to study many different extensions in the beyond standard model scenario. The main characteristics in all these models are the testability of the new physics particles or associated signals at the TeV scale. Almost all high-energy physicists are convinced that the LHC will discover new physics of some kind. Leaving aside all the theoretical arguments, it would be unprecedented
in the history of the field if the order-of-magnitude increase in available
energy did not reveal something new. It may well be that nature surprises us  and make the physics much more complicated than the simple cases discussed here.
With CERN’s Large Hadron Collider having started its Run II with 14 TeV center of mass energy with 
a substantial amount of luminosity and possibility of new colliders in the horizon, this is a very exciting time for the theorists to continue investigate the beyond standard model scenarios and refine and extend their phenomenological studies for the possible detection of possible new physics.

\nocite{*} % Use to exclude specific citations from *.bib file
\bibliography{references}
\bibliographystyle{ieeetr}%{alpha}%{ieeetr}%%
%\appendix
%\input{appendix}
\newpage
 \begin{vita}{SHREYASHI CHAKDAR}{Doctor of Philosophy}{Physics} %Creates vita
% \vitaitem{Personal Data:} Born in Darjeeling, West Bengal, India on Dec 5, 1984.
 \vitaitem{Education:} \\\\
Completed the requirements for the degree of Doctor of Philosophy with a major in Physics at Oklahoma State University in July, 2015.\\\\
Received the Master of Science in Physics at Indian Institute of Technology - Roorkee,
Roorkee, India in 2008.\\\\
Received the Bachelor of Science in Physics (Honours) at Lady Brabourne College/Calcutta University, Kolkata, India in 2006.\\
\vitaitem{Experience:} \\ 
Graduate Fellow at Kavli Institute for Theoretical Physics (KITP), University of California, Santa Barbara
in Fall 2014. 

\vitaitem{Professional Memberships:}\\
American Physical Society

  \end{vita}

\end{document}